\documentclass[11pt,a4paper]{article}

\usepackage{graphicx,color}
\usepackage{epstopdf}
\usepackage{amsmath,amsfonts,bbm}
\numberwithin{equation}{section}
\numberwithin{figure}{section}
\let\OLDthebibliography\thebibliography
\renewcommand\thebibliography[1]{
  \OLDthebibliography{#1}
  \setlength{\parskip}{0pt}
  \setlength{\itemsep}{0pt plus 0.3ex}
}

\usepackage{amsmath, amsthm,mathtools, amssymb,upgreek,slashed}

\begin{document}

%%%%%%%%%%%%%%%%%%%%%%%%%%%%%%%%%%%%%%%%%%%%%%%%
%%%%%%%%%%%%%%%%%%%%%%%%%%%%%%%%%%%%%%%%%%%%%%%%
%%%%%%%%%%%%%%%% SIMBOLI VARI %%%%%%%%%%%%%%%%%%%%%%%%
%%%%%%%%%%%%%%%%%%%%%%%%%%%%%%%%%%%%%%%%%%%%%%%%
%%%%%%%%%%%%%%%%%%%%%%%%%%%%%%%%%%%%%%%%%%%%%%%%

\font\msytw=msbm10 scaled\magstep1
\font\msytww=msbm7 scaled\magstep1
\font\msytwww=msbm5 scaled\magstep1
\font\cs=cmcsc10
\font\ottorm=cmr8

\let\a=\alpha \let\b=\beta  \let\g=\gamma  \let\d=\delta \let\e=\varepsilon
\let\z=\zeta  \let\h=\eta   \let\th=\theta \let\k=\kappa \let\l=\lambda
\let\m=\mu    \let\n=\nu    \let\x=\xi     \let\p=\pi    \let\r=r
\let\s=\sigma \let\t=\tau   \let\f=\varphi \let\ph=\varphi\let\c=\chi
\let\ps=\psi  \let\y=\upsilon \let\o=\omega\let\si=\varsigma
\let\G=\Gamma \let\D=\Delta  \let\Th=\Theta\let\L=\Lambda \let\X=\Xi
\let\P=\Pi    \let\Si=\Sigma \let\F=\Phi    \let\Ps=\Psi
\let\O=\Omega \let\Y=\Upsilon

%%%%%%%%%%%%%%%%%%%%%%%%%
\def\ins#1#2#3{\vbox to0pt{\kern-#2 \hbox{\kern#1 #3}\vss}\nointerlineskip}

\newdimen\xshift \newdimen\xwidth \newdimen\yshift

\def\vdd{{\vec d}}\def\vee{{\vec e}}\def\vkk{{\bk}}\def\vii{{\vec i}}
\def\vmm{{\vec m}}\def\vnn{{\vec n}}\def\vpp{{\vec p}}\def\vqq{{\vec q}}
\def\vxxi{{\vec \xi}}\def\vrr{{\vec r}}\def\vtt{{\vec t}}
\def\vuu{{\vec u}}\def\vvv{{\vec v}}
\def\vxx{{\xx}}\def\vyy{{\vec y}}\def\vzz{{\vec z}}
\def\un{{\underline n}} \def\ux{{\underline x}} \def\uk{{\underline k}}
\def\xxx{{\underline\xx}}\def\vxx{{\xx}} \def\vxxx{{\underline\vxx}}
\def\kkk{{\underline\kk}} \def\vkkk{{\underline\vkk}}
\def\bO{{\bf O}}\def\rr{{\bf r}} \def\bk{{\bf k}}  \def\bp{{\bf p}}
 \def\bP{{\bf P}}\def\bl{{\bf l}} 
\def\ss{{\underline \sigma}}\def\oo{{\underline \omega}}

\def\PPP{{\cal P}}\def\EE{{\cal E}}\def\cF{{\cal F}}
\def\mF{{\mathfrak{F}}}
\def\MM{{\cal M}} \def\VV{{\cal V}}\def\cB{{\cal B}}\def\cA{{\cal A}}\def\cI{{\cal I}}
\def\bV{{\bf V}_n}
\def\cE{{\cal E}}\def\si{{\sigma}}\def\ep{{\epsilon}} \def\cD{{\cal D}}\def\cG{{\cal G}}
\def\CC{{\cal C}}\def\FF{{\cal F}} \def\FFF{{\cal F}}\def\cJ{{\cal J}}
\def\cF{{\cal F}} \def\cT{{\cal T}}\def\cS{{\cal S}}\def\cQ{{\cal Q}}
\def\HHH{{\cal H}}\def\WW{{\cal W}}\def\cP{{\cal P}}
\def\TT{{\cal T}}\def\NN{{\cal N}} \def\BBB{{\cal B}}\def\III{{\cal I}}
\def\RR{{\cal R}}\def\cL{{\cal L}} \def\JJ{{\cal J}} \def\OO{{\cal O}}
\def\DD{{\cal D}}\def\AAA{{\cal A}}\def\GG{{\cal G}} \def\SS{{\cal S}}
\def\KK{{\cal K}}\def\UU{{\cal U}} \def\QQ{{\cal Q}} \def\XXX{{\cal X}}

\def\hh{{\bf h}} \def\HH{{\bf H}} \def\AA{{\bf A}} \def\qq{{\bf q}}
\def\bG{{\bf G}}
\def\BB{{\bf B}} \def\XX{{\bf X}} \def\PP{{\bf P}} \def\bP{{\bf P}} 
\def\pp{{\bf p}}
\def\vv{{\bf v}} \def\xx{{\bf x}} \def\yy{{\bf y}} \def\zz{{\bf z}}
\def\dd{{\bf d}}
\def\aaa{{\bf a}}\def\bbb{{\bf b}}\def\hhh{{\bf h}}\def\II{{\bf I}}
\def\ii{{\bf i}}\def\jj{{\bf j}}\def\kk{{\bf k}}\def\bS{{\bf S}}
\def\mm{{\bf m}}\def\Vn{{\bf n}}\def\uu{{\bf u}}\def\tt{{\bf t}}
\def\bq{{\bf q}}
\def\B{\hbox{\msytw B}}
\def\RRR{\hbox{\msytw R}} \def\rrrr{\hbox{\msytww R}}\def\mI{\hbox{\msytw I}} 
\def\rrr{\hbox{\msytwww R}} \def\CCC{\hbox{\msytw C}}\def\EEE{\hbox{\msytw E}}
\def\cccc{\hbox{\msytww C}} \def\ccc{\hbox{\msytwww C}}
\def\MMM{\hbox{\euftw M}}\font\euftw=eufm10 scaled\magstep1%
\def\NNN{\hbox{\msytw N}} \def\nnnn{\hbox{\msytww N}}
\def\nnn{\hbox{\msytwww N}} \def\ZZZ{\hbox{\msytw Z}}\def\QQQ{\hbox{\msytw Q}}
\def\zzzz{\hbox{\msytww Z}} \def\zzz{\hbox{\msytwww Z}}
\def\SSS{{\bf S}}
\def\SSSS{\hbox{\euftwww S}}
\def\1{\hbox{\msytw 1}}
\newcommand{\mR}{{\msytw R}}
\def\virg{\quad,\quad}

%%%%%%%%%%%%%%%%%%%%%%%%%%%%%%%%%%%%%%%%%%%%%%%%
%\renewcommand{\thesection}{\Roman{section}}
\renewcommand{\thesubsection}{\arabic{section}.\arabic{subsection}}

\renewcommand{\theequation}{\thesection.\arabic{equation}}

\def\\{\hfill\break}
\def\={:=}
\let\io=\infty
\let\0=\noindent\def\pagina{{\vfill\eject}}
\def\media#1{{\langle#1\rangle}}
\let\dpr=\partial
\def\sign{{\rm sign}}
\def\const{{\rm const}}
\def\tende#1{\,\vtop{\ialign{##\crcr\rightarrowfill\crcr\noalign{\kern-1pt
    \nointerlineskip} \hskip3.pt${\scriptstyle #1}$\hskip3.pt\crcr}}\,}
\def\otto{\,{\kern-1.truept\leftarrow\kern-5.truept\to\kern-1.truept}\,}
\def\defin{{\buildrel def\over=}}
\def\wt{\widetilde}
\def\wh{\widehat}
\def\to{\rightarrow}
\def\ra{\right\rangle}
\def\qed{\hfill\raise1pt\hbox{\vrule height5pt width5pt depth0pt}}
\def\Val{{\rm Val}}
\def\ul#1{{\underline#1}}
\def\lis{\overline}
\def\V#1{{\bf#1}}
\def\be{\begin{equation}}
\def\ee{\end{equation}}
\def\bea{\begin{eqnarray}}
\def\eea{\end{eqnarray}}
\def\bd{\begin{definition}}
\def\ed{\end{definition}}

\def\nn{\nonumber}
\def\pref#1{(\ref{#1})}
\def\ie{{\it i.e.}}
\def\cC{{\cal C}}
\def\lb{\label}
\def\eg{{\it e.g.}}
\def\sl{{\displaystyle{\not}}}
\def\Tr{\mathrm{Tr}}
\def\BBBB{\hbox{\msytw B}}
\def\bbb{\hbox{\msytww B}}
\def\TTT{\hbox{\msytw T}}
\def\d{\delta}
\def\bT{{\bf T}}
\def\mod{{\rm mod}}
\def\der{{\rm d}}
\def\bs{\backslash}
\newtheorem{corollary}{Corollary}[section]
\newtheorem{lemma}{Lemma}[section]
\newtheorem{conjecture}{Conjecture}[section]
\newtheorem{notation}{Notation}[section]
\newtheorem{example}{Example}[section]
\newtheorem{remark}{Remark}[section]
\newtheorem{definition}{Definition}[section]
\newtheorem{theorem}{Theorem}[section]
\newtheorem{proposition}{Proposition}[section]
\newtheorem{oss}{Remark}

%%%%%%%%%%%%%%%%%%%%%%%%%%%%%%%%%%%%%%%%%%%%%%%%%%%%%%%%%%%%%%%%%
\title{{\bf Honeycomb Hubbard Model 
at van Hove Filling}}
%\author{Zhituo Wang\\ Institute for Advanced Study in Mathematics, \\Harbin Institute of Technology\\
%Email: wzht@hit.edu.cn}

\author{Vincent Rivasseau$ ^1$, Zhituo Wang$ ^2$\\
1 Laboratoire de physique des 2 infinis Ir{\`e}ne Joliot-Curie\\
CNRS and Universit{\'e} Paris-Saclay,
91405 Orsay Cedex, France\\Email: Vincent.rivasseau@th.u-psud.fr\\
2  Institute for Advanced Study in Mathematics, \\Harbin Institute of Technology\\
Email: wzht@hit.edu.cn}

\maketitle

%%%%%%%%%%%%%%%%
\begin{abstract}
This paper is devoted to the rigorous study of the low temperature properties of the two-dimensional weakly interacting Hubbard model 
on the honeycomb lattice in which the renormalized chemical potential $\mu$ has been fixed such that the Fermi surface 
consists of a set of exact triangles. Using renormalization group analysis around the Fermi surface, we prove that this model 
is {\it not} a Fermi liquid in the mathematically precise sense of Salmhofer. 

The main result is proved in two steps. First we prove that the perturbation series for Schwinger functions as well as the 
self-energy function have non-zero radius of convergence when the temperature $T$ is above an exponentially small value, 
namely  ${T_0\sim \exp{(-C|\lambda|^{-1/2})}}$. Then we prove the necessary lower
bound for second derivatives of self-energy w.r.t. the external momentum and achieve the proof. 

\end{abstract}
%%%%%%%%%%%%%%%%
%\tableofcontents
\renewcommand{\thesection}{\arabic{section}}
\section{Introduction}

%%%%%%%%%%%%%%%%%%%%%%%%%%%%%%%%%%%%%%%%
A honeycomb monolayer of carbon atoms, known as {\it graphene} \cite{N, review1, review3, review4}, has played an important role in condensed matter research. The un-doped system has conical valence \cite{W, Feff1, Feff3} and conduction bands meeting at two different Fermi points, also called the Dirac points, and dispersion relation of the quasi- particle closely resembles the massless Dirac fermions in $2+1$ dimensions. The peculiar Fermi surface has shown to be the origin of a number of remarkable effects, such as the anomalous integer Hall effect. Theoretically, this system can be described by the $2D$ Hubbard model \cite{hubb, lieb} on the honeycomb lattice (also called the honeycomb Hubbard model) at half-filling with weak local interactions, whose rigorous construction has been achieved in \cite{GM}. The geometry of the Fermi surface as well as the physical properties change drastically with doping \cite{link, exp2, exp1}. 

In this paper we study the doped Honeycomb Hubbard model in which the value of the renormalized chemical potential $\mu$ is equal to the hopping parameter $t$, which is set to be $1$. In this setting the non-interacting Fermi surface $\cF_0$ is a collection of exact triangles in which van Hove singularities appear. Due to the lattice structure of the model, which serves as an ultraviolet cutoff, there is no ultraviolet divergence in the mathematically interesting quantities such as the Schwinger functions and the self-energy function, and we focus on the infrared analysis of this model. The achievements of the present work are the following: We establish the power counting theorem for the $2p$-point Schwinger functions, $p\ge1$ and prove that the perturbation series for the two-point many-fermion systems as well as the self-energy function have positive radius of convergence when the temperature $T$ is greater than a value ${T_0\sim \exp{(-C|\lambda|^{-1/2})}}$, where $|\lambda|\ll1$ is the bare coupling constant, $C$ is a constant which depends on the physical parameters of the model such as the electron mass, the lattice structure, etc., but not on the temperature. Then the lower bound for the second derivatives of the self-energy w.r.t. the external momentum have been established.

We believe that this paper is important 
since it addresses a doped graphene system that is of current interest in the physics community, 
and mathematically provide the first rigorous results on the non-Fermi liquid behavior on that system.

Non-Fermi liquid behaviors have also been proved in the half-filled Hubbard model on the square lattice at half-filling \cite{AMR1, AMR2}. There are important differences between the two models. First of all, the model studied in \cite{AMR1} is at half-filling, in which there are no quantum corrections to the chemical potential and the dispersion relations. This fact is due to the particle-hole symmetry, which make the
renormalization analysis much simpler. Secondly, the Schwinger functions as well as the self-energy in the current model are matrix valued functions, due to the lattice structure, which are harder to study than in \cite{AMR1, AMR2}. The fact that the Fermi surface in the current model is triangle-like but not square-like also makes the analysis more involved. 
%An important  remark is 
It is intriguing that both models exhibit non-Fermi liquid behaviors. It may indicate some universal structure in models with van Hove singularities.

The main results of this paper will be proved with the Fermionic cluster expansions and rigorous renormalization group analysis \cite{BG, FT, M2}. One major difficulty in the proof is that the non-interacting Fermi surface $\cF_0$ is deformed by interaction, and the resulting interacting Fermi surface $\cF$ is moving when the temperature changes \cite{FST1}. This shift of Fermi surface may cause divergence of many coefficients in the naive perturbation expansion. In order to solve this problem, we introduce a counter-term to the interaction potential, in such a way that the interacting Fermi surface for the {\it new model} is fixed and coincides with $\cF_0$. The inversion problem \cite{FST1, FST2, FS1, FS2}, which concerns the existence and uniqueness of the counter-term given a bare dispersion relation is not addressed in this paper. Sector analysis, the BKAR jungle formulas \cite{BK, AR} and the multi-arch expansions are the main tools that we shall employ to establish the upper and lower bounds for the Schwinger functions and the self-energy function.

%%%%%%%%%%%%%%%%%%%%%%%%%%%%%%%%%%%%%%%%%%%%%%%%%%%%%%%%%%%%%%%%%%%%%%%%%%%%%%%%%%%%%%%%
\section{The Model and Main results}
%%%%%%%%%%%%%%%%%%%%%%%%%%%%%%%%%%%%%%%%%%%%%%%%%%%%%%%%%%%%%%%%%%%%%%%%%%%%%%%%%%%%%%%
\subsection{The honeycomb lattice and the non-interacting Fermi surface}
Let $\Lambda_A=\{\xx\ \vert\ \xx=n_1\bl_1+n_2\bl_2, n_1, n_2\in\ZZZ\}\subset\RRR^2$ be the infinite triangular lattice generated by the basis vectors 
${\bl}_1=\frac12(3,\sqrt{3})$, ${\bl_2}=\frac12(3,-\sqrt{3})$. Let $\Lambda_B=\Lambda_A+\dd_i$ be the shifted triangular lattice of $\Lambda_A$ by one of the three vectors: ${\dd_1}=(1,0)$, ${\dd_2}=\frac12
(-1,\sqrt{3})$ or ${\dd_3}=\frac12(-1,-\sqrt{3})$. Due to the $\ZZZ_3$ symmetry, shifting $\Lambda_A$ by any vector $\dd_i$, $i=1,\cdots,3$ gives the same lattice. So we choose $\Lambda_B=\Lambda_A+\dd_1$, for simplicity.
The infinite honeycomb
lattice is defined as $\L=\L_A\cup L_B$. For a fixed $L\in\NNN_+$, define the finite honeycomb lattice as the torus $\Lambda_L=\L/L\L=\Lambda_{L,A}\cup \Lambda_{L,B}$, which is the union of the two sub-lattices $\Lambda_{L,A}:=\Lambda_A/L\Lambda_A$, $\Lambda_{L,B}=\Lambda_B/L\Lambda_B$, with metric $d_L:=\vert\xx-\yy\vert_{\L_L}=\min_{(n_1, n_2)\in\ZZZ^2}\vert \xx-\yy+n_1L\bl_1+n_2L\bl_2\vert$. 
%%%%%%%%%%%%%%%%%%%%%%%%%%%%%%%%%%%%%%%%%%%%%%%%%%%%%%%%%%%%%%%%%%%%%%%%%%%%%%%%%%%%%%%%
%%%%%%%%%%%%%%%%%%%%%%%%%%%%%%%%%%%%%%%%%%%%%%%%%%%%%%%%%
\begin{figure}[htp]
\centering
\includegraphics[width=.55\textwidth]{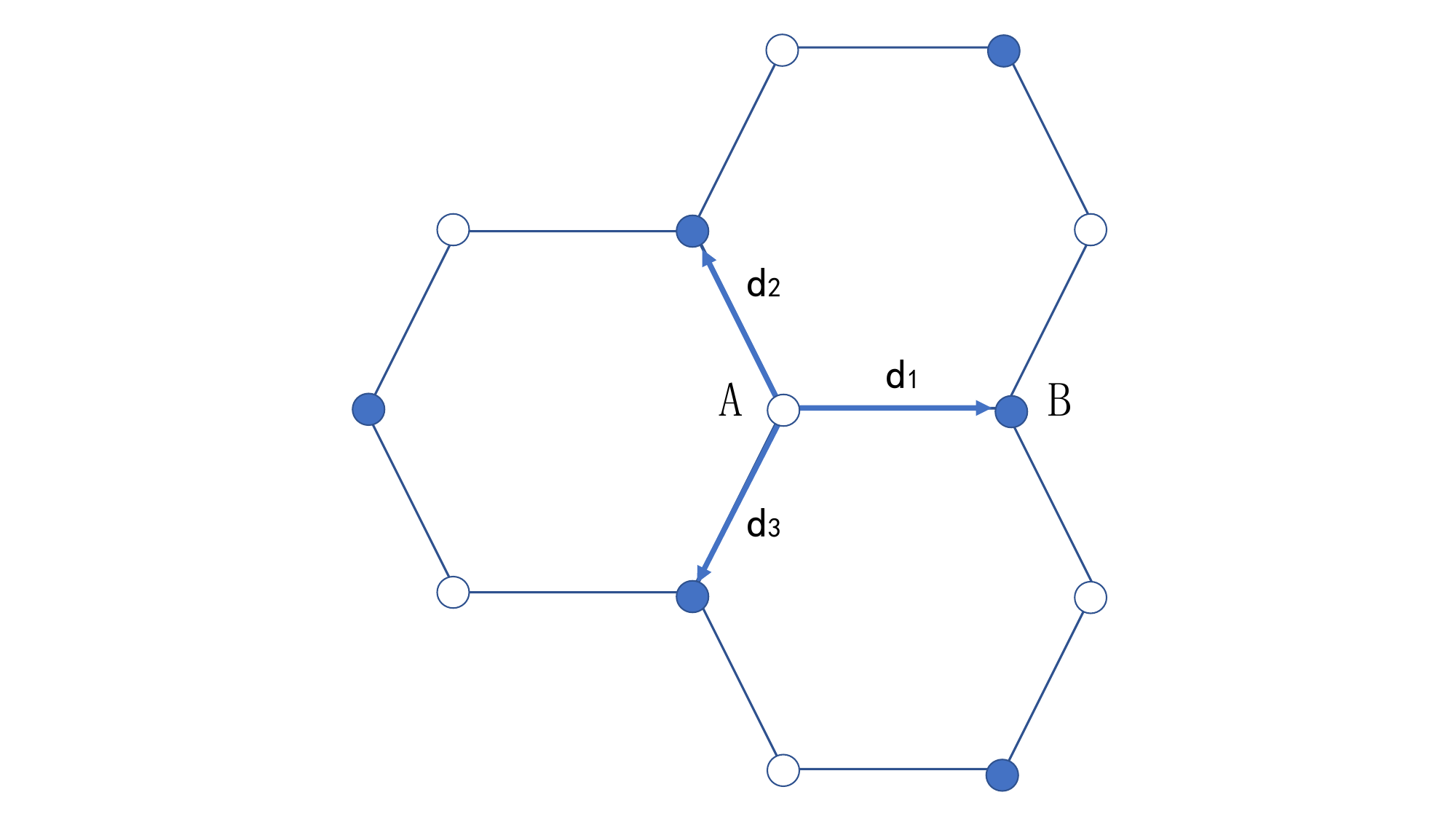}
\caption{\label{lattice}
A portion of the honeycomb lattice $\L$. The white and black dots correspond 
to the sites of the triangular sub-lattices $\L_A$ and $\L_B$, respectively.}
\end{figure}
The Fock space for the many-fermion system is constructed as follows:
let $\HHH_L=\CCC^{L^2}\otimes\CCC^2\otimes\CCC^2$ be a single-particle Hilbert space of functions {$\Psi_{\xx, \t,\a}:\L_L\times\{\uparrow,\downarrow\}\times  \{A,B\}\rightarrow\CCC$}, in which $\t\in\{\uparrow,\downarrow\}$ labels the spin index of the quasi-particle, $\a\in\{A, B\}$ distinguishes the two kinds of sub-lattices and $\xx\in \L_L$ labels the lattice point. The
normalization condition is $\Vert\Psi\Vert^2_2=\sum_{\xx,\t,\a}\vert\Psi_{\xx,\t,\a}\vert^2=1$. The Fermionic Fock space $\FFF_L$ is defined as:
\be
{\FFF_L=\CCC\oplus\bigoplus_{N=1}^{L^2}\FFF_L^{(N)},\quad \FFF_L^{(N)}=\bigwedge^N \HHH_L},
\ee
where $\bigwedge^N \HHH_L$ is the $N$-th anti-symmetric tensor product of $\HHH_L$. Let $\xi_i=(\xx_i,\t_i,\a_i)$, $i=1,\cdots,N$. Define the Fermionic operators ${\bf a}^\pm_{\xx,\t,\a}$ ({see eg. \cite{BR1}, Page 10, Example 5.2.1}) on $\FFF_L$ by:
\bea
&&({\bf a}^+_{\xx,\t,\a}\Psi)^{(N)}(\xi_1,\cdots, \xi_N)\nn\\
&&\quad\quad\quad:=\frac{1}{\sqrt{N}}\sum_{j=1}^N(-1)^j\delta_{\xx,\xx_j}\delta_{\a,\a_j}\delta_{\t,\t_j} \Psi^{(N-1)}(\xi_1,\cdots ,\xi_{j-1},\xi_{j+1},\cdots,\xi_{N}),\\
&&({\bf a}^-_{\xx,\t,\a}\Psi)^{(N)}(\xi_1,\cdots, \xi_N):= \sqrt{N+1}\ \Psi^{(N+1)}(\xx,\t,\a;\ \xi_1,\cdots,\xi_{n}),
\eea 
where $\delta_{.,.}$ is the Kronecker delta function. It is easy to find that the Fermionic operators satisfy the canonical anti-commutation relations (CAR): 
\be\{{\bf a}^+_{\xx,\t,\a}, {\bf a}^-_{\xx',\t',\a'}\}=\delta_{\xx,\xx'}\delta_{\a,\a'}\delta_{\t,\t'},\quad \{{\bf a}^+_{\xx,\t,\a}, {\bf a}^+_{\xx',\t',\a'}\}=\{{\bf a}^-_{\xx,\t,\a}, 
{\bf a}^-_{\xx',\t',\a'}\}=0.\ee 
We impose the periodic boundary conditions on these Fermionic operators: $${\bf a}^\pm_{\xx+n_1L+n_2L,\t,\a}={\bf a}^+_{\xx,\t,\a},\ \forall \xx\in\Lambda_L.$$ 
The operators ${\bf a}^\pm_{\xx,\t,A}$ and ${\bf a}^\pm_{\xx,\t,B}$ are called the Fermionic operators of type $A$ and type $B$, respectively.

The second quantized grand-canonical Hamiltonian on $\L_L$ is defined by 
\be
H_{L}(\lambda)=H^0_{L}+ V(\lambda)_{L},
\ee
in which
\bea
H^0_{L}&=&-t\sum_{\substack{\xx\in \L_{L,A}\\i=1,\cdots,3}}\sum_{\t=\uparrow\downarrow} \Big(\ 
{\bf a}^+_{\xx,\t,A} {\bf a}^-_{\xx+\dd_i, \t,B} +{\bf a}^+_{\xx+\dd_i,\t,B} {\bf a}^-_{\xx,\t,A}\ \Big)\nn\\
&&\quad\quad\quad-\mu\sum_{\substack{\xx\in \L_{L,A}}}\sum_{\t=\uparrow\downarrow}\Big(\ 
{\bf a}^+_{\xx,\t,A}{\bf a}^-_{\xx,\t,A}+{\bf a}^+_{\xx+ \dd_1,\t,B}{\bf a}^-_{\xx+\dd_1,\t,B}\ \Big),\label{hamil0}
\eea
is the non-interacting Hamiltonian; $t\in\RRR_+$ is the nearest neighbor hopping parameter and $\mu\in\RRR $ is called the {\it renormalized} chemical potential. $V(\lambda)_{L}$ is the interaction potential, to be defined later. We fix $t=1$ for the rest of this paper.

%%%%%%%%%%%%%%%%%%%%%%%%%%%
Let ${\Lambda}_L^*$ be the dual lattice of $\Lambda_L$ with basis vectors $\bG_1=\frac{2\pi}{3}(1,\sqrt{3})$, $\bG_2=\frac{2\pi}{3}(1,-\sqrt{3})$,  the first Brillouin zone is defined as 
\be\label{bril}
\DD_L:=\RRR^2/\Lambda^*_L=\big\{\bk\in\RRR^2\ \vert\ \bk=\frac{n_1}{L}\bG_1+\frac{n_2}{L}\bG_2, {n_{1,2}\in[-\frac{L}{2}, \frac{L}{2}-1]\cap\ZZZ}\big\}
.\ee
%%%%%%%%%%%%%%%%%%%%%%%%%%%%%%%%%%%%%%%%%%%%%%%%%%%%%%%%%%%%%%%%%%%%%%%
The Fourier transform for the Fermionic operators are:
\be
{\bf a}^\pm_{\xx,\t,A}=\frac{1}{|\L_L|}\sum_{\bk\in\DD_L}e^{\pm i\kk\cdot\xx}\hat {\bf a}^\pm_{\bk,\t,A},\ {\bf a}^\pm_{\xx+\dd_1,\t,B}=\frac{1}{|\L_L|}\sum_{\bk\in\DD_L}e^{\pm i\kk\cdot\xx}\hat {\bf a}^\pm_{\bk,\t,B},
\ee
in which $|\L_L|$ is the volume of $\L_L$. The inverse Fourier transform are given by:
%\be\hat {\bf a}^\pm_{\bk,\a,\t}=\sum_{\xx\in\Lambda_L}e^{\mp i\bk\cdot\xx} {\bf a}^\pm_{\xx,\a,\t},\quad\forall\ \bk\in\DD_L.
%\ee 
%%%%%%%%%%%%%%%%%%%%%%%%%%%%%%%
$\hat {\bf a}^\pm_{\bk,\t,A}=\sum_{\xx\in\Lambda_L}e^{\mp i\bk\cdot\xx} {\bf a}^\pm_{\xx,\t, A}$,
$\hat {\bf a}^\pm_{\bk,\t, B}=\sum_{\xx\in\Lambda_L}e^{\mp i\bk\cdot\xx} {\bf a}^\pm_{\xx+\dd_1,\t,B}$.
%%%%%%%%%%%%%%%%%%%%%%%%%%%%%%%%
The periodicity of ${\bf a}^\pm_{\xx,\t,\a}$ implies
$\hat {\bf a}^\pm_{\bk+n_1\bG_1+n_2\bG_2,\t,\a}=\hat {\bf a}^\pm_{\bk,\t,\a}$, and the commutation relations become:
\bea\{{\bf a}^+_{\bk,\t,\a}, {\bf a}^-_{\bk',\t',\a'}\}=|\L_L|\delta_{\bk,\bk'}\ \delta_{\a,\a'}\ \delta_{\t,\t'}
\quad \{{\bf a}^+_{\bk,\t,\a}, {\bf a}^+_{\bk',\t',\a'}\}=\{{\bf a}^-_{\bk,\t,\a}, {\bf a}^-_{\bk',\t', \a'}\}=0.
\eea
%%%%%%%%%%%%%%%%%%%%%%%%%%%%%%%%%%%%%%%%%%%%%%%%%%%%%%%%%%%%%%
It is useful to relabel the Fermionic operators of type $A$, i.e., those with $\a=A$, by $\a=1$ and the Fermionic operators of type $B$ by $\a=2$, and organize these operators into vectors. Then we can rewrite the non-interacting Hamiltonian as
\be\label{qua2}
H^0_L=-\frac{1}{|\Lambda_L|}\sum_{\kk\in\cD_L,\tau=\uparrow\downarrow}\sum_{\a,\a'=1,2}\hat {\bf a}^+_{\bk,\tau,\a}[\hat H_0(\bk)]_{\a,\a'}\hat {\bf a}_{\bk,\tau,\a'},
\ee
with matrix kernel: \be
\hat H_0(\bk)=\begin{pmatrix}\ -\mu&-\Omega^*(\bk)\\-\Omega(\bk)&-\mu\ 
\end{pmatrix},
\ee
in which $\O({\bk})=\sum_{i=1}^3 
e^{i(\dd_i-\dd_1) \bk}=1+2
e^{-i \frac32 k_1}\cos(\frac{\sqrt{3}}2 k_2)$
is called the {\it non-interacting complex dispersion relation}, and $\Omega^*(\bk)$ is the complex conjugate of $\Omega(\bk)$.
%%%%%%%%%%%%%%Measure%%%%%%%%%%%%%%%%%
\begin{lemma}[See also \cite{GM}, Lemma 1]\label{inv0}
The Hamiltonian \eqref{qua2} is invariant under the following symmetries.
\begin{itemize}
\item (a) discrete spatial rotations: ${\bf a}^\pm_{\ \bk,\t,\a}\rightarrow e^{\mp i\bk\cdot(\dd_3-\dd_1)(\a-1)}{\bf a}^\pm_{ R_{2\pi/3}(\bk),\t,\a}$, $R_\theta\in SO(2)$ is the rotation operator with $\theta\in[0,2\pi)$ independent of $\bk$.
\item (b) vertical reflections: ${\bf a}^\pm_{\ (k_1,k_2),\t,\a}\rightarrow {\bf a}^\pm_{\ (k_1,-k_2),\t,\a}$.
\item (c) interchange of particles: ${\bf a}^\pm_{\ \bk,\t,\a}\leftrightarrow {\bf a}^\pm_{\ -\bk,\t,\a'}$, for $\a\neq\a'$.
\end{itemize}
\end{lemma}
\begin{proof}
It is enough to prove that
\bea
&&\sum_{\a,\a'=1,2}\sum_{\bk\in\cD_L}\hat {\bf a}^+_{\bk,\tau,\a}[\hat H_0(\bk)]_{\a,\a'}\hat {\bf a}_{\bk,\tau,\a'}\\
&&=-\mu\sum_{\bk\in\cD_L}\big({\bf a}^+_{\bk,\tau,1}{\bf a}_{\bk,\tau,1}+{\bf a}^+_{\bk,\tau,2}{\bf a}_{\bk,\tau,2}\big)
-\sum_{\bk\in\cD_L}\Big[{\bf a}^+_{\bk,\tau,1}\Omega^*(\bk){\bf a}_{\bk,\tau,2}+{\bf a}^+_{\bk,\tau,2}\Omega(\bk){\bf a}_{\bk,\tau,1}\Big]\nn
\eea
is invariant under the transformations in $(a)-(c)$.
Since $\Omega(\bk)$ is an even function of $k_2$ and since $\cD_L$ is invariant under the transformation $k_2\rightarrow -k_2$, the conclusion of $(b)$ follows. The rotation operator in $(a)$ is $R_{2\pi/3}=\begin{pmatrix}-1/2&\sqrt3/2\\-\sqrt3/2&-1/2\end{pmatrix}$. We have
$\Omega(R^{-1}_{2\pi/3}(\bk))=e^{i(\dd_1-\dd_2)\cdot\bk}\Omega(\bk)$. Since $\cD_L$ is invariant under the rotation $R_{2\pi/3}$, we proved $(a)$. Finally, using the fact that $\Omega(\bk)=\Omega^*(-\bk)$, the conclusion of $(c)$ follows.
\end{proof}
%%%%%%%%%%%%%%%%%%%%%%%%%%%%%%%%%%%%%%%%%%%%%%
Let $T>0$ be the temperature of the system and $\beta=1/T$, the Gibbs states associated with $H_{L}$ are defined by:
\be\label{gibbs}
\langle\cdot\rangle=\Tr_{\FFF_L}\ [\ \cdot\ e^{-\beta H_{L}}]/Z_{\beta,\L_L},\ee
in which $Z_{\beta,\L_L}=\Tr_{\FFF_L}e^{-\beta H_{L}}$ is the partition function and the trace is taken w.r.t. vectors in the Fock space $\cF_L$. Define $\Lambda_{\beta, \Lambda_L}:=[-\b,\b)\times\L_L$. For $x_0\in[-\b,\b)$, the imaginary-time evolution of the Fermionic operators {\bf is} defined as ${\bf a}^\pm_{x}=e^{x^0H_L}{\bf a}^\pm_{\ \xx} e^{-x^0H_L}$,
in which $x=(x_0,\xx)\in\Lambda_{\beta, \Lambda_L}$.

The $2p$-point Schwinger functions, $p\ge0$, are (formally) defined as:
\bea\label{nptsch}
&&S_{n, \beta, L}(x_1,\e_1,\t_1,\a_1;\cdots x_{2p} ,\e_{2p},\t_{2p},\a_{2p};\lambda):=
\langle\bT\ {\bf a}^{\e_1}_{x_1,\t_1\a_1}\cdots {\bf a}^{\e_{2p}}_{x_{2p},\t_{2p},\a_{2p}}\rangle_{\beta,L}\nn\\
&&\quad\quad\quad:=\frac{1}{Z_{\beta,\Lambda_L}}\Tr_{\FFF_L} e^{-\beta H_L}\bT\{{\bf a}^{\e_1}_{({x_1^0},\xx_1),\t_1, \a_1}\cdots {\bf a}^{\e_{2p}}_{({x_{2p}^0},\xx_{2p}),\t_{2p},\a_{2p}}\},
\eea
where $\bT$ is the Fermionic time-ordering operator, defined as
\bea
&&\bT\ {\bf a}^{\e_1}_{(\xx_1, {x_1^0}),\t_1,\a_1}\cdots {\bf a}^{\e_{2p}}_{(\xx_{2p}, {x_{2p}^0}),\t_{2p},\a_{2p}}\\
&&\quad\quad\quad\quad={\rm sgn} (\pi)\ {\bf a}^{\e_{\pi(1)}}_{(\xx_{\pi(1)}, {x}_{\pi(1)}^0),\t_{\pi(1)},\a_{\pi(1)}}\cdots {\bf a}^{\e_{\pi({2p})}}_{(\xx_{\pi({2p})}, {x}_{\pi(n)}^0),\t_{\pi({2p})},\a_{\pi({2p})}},\nn
\eea
such that ${x}_{\pi(1)}^0\ge{x}_{\pi(2)}^0\ge\cdots\ge{x}_{\pi(2p)}^0$, in which $\pi$ is the permutation operator. If some operators are evaluated at equal time, the ambiguity is solved by taking the normal-ordering on these operators: putting ${\bf a}^-_{x_i,\t_i,\a_i}$ on the right of ${\bf a}^+_{x_i,\t_i,\a_i}$. 
%%%%%%%%%%%%%%%%%%%%%%%%%%%%%%%%%%%%%%%%%%%%%%%%%%%%%%%%%%%%%%%%
\subsubsection{The non-interacting Fermi surface}
\begin{figure}[!htb]
\centering
\includegraphics[scale=.26]{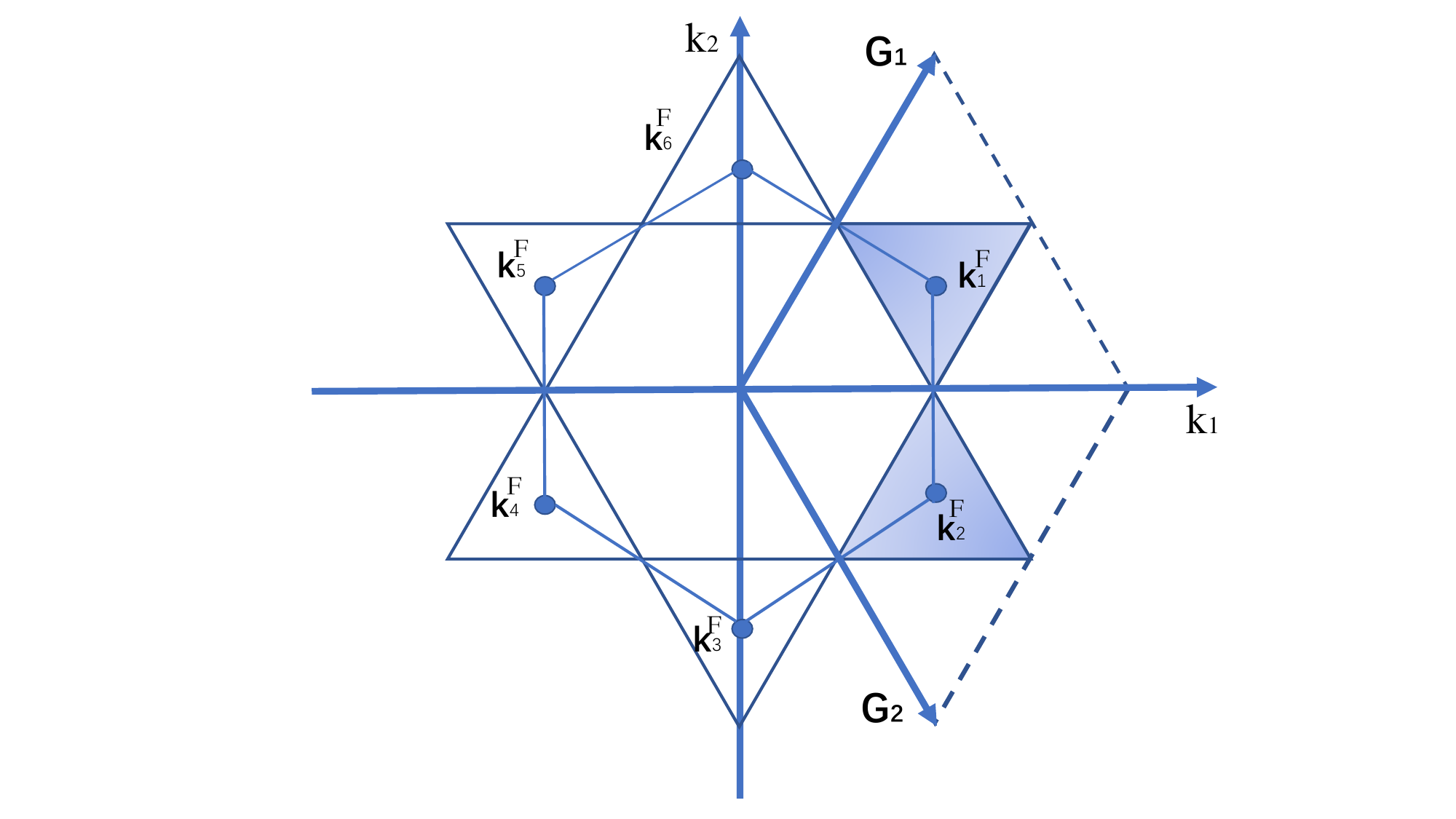}
\caption{An illustration of the Fermi triangles, among which the shaded triangles are the fundamental ones. The vertices of the Fermi triangles are the van Hove singularities. The centers of the Fermi triangles, $\kk_1^F,\cdots,\kk_6^F$, are the Fermi points. The rhombus generated by ${\bf G}_1$ and  ${\bf G}_2$ is the first Brillouin zone.}\label{fpt}
\end{figure}

The non-interacting two-point Schwinger function (also called the free propagator) is defined as:
\bea
C_{\b}(x-y)&:=&\lim_{L\rightarrow\infty}S_{2,\b,L}(x,y;0)\\
&=&\lim_{L\rightarrow\infty} {\frac{1}{\b|\L_L|}}
\sum_{k=(k_0, \kk)\in\DD_{\b, L}}e^{ik_0\cdot(x_0-y_0)+i\bk\cdot(\xx-\yy)}\hat C(k_0,\bk)\label{free2pt},
\eea
in which 
\bea\label{2ptk}
\hat C(k_0,\bk)=[-i k_0 \mI+E(\kk,\mu)]^{-1},
\eea
is the free propagator in the momentum space. The summation over the momentum $k=(k_0,\bk)$
runs over the set $\DD_{\b, L}:=\{(2n+1)\pi T,\ n\in\NNN\}\times\cD_L$, in which $k_0=(2n+1)\pi/\beta=(2n+1)\pi T$, $n\in\NNN$, are called the Matsubara frequencies. $\mI$ is the $2\times2$ identity matrix and $E(\kk,\mu)=\hat H_0(\bk)$ is called the {\it band matrix}, which is closely related to the band structure of the electrons. Inverting the denominator we obtain:
\bea\label{2ptkb}
\hat C(k_0,\bk)=\frac{1}{k_0^2+e(\bk,\mu)-2i\mu k_0} \begin{pmatrix}i k_0+\m &-\O^*(\bk) \\ -\O(\bk) &
ik_0+\m\end{pmatrix},
\eea
in which
\bea\label{band1}
e(\bk,\mu)&:=&-\det\big[ E(\kk,\mu)\big]\nn\\
&=&4\cos(3k_1/2)\cos(\sqrt{3} k_2/2)+
4\cos^2(\sqrt{3} k_2/2)+1-\mu^2.
\eea
\begin{definition}
The non-interacting Fermi surface (F.S.) is defined as:
\be{\cal F}_0:=
\{\bk=(k_1, k_2)\in \RRR^2\vert\ e(\bk,\mu)=0\}.\label{freefs}\ee
It is a one-dimensional subset of $\RRR^2$. A Fermi surface may have several connected components, each of which is called a Fermi curve (F.C.).
\end{definition}
The geometry of the Fermi surface depends crucially on the value of $\mu$: 
when $\mu=0$, the solution to the equation $e(\bk,\mu)=0$ composes of a set of points, called the {\it Fermi points} or the {\it Dirac points} (cf. eg. \cite{GM}), among which the pair $\bk^F_1=(\frac{2\pi}{3}, \frac{2\pi}{3\sqrt3})$ and $\bk^F_2=(\frac{2\pi}{3}, -\frac{2\pi}{3\sqrt3})$ are considered as the fundamental ones. When $\mu=1$, the solutions to $e(\bk,1)=0$ are the following lines:
%%%%%%%%%%%%%%%
\bea\label{sol0}
L_1&=&\{(k_1, k_2)\in\RRR^2:k_2={\sqrt{3}} k_1-\frac{4n+2}{\sqrt3}\pi,\ 
n\in\ZZZ\}, \nn\\
L_2&=&\{(k_1, k_2)\in\RRR^2:  k_2=-{\sqrt{3}} k_1+\frac{4n+2}{\sqrt3}\pi,\ 
n\in\ZZZ\},\nn\\ 
L_3&=&\{(k_1,k_2)\in\RRR^2:k_2=\pm\frac{(2n+1)\pi}{\sqrt3},\  n\in\ZZZ\},
\eea
which form a set of perfect triangles, also called the Fermi triangles. The Fermi surfaces is the union of these Fermi triangles (see Figure \ref{fpt} for an illustration of a Fermi surface composed of $6$ Fermi triangles). The following two Fermi triangles
\bea
\cF_0^+&=&\{k_2={\sqrt{3}} k_1-\frac{2\pi}{\sqrt3},\ k_1\in[\frac{2\pi}{3},\pi]\}\cup
\{ k_2=-{\sqrt{3}} k_1+\frac{2\pi}{\sqrt3},\ k_1\in[\frac{\pi}{3},\frac{2\pi}{3}]\}\nn\\
&&\quad\quad\cup \{k_2=\frac{\pi}{\sqrt3},\  k_1\in[\frac{\pi}{3},\pi]\},\\
\cF_0^-&=&\{k_2={\sqrt{3}} k_1-\frac{2\pi}{\sqrt3},\ k_1\in[\frac{\pi}{3},\frac{2\pi}{3}\}\cup
\{ k_2=-{\sqrt{3}} k_1+\frac{2\pi}{\sqrt3},\ k_1\in[\frac{2\pi}{3},\pi]\}\nn\\
&&\quad\quad\cup \{k_2=-\frac{\pi}{\sqrt3},\  k_1\in[\frac{\pi}{3},\pi]\},
\eea
centered around the Fermi points $\bk^F_1$ and $\bk^F_2$, respectively, are called {\it the fundamental Fermi triangles}. All the other Fermi triangles are considered as translations of $\cF_0^+$ and $\cF_0^-$. The vertices of the Fermi triangles are called the {\it van Hove singularities}. Lifshitz phase transitions \cite{lifshitz} may happen when the chemical potential crosses $\mu=1$, for which the geometry of the Fermi surface changes drastically: When $0<\mu<1$, the Fermi surface is a set of closed convex curves centered around the Fermi points and bordered by the Fermi triangles, and when $\mu>1$ the Fermi surfaces become concave and have more complicated geometrical properties.

%In this paper we consider the case in which the bare chemical potential is in a small neighborhood of $\mu=1$, which is the value of the renormalized chemical potential and is fixed by the counter-term.

%%%%%%%%%%%%%%%%%%%%%%%%%%%%%%%%%%%%%%%%%%%%%%%%%%%%%%%%%%%%%%%%%%%%%%%%%%%%%%%%%%%%%%%
%%%%%%%%%%%%%%%%%%%%%%%%%%%%%%%%%%%%%%%%%%%%%%%%%
\subsection{The interaction potential and the moving Fermi surface}
The many-body interaction potential for the honeycomb Hubbard model is defined as:
\bea  
V_{L}(\lambda)&=&\lambda\sum_{\substack{\xx\in \L_{L,A}\\i=1,\cdots,3}}
\Big(\ {\bf a}^+_{\xx,\uparrow,1}{\bf a}^-_{\xx,\uparrow,1}{\bf a}^+_{\xx,\downarrow,1}{\bf a}^-_{\xx,\downarrow,1}\nn\\
&&\quad\quad\quad\quad\quad+{\bf a}^+_{\xx+\dd_i,\uparrow,2}{\bf a}^-_{\xx+\dd_i,\uparrow,2}{\bf a}^+_{\xx+\dd_i,\downarrow,2}{\bf a}^-_{\xx+\dd_i,\downarrow,2}\ \Big),
 \label{hamil1}\eea
in which $\lambda\in\RRR$ is called the {\it bare coupling constant}. It is easy to prove that $V_L$
is invariant under the transformations introduced in Lemma \ref{inv0}.

If we choose the grand-canonical Hamiltonian as $\tilde H_L=H^0_L+V_L$, the interacting 2-point Schwinger function  $\tilde S^{int}_2(\l)$ (also called the interacting propagator) is defined as:
\bea\label{sfe1}
[\tilde S^{int}_2(\l,p)]_{\a\a',\tau\tau'}&:=&\langle{\bf T}\{{\bf a^-}_{p_0,{\bf p},\t,\a}{\bf a^+}_{k_0,{\bf k},\t'\a'}\}\rangle\\
&=&\delta_{\t,\t'}\delta(p-k)\Big[\big({-ik_0\mI+E(\bk,\mu)+\cE((k_0,\kk),\lambda)}\big)^{-1}\Big]_{\a,\a'},\nn
\eea
in which
\be\label{sf1}
\cE((k_0,\kk),\lambda)=\begin{pmatrix}\tilde\Sigma_{11}((k_0,\kk),\lambda)&\hat\Sigma_{12}((k_0,\kk),\lambda)\\
\hat\Sigma_{21}((k_0,\kk),\lambda)&\tilde\Sigma_{22}((k_0,\kk),\lambda)\end{pmatrix}.
\ee
is called {\it the self-energy matrix}. 
%%%%%%%%%%%%%%%%%%%%%%%%%%%%%%%%%%%%%%%%%%%%%%%%%%%%%%%%%
Since the interaction potential $V_L(\lambda)$ is invariant under the transformations of Lemma \ref{inv0}, we have 
\be\label{inv1}
\tilde\Sigma_{11}((k_0,\kk),\lambda)=\tilde\Sigma_{22}((k_0,\kk),\lambda),\quad \hat\Sigma_{12}((k_0,\kk),\lambda)=\hat\Sigma_{21}((k_0,-\kk),\lambda).\ee
The diagonal elements of \eqref{sf1} can be further
decomposed as: $
\tilde\Sigma_{11}((k_0,\kk),\lambda)=T(\lambda)+\hat\Sigma_{11}((k_0,\kk),\lambda)$, 
in which $T(\lambda)$ is independent of the external momentum $\bk$ and is called {\it the tadpole term}, and $\hat\Sigma_{11}$ is the non-local part.

Notice that the interacting propagator $\tilde S^{int}_2(\l)$ could be singular at $k_0\rightarrow0$.
%%%%%%%%%%%%%%%%%%%%%%%%%%%%%%%%%%%%%%%%%
\begin{definition}\label{intfs}
The interacting Fermi surface is defined by
\be\label{fint1}
\cF=\{\bk\vert \det (E(\bk,1)+\cE(0,\bk),\l)=0\}.
\ee
\end{definition}
Using \eqref{sf1}, we have:
\be
\cF=\Big\{\bk\vert\det \begin{pmatrix}-1+T(\lambda)+\hat\Sigma_{11}((0,\kk),\lambda)&-\Omega^*(\bk)+\hat\Sigma_{12}((0,\kk),\lambda)\\
-\Omega(\bk)+\hat\Sigma_{12}((0,-\kk),\lambda)&-1+T(\lambda)+\hat\Sigma_{11}((0,\kk),\lambda)\end{pmatrix}=0\Big\}.
\ee
Remark that, since $\cE(k,\lambda)$ is not known ahead of time, $\cF$ is also not known and is changing when $\cE$ changes. One possible way of solving this problem is to fix the interacting Fermi surface by introducing the following (nonlocal) counter-terms:
%%%%%%%%%%%%%%%%%%%%%%%%%%%%%%%%%%%%%%%%%%%
\bea
N_{L}(\lambda)&=&-\frac{1}{|\Lambda|}\sum_{\bk\in\cD_L}\sum_{\t\in \{\uparrow,\downarrow\}}\Big[\ \delta\mu(\lambda)\sum_{\a=1,2}\ {\bf a}^+_{\bk,\t,\a}{\bf a}_{\bk,\t,\a}+\sum_{\a,\a'=1,2}
\hat\nu(\bk,\lambda)_{\a\a'}{\bf a}^+_{\bk,\t,\a}{\bf a}_{\bk,\t,\a'}\Big]
\nn\\
&&:=-\frac{1}{|\Lambda|}\sum_{\bk\in\cD_L}\sum_{\t\in \{\uparrow,\downarrow\}}\sum_{\a,\a'=1,2}{\bf a}^+_{\bk,\t,\a}[\delta E(\bk,\lambda)]_{\a\a'}{\bf a}_{\bk,\t,\a'},
\eea
%%%%%%%%%%%%%%%%%%%%%%
%%%%%%%%%%%%%%%%%%%%%%%%
in which
\be
\delta E(\bk,\lambda)=\begin{pmatrix}\delta\mu(\lambda)+\hat\nu_{11}(\bk,\lambda)&\hat\nu_{12}(\bk,\lambda)\\
\hat\nu_{21}(\bk,\lambda)&\delta\mu(\lambda)+\hat\nu_{22}(\bk,\lambda)\end{pmatrix},
\ee
whose matrix elements satisfy 
\be\hat\nu_{11}(\bk,\lambda)=\hat\nu_{22}(\bk,\lambda),\quad \hat\nu_{12}(\bk,\lambda)=\hat\nu_{21}(-\bk,\lambda),
\ee 
and
\be\label{rncd}
\delta\mu(0)=0,\quad \hat\nu_{\a\a'}(\bk,0)=0,\ {\rm for}\ \a,\a'=1,2.
\ee
%%%%%%%%%%%%%%%%%%%%%%%%
The (new) grand-canonical Hamiltonian is defined by:
\be
H_L=H_L^0+V_{L}+N_{L},
\ee
and the interacting propagator is:
\bea\label{newintp}
\big(-ik_0\mI+E(\bk,1)+\delta E(\bk,\lambda)+\cE((k_0,\kk),\delta E,\lambda)\big)^{-1}.
\eea

With the introduction of the counter-terms, the singularities of the new interacting propagator \eqref{newintp} are required to coincide with the non-interacting Fermi surface $\cF_0$, which set constraints on the counter-terms, called {\it the renormalization conditions}.
We can formally expand the interacting propagator as:
\be\label{intpr0}
\sum_{n=0}^\infty\big(-ik_0\mI+E(\bk,1)\big)^{-1}\Bigg[\ \big(\delta E(\bk,\lambda)+\cE((k_0,\kk),\delta E,\lambda)\big)\big(-ik_0\mI+E(\bk,1)\big)^{-1} \Bigg]^n,
\ee
the renormalization conditions can be formulated as:
\begin{definition}[The renormalization conditions]\label{conj1}\\
\begin{itemize}
\item (a) The numerator in \eqref{intpr0} vanishes on the Fermi surface:
\be\label{rncd1}
\delta E(\bk,\lambda)+\cE((0,\kk),\lambda)=0,\ {\forall}\ \bk\ {\rm with}\ e(\bk,1)=0,
\ee
\item (b) the ratio
\be\label{rncd2}
\big(\delta E(\bk,\lambda)+\cE((k_0,\kk),\delta E,\lambda)\big)\big({-ik_0\mI+E(\bk,1)}\big)^{-1}
\ee
is locally bounded for all $(k_0,\bk)\in\cD_{\b,L}$, up to a zero-measure set.
\end{itemize}
\end{definition}

Define the projection operator $P_F$ which maps each $\bk\in\cD_L$ to a unique $P_F\bk\in\cF_0$. Using the explicit expressions for $\delta E$ and $\cE$, Formula \eqref{rncd1} can be written as:
\bea
&&\delta\mu(\lambda)+T(\lambda)=0,\ {\rm and}\label{rncd3}\\ 
&&\hat\nu_{\a\a'}(P_F\bk,\lambda)+\hat\Sigma_{\a\a'}((0,P_F\bk),\delta E,\lambda )=0,\ \a,\a'=1,2.\label{rncd4}
\eea
%%%%%%%%%%%%%%%%%%%%%%%%%%%%%%%%%%%%%%%%%%%%%%%%%%%%%%
Due to the symmetry properties of the self-energy matrix (cf. \eqref{inv1}), \eqref{rncd4} reduces to:
\be\label{rncd5}
\hat\nu_{11}(P_F\bk,\lambda)+\hat\Sigma_{11}((0,P_F\bk),\delta E,\lambda )=0,\quad \hat\nu_{12}(P_F\bk,\lambda)+\hat\Sigma_{12}((0,P_F\bk),\delta E,\lambda )=0.
\ee
%%%%%%%%%%%%%%%%%%%%%%%%%%%5
\begin{remark}
It is important to remark that, since the set of counter-terms $\delta E$ that satisfy condition $(a)$ maybe highly non-trivial, we have indeed defined a class of honeycomb-Hubbard models whose interacting Fermi-surfaces are fixed and coincide with $\cF_0$.
\end{remark}
%%%%%%%%%%%%%%%%%%%%%%%%%%%%%%%%%%%%%%%%%%%%%%%%%%
We have the following theorem concerning the counter-terms, which will be proved in Section \ref{contthm}.
\begin{theorem}\label{conj2}
There exists a counter-term matrix $\delta E(\bk,\lambda)$ such that the renormalization conditions introduced in Definition \ref{conj1} can be satisfied. The tadpole counter-term $\delta\mu(\l)$ is a bounded function of $\lambda$. The counter-terms $\hat\nu_{\a\a'}(\bk,\l)$, $\a,\a'=1,2$, are bounded and uniformly $C^{1}$ but not uniformly $C^2$ in the external momentum $\bk$.
\end{theorem}

Notice that we can combine the quadratic terms in $H^0_L$ and $N_L$:
\bea
-\frac{1}{|\Lambda|}\sum_{\bk\in\cD_L}\sum_{\t\in \{\uparrow,\downarrow\}}\sum_{\a,\a'=1,2}{\bf a}^+_{\bk,\t,\a}[ E_{bare}(\bk,\lambda)]_{\a\a'}{\bf a}_{\bk,\t,\a'},
\eea
in which the kernel matrix
\be\label{bme}
E_{bare}=\begin{pmatrix}-1+\delta\mu(\lambda)+\hat\nu_{11}(\bk,\lambda)&-\Omega^*(\bk)+\hat\nu_{12}(\bk,\lambda)\\
-\Omega(\bk)+\hat\nu_{12}(-\bk,\lambda)&-1+\delta\mu(\lambda)+\hat\nu_{11}(\bk,\lambda)\end{pmatrix}
\ee
is called {\it the bare band matrix}, $\mu_{bare}=1-\delta\mu(\lambda)$ is called the bare chemical potential. The Hamiltonian for the {\it new model} can be considered as
the one with band matrix $E_{bare}$ and interaction potential $V_L(\lambda)$. 
%%%%%%%%%%%%%%%%%
%{\color{red} rewrite}
%%%%%%%%%%%%%%%%

Before proceeding, let us recall the Salmhofer's criterion on the Fermi liquid \cite{salm} at equilibrium:
\begin{definition}[Salmhofer's criterion]\label{salmc}
A $2$-dimensional many-fermion system at positive temperature is a Fermi liquid if the thermodynamic limit of the momentum space Green's functions exists for $|\lambda|<\lambda_0(T)$ and if there are constants $C_0, C_1, C_2>0$ independent of $T$
and $\lambda$ such that the following holds. (a) The perturbation expansion for the momentum space self-energy $[\hat\Sigma(k,\lambda)]_{\a\a'}$, $\a,\a'=1,2$, converges for all $(\lambda,T)$ with $|\lambda\log T|<C_0$. (b) The self-energy $[\hat\Sigma(k, \lambda)]_{\a\a'}$ satisfies the following regularity conditions:
\begin{itemize}
\item $[\hat\Sigma(k, \lambda)]_{\a\a'}$ is twice differentiable in $k_0$, $k_+,k_-$ and
\be
\Vert[\partial_{k_{i}k_{j}}^2\hat\Sigma(k, \lambda)]_{\a\a'}\Vert_{L^\infty}\le C_1,\ \ i=0,\pm, \ j=0,\pm,\ \a,\a'=1,2,
\ee
\item The restriction of the self-energy on the Fermi surface $\cF_0$ is $C^{2}$ differentiable w.r.t. the momentum, 
 and
\be
\Vert[\partial_{k_{i}k_{j}}^2\hat\Sigma(k, \lambda)]_{\a\a'}\vert_{k_0=0,k_{\pm\in\cF_0}}\Vert_{L^\infty}\le C_2,\ i=0,\pm,\ j=0,\pm, \ \a,\a'=1,2.
\ee
\end{itemize}
\end{definition} 
\vskip.3cm

\subsection{The main result}
The most interesting quantities in this model are the connected Schwinger functions $S^c_{2p,\beta}(\lambda)$, $p\ge0$, and the self-energy function $\Sigma(\lambda)$ in the {\it thermodynamic limit} $L\rightarrow\infty$ or $\L_L\rightarrow\L$. A fundamental mathematical problem is whether such quantities are well defined. 
In this paper we will provide a positive answer to this problem, and study the analytic properties of the connected Schwinger function  $S^c_{2p,\beta}(\lambda)$ for $p\ge1$ and the self-energy function, in the thermodynamic limit. The main results are summarized in the following theorem (see also Theorem \ref{tpc}, Theorem \ref{cth1}, Theorem \ref{mqua}, Theorem \ref{maina}, Theorem \ref{mainb} and Theorem \ref{thmain2} for the precise presentation of the main results).
%%%%%%%%%%%%%%%%%%%%%%%%%%%%%%%%%%%%%%%%%%%%%%%%%%%%%%%%%%%%%%%%%%%%%%%%%%%%%%%%%%%%%%%
%%%%%%%%%%%%%%%%%%%%%%%%%%%%%%%%%%%%%%%%%%%%%%%%%%%%%%%%%%%%%%%%%%%%%%%%%%%%%%%%%%%%%%%%%555
\begin{theorem}\label{mainthm}
Consider the doped honeycomb Hubbard model with renormalized chemical potential $\mu=1$ at positive temperature $0<T\ll1$, which corresponding to the far-infrared region of the model. There exists a counter-term matrix $\delta E$ obeying the renormalization conditions (cf. Definition \ref{conj1}), such that, after taking the thermodynamic limit $L\rightarrow\infty$, the perturbation series of the connected $2p$-point Schwinger functions, $p\ge1$, as well as the self-energy have positive radius of convergence in the set $\RR_T:=\{\lambda\in\RRR\ |\ \lambda<C/|\log T|^2\}$, in which $0<C<0.01$ is a positive constant which depends on the physical parameters of the model but not on $T$ and $\lambda$. The first derivative of the self-energy w.r.t. the external momentum is uniformly bounded but the second derivatives of the self-energy w.r.t. the external momentum are divergent for $T\rightarrow0$ (cf. Theorem \ref{thmain2}). Therefore this model doesn't satisfy Salmhofer's criterion and the ground state of this model is not a Fermi liquid.
\end{theorem}
%%%%%%%%%%%%%%%%%%%%%%%%%%
These results will be proved with rigorous renormalization group analysis. In the first step, we shall express the Schwinger functions in terms of Grassmann functional integrations, which are more suitable for the multi-scale analysis. 
%%%%%%%%%%%%%%%%%%%%%%%%%%%%%%%%%%%%%%%%%%%%%%%%%%%%%%%%%%%%%%%%%%
\section{The Multi-scale Analysis}
\subsection{The Berezin integrals}
%\begin{notation}
%Instead of using the labeling $A$ and $B$ for the two types of quasi-particles, we introduce the new labeling as $A=1$ and $B=2$.
%\end{notation}
The Berezin integrals are linear functionals on the Grassmann algebra ${\bf Gra}$, generated by the Grassmann variables $\{\hat\psi^\e_{k,\t,\a}\}^{\t=\uparrow\downarrow;\a=1,2;\epsilon=\pm}_{k\in\DD_{\b,L}}$, 
which satisfy the periodic condition in the momentum variables: $\hat\psi^\e_{k_0,\bk+n_1{\bf G}_1+n_2{\bf G}_2,\t,\a}=\hat\psi^\e_{k_0,\bk,\t,\a}$, but anti-periodic condition in the frequency variable: $\hat\psi^\e_{k_0+\beta,\bk,\t,\a}=-\hat\psi^\e_{k_0,\bk,\t,\a}$.
The product of ${\bf Gra}$ is defined by: $\hat\psi^\e_{k,\t,\a}\hat\psi^{\e'}_{k',\t',\a'}=-\hat\psi^{\e'}_{k',\t',\a'}\hat\psi^\e_{k,\t,\a}$, for $(\e,\t,\a,k)\neq(\e',\t',\a',k',)$ and $(\hat\psi^\e_{k,\t,\a})^2=0$. Let $D\psi=\prod_{k\in\DD_{\b, L}, 
\t=\pm, \a=1,2}d\hat\psi_{k,\t,\a}^+
d\hat\psi_{k,\t, \a}^-$ be the measure of the Berezin integral,
$Q( \hat\psi^-, \hat\psi^+)$ be a monomial function of $\hat\psi_{k,\t,\a}^-, \hat\psi_{k,\t,\a}^+$. The Berezin integral $\int Q D\psi$ is defined to be $1$ for $Q( \hat\psi^-, \hat\psi^+)=\prod_{k\in\DD_{\b, L}, 
\t=\pm, \a=1,2} \hat\psi^-_{k,\t,\a} \hat\psi^+_{k,\t,\a}$, up to a permutation of the variables, and $0$ otherwise. The Grassmann differentiation is defined by 
${\partial_{ \hat\psi^\e_{k,\t,\a}}}{ \hat\psi^{\e'}_{k',\t',\a'}}=\delta_{k,k'}\delta_{\t,\t'}\delta_{\a,\a'}\delta_{\e,\e'}$, which also satisfy the anti-commutation relation: ${\partial_{ \hat\psi^\e_{k,\t,\a}}}\partial_{ \hat\psi^{\e'}_{k',\t',\a'}}=-\partial_{ \hat\psi^{\e'}_{k',\t',\a'}}{\partial_{ \hat\psi^\e_{k,\t,\a}}}$ for $(\e,\t,\a,k)\neq(\e',\t',\a',k',)$, and equals to zero otherwise.

The {\it Grassmann Gaussian measures} $P(d\psi)$ with covariance $\hat C(k)$ is defined as:
\be
P(d\psi) = (\det \NN)^{-1} D\psi \cdot\;\exp \Bigg\{-{\frac{1}{
\b|\L_L|}} \sum_{k\in\DD_{\b, L},\t={\uparrow\downarrow}, \a=1, 2 } 
\hat\psi^{+}_{k,\t, \a}{\hat C({k})}^{-1}\hat\psi^{-}_{k,\t,\a}\Bigg\}\;,
\label{ggauss}\ee
where
\be
\NN=\prod_{\kk\in\DD_L,\t={\uparrow\downarrow}}{\frac{1}
{\b|\L_L|}}
\begin{pmatrix}-i k_0-1 & -\O^*(\bk) \\ -\O(\bk) &
-ik_0-1\end{pmatrix},\label{norma}
\ee
is the normalization factor. The Grassmann fields are defined as:
\be
\psi^\pm_{x,\t,\a}=\sum_{k\in\DD_{\b, L}}
e^{\pm ikx}\hat\psi^\pm_{k,\t,\a},\ \ x\in\Lambda_{\beta,L}.
\ee

The interaction potential becomes:
\bea 
&&\VV_L(\psi,\lambda)=
\l\sum_{\a,\a'=1,2}\ \int_{\Lambda_{\beta,L}} d^3x \ \psi^+_{x,\uparrow,\a}
\psi^-_{x,\uparrow,\a'}\psi^+_{x,\downarrow,\a}
\psi^-_{x,\downarrow,\a'}\label{potx}\\
&&\quad\quad+\frac{1}{\b|\Lambda_L|}\sum_{k\in\DD_{\b, L}}\sum_{\a,\a'=1,2}\sum_{\tau=\uparrow, \downarrow}[\delta\mu(\lambda)\delta_{\a\a'}+\nu_{\a\a'}(\kk,\lambda)]
\hat\psi^+_{k,\t,\a}\hat\psi^-_{k,\t,\a'}\nn
\eea
where $\int_{\Lambda_{\beta,L}} d^3x:=\int_{-\beta}^\beta dx_0\ \sum_{\xx\in\L_L}$ is a short-handed notion for the integration and sum. Define the non-Gaussian measure $P^I_L(d\psi):=P(d\psi)e^{-\VV_L(\psi)}$ over ${\bf Gra}$ and let $P^I(d\psi)=\lim_{L\rightarrow\infty}P^I_L(d\psi)$ be the limit of the sequence of measures indexed by $L$ (in the topology of weak convergence of measures), then we can easily prove that these non-Gaussian measure are invariant under the transformations introduced in Lemma \ref{inv0}. 
The Schwinger functions are defined as the moments of the measure $P^I(d\psi)$:
\bea
&&S_{2p,\b}(x_1,\t_1,\e_1,\a_1,\cdots,x_{2p},\t_{2p},\e_{2p},\a_{2p};\l)\nn\\
&&\quad\quad\quad\quad\quad\quad\quad\quad\quad=\lim_{L\rightarrow\infty}\frac{\int\psi^{\epsilon_1}_{x_1,\t_1,\a_1}\cdots \psi^{\epsilon_{2p}}_{x_{2p},\t_{2p},\a_{2p}} P(d\psi)e^{-\VV_L(\psi,\lambda)}}{\int P(d\psi)e^{-\VV_L(\psi,\lambda)}}\nn\\ 
&&\quad\quad\quad\quad\quad\quad\quad\quad\quad=:\frac{\int\psi^{\epsilon_1}_{x_1,\t_1,\a_1}\cdots \psi^{\epsilon_{2p}}_{x_{2p},\t_{2p},\a_{2p}} P(d\psi)e^{-\VV(\psi,\lambda)}}{\int P(d\psi)e^{-\VV(\psi,\lambda)}}.
\eea
We assume in the rest of this paper that the thermodynamic limit has been already taken and will drop the parameter $L$. This assumption is justified by rigorous construction of the connected Schwinger functions in the limit $L\rightarrow\infty$.
%%%%%%%%%%%%%%%%%%%%%%Now The Fermi surfaces%%%%%%%%%%%%%%%%%%%%%%%%%%%%%%%%%%%%
\subsection{Scale Analysis}
The lattice structure plays the role of the short-distance cutoff for the spatial momentum, so that the ultraviolet behaviors of the Schwinger functions are rather trivial. The two-point Schwinger function 
is not divergent but has a discontinuity at $x_0=0$, $\xx=0$ \cite{BG1}.\footnote {Although summation over all scales of the tadpole terms is not absolutely convergent for $k_0\rightarrow\infty$, this sum can be controlled by using the explicit expression of the single scale propagator.} So we omit the ultraviolet analysis but introduce a suitable ultraviolet (UV) cutoff function $U(\bk)$, $\bk\in \RRR^2$, which is smooth and compactly supported. This can keep the momentum $\bk$ bounded. We consider only the far infrared behaviors, which correspond to the cases of $T\ll1$ and the momenta getting close to the Fermi surface. It is mostly convenient to choose the infrared cutoff functions as the Gevrey class functions, defined as follows.
\begin{definition}\label{gev}
Given $\DD\subset\RRR^d$ and $h>1$, the Gevrey class $G^h_0(\DD)$ of functions of index $h$ is defined as the set of smooth functions $\phi\in\cC^\infty(\DD)$ such that for every compact subset $\KK\subset\DD$, there exist two positive constants $A$ and $\g$, both depending on $\phi$ and $\KK$, satisfying:
\bea
\max_{x\in \KK}|\partial^\alpha \phi(x)|\le A\g^{-|\alpha|}(|\alpha!|)^h,\ \alpha\in\ZZZ^d_+,\ |\alpha|=\alpha_1+\cdots+\alpha_d.
\eea
The Gevrey class of functions with compact support is defined as: 
$G_0^h(\DD)=G^h(\DD)\cap C^\infty_0(\DD)$.
The Fourier transform of any $\phi\in G_0^h$ satisfies
\be
\max_{k\in\RRR^d}|\hat \phi(k)|\le Ae^{-h(\frac{\g}{\sqrt{d}}|k|)^{1/h}}.
\ee
\end{definition}

\begin{figure}[htp]
\centering
\includegraphics[width=.42\textwidth]{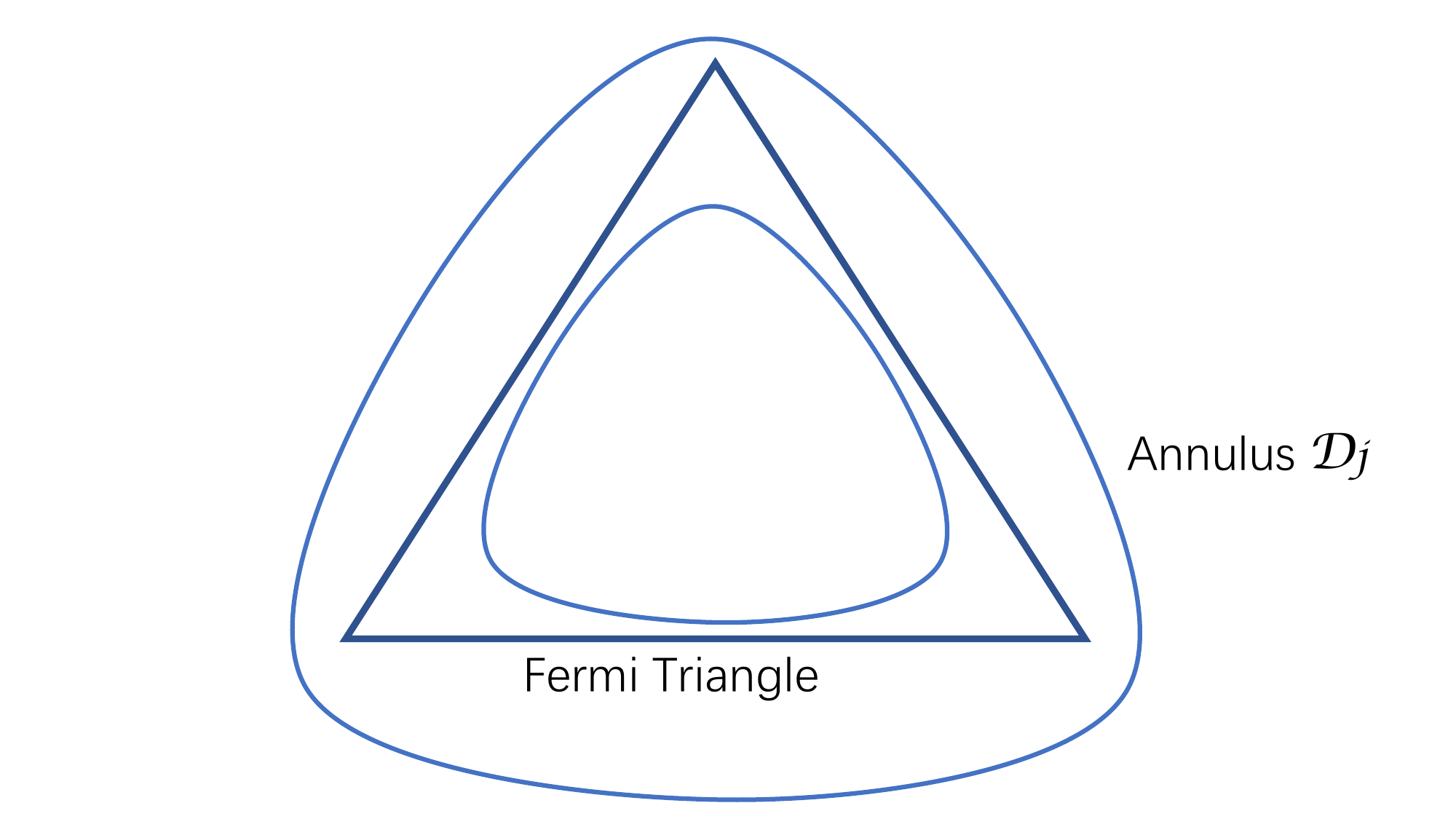}
\caption{\label{annulus}
An annulus $D_j$, $j\ge1$, surrounding a Fermi triangle.}
\end{figure}

Let $\chi\in G^h_0(\RRR)$, define:
\be
\chi(t)=\chi(-t)=
\begin{cases}
=0\ ,&\quad {\rm for}\quad  |t|>2,\\
\in(0,1)\ ,&\quad {\rm for}\quad  1<|t|\le2,\\
=1,\ &\quad {\rm for}\quad  |t|\le 1. 
\end{cases}\label{support}
\ee
Given any fixed constant $\gamma\ge10$, define the following partition of unity:
\bea\label{part1}
1&=&\sum_{j=0}^{\infty}\chi_j(t),\ \ \forall t\neq 0;\\
\chi_0(t)&=&1-\chi(t),\ 
\chi_j(t)=\chi(\gamma^{2j-2}t)-\chi(\gamma^{2j}t),\ {\rm for}\ j\ge1.\nn
\eea
The support of the cutoff function $\chi_j$, $j\ge1$, is an annulus (See Figure \ref{annulus} for an illustration):
\be\label{multi1}
\cD_j=\Big\{k=(k_0,\kk) \vert\g^{-2j-2}\le 4k_0^2+e^2(\kk,1)\le 2\g^{-2j}\Big\}.\ee

Define also 
$$\chi^{(\le j)}(t)=\sum_{i=0}^j\chi_{i}(t),\ \chi^{(> j)}(t)=\sum_{i=j+1}^\infty\chi_{i}(t).$$
%%%%%%%%%%%%%
%The infrared propagators are defined as follows.
%%%%%%%%%%
\begin{definition}
Let $j_0\gg1$ be a positive integer, the free propagators in the far infrared region are defined as:
\bea\label{irprop}
\hat C^{ir}(k)_{\a\a'}&=&\hat C(k)_{\a\a'}[U(\bk)-\chi^{(\le j_0)}(4k_0^2+e^2(\kk,1))]\nn\\
&=&\sum_{j=j_0}^\infty \ \hat C_j(k)_{\a\a'},\ \a,\a'=1,2,\\
\hat C_j(k)_{\a\a'}&=&\hat C(k)_{\a\a'}\cdot \chi_j[4k_0^2+e^2(\kk,1)].\label{irprop1}
\eea
The number $j_0$ serves as a UV cutoff index for $k_0$ and is also called the infrared threshold.
\end{definition}
\begin{remark}
Remark that, it is always possible to choose a proper $U(\bk)$ such that $U(\bk)\ge \chi^{(\ge j)}[4k_0^2+e^2(\kk,1)],$ for all $j\ge0$ and $k=(k_0,\bk)\in\RRR\times \RRR^2$. 
\end{remark}
Now we consider the sliced propagators $\hat C_j(k)$ with $0\le j\le j_0$, we have:
\begin{proposition}
The $p$-th power of the sliced propagator, $[\hat C_j]^p$, is integrable for $0\le j\le j_0$ and $p\ge1$.
\end{proposition}
\begin{proof}
Since the integration domain of $[\hat C(k)]^p$ is bounded, and the denominator of $\tilde C(k)$ is strictly bounded away from zero in the integration domain, the conclusion follows.
\end{proof}
%This proposition states that the interacting Schwinger functions are also well defined for $0\le j\le j_0$. 
%%%%%%%%%%%%%%%%%%
\begin{definition}\label{wholeprop}
Define the infrared cutoff index $j_{max}=\EEE(\tilde j_{max})$, in which $\tilde j_{max}$ is the solution to the equation $\gamma^{\tilde j_{max}-1}= 1/\sqrt2\pi T$ and
$\EEE(\tilde j_{max})$ is the integer part of $\tilde j_{max}$. Then the infrared propagator with cutoff index $j_{max}$ is defined as:
\bea\label{irpropm}
\hat C^{ir,j_{max}}(k)_{\a\a'}=\hat C(k)_{\a\a'}[U(\bk)-\chi^{(\le j_0)}(4k_0^2+e^2(\kk,1))-\chi^{(> j_{max})}(4k_0^2+e^2(\kk,1))].\nn\\
\eea
\end{definition}
%%%%%%%%%%%%%%%%%%%%%%%%%%%%%%%%
It is useful to rewrite the propagator $\hat C(k)$ in \eqref{2ptkb} (with $\mu=1$) as a product of a scalar function $\tilde C(k)=[-2ik_0+e(\kk,1)+k^2_0]^{-1}$ with the $2\times2$ matrix $A=A(k, 1)$. We have:
%%%%%%%%%%%%%%%%%
% $\hat C(k)=\tilde C(k) A(k)$, in which $\tilde C(k)=[-2ik_0+e(\kk,1)+k^2_0]^{-1}$ and $A(k)$
%
%\be\label{redprop}\tilde C(k):=\frac{1}{-2ik_0+e(\kk,1)+k^2_0},
%\ee 
%and 
%\be\label{freep3}
%A(k)=\begin{pmatrix}i k_0+1&-\O^*(\bk) \\ -\O(\bk) &
%ik_0+1\end{pmatrix}.
%\ee
%%%%%%%%%%%%
\begin{proposition}\label{mat0}
Let $A_{\a\a'}$, $\a,\a'=1,2$ be the matrix elements of $A$. Let $j\in[j_0, j_{max}]$ be any scale index and $\gamma\ge10$ be a fixed constant such that $\g^{-2j-2}\le 4k_0^2+e^2(\kk,1)\le 2\g^{-2j}$, then there exist two constants $K$, $K'$ which are independent of the scale index $j$ and satisfy  $0.9\le K<K'\le 2$, such that 
\be\label{matele2}
K\le\vert A_{\a\a'}\vert\le K',\  \a,\a'=1,2\ .
\ee 
\end{proposition} 
\begin{proof}
%%%%%%%%%%%%%%%%%%%%%%%%%%%%%%%%%
%It is easy to find that {\bf the support} of the cutoff function $\chi_j$ at $j\ge j_0$ is the annulus:
%\be\label{multi1}
%\cD_j=\Big\{k=(k_0,\kk) \vert\g^{-2j-2}\le 4k_0^2+e^2(\kk,1)\le 2\g^{-2j}\Big\},\ee
%which implies that
%\be\frac14 \g^{-2j-2}\le k_0^2\le \frac12\g^{-2j},\label{cond1}\ee
%\be
% \frac12\g^{-j-1}\le \vert e(\kk,1)\vert\le \frac{\sqrt2}{2}\g^{-j}
%\label{cond0}.
%\ee
Consider first the elements $\vert A_{11}\vert=\vert A_{22}\vert=\vert ik_0+1\vert=(1+k_0^2)^{1/2}$. Choosing $j_0=1$ and by \eqref{multi1}, we have $1<(1+k_0^2)^{1/2}<2$. Now we consider the elements $\vert A_{12}\vert=\vert A_{21}\vert=\vert \O_0(\kk)\vert=(1+e(\kk,1))^{1/2}$ (cf. \eqref{band1}). By \eqref{multi1}, we can easily find that $0.9\le(1+e(\kk,1))^{1/2}\le2$. So we can always choose $K$ and $K'$ which satisfy \eqref{matele2}.
\end{proof}
%%%%%%%%%%%

\begin{lemma}\label{bdsp}
Let $\hat C_j(k)_{\a\a'}$, $\a,\a'=1,2$, be any matrix element of the momentum space free propagator at slice $j$. There exists a positive constant $K$, which is independent of the scale index, such that
\be\label{tad1}
\sup_{k\in \cD_j}\vert\hat C_j(k)_{\a\a'}\vert\le K\g^{j}.
\ee
\end{lemma}
\begin{proof}
Using the definition of the support function $\chi_j$, and by \eqref{irprop}, we have
\be
\sup_{k\in \cD_j}\vert \tilde C_j(k)_{\a\a'}\vert\le K'\g^{j}\ ,
\ee
for certain positive constant $K'$ independent of the scale index. By Proposition \ref{mat0}, the result of this lemma follows.
\end{proof}
%%
%%%%%%%%%%%%%%%%%%%%%%%%%%%%%%%%%%%%%%%%%%%%%%%%%%%%%%%%%%%%%%%%%%%%%%%%%%%%%%%%%%%%%%%%%%%%5
\subsection{Sectors and angular analysis}
\begin{figure}[htp]
\centering
\includegraphics[width=.55\textwidth]{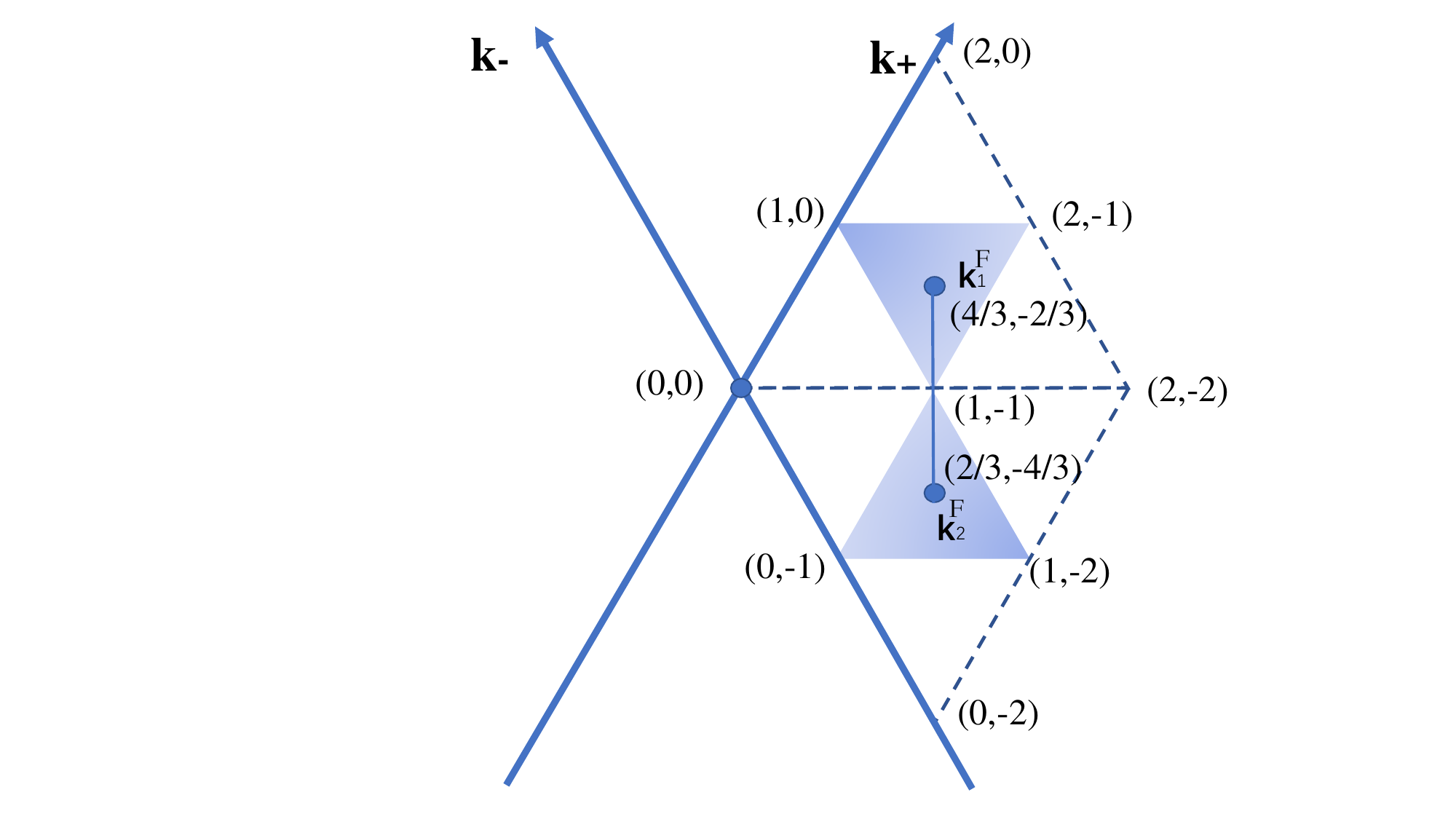}
\caption{\label{kpm}
The Brillouin zone $\tilde\cD_1$ in the coordinate system $(k_+,k_-)$. The vertices of the Fermi triangles are the van Hove singularities.}
\end{figure}

Due to the $\ZZZ^3$ symmetry of the Fermi triangle (see Figure \ref{figsec}), it is convenient to introduce a new basis $(e_+, e_-)$:
\be
e_+=\frac{\pi}{3}(1, \sqrt3),\quad e_-=\frac{\pi}{3}(-1, {\sqrt3}),
\ee
which is neither orthogonal nor normal. 
%%%%%%%%%%%%%
%The Jacobian for this transformation is $J=\frac{2\pi^2}{3\sqrt3}$. 
%%%%%%%%%%%%%%%
Let $(k_+,k_-)$ be the coordinates in the new basis, the transformation law is
\be\label{cotrans}
k_1=\frac{\pi}{3}(k_+-k_-),\quad k_2=\frac{\pi}{\sqrt3}(k_++k_-).
\ee
In the new coordinate system, the first Brillouin zone, denoted by $\tilde\cD_1$, becomes a rescaled rhombus in which
$k_+\in[0,2]$ and $k_-\in[-2,0]$. See Figure \ref{kpm} for an illustration. Formula \eqref{band1} can be rewritten as:
\be\label{band3}
e(\bk,1)=8\cos\frac{\pi(k_++k_-)}{2} \cos\frac{\pi k_+}{2} \cos\frac{\pi k_-}{2}.
\ee
The edges of the Fermi triangles are given by the equations $k_+=\pm1$, $k_-=\pm1$ and $\frac{k_++k_-}{2}=\pm1$, respectively.

\begin{figure}[!htb]
\centering
\includegraphics[scale=.3]{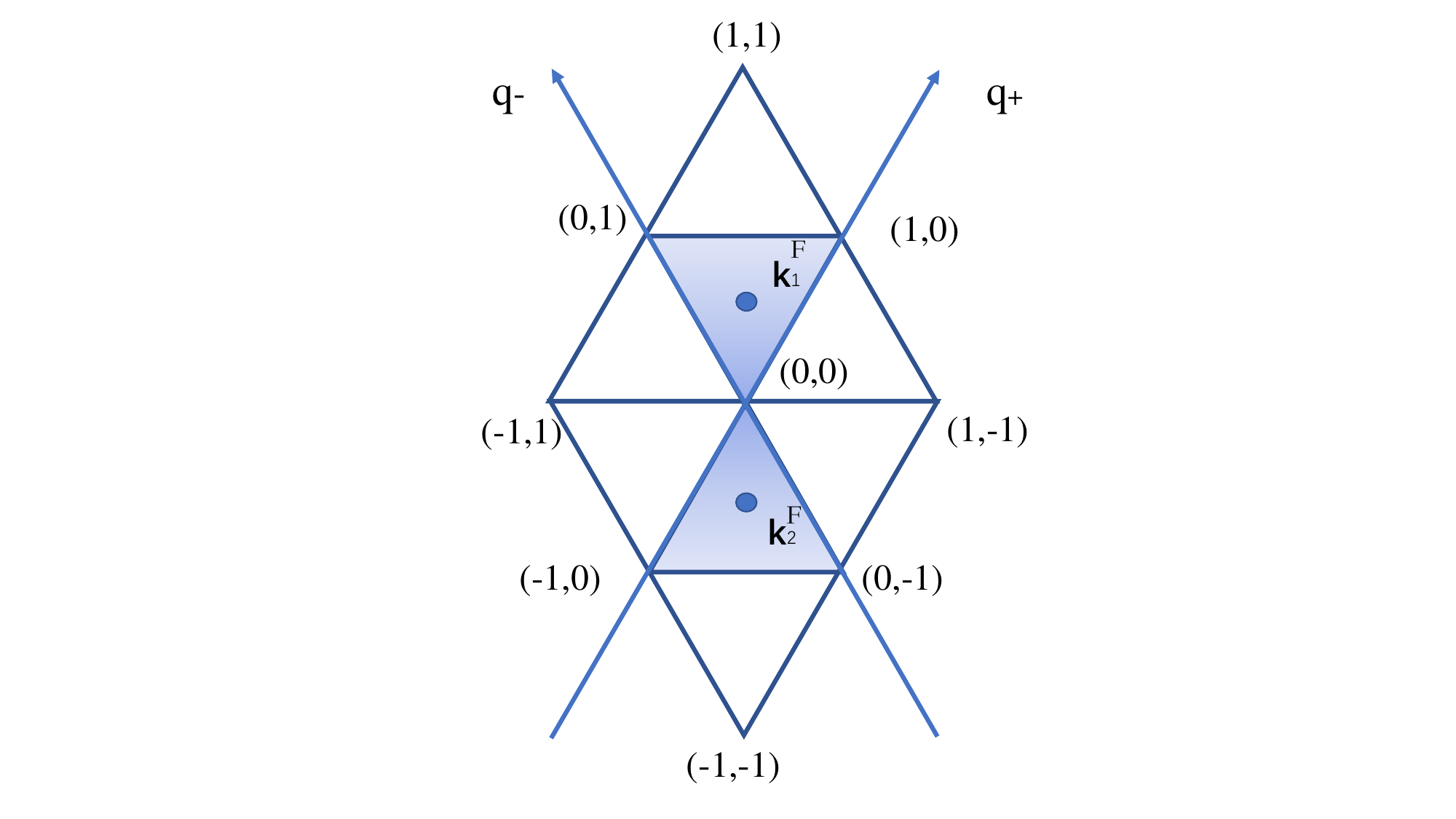}
\caption{An illustration of the first Brillouin zone in the basis $(e_+,e_-)$ with the rescaled coordinates. The vertices of the Fermi triangles are the van Hove singularities. }\label{fpt1}
\end{figure}

Now we consider the variables $(x_+,x_-)$ dual to the lattice momentum $(k_+,k_-)$. Without losing generality, we can pick up any lattice point of type $A$ and study the corresponding change of coordinates. Coordinate transforms for points of type $B$ are the same (modulo a shift of the origin), due to the periodic structure on the lattice. Consider the lattice point $\xx=n_1{\bl}_1+n_2{\bl}_2$, in which ${\bl}_1=\frac12(3,\sqrt{3})$, ${\bl_2}=\frac12(3,-\sqrt{3})$ and $n_1, n_2\in\ZZZ$. We have
$x_1=\frac32(n_1+n_2)$, $x_2=\frac{\sqrt3}{2}(n_1-n_2)$. The coordinate transformation in the direct space corresponding to \eqref{cotrans} is:
\bea\label{coor2}
x_+=\pi\cdot(\frac{x_1}{3}+\frac{x_2}{\sqrt3})=\pi n_1,\quad
x_-=\pi\cdot(-\frac{x_1}{3}+\frac{x_2}{\sqrt3})=-\pi n_2,\quad  n_1, n_2\in\ZZZ.\nn
\eea
The free propagator can be written as
\bea\label{prob2x}
C(x)=C(x_0,x_+,x_-)=\frac{2\pi^2}{3\sqrt3}\int_{\TTT_\beta} dk_0 \int_{\tilde\cD_1} dk_+dk_-\hat C(k_0,k_\pm)e^{ik_0x_0+ik_+x_++k_-x_-},
\eea
\bea\label{prop2k}
\hat C(k_0,k_\pm)=\frac{1}{-2i k_0+ e(\bk,1)+k_0^2} \begin{pmatrix}i k_0+1 &-\tilde\O^*(\bk) \\ -\tilde\O(\bk) &
ik_0+1\end{pmatrix},
\eea
in which $\tilde\Omega(\bk)=1+2e^{-i\frac{\pi}{2}(k_+-k_-)}\cos \frac{\pi}{2}(k_++k_-)$ and the prefactor $\frac{2\pi^2}{3\sqrt3}$ is the Jacobian of the transform $(k_1,k_2)\rightarrow(k_+,k_-)$. The integral $\int dk_0$ is the discrete sum $2\pi T\sum_{n\in\ZZZ}(2n+1)\pi T$ and the integration $\int dk_+dk_-$ is constrained in $\tilde\cD_1$. Define the quasi-momentum $q_\pm$ by
\bea\label{qa}
q_\pm=
\begin{cases}
k_\pm-1,&{\rm for}\quad k_\pm\ge0,\\
k_\pm+1,& {\rm for}\quad k_\pm\le0,
\end{cases}
\eea
then \eqref{band3} becomes
\be\label{band4}
e({\bf q},1)=-8\cos\frac{\pi(q_++q_-)}{2} \sin\frac{\pi q_+}{2} \sin\frac{\pi q_-}{2}.
\ee
The first Brillouin zone with new coordinates $(q_+,q_-)$, denoted by $\tilde\cD$, is a rhombus centered around $(0,0)$ but with rescaled coordinates, see Figure \ref{fpt1} for an illustration. The fundamental Fermi triangles are defined by the equations $q_+=0$, $q_-=0$ and $q_++q_-=\pm1$. 
%%%%%%%%%%%%%%%%%%%%%%%%%%%%%%%%%%%%%%%%%%%%%%%%%%%%%%%%%%%%%%%

Remark that, we have 
$k_+\ge0$ and $k_-\le0$ in $\tilde\cD_1$, and the quasi-momenta are given by 
\be
q_+=k_+-1,\quad\quad q_-=k_-+1.
\ee
Define $\tilde\cD_\b=\TTT_\beta\times\tilde\cD$, the free propagator can be rewritten as:
\bea\label{prop2qx}
C(x_0,x_+,x_-)=\frac{2\pi^2}{3\sqrt3}\int_{\tilde\cD_\b} dk_0 dq_+dq_-\hat C(k_0,q_+,q_-)e^{ik_0x_0+(q_++1) x_+ +(q_--1) x_-},
\eea
in which
\bea\label{prop2q}
\hat C(k_0,q_+,q_-)&=&\frac{1}{-2i k_0+ e(\bq,1)+k_0^2} \begin{pmatrix}i k_0+1 &-\tilde\O^*(\bq) \\ -\tilde\O(\bq) &
ik_0+1\end{pmatrix},\\
\tilde\Omega(\bq)&=&1-2e^{-i\frac{\pi}{2}(q_+-q_-)}\cos \frac{\pi}{2}(q_++q_-).\nn
\eea

\begin{definition}[UV cutoff for the quasi-momenta]\label{uvq}
In the new coordination, we define the UV cutoff function $U(k_+,k_-)=U(q_++1,q_--1)$ as
\be
U(q_++1,q_--1)=\begin{cases} 0,\quad {\rm for}\quad \vert q_+\vert>1,\ \vert q_-\vert>1,\\
1,\quad {\rm for}\quad \vert q_+\vert\le1,\ \vert q_-\vert\le1, \end{cases}
\ee
and the infrared propagator (cf. Definition \ref{wholeprop}) becomes:
\bea\label{irpropmq}
&&\hat C^{ir,j_{max}}(k_0,q_+,q_-)_{\a\a'}=\hat C(k_0,q_+,q_-)_{\a\a'}\ \Big[\ U(q_+,q_-)\\
&&\quad\quad\quad\quad\quad-\chi^{(\le j_0)}(4k_0^2+e^2(\bq,1))-\chi^{(> j_{max})}(4k_0^2+e^2(\bq,1))\ \Big].\nn
\eea

\end{definition}
%in which $\tilde\Omega(\bq)=1-2e^{-i\frac{\pi}{2}(q_+-q_-)}\cos \frac{\pi}{2}(q_++q_-)$. 

%%%%%%%%%%%%%%%%%%%%%%%%%%%%%%%%%%%%%%%%%%%%%%%%%%%%%%%%%%%%%%%%%%%%%%%%%%%%%%%%
%%%%%%%%%%%%%%%%%%%%%%%%%%%%%%%%
%%We have important mistakes
%%%%%%%%%%%%%%%%%%%%%%%%%%%%
Remarked that, since we are mainly interested in the far-infrared behaviors of the free propagators, which corresponds to the cases of $j\gg1$, we have $\vert k_0^2\vert\ll\vert k_0\vert$, and we can drop the term $k_0^2$ in the denominator of \eqref{prop2q}. By Formula \eqref{multi1}, the size of $k_0^2+e^2(\bk,1)$ is bounded by
$O(1)\gamma^{-2j}$, at any scale index $j\ge1$. But the size of $e^2(\bk,1)$, which can be of order $\gamma^{-2i}$ with $i\ge j$, is not fixed. In order to obtain the optimal decaying bounds for the direct-space propagators, we need to control $e^2(\bk,1)$ by further introducing cutoff functions for the spatial momentum.
\begin{definition}
Define the factors $\{t^{(a)}\}$, $a=1,2,3$, by
\be\label{threefactors}
t^{(1)}=4\cos^2\frac{\pi k_+}{2},\ t^{(2)}=4\cos^2\frac{\pi k_-}{2},\ t^{(3)}=4\cos^2\frac{\pi(k_++k_-)}{2},
\ee
or in terms of coordinates $q^{(a)}$, by
\be\label{threefactorsq}
t^{(1)}=4\sin^2\frac{\pi q_+}{2},\ t^{(2)}=4\sin^2\frac{\pi q_-}{2},\ t^{(3)}=4\cos^2\frac{\pi(q_++q_-)}{2}.
\ee
Correspondingly, define the three local coordinates $\{k^{(a)}\}$, $a=1,2,3$, by
\be
k^{(1)}=k_+,\ k^{(2)}=k_{-},\ k^{(3)}=k_++k_-,
\ee
and the coordinates of the quasi-momentum $\{q^{(a)}\}$, $a=1,2,3$, by
\be
q^{(1)}=k_+\pm1,\ q^{(2)}=k_-\pm1,\ q^{(3)}=q_++q_-.
\ee
\end{definition}
%%%%%%%%%%%%%%%%%%%%%%%%%%%%%%%%%%%%%%%%%
%Since the factors in \eqref{threefactors} or \eqref{threefactorsq} are highly nonlinear in $k^{(a)}$ 
%or $q^{(a)}$, for $k^{(a)}$ not very close to $\pm1$, or for $q^{(a)}$ not very close to $0$, instead of slicing the quasi momentum $k^{(a)}$ or the quasi-momentum $q^{(a)}$ directly, we choose to slice the functions in \eqref{threefactors} or \eqref{threefactorsq}. The support of the cutoff functions are called the {\it sectors} \cite{FMRT}. 
%%%%%%%%%%%%%%%%%%%%%%%%%%%%%%%%%%%%%%%%%%
%%%%%%%%%%Lifshitz transition

\begin{definition}[Partition of unity]
To each factor $t^{(a)}$ of \eqref{threefactors} or \eqref{threefactorsq}, $a=1, 2, 3$, we introduce a set of indices $s^{(a)}\in\{0,1,\cdots, j\}$ and define the following functions of partition of unity:
\bea\label{secf}
1=\sum_{s^{(a)}=0}^{j}v_{s^{(a)}}(t^{(a)}),\quad \begin{cases} v_0(t^{(a)})=1-\chi(\g^2t^{(a)}),\quad\ {\rm for}\quad s^{(a)}=0,\\
v_{s^{(a)}}(t^{(a)})=\chi_{s^{(a)}+1}(t^{(a)}),\quad {\rm for}\quad 1\le s^{(a)}\le j-1, \\
v_j(t^{(a)})=\chi(\g^{2j}t^{(a)}),\quad\quad\quad {\rm for}\quad s^{(a)}=j.
\end{cases}
\eea
\end{definition}
%%%%%%%%%%%
%%%%%%%%%%%%%%%%%%%%%%%%%%%%%%%%%%%%%%
%Remark that, for a vector $\bk\in\RRR^2$, we need a pair of sector indices to indicate the regions it belongs to. 
%%%%%%%%%%%%%%5
%Then we can define the sectorized propagators, as follows.
\begin{definition}
Let $t^{(a)}$ and $t^{(b)}$, $a,b=1,2,3$, $a\neq b$, be the factors defined by \eqref{threefactors} or \eqref{threefactorsq}, whose values are close to zero. 
The free propagator of scale index $j$ can be decomposed as:
\be\label{sec0}
\hat C_j(k)=\sum_{\s=(s^{(a)}, s^{(b)})}\hat C_{j,\s}(k),
\ee
in which
\be
\hat C_{j,\s}(k)=\hat C_j(k)\cdot v_{s^{(a)}}[t^{(a)}]\ v_{s^{(b)}}[t^{(b)}].\label{sectz}
\ee
is called a sectorized propagator. $\s=(s^{(a)},s^{(b)})$, in which $s^{(a)},s^{(b)}=0,1,\cdots, j$, is called the sector indices at scale $j$ and the summation runs over all such sector indices. The support of $\hat C_{j,\s}(k)$, denoted by $\Delta^j_{s^{(a)},s^{(b)}}$, is called a sector \cite{FMRT} with scale index $j$ and sector indices $\s=(s^{(a)},s^{(b)})$.
\end{definition}

Notice the three cosine functions in \eqref{band3} are not independent of each other. We have:
\begin{lemma}\label{bdcos}
Let $s\ge2$. If $\vert\cos\frac{\pi k_+}{2}|\le\gamma^{-s}$ and $|\cos\frac{\pi k_-}{2}|\le \sqrt{2}/2$, then there exists some strictly positive constant $c'>0$ such that 
\be\label{bd3f}
c'\le\vert\cos\frac{\pi(k_++k_-)}{2}\vert\le1.
\ee
\end{lemma}
%%%%%%%%%%%%%%%%%%%%%%%%%%%
\begin{proof}
By trigonometrical formula 
$\cos\frac{\pi(k_++k_-)}{2}=\cos\frac{\pi k_+}{2} \cos\frac{\pi k_-}{2}-\sin\frac{\pi k_+}{2} \sin\frac{\pi k'_-}{2}$,
we have 
\bea
|\cos\frac{\pi(k_++k_-)}{2}|&\ge& \frac{\sqrt2}{2}[1-\frac{\gamma^{-2s}}{2}-\g^{-s}+O(\g^{-4s})]\nn\\
&\ge& \frac{\sqrt2}{2}(1-2\g^{-2}),
\eea
for $\g\ge10$.
Choosing $c'=\frac{\sqrt2}{2}(1-2\g^{-2})$, which is strictly greater than zero, we proved this lemma. We can prove similar results for the three factors in \eqref{threefactorsq}.
\end{proof}
This lemma states that, if any two factors among the three, say, $t^{(1)}$ and $t^{(2)}$, are close to zero, then the third factor $t^{(3)}$ must be strictly bounded away from zero. Therefore, we only need to introduce two sector indices $(s^{(1)}, s^{(2)})$ to control the size of $e(\bk,1)$. Remark that the constant $\sqrt{2}/2$ in Lemma \eqref{bdcos} can be replaced by some other positive constant that is strictly smaller than one.
%%%%%%%%%%%%%%%%%%%%%%%%%%%%%%%%%%%%%%%%%%%%%%%%%%%%%%%%%%%%%%%%%%%%%%%%%%
\subsection{Constraints on the sector indices}
\begin{figure}[htp]
\centering
\includegraphics[width=.6\textwidth]{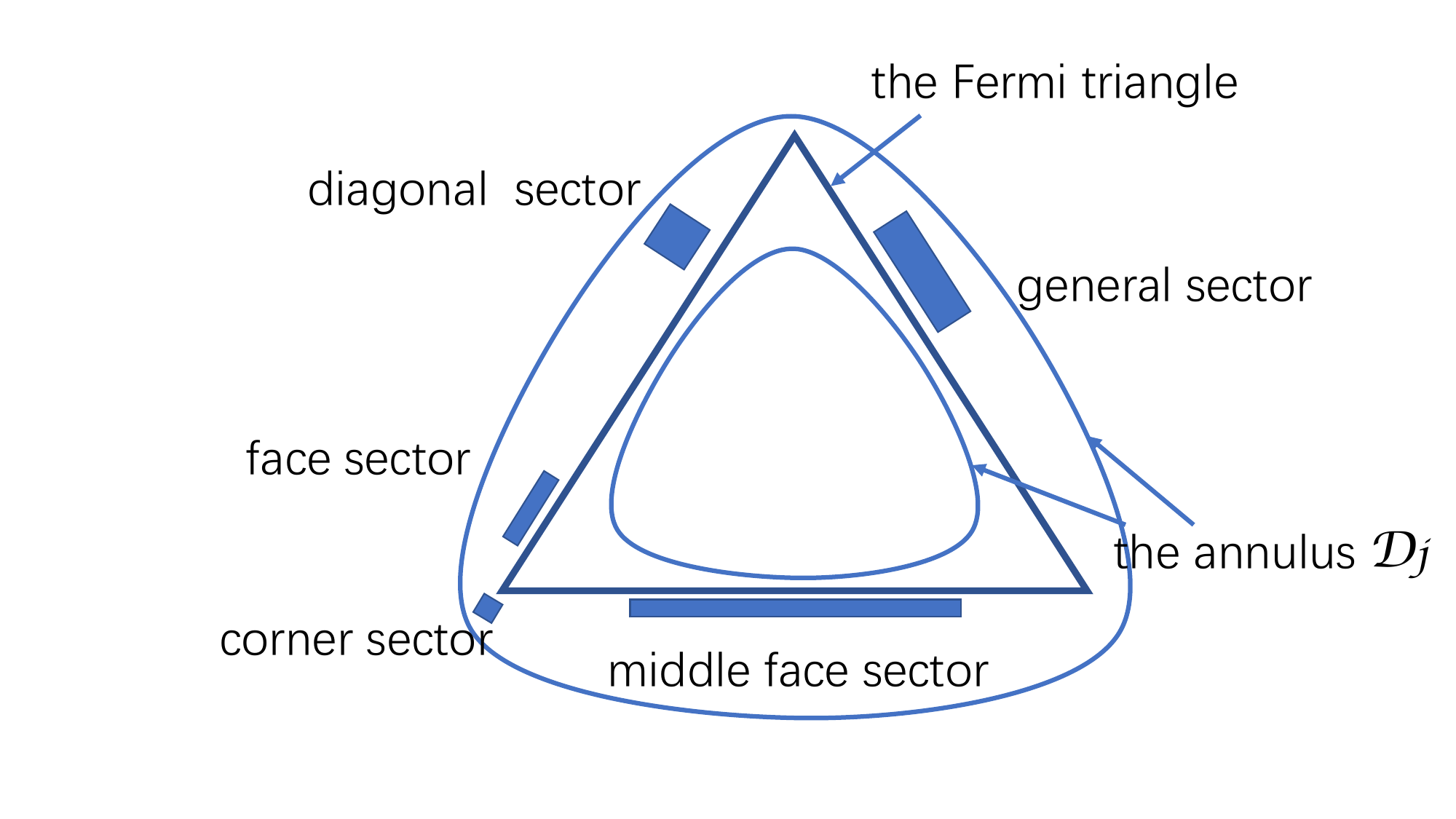}
\caption{\label{figsec}An illustration of the various sectors.
}
\end{figure}

In this part we consider the possible constraints on the sector indices. The first one is the following: 
\begin{lemma}\label{spm}
Let $j$ be any scale index. Let $t^{(a)}$ and $t^{(b)}$, $a,b=1,2,3$, $a\neq b$, be the factors that are close to zero and
$s^{(a)}$, $s^{(b)}$ be the corresponding sector indices, then the possible values of the sector indices $s^{(a)}$ and $s^{(b)}$ must satisfy:
\be
s^{(a)}+s^{(b)}\ge j-1.
\ee
\end{lemma}
\begin{proof}
By Formulas \eqref{secf} and \eqref{bd3f}, we have
\be\label{compare1}
4\g^{-2s^{(a)}}\g^{-2s^{(b)}}\ge e^2(\kk,1)=t^{(1)}\cdot t^{(2)}\cdot t^{(3)}\ge 4c'^2\gamma^{-2s^{(a)}-2}\gamma^{-2s^{(b)}-2},
\ee
in which $c'=\frac{\sqrt2}{2}(1-2\g^{-2})$.
In order that the sliced propagator is non-vanishing, $e^2(\kk,1)$ must obey Formulas \eqref{multi1}, so we have
\be\label{compare2}
\frac12\g^{-2j}\ge e^2(\kk,1)\ge\frac14\g^{-2j-2}
\ee
In order that \eqref{compare1} be consistent with \eqref{compare2}, we have
\be
4c'^2\gamma^{-2s^{(a)}-2}\gamma^{-2s^{(b)}-2}\le \frac12\g^{-2j},
\ee 
So we obtain:
\be
s^{(a)}+s^{(b)}\ge j-2+\log_\gamma [4(1-2\g^{-2})].
\ee
Since the sector indices are integers and $0<\log_\gamma [4(1-2\g^{-2})]<1$ for $\g\ge 10$, we have
\be
s^{(a)}+s^{(b)}\ge j-1.
\ee
\end{proof}
This lemma also put constraints on the shapes of the sectors. In order to better understand the geometry of the sectors, we introduce the following definitions (see Figure \ref{figsec} for an illustration):
\begin{definition}
A face $f^{(a)}$, $a=1,2,3$, is defined as the region close to the Fermi triangle, in which the factor $t^{(a)}$ takes values in the neighborhood of zero; a corner $I^{(ab)}$, $a,b=1,2,3$, is defined as the region close to the Fermi triangle, in which two factors $t^{(a)}$ and $t^{(b)}$ take values in a neighborhood of zero. 
E.g., the face $f^{(1)}$ is the region for which $t^{(1)}$ is close to zero and the corner $I^{(23)}$ is the region for which both $t^{(2)}$ and $t^{(3)}$ are close to zero.
We introduce also the following notions for the sectors at scale $j$:
\begin{itemize}

\item {the sectors with sector indices $(s,j)$ and $(j,s)$, with $j>s$, are called the face sectors, in particular, the sectors with sector indices $(0,j)$ and $(j,0)$ are called the middle-face sectors.}

\item the sectors with sector indices $(j,j)$ are called the corner sectors.
\item the sectors with sector indices $(s,s)$, with $(j-1)/2\le s<j$, are called the diagonal sectors.
\item other sectors are called the general sectors. 
\end{itemize}
\end{definition}

Now we consider the possible constraints on the sector indices placed by conservation of momentum. Let $\s=(s^{(a)},s^{(b)})$, $a\neq b$, $a,b=1,2,3$, be the sector indices for the momentum $\bk$. In order that $\hat C_{j,\s}(k)\neq0$, $k=(k_0,\bk)$ must satisfy the following bounds:
\be
\frac{1}{4}\gamma^{-2j-2}\le k^2_0\le \frac12 \g^{-2j},
\ee
and
\bea\label{supp1}
\begin{cases}
\g^{-2}\le t^{(a)}\le 1,&{\rm for}\ s^{(a)}=0,\\
\g^{-s^{(a)}-1}\le t^{(a)}\le\sqrt2 \g^{-s^{(a)}},& {\rm for}\  1\le s^{(a)}\le j-1,\\
 t^{(a)}\le\sqrt2 \g^{-j},&{\rm for}\  s^{(a)}=j.
\end{cases}
\eea
When $k^{(a)}$, $a=1,\cdots,3$, is close to $1$, or equivalently when the quasi-momentum $q^{(a)}$ is close to zero, the constraints in \eqref{supp1} can be formulated as:
\bea
\begin{cases}
1/{\pi \g}\le\vert q^{(a)}\vert\le 1,\quad\quad\quad\quad\quad\quad\quad {\rm for}\quad s^{(a)}=0,\\
{\g^{-s^{(a)}-1}}/{\pi }\le\vert q^{(a)}\vert\le\frac{\sqrt2}{\pi} \g^{-s^{(a)}},\quad\ {\rm for}\quad  1\le s^{(a)}\le j-1,\\
\vert q^{(a)}\vert\le\frac{\sqrt2}{\pi} \g^{-j},\ \quad\quad\quad\quad\quad\quad\quad\quad {\rm for}\quad  s^{(a)}=j.
\end{cases}\label{supp2}
\eea
%for $a=1,2,3$.
\vskip.5cm

Let $k_i=(k_{i,0},\bk_i)$, $i=1,\cdots,4$ be the four momenta entering or exiting the vertex $v$. By conservation of momentum and the periodic boundary condition (c.f. Formula \eqref{bril}), we have:
%%%%%%%%%%%%%%%%
\bea\label{com1}
\sum_{i=1}^4k_{i,0}=0,\quad \sum_{i=1}^4k_{i,1}=\frac{2\pi}{3}n_1,\quad \sum_{i=1}^4k_{i,2}=\frac{2\pi}{\sqrt3}n_2,\ n_1, n_2\in\ZZZ,
\eea
%%%%%%%%%%%%%%%%%
in which the last two equations can be rewritten in the new coordination system as:
\be\label{com11}
\sum_{i=1}^4k_{i,+}=n_+,\quad \sum_{i=1}^4k_{i,-}=n_-,
\ee
%%%%%%%%%%%
%\bea
%&&k_{1,+}+k_{2,+}+k_{3,+}+k_{4,+}=2n_+,\label{com11}\\
%&&k_{1,-}+k_{2,-}+k_{3,-}+k_{4,-}=2n_-,\label{com22}
%\eea
%%%%%%%%%%%%
where $n_+=n_1+n_2$ and $n_-=-n_1+n_2$. So that $n_+$ and $n_-$ have the same parity. Adding up the two equations of \eqref{com11}, we obtain:  
\be\label{com33}
\sum_{i=1}^4k_{i}^{(3)}=n_0,
\ee
%%%%%
in which $n_0=n_++n-$. Since $n_+$ and $n_-$ have the same parity, $n_0$ is always an even integer.
%%%%%%%%%
In terms of the quasi-momentum $q_\pm=k_\pm\pm1$, \eqref{com1}-\eqref{com33} can be written as
\be\label{coq1}
\sum_{i=1}^4q_{i,+}=m_+,\quad \sum_{i=1}^4q_{i,-}=m_-,\quad \sum_{i=1}^4q_{i}^{(3)}=m_0.
\ee
Since even sums of $\pm1$ are still even numbers, the integers $m_+$ and $m_-$ still have the same parity, and $m_0=m_++m_-$ is an even integer. 
%%%%%%%%%%%%%%%%%%%%%%%%%%%%%%%%%%%%%%%%%%%%
Now we rename the indices $m_\pm, m_0$ by $m^{(a)}$, $a=1,\cdots,3$, with $m^{(1)}=m_+$, $m^{(2)}=m_-$ and $m^{(3)}=m_0$. We have the following lemma concerning the sector indices which depend on the value of $m^{(a)}$. Consider first the case of $m^{(a)}=0$, we have:

%%%%%%%%%%%%%%
\begin{lemma}\label{secmain}
Let $q_{i}^{(a)}$, $i=1,\cdots,4$, $a=1,\cdots,3$, be the quasi-momenta entering or exiting the vertex $v$ and let $j_i$, $s_{i}^{(a)}$ be the corresponding scale indices and sector indices. Let $s_{1}^{(a)}, s_{2}^{(a)}$ be the two indices with smallest values among all the sector indices such that $s_{1}^{(a)}\le s_{2}^{(a)}$. Let $\g\ge10$ be a fixed constant. If $m^{(a)}=0$, we have the following constraints concerning the possible values of $s_{1}^{(a)}$ and $s_{2}^{(a)}$: either $|s_{1}^{(a)}-s_{2}^{(a)}|\le1$, or $s_{1}^{(a)}=j_1$, in which $j_1$ is strictly smaller than $j_2, j_3$ or $j_4$. We have exactly the same results for the sector indices $s_{i}^{(b)}$, $i=1,\cdots,4$, $b=1,\cdots,3$, $b\neq a$.
\end{lemma}
%%%%%%%%%%%
\begin{proof}
First of all we consider the possible constraints for the quasi momentum $q_{i}^{(1)}=q_{i,+}$ and $q_{i}^{(2)}=q_{i,-}$, $i=1,\cdots,4$. We can always arrange the sector indices in the order $s_{1,+}\le s_{2,+}\le s_{3,+}\le s_{4,+}$, and the scale indices in the order $j_{1}\le j_{2}\le j_{3}\le j_{4}$. Then either $s_{1,+}<j_1$ or $s_{1,+}=j_1$. For both cases we have (cf. Formula \eqref{supp2}) $|q_{i,+}|\le\frac{\sqrt2}{\pi}\g^{-s_{2,+}}$, for $i=2,3,4$, and $|q_{1,+}|\ge{\g^{-s_{1,+}-1}}/{\pi }$.
In order that the equation $q_{1,+}+q_{2,+}+q_{3,+}+q_{4,+}=0$ holds, we must have
\be
{\g^{-s_{1,+}-1}}/{\pi }\le \frac{3\sqrt2}{\pi}\g^{-s_{2,_+}},
\ee
which implies
\be
s_{2,+}\le s_{1,+}+1+\log_\g (3\sqrt2).
\ee
For any $\g\ge 10$, we have $0<\log_\g (3\sqrt2)<1$, . Since $s_{1,+}$ and $s_{2,+}$ are integers, we have $|s_{2,+}-s_{1,+}|\le1$. Following the same arguments we can prove the same result for sectors in the $"-"$ direction. Using the fact that $q_i^{(3)}=q_{i,+} +q_{i,-}$ and the fact that this constraints are valid for both $q_{i,+}$ and $q_{i,-}$, they are also valid for $q_i^{(3)}$. Hence we conclude this proposition.
\end{proof}
%%%%%%%%%%%%%%%
Now we consider the case of $m^{(a)}\ge1$, $a=1,\cdots,3$.
\begin{lemma}
%%%%%%%%%
Let $q_{i}^{(a)}$, $i=1,\cdots,4$, $a=1,2$ or $3$, be the quasi-momenta entering or exiting the vertex $v$ and $j_i$, $s_{i}^{(a)}$ be the corresponding scale indices and sector indices.
Let $\g\ge10$ be a fixed constant. If $\vert m^{(a)}\vert\ge2$, then $s^{(a)}_{i}=0$ for two or more labeling indices $i$. If $m^{(a)}=\pm1$, then we only have the following two cases: (i) $s^{(a)}_{i}=0$ for two or more labeling indices $i$; (ii) There exists exactly one quasi momentum which equals to $\pm1$. In case (ii), we can assume, without losing generality that $q^{(a)}_1=1$ and assume that $s_{2}^{(a)}\le s_{3}^{(a)}\le s_{4}^{(a)}$. Then either $|s_{2}^{(a)}-s_{3}^{(a)}|\le1$, or $s_{2}^{(a)}=j_2$, in which $j_2$ is strictly smaller than $j_3$ and $j_4$.
\end{lemma}

\begin{proof}
We prove first the case of $m^{(1)}=m_+$. Since $\vert q_{i,+}\vert\le1$, with $i=1,\cdots,4$, we have $\vert m_+\vert\le4$. 
\begin{itemize}
\item
For the case of $\vert m_+\vert=4$, we have $\vert q_{i,+}\vert=1$ for all $i$, which implies that $s_{i,+}=0$, for all $i$. 

\item For the case of $\vert m_+\vert=2$ or $3$, suppose that $s_{i,+}\neq0$ for the three labellings $i=1,2,3$ and $s_{4,+}=0$ . By Formula \eqref{supp2} we have $\vert q_{i,+}|\le\frac{ \sqrt2}{\pi}\g^{-1}$ for $i=1,2,3$, so that
\be\label{qplus}
q_{1,+}+q_{2,+}+q_{3,+}+q_{4,+}\le \frac{3\sqrt2}{\pi}\g^{-1}+1<2,\quad {\rm for}\quad \g\ge 10,
\ee
which contradicts \eqref{coq1}.
\item Now suppose that $m_+=\pm1$. Consider first the case of $q_{1,+}+q_{2,+}+q_{3,+}+q_{4,+}=1$. Suppose that $s_{1,+}=0$ but
$q_{1,+}\neq1$, then since $\vert q_{i,+}\vert<0.1$ for $s_{i,+}\ge1$ (cf. \eqref{supp1}), there exists 
at least one more sector such that $s_{i,+}=0$, for $i=2,3$ or $4$. So we proved case $(i)$. If $q_{1,+}=1$, then \eqref{coq1} becomes $q_{2,+}+q_{3,+}+q_{4,+}=0$. The constraints for the sector indices are very similar to that in Lemma \ref{secmain}, except that we have one momentum less. The proof for this part is the same as in Lemma \ref{secmain} so we don't repeat it here.
\end{itemize} 
With exactly the same method we can prove the case for $m_-$. Using the fact that $q^{(3)}_i=q_{1,+}+q_{1,-}$, we can easily prove that the same constraints are also valid for $q^{(3)}_i$, $i=1,\cdots,4$. This concludes the Lemma.
\end{proof}

%%%%%%%%%%%%%%%%%%%%%%%%%%%%%%%%%%%%%%%%%%%%%%%%%%%
\subsection{Decaying properties of the sectorized propagators in the direct space}
In this section we study the decay of the free propagators in the direct space.
First of all, define the dual coordinates to the momentum $k^{a}$, $a=1,2,3$, as: 
\be
x^{(1)}:=x_+=\pi(x_1/3+x_2/\sqrt3), x^{(2)}:=x_-=\pi(-x_1/3+x_2/\sqrt3), x^{(3)}=x_++x_-.
\ee
\begin{notation}
In the following we shall use $(k^{(a)},k^{(b)})$ and $(q^{(a)},q^{(b)})$, $a,b=1,2,3$, $a\neq b$, to indicate the momentum and quasi momentum with sector indices $(s^{(a)},s^{(b)})$, whose dual variables are $(x^{(a)},x^{(b)})$.
\end{notation}

For a sector with scale index $j$ and sector indices $\sigma=(s^{(a)},s^{(b)})$, it is useful to introduce a new index, $l(\s)$, which describes the distance of this sector to the Fermi surface, defined as $l(j,\s)=s^{(a)}+s^{(b)}-j+1$. It is also called the $depth$ of a sector
and denoted by $l$ when the scale indices and the sector indices are clear from the context. 
%%%%%%%%%%%%%%%%%%%%%%%%%%%%%%%%%5with $x=(x_0,x_+,x^{(b)})$,
\begin{lemma}\label{bdx1}
Let $[C_{j,\sigma}(x-y)]_{\a\a'}$ be the Fourier transform of $[\hat C_{j,\sigma}(k_0,k^{(a)},k^{(b)})]_{\a\a'}$, (c.f. Eq. \eqref{sectz}). There exist model dependent constants $K$ and $c$ such that for all $j\ge j_0$ and $\sigma=(s^{(a)},s^{(b)})$, $0\le s^{(a)},s^{(b)}\le j$, the following bound holds:
%\be\label{decay1}
%|C_{j,\sigma}(x-y)]_{\a\a'}|=\sup_{x,y\in\L}\vert\ [C_{j,\sigma}(x-y)]_{\a\a'}\vert\le O(1) \g^{-j-l}\ e^{-c[d_{j,\s}(x,y)]^\a},
%\ee
%%%%%%%%%%
\be\label{decay1}
\vert[C_{j,\sigma}(x-y)]_{\a\a'}\Vert_{L^\infty}\le K \g^{-j-l}\ e^{-c[d_{j,\s}(x,y)]^\a},
\ee
%%%%%%%%%%
where 
\be\label{dist0}
d_{j,\s}(x,y)=\g^{-j}\vert x_0-y_0\vert+\g^{-s^{(a)}}\vert x^{(a)}-y^{(a)}\vert+\g^{-s^{(b)}}\vert x^{(b)}-y^{(b)}\vert,
\ee
$\a=1/h\in(0,1)$ is the index characterizing the Gevrey class of functions (cf. Definition \ref{gev}). \footnote{The interested readers who are familiar with the sectors for strictly convex Fermi surfaces (cf. \cite{DR1}, \cite{DR2}, \cite{BGM2}, \cite{FKT} are invited to compare the different decaying properties.}.
\end{lemma}
%%%%%%%%%%%%%%%%%%%%%%%%%%%%%%%%%%%%%%%%%%%%%%%55
\begin{proof}
By Lemma \ref{mat0} we know that each matrix element of the sectorized propagator is dominated by $\tilde C(k)$ times a uniform constant, which can be ignored for the moment. Let $\tilde C_{j,\sigma}(x-y)$ be the Fourier transform of $\tilde C_{j,\sigma}(k)$, then it is enough to prove that:
\be\label{decay2}
\Vert\tilde C_{j,\sigma}(x-y)\Vert_{L^\infty}\le K \g^{-j-l}\ e^{-c[d_{j,\s}(x,y)]^\a}.
\ee
This is essentially Fourier analysis and integration by parts. Using the fact that $\tilde C_{j,\sigma}(k)=\tilde C_{j,\sigma}(k_0,q^{(a)}\pm1,q^{(b)}\pm1)$, the integral $\int dk_0dq^{(a)}dq^{(b)}$ constrained to a sector is bounded by $\g^{-j}\cdot \g^{-s^{(a)}}\cdot\g^{-s^{(b)}}$, while the integrand is bounded by $\frac{1}{\gamma^{-j}}$. So we obtain the pre-factor of \eqref{decay2}. Let $\frac{\partial}{\partial k_0}f=(1/2\pi T)[f(k_0+2\pi T)-f(k_0)]$ be the difference operator.  To prove the decaying behavior of \eqref{decay2}, it is enough to prove that
\bea\label{decay4}
\Vert\frac{\partial^{n_0}}{\partial k_0^{n_0}}\frac{\partial^{n^{(a)}}}{\partial (q^{(a)})^{n^{(a)}}}\frac{\partial^{n^{(b)}}}{\partial (q^{(b)})^{n^{(b)}}} \tilde C_{j,\s}(k_0, q^{(a)},q^{(b)})\Vert_{L^\infty}
\le K^n\g^{jn_0}\g^{s^{(a)}n^{(a)}}\g^{s^{(b)}n^{(b)}}(n!)^{1/\alpha},\nn
\eea
where $n=n_0+n^{(a)}+n^{(b)}$. By \eqref{supp2}, we can easily prove that there exists some constant $K_1$ such that $\Vert\frac{\partial}{\partial q^{(b)}}v_{s^{(b)}}[\cos^2(q^{(b)}\pm1)\pi/2]\Vert\le K_1\g^{s^{(b)}}$; when the operator $\frac{\partial}{\partial q^{(b)}}$ acts on $\chi_j[k_0^2+e^2(q^{(a)},q^{(b)},1)]$, the resulting term is simply bounded by $\g^{2j-2s^{(a)}-s^{(b)}}$; when $\frac{\partial}{\partial q^{(b)}}$ acts on $[-2ik_0-e(q^{(a)},q^{(b)},1)+k_0^2]^{-1}$, the resulting term is bounded by $\g^{j-s^{(a)}}$. Using the constraint $s^{(a)}+s^{(b)}\ge j-1$, we find that each of the three factors is bounded by $K_2\g^{s^{(b)}}$, for some positive constant $K_2$. When $\frac{\partial}{\partial q^{(b)}}$ acts on a factor $\cos(q^{(b)}\pm1)\pi/2$, which is generated in the previous derivations, it costs a factor $K_3\g^{s^{(b)}}$. Similarly, each $\frac{\partial}{\partial q^{(a)}}$ acting on $\tilde C_{j,\s}(k_0,q^{(a)},q^{(b)})$ is bounded by $K\g^{s^{(a)}}$. Finally, each derivation $\frac{\partial}{\partial k_0}$ on the propagator results in a factor $\g^j$.
The factor $(n!)^{1/\alpha}$ comes from derivations on the compact support functions, which are Gevrey functions of order $\alpha$. When $j=j_{max}$, the propagator decays only in the $x_0$ direction but not in the $x^{(a)}$ or $x^{(b)}$ direction. Let $K$ be the product of the positive constant $K_1$, the result of this Lemma follows.
\end{proof}
Then we have the following lemma:
%%%%%%%%%%%%%%%%%%%%%%%%%%%%%%%%%%%%%%%%%%%%%%%%%%%%%%%%%%%555
\begin{lemma}
The $L^1$ norm of $[C_{j,\sigma}(x)]_{\a\a'}$, $x\in\Lambda_{\beta}$, $\a,\a'=1,2$ is bounded as follows: 
\be\label{tad2}
\Big\Vert\ [C_{j,\sigma}(x)]_{\a\a'}\ \Big\Vert_{L^1}\le O(1)\g^{j}.
\ee
\end{lemma}
\begin{proof}
This lemma can be proved straightforwardly using Lemma \ref{bdx1}. We have
\bea
&&\Big\Vert\ [C_{j,\sigma}(x)]_{\a\a'}\ \Big\Vert_{L^1}\quad\le O(1)\ \int_{\Lambda_{\beta, L}} dx_0 dx^{(a)} dx^{(b)} \ \Big|\tilde C_{j,\sigma}(x)\Big|\nn\\
&&\quad\le O(1)\g^{-j-l}\g^{(j+s^{(a)}+s^{(b)})}\le O(1)\g^{j}.
\eea
\end{proof}
Remark that, comparing to the $L^\infty$ norm for a sliced propagator $[C_{j,\sigma}(x)]_{\a\a'}$, 
a factor $\g^{2j+l}$ is lost when taking the $L^1$ norm. In order to capture this scaling property it is convenient to define a new scale index:
\begin{definition}\label{indexr}
Define a index $r=\EEE(j+l/2)$, where $\EEE(\cdot)$ takes the integer value of its variable. We have $r\ge0$ and $r_{max}(T):=\EEE(1+\frac{3}{2}\ j_{max}(T))$. We call this new scale index an $r$-index, or simply a scale index $r$. Correspondingly, we have the following decomposition for the propagator:
\be
\hat C(k_0,q^{(a)},q^{(b)})=\sum_{r=0}^{r_{max}(T)}\sum_\s\hat C_{r, \s}(k_0,q^{(a)},q^{(b)}).
\ee
In terms of the $r$-index, the sector $\Delta^j_{s^{(a)},s^{(b)}}$ is also denoted by $\Delta^r_{s^{(a)},s^{(b)}}$.
\end{definition}
Since $|x-\EEE(x)|\le1$, $\forall x\in\RRR$, we shall simply drop the integer part $\EEE(\cdot)$ in the future sections. The four indices $j$, $s^{(a)}$, $s^{(b)}$ and $r$ are related by the relation $r=j+\frac l2=\frac{j+s^{(a)}+s^{(b)}+1}{2}.$ Then the constraints for the sector indices
$s^{(a)}+s^{(b)}\ge j-1$, $0\le s_{\pm}\le j$, $0\le j\le j_{max}$
becomes $s^{(a)}+s^{(b)}\ge r-1$, $0\le s^{(a)},s^{(b)}\le r$ and $0\le r\le r_{max}=3j_{max}/2$.
The depth index can be expressed as $l=2(s^{(a)}+s^{(b)}-r+1)$.

%%%%%%%%%%%%%%%%%%%%%%%%%%%%%%%%%%%%%%%%%%%%%%%%%%%%%%%%%%%%%%%%%%%%%%%%%%%%%%%%%%%%
\section{The perturbation expansion}
\subsection{The BKAR jungle formula and the power-counting theorem}
In this section we study the perturbation expansion for the Schwinger functions. It is most conveniently to label the perturbation terms by graphs \cite{RW1}. Before proceeding, let us recall some notations in graph theory.
\begin{definition}(cf., eg. \cite{tutte})\label{defgraph}
Let $n\ge 1$, be an integer, $I_n=\{1,\cdots,n\}$, $\cP_n=\{\ell=(i,j), i, j\in I_n, i\neq j\}$ be the set of unordered pairs in $I_n$. A graph $G=\{V_G, E_G\}$ of order $n$ is defined as a set of 
vertices $V_G=I_n$ and of edges $E_G\subset\cP_n$, whose cardinalities are noted by $|V_G|$ and $|E_G|$, respectively. A graph $G'=\{V_{G'}, E_{G'}\}\subset G$ is called a subgraph of $G$ if $V_{G'}\subset V_G$ and $E_{G'}\subset E_G$. It is called a connected component if $G'$ is connected, i.e. there exists a non-empty set of edges connecting any pair of vertices of $G'$. A half-edge, which corresponds to an external field and is noted by $(i,\cdot)$, is an object such that each pair of them form an edge: $[(i,\cdot), (j, \cdot)]=(i,j)$, for $i,j\in I_n$. A graph $G$ with half-edges is also called a decorated graph or an extended graph.  
\end{definition}
\begin{definition}
A forest $\cF$ is a graph which contains no loops, i.e. no subset\\ $L=\{(i_1,i_2), (i_2, i_3), \cdots, (i_k, i_1)\}\subset E_{\cF}$ with $k\ge3$. An edge in a forest is also called a tree line and an edge in $L$ is called a loop line. A maximally connected component of $\cF$ is called a tree, noted by $\cT$. A tree with a single vertex is allowed. $\cT$ is called a spanning tree if it is the only connected component of $\cF$.
\end{definition}

Recall that a general $2p$-point Schwinger function at temperature $\b=1/T$ is defined as:
\bea
&&S_{2p,\b}(\lambda; x_1,\t_1,\a_1;\cdots,x_{2p},\t_{2p},\a_{2p},\l)\\
&&=\frac{1}{Z}\int d\mu_C(\bar\psi,\psi)\Big[ \prod_{i=1}^p\prod_{\e=\pm,\a_i\t_i}\psi^{\e_i}_{\tau_i, \a_i}(x_i)\ \Big]\ \Big[
\prod_{i=p+1}^{2p}\prod_{\e_i=\pm,\a_i,\t_i}\psi^{\e_i}_{\tau_i,\a_i}(x_i)\Big]e^{- {\VV}(\bar\psi,\psi)},\nn
\eea
where
\be
Z=\int d\mu_C(\bar\psi,\psi)e^{-\VV(\psi,\bar\psi)}
\ee
is the partition function. 
\begin{remark}\label{rmkindex}
Remark that, by Proposition \ref{mat0}, each matrix element of the propagator is bounded by an absolute positive constant. So we can simply replace each matrix element $\hat C_{\a\a'}$ by $\tilde C$ times a constant. In order to estimate the upper bound for $S_{2p,\b}$, $p\ge2$, it is enough to estimate the upper bound of any matrix element $[S_{2p,\b}]_{\a\a'}$. In order to simplify the notation, we shall drop the matrix indices of the Schwinger functions in the rest of this section and in Section $5$, and write $[S_{2p,\b}]_{\a\a'}$ as $S_{2p,\b}$. We will restore the matrix indices when we consider the renormalization for the two-point Schwinger function and the self-energy function. 
\end{remark}
Let $\{\xi^x_i=(x_{i},\t_{i},\a_{i})\}$, $\{\xi^y_i=(y_{i},\t_{i},\a_{i})\}$ and $\{\xi^z_i=(z_{i},\t_{i},\a_{i})\}$ be set of indices associated with the Grassmann variables $\{\psi^\e_{\tau_i,\a_i}(x_i)\}$, $\{\psi^\e_{\tau_i,\a_i}(y_i)\}$ and $\{\psi^\e_{\tau_i,\a_i}(z_i)\}$, respectively. Expanding the exponential into power series and performing the Grassmann integrals, we obtain:
\bea\label{schw1}
&&S_{2p}(\lambda; x_1,\t_1,\a_1;\cdots,x_{2p},\t_{2p},\a_{2p})\\
&=&\sum_{N=1}^{\infty}\sum_{n+n_1+n_2=N}\frac{\l^n}{n!}\frac{(\delta\mu(\lambda))^{n_1}}{n_1!}\int_{{(\L_{\beta,L})}^{n+n_1}}d^3y_1\cdots d^3y_{n+n_1}\nn\\
&&\int_{{(\L_{\beta,L})}^{2n_2}} d^3z_{1}\cdots d^3z_{2n_2}\prod_{i,j=1}^{2n_2}\nu(\zz_i-\zz_j)\delta(z_{i,0}-\delta z_{j,0})\nn\\
&&\sum_{\underline{\t},\underline{\a}}\Bigg\{\begin{matrix}&\xi^x_1,&\cdots, \xi^x_p,&\xi^z_1,&\cdots, \xi^z_{n_2},&\xi^y_1,\cdots,\xi^y_n\\
&\xi^x_{p+1},&\cdots, \xi^x_{2p},&\xi^z_{n_2+1},&\cdots,\xi^z_{2n_2},&\xi^y_1,\cdots, \xi^y_n
\end{matrix}\Bigg\},\nn
\eea
where $N=n+n_1+n_2$ is the total number of vertices, $n$ is the number of interaction vertices, to each of which is associated a (bare) coupling constant $\lambda$, and $n_1$ is the number of two-point vertices, each of which is associated with a bare chemical potential counter-term $\delta\mu(\l)$. $\nu(\zz_i-\zz_j)=\frac{1}{\vert\L_L\vert}\int d\bk e^{i\kk\cdot(\zz_i-\zz_j)}\hat\nu(\bk)$ is counter-term in the direct space. The two-point vertices are also called {\it the counter-term vertices}. We have used Cayley's notation (c.f. \cite{rivbook}) for determinants:
\bea
\Bigg\{\begin{matrix}\xi^x_i\\
\xi^y_j \end{matrix}\Bigg\}=
\Bigg\{\begin{matrix}
x_{i,\t_i,\a_i}\\ y_{j,\t_j,\a_j}
\end{matrix}\Bigg\}=\det\Big[\ \delta_{\t_i\t_j}[C_{j,\t}(x_i-y_{j})]\ \Big]_{\a_i,\a_j}.
\eea
%%%%%%%%%%%%%%%%%%
Remark that the perturbation series would be divergent if we fully expand the determinant \cite{RW1}. Instead, we can only partially expand the determinant such that the expanded terms are labeled by forest graphs, which don't proliferate very fast \cite{RW1}. In order to make the partial expansions consistent with the multi-slice analysis, to each tree line in the forest, which corresponds to a sectorized propagator, we associate a scale index $r=j+l/2$, and we arrange the set of tree lines according to the increasing order of $r$. The set of forests with labeling forms a layered object, called {\it a jungle} graph. The canonical way of generating the jungles in perturbation theory is the BKAR jungle formula (see \cite{AR}, Theorem $IV.3$):
\begin{theorem}[The BKAR jungle Formula.]\label{ar1}
Let $n\ge1$ be an integer, $I_n=\{1,\cdots, n\}$ be an index set and $\cP_n=\{\ell=(i,j), i, j\in I_n, i\neq j\}$. Let $\cF$ be a forest of order $n$ and $\cS$ be the set of smooth functions from $\RRR^{\cP_n}$ to an arbitrary Banach space. Let ${\bf x}=(x_\ell)_{\ell\in\cP_n}$ be an arbitrary element of $\RRR^{\cP_n}$ and ${\bf 1}\in \RRR^{\cP_n}$ be the vector with every entry equals $1$. Then for any $f\in \cS$, we have:
\be\label{BKAR}
f({\bf 1})=\sum_{\cJ=(\cF_1\subset\cF_1\cdots\subset\cF_{m})}\Big(\int_0^1\prod_{\ell\in\cF_m} dw_\ell\Big)\Bigg(\prod_{k=1}^m\Big(\prod_{\ell\in\cF_k\setminus\cF_{k-1}}\frac{\partial}{\partial {x_\ell}}\ \Big)\Bigg)\ f[X^\cF(w_\ell)],
\ee
%%%%%%%%%%%%%%%%%%%%%%%%%%%%%%%%%%%%%%
where the summation runs over all jungles $\{\cJ=({\cF}_0\subset{\cF}_1\cdots\subset{\cF}_{m}={\cF})\}$, in which the last forest $\cF=\cF_m$ is a spanning forest of the fully expanded graph $G$ with $n$ vertices. $\cF_{0}:={\bf V_n}$ is the completely disconnected forest of $n$ connected components, each of which corresponds to the interaction vertex $\VV(\psi,\lambda)$ ( cf. Formula \eqref{potx} ). $X^\cF(w_\ell)$ is a vector ${(x_\ell)}_{\ell\in \cP_n}$, whose elements $x_\ell= x_{ij}^\cF(w_\ell)$ are defined as follows:
%%%%%%%%%%%%%%%%%%%%%%%%%%%%%%%%%%%%%%%
%where the sum over $\cF$ runs over all forests with $n$ vertices, including the empty one which has no edges; $\cJ=({\cF}_0\subset{\cF}_1\cdots\subset{\cF}_{r_{max}}={\cF})$ is a layered object of forests $\{\cF_0,\cdots,\cF_{r_{max}}\}$, also called a jungle, in which the last forest $\cF=\cF_{r_{max}}$ is a spanning forest of the fully expanded graph $G$ containing $n$ vertices and $2p$ external edges. $\cF_{0}:={\bf V_n}$ is the completely disconnected forest of $n$ connected components, each of which corresponds to the interaction vertex $\VV(\psi,\lambda)$ ( cf. Formula \eqref{potx} ). $X^\cF(w_\ell)$ is a vector ${(x_\ell)}_{\ell\in \cP_n}$ with elements $x_\ell= x_{ij}^\cF(w_\ell)$, which are defined as follows:
%%%%%%%%%%%%%%%%%%%%%%%%%%%%%%
\begin{itemize}
\item $x_{ij}^\cF=1$, if $i=j$, 
\item $x_{ij}^\cF=0$, if $i$ and $j$ are not connected by $\cF_k$,
\item $x_{ij}^\cF=\inf_{\ell\in P^{\cF}_{ij}}w_\ell$, if $i$ and $j$ are connected by the forest $\cF_k$ but not $\cF_{k-1}$, where $P^{\cF_k}_{ij}$ is the unique path in the forest that connects $i$ and $j$,
\item $x_{ij}^\cF=1$, if $i$ and $j$ are connected by $\cF_{k-1}$.
\end{itemize}
\end{theorem}
We obtain: 
\bea\label{rexp1}
S_{2p,\b}(\lambda)&=&\sum_{N=n+n_1+n_2} S_{2p,N,\b},\\
S_{2p,N,\b}&=&\frac{1}{n!n_1!n_2!}\sum_{\{\underline{\tau}\}}\sum_{\cJ_{r_{max}}}\prod_v\int_{\Lambda_{L,\beta}} d^3x_v\lambda^n (\delta\mu(\lambda))^{n_1} (\nu)^{n_2}\nn\\
&&\quad\quad\cdot\prod_{\ell\in\cF}\int dw_\ell
C_{r,\sigma_\ell}(x_\ell,x'_\ell)\det[C_{r,\sigma}(w)]_{left}\ .\label{rexp11}
\eea
in which the sum runs over all jungles $\{\cJ_{r_{max}}=({\cF}_0\subset{\cF}_1\cdots\subset{\cF}_{r_{max}})\}$, $\e(\cJ_{r_{max}})$ is a product of the factors $\pm1$ along the jungle. $\det[C_{r,\sigma}(w)]_{left}$ is the determinant for the remaining $2(n+1)\times 2(n+1)$ dimensional square matrix, which has the same form as \eqref{rexp1}, but is multiplied by the interpolation parameters $\{w_\ell\}$. So it is still a Gram matrix, whose elements are pairs of Grassmann fields and anti-fields that don't form tree propagators. Let $r_f$ be the $r$-index of a field or anti-field $f$, then the $(f,g)$ entry of the determinant reads:
\bea\label{intc1}
C_{r,\s}(w)_{f,g}=\delta_{\t(f)\t'(g)}\sum_{v=1}^n\sum_{v'=1}^n\chi(f,v)\chi(g,v') x^{\cF,r_f}_{v,v'}(\{w\})C_{r,\t(f),\s(f)}(x_v,x_{v'}),
\eea
where $[x^{\cF,r_f}_{v,v'}(\{w\})]$ is an $n\times n$ dimensional positive matrix, whose elements are defined in the same way as in \eqref{BKAR}:
\begin{itemize}
\item If the vertices $v$ and $v'$ are not connected by $\cF_r$, then $x^{\cF,r_f}_{v,v'}(\{w\})=0$,
\item If the vertices $v$ and $v'$ are connected by $\cF_{r-1}$, then $x^{\cF,r_f}_{v,v'}(\{w\})=1$,
\item If the vertices $v$ and $v'$ are connected by $\cF_r$ but not $\cF_{r-1}$, then  $x^{\cF,r_f}_{v,v'}(\{w\})$ is equal to the infimum of the $w_\ell$ parameters for $\ell\in\cF_r/\cF_{r-1}$ which is in the unique path connecting the two vertices. The natural convention is that $\cF_{-1}=\emptyset$ and that $x^{\cF,r_f}_{v,v}(\{w\})=1$.
\end{itemize}

\begin{figure}[htp]
\centering
\includegraphics[width=0.53\textwidth]{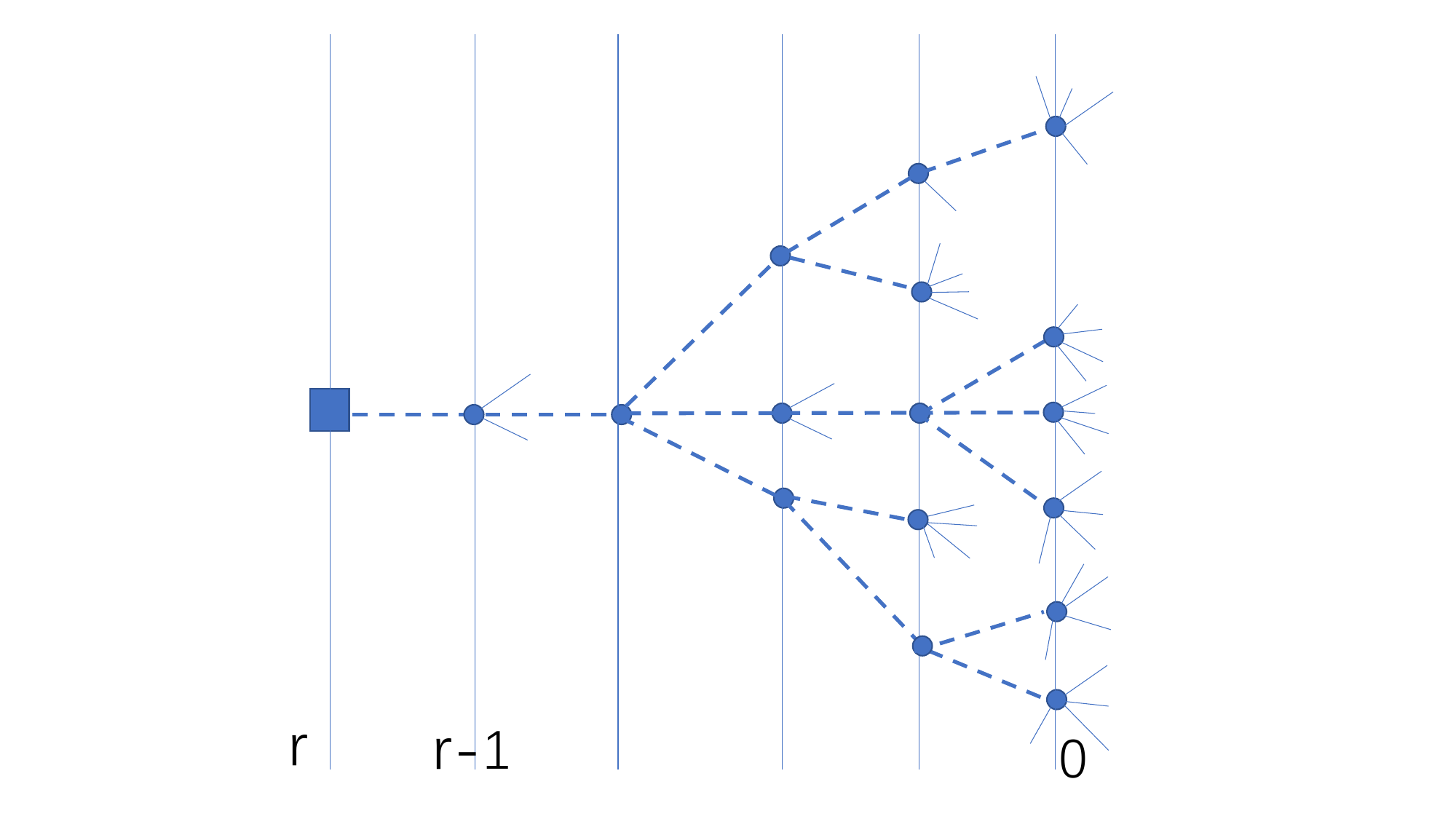}
\caption{\label{gn1}A Gallavotti-Nicol\`o tree. The round dots
are the nodes and bare vertices and the big square is the root. They are arranged according to the 
scale index $r$, such that nodes of the same $r$-index belong to the same thick line.
The thin lines are the external fields, the dash lines indicate the inclusion relations.
}
\end{figure}

\begin{figure}[htp]
\centering
\includegraphics[width=0.45\textwidth]{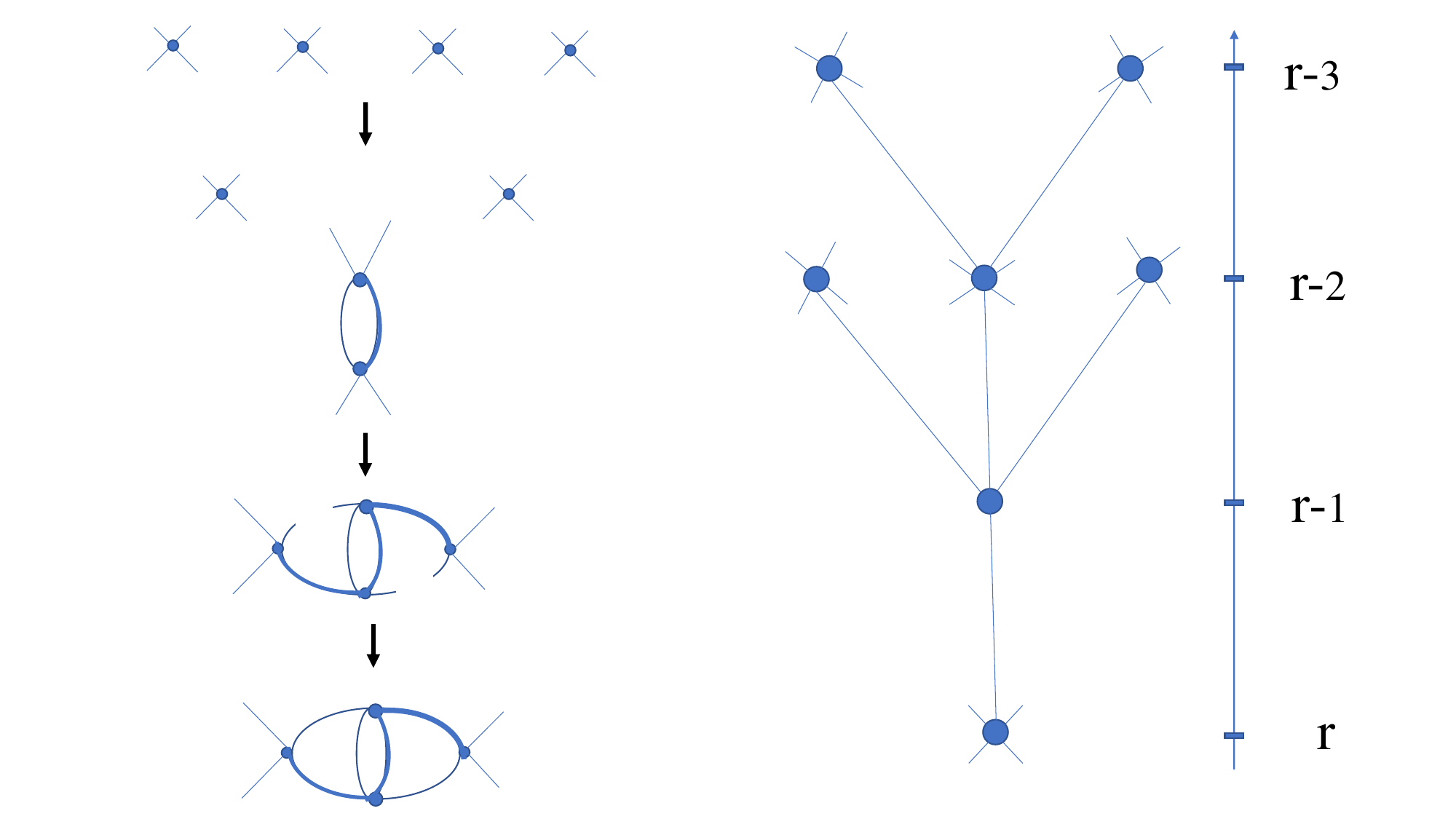}
\caption{\label{gn2}Grouping of a quadruped Feynman graphs into the GN tree. %The figure on the l.h.s. shows a quadruped Feynman graph with its subgraphs. The thick lines are %the propagators in the spanning tree. The Figure on the r.h.s. shows the corresponding GN tree.
}
\end{figure}

Taking the logarithm on $S_{2p,\b}$, we obtain the {\it connected} $2p$-point Schwinger function $S^c_{2p,\b}$:
\bea\label{rexp2}
S^c_{2p,\b}&=&\sum_{N=n+n_1+n_2}S^c_{2p,N,\b},\\
S^c_{2p,N,\b}&=&\frac{1}{n!n_1!n_2!}\sum_{\{\underline{\tau}\}}\sum_{\cJ_{r_{max}}'}\prod_v\int_{\L_{\b,L}} d^3x_v\lambda^n  (\delta\mu)^{n_1}[\nu]^{n_2}\nn\\
&&\quad\cdot\prod_{\ell\in\cT}\int dw_\ell C_{r,\sigma_\ell}(x_\ell,x'_\ell){\det}[ C_{r,\sigma}(w)]_{left}\ ,\label{rexp3}
\eea
in which $S^c_{2p,\b}$ has almost the same structure as $S_{2p,\b}$ in \eqref{rexp11}, except that the summation over jungles is restricted to the ones $\cJ_{r_{max}}'=({\cF'}_0\subset{\cF'}_1\cdots\subset{\cF'}_{r_{max}}={\cT})$, in which the final layered forest is a spanning tree $\cT$ with $n$ vertices. Without losing generality, suppose that a forest $\cF'_r$ contains $c(r)\ge1$ trees, noted by $\cT_r^k$, $k=1,\cdots, c(r)$, and a link in $\cT_r^k$ is noted by $\ell(T)$. To each $\cT_r^k$ we introduce an extended graph (cf. Definition \ref{defgraph}) $G^k_r$, which contains $\cT_r^k$ as a spanning tree and contains a set of half-edges, $e(G^k_r)$, such that the cardinality of $e(G^k_r)$, denoted by $|e(G^k_r)|$, is an even number. By construction, the scale index of any external field, $r_f$, is greater than $r_{\ell(T)}$. The graphical structure of a component $G^k_r$ is highly nontrivial: besides a tree structure $\cT_r^k$, it contains also a set of internal fields (which would form loop lines if they were fully expanded) which still form a determinant. Each connected component $G^k_r$ is contained in a unique connected component with a lower scale index $r$. The inclusion relation of the graphs $G^k_r$, $r=0,\cdots,r_{max}$, has a tree structure, called the Gallavotti-Nicol\`o tree.
\begin{definition} [cf. \cite{GN}]
A Gallavotti-Nicol\`o tree (GN tree for short) $\cG^{r_{max}}$ is an abstract tree graph in which the vertices, also called the nodes, correspond to the extended graphs $G^k_r$, $r=0,\cdots, r_{max}$, $k=1,2,\cdots, c(r)$, and the edges are the inclusion relations of these nodes. The node $G_{r_{max}}$, which corresponds to the full Feynman graph $G$, is called the root of $\cG^{r_{max}}$. Obviously each GN tree has a unique root. The bare nodes of the GN tree, which form the set ${\bf V_N}=\cF_0$, are called the leaves. The cardinality of the set of bare nodes, $|{\bf V_N}|$, is called the order of $\cG$. A GN tree of order $N$ is also noted as $\cG^{r_{max}}_N$.
\end{definition}
An illustration of a GN tree with $16$ nodes and $8$ bare nodes is shown in Figure \ref{gn1}. Figure \ref{gn2} is an illustration of the grouping the subgraphs of a Feynman graph into the corresponding GN tree. The readers that are not familiar with the GN trees are invited to consult \cite{BG1} or \cite{rivbook} for more details. Now we consider the amplitudes of the connected Schwinger functions. 
%%%%%%%%%%%%%%%%%%%%%%%
%%%%%%%%%%%%%%%%%%%%%%%%%%%%
%%%%%%%%%%%%%%%%
%%%%%%%%%%%%%%%%%%%%%%
\begin{remark}\label{ctbd}
Remark that since the counter-terms are quadratic in the two external fields, they only
contribute the factors $|\delta\mu|^{n_1}$ and $|\nu_{\a\a'}|^{n_2}$ (though the latter is non-local)  to the amplitudes of the correlation functions. As will be proved in Theorem \ref{flowmu} and \ref{mainc}, these counter-term are simply bounded by some absolute positive constants hence are not essential for the power-counting. So we drop these counter-terms in the rest of this section, just for simplicity. We will retain them in future sections when we study the renormalization of the two-point function.
\end{remark}
%%%%%%%%%%%%%%%%%%%%%
%We can express the $\delta$ functions in \eqref{form}, which encode the constraints from the conservation of momentum, as the Fourier transform of an oscillation factor. Then if we naively bound the delta functions with the Gram-Hadamard inequality, the Fourier oscillation factors will be simply bounded by one and constraints from the conservation of momentum will be lost. In order to solve this problem, we introduce an indicator function $\chi_{i}(\{\si\})= \chi (\sigma^1_i, \sigma^2_i, \sigma^3_i, \sigma^4_i)$ at each vertex in the graph, defined as follows: $\chi_{i}(\{\si\})$ equals to $1$ if the sector indices $\{\si\}$ satisfy the constraints in Lemma \ref{secmain}, and equals to $0$ otherwise. 

%%%%%%%%%%%%%%%%%%%%%%%
The $2p$-point Schwinger function can be written as:
\bea\label{s2pa}
S^c_{2p,\b} &=&\sum_{n}S^c_{2p,n,\b}\lambda^n,\\
S^c_{2p,n,\b}&=&{\frac{1}{n!}}{\frac{1}{n_1!}}{\frac{1}{n_2!}}\sum_{ \{\underline{\t}\},\cG^{r_{max}}, \cT} \ 
\sum_{\cJ_{r_{max}}',\{\sigma\}}
\epsilon (\cJ') \prod_{j=1}^{n}   \int d^3x_{j}  \delta(x_1)
\prod_{\ell\in \cT} \int_{0}^{1} dw_{\ell}
C_{r_{\ell},\si_{\ell}} 
(x_{\ell}, \bar x_{\ell}) \nonumber\\
&&\quad \prod_{i=1}^{n} 
\chi_{i}(\si)    
\det[C_{r,\sigma}(w)]_{left} \ ,\label{form} 
\eea
%%%%%%%%%%%%%%%%%%
in which the function $\chi_{i}(\si)= \chi (\sigma^1_i, \sigma^2_i, \sigma^3_i, \sigma^4_i)$ associated to each vertex $i=1,\cdots, n$ of the graph is the indicator function, defined as follows: $\chi_{i}(\si)$ equals to $1$ if the sector indices $\si$ satisfy the constraints in Lemma \ref{secmain}, and equals to $0$ otherwise. This function describes the conservation of momentum at each vertex. 
%%%%%%%%%
%Constraints on the sector indices placed by the GN tree structure also need to be considered. 
%%%%%%%
%%%%%%%%%%%%%%%%%%%%%%%%%%%%%%%%%%%%%%%%%%%%%%%5

Notice that each matrix element \eqref{intc1} of $\det[C_{r,\sigma}(w)]_{left}$ can be written as an inner product of two vectors:
\be
C_{r,\s}(w)_{f,g;\tau,\tau'}=( e_\tau\otimes A_f(x_v, ),  e_{\tau'}\otimes B_g(x_{v'},))\ ,
\ee
in which the unit vectors $e_{\uparrow}=(1,0)$, $e_{\downarrow}=(0,1)$ are the spin variables and
\bea
A_f&=&\sum_{k\in\DD_{\beta, \L}}\sum_{v=1}^n\chi(f,v)[x^{\cF,r_f}_{v,v'}(\{w\})]^{1/2}e^{-ik\cdot x_v}\cdot\\
&&\quad\quad\quad\cdot\Big[\ \chi_j[k_0^2+e^2(\kk,1)]\cdot
v_{s^{(a)}}[\cos^2(k^{(a)}/2)]\cdot v_{s^{(b)}}[\cos^2(k^{(b)}/2)]\ \Big]^{1/2},\nn
\eea
\bea
B_g&=&\sum_{k\in\DD_{\beta, \L}}\sum_{v'=1}^n\chi(g,v')[x^{\cF,r_f}_{v,v'}(\{w\})]^{1/2}e^{-ik\cdot x_{v'}}\hat C(k)_{\tau_\ell,\sigma_\ell}\\
&&\quad\quad\quad\cdot\Big[\ \chi_j[k_0^2+e^2(\kk,1)]\cdot
v_{s^{(a)}}[\cos^2(k^{(a)}/2)]\cdot v_{s^{(b)}}[\cos^2(k^{(b)}/2)]\ \Big]^{1/2}\nn,
\eea
are vectors in the Hilbert space such that
\bea
\Vert A_f\Vert^2_{L^2}\le \sqrt{K}\g^{-j_f/2-s_{f,+}/2-s_{f,-}/2},\quad \Vert B_f\Vert^2_{L^2}\le \sqrt{K}\g^{-s_{f,+}/2-s_{f,-}/2+j_f/2},
\eea
for some positive constant $K$ that is independent of $r$ and $l$. By Gram-Hadamard's inequality \cite{Le, GK} and using the fact that $l_f=s_{f}^{(a)}+s_{f}^{(b)}-j+1$, $a\neq b$,
we have
\be
|\det(A_f, B_g)|\le\prod_{f}\Vert A_f\Vert_{L^2}\cdot \Vert B_f\Vert_{L^2}\le K\prod_{f}\g^{-(j_f+l_f)/2}.
\ee
So we have:
\be
\det\nolimits_{{ left}}\le K^n\prod_{f\ left}\g^{-(j_f+l_f)/2}= K^n\prod_{f\ left}\g^{-r_f/2-l_f/4}.\ee
Integrating over the all position variables except the fixed one,
$x_1$, we obtain the following bound:
\be | S^c_{2p,n,\b} | \le {\frac{K^n}{n!}}
\sum_{\{\underline\tau\}, \cG^{r_{max}}, \cT}\ \sum'_{\{\si \}} 
\prod_{i=1}^{n} \chi_{i}(\si) 
\prod_{\ell \in \cT} \g^{2r_{\ell}}
\prod_{f} \g^{-r_f/2-l_f/4}, \label{absol1}
\ee
in which the last product runs over all the $4n$ fields
and anti-fields, and the summation $\sum'$ means that we have taken into account the constraints on the sector indices among the different connected components in the GN tree. We have the following lemma concerning the last two terms in \eqref{absol1}:
\begin{lemma}\label{indmain}
Let $c(r)$ be the number of connected components at level $r$ in the GN tree, we have the following inductive formulas:
\bea &&\prod_{f} \g^{-r_{f}/2}= \prod_{r=0}^{r_{max}}\ \prod_{k=1}^{c(r)} \g^{-|e(G_r^k)|/2}\ ,
\label{induc1}\\
&&\prod_{\ell \in \cT} \g^{2r_{\ell}}=\g^{-2r_{max}-2}
\prod_{r=0}^{r_{max}}\ \prod_{k=1}^{c(r)}  \g^{2}\ .
\label{induc2}
\eea
\end{lemma}
\begin{proof}
Both formulas can be proved by induction. We prove \eqref{induc1} first. Consider a graph $G$ with $n$ vertices. Let $N_f$ be the set of fields (half-edges) in $G$ and $n(r,f)$ be the number of fields such that $r_f=r$. Then the l.h.s. of \eqref{induc1} is equal to
$\g^{-\frac12\sum_{f\in N_f} r_{f}}=\g^{-\frac12[\ \sum_{r=0}^{r_{max}}\ n(r,f)\cdot r\ ] }$.
Let $\cup_{k=1}^{c(r)}e(G_r^k)$ be the set of external fields of any connected component in the GN tree of scale index $r$. Obviously we have $|e(G_r)|=\sum_{k=1}^{c(r)}\  |e(G_r^k)|$. Then
the r.h.s. of \eqref{induc1} is equal to $\g^{-\frac12[\ \sum_{r=0}^{r_{max}}\  |e(G_r)|\ ]}$.
Fix an external field $f_e$ of a connected component in $\cG^{r_{max}}$ such that $r_{f_e}=r\ge1$, which means that $f_e$ is an external field of $G_{\le r-1}$, but an internal field of $G_{r}$, hence $f_e\in e(G_0)\cap\cdots\cap e(G_{r-1})$. So we have
$ \sum_{r=0}^{r_{max}} |e(G_r)|= \sum_{r=0}^{r_{max}}\ r\cdot n(f_e,r)$, where $n(f_e,r)$ is the number of external fields at level $r$, and
\be
\prod_{r=0}^{r_{max}}\ \prod_{k=1}^{c(r)} \g^{-|e(G_r^k)|/2}=\g^{-\sum_{r=0}^{r_{max}}\ r\cdot n(f,r)/2}=\prod_{f} \g^{-r_{f}/2}\ .
\ee
Now we consider \eqref{induc2}. Suppose that the spanning tree $\cT$ has $k$ edges $\ell_1, \ell_2,\cdots, \ell_k$, each is assigned a scale index $r_{\ell_i}$, $i=1,\cdots,k$. We can always order the tree lines according to the order $r_{\ell_1}\le r_{\ell_2}\cdots\le r_{\ell_k}\le r_{max}$. Let $n(\ell,r)$ be the number of tree lines in $\cT$ whose scale indices are equal to $r$, then the l.h.s. of \eqref{induc2} is equal to
$\g^{2\sum_{\ell\in\cT}r_\ell}=\g^{2\sum_{r=0}^{r_{max}}\ r\cdot n(l,r)}$, and the r.h.s. of \eqref{induc2} is equal to 
\bea\label{induc2r}
&&\g^{-2r_{max}-2}\ 
\prod_{r=0}^{r_{max}}\ \g^{2c(r)}=
\g^{2 \sum_{r=0}^{r_{max}}\ \big[c(r)-1\big]}\ .
\eea
Since $c(r_{max})=1$, we have
\be
c(r)-1=n(\ell,r+1)+n(\ell,r+2)+\cdots n(\ell,r_{max}),
\ee
where $n(\ell,i)=c(i)-c(i+1)$, and
\be
\sum_{r=0}^{r_{max}}\ \big[c(r)-1\big]=1\cdot n(\ell,1)+2\cdot n(\ell,2)+\cdots +r_{max}\cdot  n(\ell,r_{max}),
\ee
which means that
\be
\g^{2 \sum_{r=0}^{r_{max}}\ \big[c(r)-1\big]}=\g^{2 \sum_{r=0}^{r_{max}}\ r\cdot n(\ell,r)}=\prod_{\ell\in\cT}\ \g^{2r_\ell}\ .
\ee
This concludes the lemma.
\end{proof}
\begin{theorem}[The power counting theorem]\label{tpc}
There exists a positive constant $K$ which may depend on the physical parameters of the model but is independent of the scale indices and the coupling constant $\l$, such that the connected, $2p$-point Schwinger functions, with $p\ge0$, satisfy the following bound: 
\be\label{pc1} | S^c_{2p,n,\b} | \le {\frac{K^n\l^n}{n!}}
\sum_{\{\underline\tau\}, \cG^{r_{max}}, \cT}\ \sum'_{\{\si \}}\prod_{i=1}^{n}\ \big[\chi_{i}(\si)\g^{-(l_i^1 + l_i^2 + l_i^3 + l_i^4)/4}\big]\ 
\prod_{r=0}^{r_{max}}\prod_{k}  \g^{2-e(G_r^k)/2}\ .
\ee
So that the two-point functions are relevant, the four point functions are marginal and the Schwinger functions with external legs $2p\ge6$ are irrelevant.
\end{theorem}
%%%%%%%%%%%%%%%%%%%%%%%%%%%%
\begin{proof}
Writing the product $\prod_f\g^{-l_f/4}$ in Formula \eqref{absol1} as $\prod_{i=1}^{n}e^{-(l_i^1 + l_i^2 + l_i^3 + l_i^4)/4}$, taking into account the conservation of momentum at each vertex $i$ and using Lemma \ref{indmain}, the result of this theorem follows.
\end{proof}
Then we need to consider the summation over the sector indices , which is easily getting unbounded if we don't take into account the constraints placed by the conservation of momentum. This is the so-called sector counting problem and will be studied in the next subsection.
\subsection{The sector counting lemma}
%%%%%%%%%%%%%%%%%%%%%%%%%%%%%%%%%%%%%%%%%%%%%%%%%%%%%%%%%%%%%
\begin{lemma}[Sector counting lemma for a single bare vertex]\label{sec1}
Let the four fields (half-lines) attached to a vertex be $f_1,\cdots, f_4$ with scale indices $j_1,\cdots, j_4$ and sector indices $\si_1=(s_{1}^{(a)}, s_{1}^{(b)}),\cdots,\si_4=(s_{4}^{(a)}, s_{4}^{(b)})$. Let the $r$-indices associated to the four fields be assigned as
$r_{f_1}=r_{f_2}=r_{f_3}=r$ and $r_{f_4}>r$. Then there exists a positive constant $K$, which is independent of the scale indices, such that for fixed $\si_4$, we have
\be\label{ss1}
\sum_{\si_1, \si_2, \si_3} \chi (\si_1, \si_2, \si_3, \si_4) 
\gamma^{-(l_1+l_2+l_3 )/4} \le K.r\ .\ee
\end{lemma}
\begin{proof}
This lemma has been proved in \cite{Riv} for a similar setting. Here we present a shorter proof, for reader's convenience. Let the four fields hooked to a vertex $i$ in a node $G_r^k$ be
$f_1,\cdots, f_4$, whose sector indices are $\s_1,\cdots,\s_4$ and depth indices are $l_i^1,\cdots, l_i^4$. Remark that, among the four fields, we can always choose one, say, $f_4$, as the {\it root} field. Then $r_4$ is greater than all the other indices $r_1=r_2=r_3=r$. We can always organize the sector indices $\s_1=(s_{1}^{(a)},s_{1}^{(b)}),\cdots, \s_3=(s_{3}^{(a)},s_{3}^{(b)})$ such that $s_{1}^{(a)}\le s_{2}^{(a)}\le s_{3}^{(a)}$ and $s_{1}^{(b)}\le s_{2}^{(b)}\le s_{3}^{(b)}$. Then, by Lemma \ref{secmain}, either $\s_1$ collapses with $\s_2$, or one has $s_{1}^{(a)}=s_{1}^{(b)}=j_1$, such that $j_1<\min\{j_2,\cdots,j_4\}$. 
So we only need to consider the following possibilities:
\begin{itemize}
\item if $\sigma_1\simeq\sigma_2$, we have $s_{2}^{(a)}= s_{1}^{(a)}\pm1$ and $s_{2}^{(b)}= s_{1}^{(b)}\pm1$. The depth indices are arranges as $l_1\le l_2\le l_3$. Then the l.h.s. of \eqref{ss1} is bounded by
\be
\sum_{\si_1, \si_3}
\g^{-(2l_1+l_3 )/4}= \sum_{\si_1}\g^{-l_1/2}\sum_{\si_3}
\g^{-l_3 /4}\ ,
\ee
Using the fact that $r_k=j_k+l_k/2$ and $l_k=s_{k}^{(a)}+s_{k}^{(b)}-j_k+1$, we have
$l_k=2(s_{k}^{(a)}+s_{k}^{(b)}-r_k+1)$, for $k=1,\cdots,3$. 
For fixed $s_1=(s_{1}^{(a)}, s_{1}^{(b)})$, summation over $\s_3=(s_{3}^{(a)}, s_{3}^{(b)})$ can be bounded as follows:
\bea
&&\sum_{\si_3=(s_{3}^{(a)},s_{3}^{(b)})}\g^{-l_3 /4} = \sum_{\si_3=(s_{3}^{(a)},s_{3}^{(b)})} \g^{-(l_3-l_1) /4}
\g^{-l_1/4}\\
&&\le \sum_{s_{3}^{(a)}\ge s_{1}^{(a)}}\g^{-(s_{3}^{(a)}-s_{1}^{(a)})/2}\sum_{s_{3}^{(b)}\ge s_{1}^{(b)}}\g^{-(s_{3}^{(b)}-s_{1}^{(b)}) /2}\g^{-l_1/4} \le K_1\cdot  \g^{-l_1/4}\ ,\nn
\eea
for some positive constant $K_1$ which is independent of the scale indices.
Now we consider the summation over $\s_1$. By the constraint $s_{1}^{(a)}+s_{1}^{(b)}\ge r-1$ and take into account the factor $\g^{-l_1/4}$ from the above formula, we have:
\bea
\sum_{\si_1}\g^{-l_1/2}\cdot \g^{-l_1/4}&\le&\g^{3r/2} \sum_{s_{1}^{(a)}=0}^{r}\g^{-3s_{1}^{(a)}/2}\sum_{s_{1}^{(b)}=r-2-s_{1}^{(a)}}^{r}\g^{-3s_{1}^{(b)}/2}\\
&&\le \g^{3r/2}\sum_{s_{1}^{(a)}=0}^{r}\g^{-3s_{1}^{(a)}/2}\g^{-3r/2+3s_{1}^{(a)}/2}
\le K_2\cdot r\ ,\nn
\eea
in which $ K_2$ is another positive constant. By choosing $K=K_1\cdot K_2$ we proved the lemma for this case.
\item if $j_1=s_{1}^{(a)}=s_{1}^{(b)}$ and is the smallest index among $(j_1,j_2,j_3,j_4)$, we have $l_1=j_1+1$.
Summing over $\s_1$ is simply bounded by $\g^{-j_1/4}$ and summation over $\s_3\ge \s_2$ is bounded by a positive constant. Finally summation over $\s_2$ gives the factor $r$. So there exists a positive constant $K$, which is also independent of all the scale indices, such that the l.h.s. of \eqref{ss1} is bounded by $K\cdot\g^{-j_1/4}r\le K\cdot r$.
\end{itemize}
\end{proof}
%%%%%%%%%%%%%%%%%%%%%%%%%%%%%%%%%%%%%%%%%%%%%%%%%%%%%%%%%%%%%%%%%%%%%%%%%%%%%%
\section{The convergent contributions to the Schwinger functions}
\subsection{More notations about the Gallavotti-Nicol\`o trees}
Before proceeding, let us introduce the following notations concerning some specific Gallavotti-Nicol\`o trees.
\begin{definition}[Biped trees]
Let $\cG^{r_{max}}$ be a Gallavotti-Nicol\`o tree in which the scale index of the root is $r_{max}$, and $G_r^k$ be a node in $\cG^{r_{max}}$. Let $e(G_r^k)$ be the set of external fields of $G_r^k$ whose cardinality is denoted by $|e(G_r^k)|$. A biped $b$ is a node in $\cG^{r_{max}}$ such that $|e(G_r^k)|=2$. The set of all bipeds is denoted by $\cB:=\{G_r^k,\ r=0,\cdots, r_{max}; k=1,\cdots, c(r)\ \big|\ |e(G_r^k)|=2\}$. A biped tree $\cG^{r_{max}}_\cB$ is defined as a subgraph of a Gallavotti-Nicol\`o tree in which the set of its nodes, noted by $V(\cG^{r_{max}}_\cB)$, consists of the following elements: i) the bare nodes $\VV$ of $\cG^{r_{max}}$, ii), the bipeds $b$ and iii) the root which corresponds to the complete graph $G$. The edges of $\cG^{r_{max}}_\cB$ are the natural inclusion relations for the nodes in $V(\cG^{r_{max}}_\cB)$. 
\end{definition}
\begin{definition}
Let $b\in\cB$ be a biped. The set of external fields of $b$ is denoted by $e_b=\{\bar\psi_b,\psi_b\}$, and the set of external fields of $\cG^{r_{max}}_\cB$ is denoted by ${\cal EB}$. We have ${\cal EB}:=\big(\cup_{b\in B}\ e_b\big)\setminus e(G)$, where $e(G)$ is the set of external fields of the complete graph $G$.
\end{definition}
Similarly, define the quadruped Gallavotti-Nicol\`o trees as:
\begin{definition}
A quadruped $Q$ is a node of a Gallavotti-Nicol\`o tree $\cG^{r_{max}}$ which has four external fields. The set of all quadrupeds in $\cG^{r_{max}}$ is noted by $\cQ$. A quadruped GN tree $\cG^{r_{max}}_\cQ$ is defined as a subgraph of $\cG^{r_{max}}$ whose set of nodes, denoted by $V(\cG^{r_{max}}_\cQ)$, composes of the following elements: the bare nodes of $\cG^{r_{max}}$, the quadruped $\cQ$ and the root of $\cG^{r_{max}}$ which is the complete graph $G$ such that $|e(G)|=4$. The edges of $\cG^{r_{max}}_\cQ$ are the inclusion relations of its nodes. The set of external fields associated to $q$ is noted by $e_q$, and the set of external fields of $\cQ$ is noted by ${\cal EQ}$. We have ${\cal EQ}=(\cup_{Q\in\cQ}e_Q)\setminus e(G)$. 
\end{definition}
%Remark that both the bare vertices and the root can be considered as quadrupeds. 

\begin{definition}\label{gn3}
A convergent Gallavotti-Nicol\`o tree $\cG^{r_{max}}_{\cC}$ is a subgraph of a Gallavotti-Nicol\`o tree which doesn't contain the nodes of $\cB$ or $\cQ$. In other words, the set of nodes in $\cG^{r_{max}}_{\cC}$ is given by $V(\cC):=\big\{ G_{r}^{k},\ r=0,\cdots, r_{max},\ k=1\cdots c(r)\big\vert\ |e( G_{r}^{k})|\ge 6\big\}$ and the edges are the natural inclusion relations of the nodes.
%%%%%%%%%%%%%%%%%%
\end{definition}
Correspondingly, we have the following definitions for the Schwinger functions.
\begin{definition}
Let $\{\cG^{r_{max}}_{\cC}\}$ be the set of convergent GN trees of root scale index $r_{max}$. The corresponding connected Schwinger functions, denoted by $S^c_{\cC,2p,\b}$, with $p\ge3$, are called the convergent Schwinger functions. They are the contributions to the Schwinger functions $S_{2p}$ from the convergent graphs. Similarly, $S^c_{\cQ,\b}$ is defined as the set of quadruped Schwinger functions whose GN trees are the quadrupeds $\{\cG^{r_{max}}_\cQ\}$. Finally, define the set of Schwinger functions $S^c_{\cB,\b}$ corresponding to the biped GN trees $\{\cG^{r_{max}}_{\cB}\}$, as the set of biped Schwinger functions.
\end{definition}
In the rest of this section, we shall construct the connected $2p$-point Schwinger functions for $p\ge2$, which include the convergent Schwinger functions and the quadruped ones, and study their analytic properties. The connected biped Schwinger functions will be studied in the next section.
%%%%%%%%%%%%%%%%%%%%%%%%%%%%%%%%%%%%%%%%%%%%%%%%%%%%%%%%%%%%%%%%%%%%%%%%%%%
\subsection{The $2p$-point Schwinger functions with $p\ge3$.}
The perturbation series of $S^c_{\cC,2p}$ can be written as
\be
S^c_{\cC,2p,\b}(\lambda)=\sum_n\l^n S_{\cC,2p,n},\ee
\be\label{conv1}
S_{\cC,2p,n} = {\frac{1}{n!}}\sum_{{\cal B} = \emptyset,
{\cal Q}=\emptyset \atop  \{\underline\tau\}, \cJ_{r_{max}}' }
\sum'_{\{\si \}} \ep (\cJ')\prod_{v} \int_{\Lambda_{\beta}} dx_{v} 
\prod_{\ell\in \cT} \int_{0}^{1} dw_{\ell}
C_{r_\ell,\si_{\ell}} (x_{\ell}, y_{\ell})
[\det C_{r,\si}(w)]_{{ left}}.
\ee
We have the following theorem:
\begin{theorem}[The Convergent contributions]\label{cth1}
There exists a positive constant $C_1$ independent of the scale index and $\l$, such that the connected Schwinger functions $S^c_{\cC, 2p,\b}(\lambda)$, $p\ge3$, are analytic functions of $\lambda$, for $|\lambda\log T|\le C_1 $.
\end{theorem}
\begin{proof}
The proof follows closely \cite{Riv}. Here we try to make the proof simpler and more pedagogical. 
Formula \eqref{conv1} can be written as:
\bea\label{conv2}
S_{\cC,2p,n} &=& {\frac{1}{n!}}\sum_{\{G^k_r, {r=0},\cdots,r_{max}; {k=1},\cdots,c(r)\},\atop{ {\cal B} = \emptyset,
{\cal Q}=\emptyset}}\sum_{\underline\tau}
\sum_{\{\si \}}' \ep (\cJ)\nn\\
&&\quad\quad \prod_{v} \int_{\Lambda_{\beta}} dx_{v} 
\prod_{\ell\in \cT} \int_{0}^{1} dw_{\ell}
C_{r_{\ell},\si_{\ell}} (x_{\ell}, y_{\ell})
[\det C_{r,\s}(w)]_{left},
\eea
for some positive constant $K$. The summation over $\cJ_{r_{max}}'$ has also been written in a more explicit from. By Theorem \ref{tpc}, we have:
\bea\label{conv3}
|S_{\cC,2p,n}| &\le& 
{\frac{K^n}{n!}}
\sum_{\{G^k_r, {r=0},\cdots,r_{max}; {k=1},\cdots,c(r)\},\atop{ {\cal B} = \emptyset,
{\cal Q}=\emptyset}}\sum_{\underline\tau,\cT} \sum'_{\{\si \}}\prod_{i=1}^{n}\ \Big[\ \chi_{i}(\si)e^{-(l_i^1 + l_i^2 + l_i^3 + l_i^4)/4}\ \Big]\nn\\ 
&&\quad\cdot\prod_{i=1}^n \g^{-[r_i^1+r_i^2+r_i^3+r_i^4]/6}
\label{cpt1},
\eea
in which we have used the fact that $2-|e(G^k_r)|/2\le-|e(G^k_r)|/6$, for $|e(G^k_r)|\ge 6$, and the fact that $$\prod_{r=0}^{r_{max}}\g^{-|e(G^k_r)|/6}=\prod_{i=1}^n \g^{-[r_i^1+r_i^2+r_i^3+r_i^4]/6}.$$

Now we consider summation over the sector indices in \eqref{conv3}.
%%%%%%%%%%%%%%%%%%%%%%%%
%\be\label{secsumcov}
%\sum'_{\{\si \}}\prod_{i=1}^{n}\ \Big[\ \chi_{i}(\si)e^{-(l_i^1 + l_i^2 + l_i^3 + l_i^4)/4}\ \Big]\prod_{i=1}^n \g^{-[r_i^1+r_i^2+r_i^3+r_i^4]/6}.\ee
%%%%%%%%%%%%%%%%%%%%%%%%%
Remark that, since the scale indices for the four fields attached to a vertex are not necessarily the same, we can't apply Lemma \ref{sec1} directly. Let us fix a Fermionic field (half-edge) with maximal index $r$, which eventually goes to the root. This field can be chosen as $f_4$ with scale index $r_4$, without losing generality. The constraint from conservation of momentum implies that either the two smallest sector indices among the four, chosen as $s_{1}^{(a)}$ and $s_{2}^{(a)}$, are equal (modulo $\pm1$), or the smallest sector index $s_{1}^{(a)}$ is equal to the $j_1$, which is the smallest scale index.
Then summing over the sector
indices $s_{1}^{(a)},s_{2}^{(a)}$ is bounded by the factor $\bar r:=\max \{r_i^1, r_i^2\}$. Now we consider the summation over the sectors $(s_{3}^{(a)},s_{3}^{(b)})$, for which we obtain (see the proof of Lemma \ref{sec1})
\be
\sum_{s_{3}^{(a)},s_{3}^{(b)}}\g^{-l_i^3/4}\le K_1.r_i^3,
\ee
for some positive constant $K_1$. In total, we lose a factor $K_1 \bar r\cdot r_i^3$ at each vertex $i=1,\cdots, n$. Summation over the sectors which are not the root sectors is bounded by
\be
\prod_{i=1}^n\sum_{r_i^1,\cdots, r_i^{4}=0}^{r_{max}}[\bar r \g^{-r_i^1/6}\g^{-r_i^2/6}\cdot r_i^3\g^{-r_i^3/6}\g^{-r_i^4/6}]\le K_2.
\ee 
Finally, we have to sum over the sector indices for the root fields, one for each vertex $i$. This summation is bounded by $r_{max}=3j_{max}/2=3|\log T|/2$. Summing over all the GN trees and
spanning trees cost a factor $K_3^n n!$, where $K_3$ is certain positive constant (see \cite{Riv} for the detailed proof of this combinatorial result.). Choosing the positive constant $K_0=K\cdot K_1\cdot K_2\cdot K_3$, we have
\be
|S_{\cC,2p,n}|\le \sum_{n=0}^\infty K_0^n|\lambda|^n|\log T|^n.
\ee
Obviously the above series is convergent for $K_0|\lambda| |\log T|<1$. Let $C_1$ be a positive constant that is strictly smaller than $1/K_0$, we conclude this theorem.
\end{proof}

%%%%%%%%%%%%%%%%%%%%%%%%%%%%%%%%%%%%%%%%%%%%%%%%%%%%%%%%%%%%%%%%%%%%%%%%
\subsection{The quadruped Schwinger functions}
In this part we consider the connected quadruped Schwinger functions $S^c_{\cQ,\b}$. Since both the
bare vertex and the general quadruped $Q$ have four external fields, we introduce to each bare vertex an indicator function $\chi_i(\{\sigma\})$ and to each quadruped $Q$ an indicator function $\chi_Q(\{\sigma\})$. The latter indication function is defined as follows: $\chi_Q(\{\sigma\})$ is equal to $1$ if the sector indices $\{\sigma\}=\{\s_Q^1,\cdots,\s_Q^4\}$ of the external fields of $Q$ satisfy the constraints in Lemma \ref{secmain}, and is equal to $0$ otherwise. Then the quadruped Schwinger functions can be written as
\bea\label{conv22}
S^c_{\cQ,\b}(\lambda)&=&\sum_{n=0}^\infty \lambda^n S_{\cQ,n},\\
S_{\cQ,n} &=& {\frac{1}{n!}}\sum_{\cG^{r_{max}}_\cQ,{\cal{EQ}},\cJ_{r_{max}}'}\sum_{\underline\tau}
\sum_{\{\si \}}' \ep (\cJ')\prod_{i=1}^n\chi_i(\{\sigma\})\prod_{Q\in\cQ}\chi_Q(\{\sigma\})\nn\\
&&\quad\quad \prod_{v} \int_{\Lambda_{\beta}} dx_{v} 
\prod_{\ell\in \cT} \int_{0}^{1} dw_{\ell}
C_{r_\ell,\si_{\ell}} (x_{\ell}, y_{\ell})
[\det C_{r,\si}(w)]_{left}.
\eea
We have the following theorem:
\begin{theorem}\label{mqua}
Let $T=1/\b>0$ be the temperature and $S^c_{\cQ,\b}(\lambda)$ be the connected quadruped Schwinger function. There exists a constant $C$, which may depend on the model but is independent of the scale indices, such that the perturbation series for $S^c_{\cQ,\b}(\lambda)$ is convergent in the domain $\{\lambda\in\RRR\vert|\lambda|<C/|\log T|^2\}$.
\end{theorem}
Remark that a theorem similar to this one has been already proved in \cite{Riv}, for a different setting. We present here a more pedagogical proof, for reader's convenience.
Before proceeding, we introduce the following definitions.
%%%%%%%%%%%%%%%%%%%%%%%%%%%%%%%%%%%%%%%%%%%%%%%%%%%%%%%%%%
\begin{definition}[The maximal sub-quadruped \cite{Riv}]
Let $Q\in\cG^{r_{max}}_\cQ$ be a quadruped which is not a leaf. By the Gallavotti-Nicol\`o tree structure of $\cG^{r_{max}}_\cQ$, $Q$ must be linked directly to a set of quadrupeds $\{Q'_1,\cdots,Q'_{d_Q}\}$, $d_Q\ge1$, called the maximal sub-quadrupeds of $Q$. These sub-quadrupeds could be either the bare vertices or some general quadrupeds.
\end{definition}
\begin{remark}
Remark that a maximal sub-quadruped $Q'$ of $Q$ may still contain some sub-quadrupeds $Q''_1,\cdots, Q''_{d(Q')}$. The inclusion relation between $Q$ and $Q''$ is not an edge of the quadruped Gallavotti-Nicol\`o tree.
\end{remark}
Now we consider summation over sector indices for a quadruped. We have: 
\begin{lemma}[Sector counting lemma for quadrupeds]\label{secqua}
Let $Q$ be a quadruped of scale indices $r$, which is linked to $d_Q$ maximal sub-quadrupeds ${Q'_1,\cdots,Q'_{d_Q}}$. Let the external fields of $Q$ be ${f_Q^1,\cdots,f_Q^4}$, with scale indices ${r_Q^1,\cdots,r_Q^4 }$ and sector indices ${\s_Q^1,\cdots,\s_Q^4}$, respectively. Let the external fields of $Q'_v$, $v=1,\cdots Q_d$, be ${f_v^1,\cdots,f_v^4 }$, with scale indices ${r_v^1,\cdots,r_v^4 }$ and sector indices ${\s_v^1=(s_{v,1}^{(a)},s_{v,1}^{(b)}),\cdots,\s_v^4=(s_{v,4}^{(a)},s_{v,4}^{(b)} ) }$, respectively. Let $\chi_v(\{\sigma_v\})$ be the characteristic function at the sub-quadruped $Q'_v$, with $v=1,\cdots, d_Q$, we have
\bea
\sum_{\{\sigma_1\},\cdots,\{\sigma_{d_Q}\}} \prod_{v=1}^{d_Q}\chi_v(\{\sigma_v\})\chi_Q(\{\sigma\})e^{-[l_v^1+l_v^2+l_v^3+l_v^4]/4}\le K_1^{d_Q-1} {r}^{d_Q-1},
\eea
for some positive constant $K$.
\end{lemma}

\begin{proof}
Let $Q$ be a quadruped and $\cT$ a spanning tree of $G$, then $\cT_Q=\cT\cap Q$ is the set of tree lines in $Q$. Among all the internal fields contained in $Q$, we fix the root field with the highest scale index $r_Q$, denoted by $f_{r_Q}$. Since a root field also belongs to some maximal sub-quadruped of $Q$, we fix also a root field for each of the $d_Q$ sub-quadrupeds. Define the external vertices of $Q$ as the set of maximal sub-quadrupeds $Q'$ to which the external fields of $Q$ are hooked. So there can be at most four of them. We call a field a tree field if, when contracted with another field, a tree line of $\cT_Q$ can be formed. We consider the constraints on the sector indices for the maximal sub-quadrupeds, starting from an external sub-quadruped $Q'_1$, which contains at least one external field, to the next maximal sub-quadruped. 
By conservation of momentum, whenever two tree fields of an external quadruped are fixed, the last field in that external quadruped is also determined. In this way we find that the number of pairs of sector indices to be determined is equal to the number of tree lines in $Q$ connecting the maximal sub-quadrupeds, which is $d_Q-1$. 
%%%%%%%%%
%%%%%%%%%%
%%%%%
%%%%%%%%
Since summing over each pair of sector indices for a root field is bounded by $\sum_{(s_{v}^{(a)},s_{v}^{(b)})}\g^{-l/4 }\le K_1.r_v$ (cf. Lemma \ref{sec1} ),
in which $r_v\le r_Q\le r$, we obtain
\bea
\sum_{\{\sigma_1\},\cdots,\{\sigma_{d_Q}\}} \prod_{v=1}^{d_Q}\chi_v(\{\sigma_v\})\chi_Q(\{\sigma\})e^{-[l_v^1+l_v^2+l_v^3+l_v^4]/4}\le
(K_1.r_v)^{d_Q-1}\le K_1^{d_Q-1} {r}^{d_Q-1}.
\eea
Thus we conclude this lemma.
\end{proof}

%%%%%%%%%%%%%%%%%%%%%%%%%%%%%%%%%%%%%%%%%%%%%%%%%%%%%%%%%%%%%%%%%%%%%%%%%%%%%%%%

\begin{proof}[Proof of Theorem \ref{mqua}]
In order to sum over all the quadruped trees, it is useful to keep the GN tree structure explicit and write a quadruped tree as $\cG^{r_{max}}_\cQ=\{Q_r^k, r=0,\cdots, r_{max}, k=1,\cdots, c(r)\}$. The quadruped Schwinger functions satisfy the following bound (cf. Formula \eqref{cpt1}):
\bea
|S_{\cQ,n}| &\le& 
{\frac{K^n}{n!}}
\sum_{\{Q^k_r, {r=0},\cdots,r_{max}; {k=1},\cdots,c(r)\},\atop{ {\cal B} = \emptyset}}\sum_{\underline\tau,\cT} \sum'_{\{\si \}}\prod_{Q\in\cQ}\chi_Q(\{\sigma\})\prod_{i=1}^{n}\ \Big[\chi_{i}(\si)e^{-[l_i^1 + l_i^2 + l_i^3 + l_i^4]/4}\Big]\nn\\
&&\quad\cdot\prod_{r=0}^{r_{max}}\g^{2-|e(G^k_r)|/2}\ .\label{convq}
\eea
Now we sum over all the sector indices, from the leaves of a quadruped tree to the root. The first quadruped $Q_0$ that we encounter contains the bare vertices as the maximal sub-quadrupeds. The second quadruped $Q_1$ contains the quadruped $Q_0$, some other quadrupeds at the same scale index than $Q_0$, and bare vertices. More quadrupeds will be encountered when we are going towards the root. In this process, we apply Lemma \ref{secqua} to each quadruped $Q$ that we meet, until we arrive at the root node of $\cG^{r_{max}}_\cQ$. Then there exists a constant $K_2$ independent of the scale index such that:
\bea
|S_{\cQ,n}| &\le& \prod_{Q\in\cQ}\Big[K_2^{d_Q} \sum_{r=0}^{r_{max}} {r}^{d_Q-1}\Big]
\le \prod_{Q\in\cQ}K_3^{d_Q}|\log T|^{d_Q},
\label{convq2}
\eea
in which $K_3=3 K_2\cdot K/2$ is another positive constant, hence is also independent of the scale index; The sum over scale indices in $[\cdots]$ means that we sum over the root scale indices for each quadruped $Q$, from $0$ to $r_{max}=3|\log T|/2$. We have used the fact that the number of quadruped trees with $n$ vertices is bounded by $c^n n!$ (see \cite{Riv}), which is a variation of Cayley's theorem concerning the number of labeled spanning trees with fixed vertices.
Using following well-known formula 
\be\label{ind4p}\sum_{Q\in \cQ} d_Q=|\cQ|+n-1\le 2n-2,
\ee 
where $| \cQ|$ is the cardinality of the set of quadrupeds $\cQ$, for which we have $| \cQ|\le n-1$,
we can prove the following bound:
\be
|S_{\cQ,n}|\le\prod_{Q\in  \cQ}K_3^{d_Q}|\log T|^{d_Q}\le K_3^{2n-2}|\log T|^{2n-2},
\ee
and
\be
|S^c_{\cQ,\b}(\lambda)|\le\sum_{n=0}^\infty K_3^{2n-2}\cdot\lambda^n\cdot|\log T|^{2n-2}.
\ee
Let $0<C\le1/K_3$ be some constant, define 
\be\label{adoq}
\RR^\cQ_T:=\{\lambda\ \vert |\lambda\log^2T|<C \},
\ee 
then the perturbation series of $S_{\cQ}(\lambda)$ is convergent for $\lambda\in\RR^\cQ_T$.
We conclude this theorem.
\end{proof}
%%%%%%%%%%%%%%%%%%%%%%%%

\begin{remark}\label{rmtad}
Obviously $\RR^{\cQ}_T\subset\RR^{c}_T$. This fact also set a constraint to the analytic domains for the biped Schwinger functions. Define the analytic domain for the two-point Schwinger functions by $\RR_T$, then we have 
\be
\RR_T=\RR_T\cap\RR^{c}_T\cap \RR^{\cQ}_T\subseteq\RR^{\cQ}_T.
\ee 
Therefore, in order that the perturbation series of the connected $2p$-point Schwinger functions, $p\ge1$, to be convergent, the bare coupling constant $\lambda$ should satisfy 
\be
|\lambda|<C/{j^2_{max}}=C/{\log^2 T}.
\ee
\end{remark}
%%%%%%%%%%%%%%%%%%%%%%%%%%%%%%%%%%%%%%%%%%%%%%%%%%%%%%%%%%%%%%%%%%%%%%%%%

%%%%%%%%%%%%%%%%%%%%%%%%%%%%%%%%%%%%%%%%%%%%%%%%%%%%%%%%%%%%%%%%%%%%%%%%%%%%%%%%%%%%%%%%%%
\section{The $2$-point Functions}
In this section we study the connected $2$-point Schwinger function and the self-energy function. While the perturbation series for the former are labeled by connected graphs, the ones for the latter are labeled by the one-particle irreducible graphs (1PI for short), which are the graphs that can't be disconnected by deleting one edge. Since the renormalization conditions are not the same for different matrix elements of the $2$-point Schwinger function and the self-energy function, we shall keep explicit the matrix elements of these functions. A matrix element of a connected $2$-point Schwinger function $S^c_{2,\b}(y,z)$, in which $y$ and $z$ are the coordinates of the two external fields, will be denoted by $[S^c_{2,\b}(y,z)]_{\a\a'}$, $\a,\a'=1,2$. Matrix elements of the self-energy function are denoted in the same way. Using the BKAR tree formula (\cite{RW1}, Theorem $4.1$) and organizing the perturbation terms according to the GN trees, we can write $S^c_{2,\b}(y,z)$ as:
\bea\label{consch}
&&[S^c_{2,\b}(y,z)]=\sum_{n=0}^\infty\frac{\lambda^{n+2}}{n!}\int_{({\Lambda_{\beta}})^n} d^3x_1\cdots d^3x_n\sum_{\cG^{r_{max}}}\sum_{\cG^{r_{max}}_\cB}\sum_{{\cal EB}}\sum_{\{\sigma\}}\sum_{\cT, \underline\tau}\Big(\prod_{\ell\in\cT}\int_{0}^1 dw_\ell\Big)\nn\\
&&\quad\quad\quad\quad\cdot \Big[\prod_{\ell\in\cT}C(f_\ell,g_\ell)\Big]\cdot\det\Big(C(f,g,\{w_\ell\})\Big)_{left},
\eea
in which $\cG^{r_{max}}_\cB$ is a biped GN tree (cf. \cite{RW1}, Definition $5.1$) and $\cT$ is a spanning tree in the root graph of $\cG^{r_{max}}_\cB$.

It is well known that (see eg. \cite{iz}, page 290) the self-energy $\Sigma(y,z)$ can be obtained by Legendre transform on the generating functional for $S^c_{2,\b}$. In terms of Feynman graphs this corresponds to replacing the connected graphs labeling the connected functions by the 1PI graphs (which are also called the two-connected graphs in Graph theory). Let $\{\Gamma\}$ be the set of 1PI graphs over the $n+2$ vertices, the self-energy is defined by:
\bea\label{selfeng}
&&\Sigma(y,z,\lambda)=\sum_{n=0}^\infty\frac{\lambda^{n+2}}{n!}\int_{({\Lambda_{\beta}})^n} d^3x_1\cdots d^3x_n\sum_{\cG^{r_{max}}}\sum_{\cG^{r_{max}}_\cB}\sum_{{\cal EB}}\sum_{\{\sigma\}, \underline\tau}\sum_{\{\cT\}}\sum_{\{\Gamma\}}\\
&&\quad \Big(\prod_{\ell\in\cT}\int_{0}^1 dw_\ell\Big)\cdot \Big[\prod_{\ell\in\cT}C(f_\ell,g_\ell)\Big]\Big[\prod_{\ell\in\Gamma\setminus\cT}C(f_\ell,g_\ell)\Big]
\cdot\det\Big(C(f,g,\{w_\ell\})\Big)_{left,\Gamma}.\nn
\eea 
Through the Fourier transform
\be
[\Sigma_2(y,z,\lambda)]_{\a\a'}=\int_{\cD_{\b,L}} dk\ [\hat\Sigma(k,\lambda)]_{\a\a'}e^{ik(y-z)},\ \a,\a'=1,2,
\ee
we define the self-energy function in the momentum space $\hat\Sigma(k,\lambda)$. 
Remark that the above expressions are still formal, as summation over the 1PI graphs could be unbounded. The canonical way of generating the 1PI graphs without divergent combinatorial factors is called the multi-arch expansion, which will be introduced shortly. The construction of the 2-point Schwinger functions and the self-energy requires renormalization theory, which will be introduced in the next subsection.

%%%%%%%%%%%%%%%%%%%%%%%%%%%%%%%%%%%%%%%%%%%%%%%%%%%%%%%%%%%%%%%%%%%%%%%%%%%%%%%%%%%%%%%%
\subsection{Localization of the two-point functions}
The localization of the two-point Schwinger function  is naturally defined in the momentum space.
Let $\hat S_2(p)$ a two-point function with external momentum $p$.
Suppose that the internal momentum of the lowest scale belongs to the sector with scale index $j_r$ and sector indices $(s^{(a)}_{j_r}, s^{(b)}_{j_r})$ while the external momentum $p$ belongs to the sector with scale index $j_e$ and sector indices $(s^{(a)}_{j_e}, s^{(b)}_{j_e})$. The localization operator is defined as:
\be
\tau\hat S_2(p)=\sum_{j=1}^\infty\sum_{\sigma=(s^{(a)},s^{(b)})}\chi_j(4p_0^2+e^2(\bp,1))\cdot v_{s^{(a)}}[t^{(a)}(\bp)]
\cdot v_{s^{(b)}}[t^{(b)}(\bp)]\cdot\hat S_2 (2\pi T, \bk_F),
\ee
%%%%%%%%%%%%%%%%%%%%%%%%%%%%%%%%%%%%%%%%%%%%%%%%%%%%%%%%%%%%%%%%%%%%%%%%%%%%%%%%%%%
in which $\kk_F=P_F(\bp)$. Notice that $\hat S_2(2\pi T,\kk_F)$ is not a constant on $\cF_0$ but depends non-trivially on $\kk_F$. In order to establish the non-perturbative bound, it is important to perform the localization in the direct space. The corresponding localization operation, noted by $\tau^*$, is defined by the Fourier transform as follows. Consider the integral
\bea\label{rn2pt0}
I=\int_{\cD_{\beta,L}\times\cD_{\beta,L}} dp dk\ \hat S_{2}
(p)\hat C(k)\hat R(p,k,P_e),
\eea
in which $\hat C(k)=\sum_j\sum_{\sigma=(s^{(a)},s^{(b)})}\hat C_{j,\sigma}(k)$ is a $2\times2$ matrix (cf. \eqref{prob2x}), $\hat R(p,k,P_e)=\bar R(p,P_e)\delta(p-k)$ is also a $2\times2$ matrix, in which $\bar R(p,P_e)$ is a function of $p$ and external momentum $P_e$. Define the localization operator $\tau$ on $I$ by:
\be
\tau I=\int dp dq\ \hat S_{2}
(k_F)\hat C(q) \hat R(p,k,P_e),
\ee
and the remainder term by $\hat R I=(1-\tau)I$, in which $\hat R:=(1-\tau)$ is called the remainder operator. The direct space representation of $I$ is given by:
\bea\label{rn2pt1}
\tilde I=\int dy dz\ S_{2} (x,y)\ C(y,z)R(z,x,P_e),
\eea
which is indeed independent of $x$, due to translational invariance. Then the operators $\tau$ and $\hat R:=(1-\tau)$ induce the actions $\tau^*$ and $\hat R^*:=(1-\tau^*)$ in the direct space. The localized term is
%\be
%[\tau^*\tilde I]_{\a\a'}=\int dy dz\ \sum_{\a_1,\a_2=1}^2[S_{2}(x,y)]_{\a\a_1}[e^{ik^0_F(x_0-y_0)+i\kk_F\cdot (\xx-\yy)} [C(x,z)]_{\a_1\a_2}][R(z,x,P_e)]_{\a_2\a'}.
%\ee
%%%%%%%%%
\be
[\tau^*\tilde I]_{\a\a'}=\int dy dz e^{ik^0_F(x_0-y_0)+i\kk_F\cdot (\xx-\yy)} \Big[S_{2}(x,y)C(x,z)R(z,x,P_e)\Big]_{\a\a'}.
\ee
%%%%%%%
Comparing with \eqref{rn2pt1} we find that the localization operator moves the starting point $y$ of the free propagator to the localization point $x$, with the compensation of a phase factor:
\be
[\tau^*C(y,z)]_{\a\a'}=e^{i2\pi T(x_0-y_0)+i\kk_F\cdot (\xx-\yy)} [C(x,z)]_{\a\a'}.
\ee
The  remainder term is:
%%%%%%%%%%%%%%%%%%%%%%%%%%%%%%%%%%%%%%%%%%%
\bea\label{rmdi}
[\hat R^* I]_{\a\a'}&=&\int dy dz\sum_{\a_1,\a_2=1}^2\ [S_{2}(x,y)]_{\a\a_1}\\
&&\cdot[C(y,z)-e^{ik^0_F(x_0-y_0)+i\kk_F\cdot (\xx-\yy)} C(x,z)]_{\a_1\a_2}[R(z,x,P_e)]_{\a_2,\a'},\nn
\eea
%%%%%%%%%%%%%%%%%%%%%%%%%%%%%%%%%%%%%%%%%%%%%%%%%%%%%%%%%%%%
in which
\bea
&&[C(y,z)-e^{ik^0_F(x_0-y_0)+i\kk_F\cdot (\xx-\yy)} C(x,z)]_{\a_1\a_2}\\
&&=\int_0^1 dt(y_0-x_0)\Big[\frac{\partial}{\partial x_0}C((ty_0+(1-t)x_0,\yy),z)\Big]_{\a_1\a_2}\nn\\
&&+\frac12\sum_{a,b=1\cdots3}(y^{(a)}-x^{(a)})(y^{(b)}-x^{(b)})\Big[\partial_{x^{(a)}}\partial_{{x}^{(b)}}
C((x_0,\xx),z)\Big]_{\a_1\a_2}\nn\\
&&+\int_0^1 dt(1-t)\sum_{a,b=1\cdots3}(y^{(a)}-x^{(a)})(y^{(b)}-x^{(b)})\Big[\partial_{y^{(a)}}\partial_{{y}^{(b)}}
C((x_0, t\yy+(1-t)\xx),z)\Big]_{\a_1\a_2}\nn\\
&&+[C((x_0,\xx),z)]_{\a_1\a_2}[1-e^{ik^0_F(x_0-y_0)+i\kk_F\cdot (\xx-\yy)}]\nn.
\eea
The terms in  the last line means that there exists an additional propagator $ C(x,z)$ attached to the biped graph and the new graph has three external lines. So it is no more linearly divergent and we gain a convergence factor in the power counting. Now we consider the other terms.
%%%%%%%%%%%%%%%
Suppose that the lowest internal momentum of the connected two-point function $S_{2}(x,y)$ belongs to the sector $\Delta^{j_r}_{{s^{(a)}_{j_r},s^{(b)}_{j_r}}}$ with scale index ${j_r}$ and sector indices $({{s^{(a)}_{j_r},s^{(b)}_{j_r}}})$, and the external momentum belongs
to the sector $\Delta^{j_e}_{{s^{(a)}_{j_e},s^{(b)}_{j_e}}}$,
then there exists a constant $K_1$, $K_2$ such that
\bea\label{rmdx}
&&|y_0-x_0|\le O(1)\gamma^{j_r},\  |y^{(a)}-x^{(a)}|\cdot|y^{(b)}-x^{(b)}|\le\g^{s_{j_r}^{(a)}+s_{j_r}^{(b)}},\nn\\
&&\Vert[\partial_{x_0}C((x_0,\yy),z)]_{\a_1\a_2}\Vert_{L^\infty}\le K_1\g^{-j_e}\Vert[C((x_0,\yy),z)]_{\a_1\a_2}\Vert_{L^\infty},\nn\\
&&\ \Vert[\partial_{x^{(a)}}\partial_{{x}^{(b)}}C((x_0,\xx),z)]_{\a_1\a_2}\Vert_{L^\infty}\le K_2\g^{-s_{j_e}^{(a)}-s_{j_e}^{(b)}}.\nn
\eea
Since the perturbation terms are organized according to the Gallavotti-Nicol\`o tree structure, we have $j_r\le j_e$. As will be proved in Section $7.2$, we can always choose the optimal internal propagators (rings propagators) such that $s_{j_e}^{(a)}+s_{j_e}^{(b)}\ge s_{j_r}^{(a)}+s_{j_r}^{(b)}$. Hence we gain the convergence factor
$\g^{-(j_e-j_r)}$ and $\g^{-[(s_{j_e}^{(a)}+s_{j_e}^{(b)})-(s_{r_0}^{(a)}+s_{r_0}^{(b)})]}$.
%%%%%%%%%%%%%%%%%%%%%%%%%%%%%%%%%%%%%%%%%%%%%%%%%%%%%%%%%%%%

Now we consider the localization for the self-energy $\hat \Sigma(k_0,\bk,\l)$. Again, we suppose that the lowest internal momentum belongs to the sector $\Delta^{j_r}_{{s^{(a)}_{j_r},s^{(b)}_{j_r}}}$ and the external momentum $(k_0,\bk)$ belongs to the sector $\Delta^{j_e}_{{s^{(a)}_{j_e},s^{(b)}_{j_e}}}$. We have:
\bea
[\tau\hat\Sigma(k_0,\bk,\l)]_{\a\a'}&=&\sum_{j=1}^{j_{max}}\sum_{\sigma=(s^{(a)},s^{(b)})}\chi_j(4k_0^2+e^2(\bk,1))\\
&&\quad\quad\cdot v_{s^{(a)}}[t^{(a)}(\bk)]
\cdot v_{s^{(b)}}[t^{(b)}(\bk)]\cdot[\Sigma (k_F^0, P_{F}\bk,\l)]_{\a\a'},\nn
\eea
in which $(k_F^0, P_{F}\bk)$ is the Fermi momentum at which we perform the renormalization.
Let $\hat \Sigma^{j_e,{s_{j_e}^{(a)},s_{j_e}^{(b)}}}_{\a\a'}(k_0,\bk)$, $\a,\a'=1,2$, be a matrix element of the self-energy function in which the external momentum $(k_0,\bk)$ is constrained in the sector $\Delta^{j_e}_{{s^{(a)}_{j_e},s^{(b)}_{j_e}}}$, we have:
%\bea\label{rmd11}
%&&\hat R\hat \Sigma(p_0,\bp)_{s^{(a)},s^{(b)}}:=(1-\tau)\hat \Sigma(k_0,\bk)_{s^{(a)},s^{(b)}}\\
%&=&\hat \Sigma(p_0,\bp)_{s^{(a)},s^{(b)}}-\hat \Sigma(k_F^0,\bp)_{s^{(a)},s^{(b)}}+\hat \Sigma(k_F^0,\bp)_{s^{(a)},s^{(b)}}-\hat \Sigma(k_F^0,\bk_F)_{s^{(a)},s^{(b)}}\nn\\
%&=&\int_0^1 dt (p_0-k_F^0)\frac{\partial}{\partial p_0(t)}\hat \Sigma(k_F^0+t(p_0-k_F^0),\bp)_{s^{(a)},s^{(b)}}\nn\\
%&+&\int_0^1 dt(1-t)(p^{(a)}-k_{F}^{(a)})(p^{(b)}-k_{F}^{(b)})\frac{\partial^2}{\partial p^{(a)}\partial p^{(b)}}\hat \Sigma(k_F^0,\bk_F+t(\bp-\bk_F))_{s^{(a)},s^{(b)}}\nn\ ,
%\eea
%%%%%%%%%%%%%%%%%%%%%%%%%%%%%%%%%%%%%%%%
\bea\label{rmd11}
&&\hat R\hat \Sigma^{j_e,{s_{j_e}^{(a)},s_{j_e}^{(b)}}}_{\a\a'}(k_0,\bk):=(1-\tau)\hat \Sigma^{j_e,{s_{j_e}^{(a)},s_{j_e}^{(b)}}}_{\a\a'}(k_0,\bk)\\
&=&\int_0^1 dt (k_0-k_F^0)\frac{\partial}{\partial k_0(t)}\hat \Sigma^{j_e,{s_{j_e}^{(a)},s_{j_e}^{(b)}}}_{\a\a'}(k_F^0+t(k_0-k_F^0),\bk)+\sum_{a,b=1}^3\int_0^1 dt(1-t)\nn\\
&\cdot&(k^{(a)}-(P_{F}\bk)^{(a)})(k^{(b)}-(P_{F}\bk)^{(b)})\frac{\partial^2}{\partial k^{(a)}(t)\partial k^{(b)}(t)}\hat \Sigma^{j_e,{s_{j_e}^{(a)},s_{j_e}^{(b)}}}_{\a\a'}(k_F^0,P_{F}\bk+t(\bk-P_{F}\bk))\nn\ ,
\eea
%%%%%%%%%%%%%%%%%%%%%%%%%%%%%%%%%%%%%%%%
in which $k_0(t)=k_F^0+t(k_0-k_F^0)$ and $k^{(a)}(t)=(P_{F}\bk)^{(a)}+t(k^{(a)}-(P_{F}\bk)^{(a)})$. 
We have $|k_0-k_F^0|\sim \gamma^{-j_e}$, $\Vert\partial_{ k_0(t)}\hat \Sigma^{j_e,{s_{j_e}^{(a)},s_{j_e}^{(b)}}}_{\a\a'}(k_F^0,\bk)\Vert\sim\gamma^{j_r}\Vert\hat \Sigma^{j_e,{s_{j_e}^{(a)},s_{j_e}^{(b)}}}_{\a\a'}(k_F^0,\bk)\Vert$.
Thus we obtain
\be\label{rmd12}
\Vert(k_0-k_F^0)\frac{\partial}{\partial k_0}\hat \Sigma^{j_e,{s_{j_e}^{(a)},s_{j_e}^{(b)}}}_{\a\a'}(k_F^0,\bk)\Vert\le K_1\gamma^{-(j_e-j_r)}\Vert\hat \Sigma^{j_e,{s_{j_e}^{(a)},s_{j_e}^{(b)}}}_{\a\a'}(k_F^0,\bk)\Vert,
\ee
and
\bea\label{rmd13}
&&\Vert(k^{(a)}-(P_{F}\bk)^{(a)})(k^{(b)}-(P_{F}\bk)^{(b)})\frac{\partial^2}{\partial k^{(a)}(t)\partial k^{(b)}(t)}\hat \Sigma^{j_e,{s_{j_e}^{(a)},s_{j_e}^{(b)}}}_{\a\a'}(k_F^0,\bk)\Vert\\
&&\quad\quad\quad\le
K_2\g^{-[(s_{j_e}^{(a)}+s_{j_e}^{(b)})-(s_{j_r}^{(a)}+s_{j_r}^{(b)})]}\Vert\Sigma^{j_e,{s_{j_e}^{(a)},s_{j_e}^{(b)}}}_{\a\a'}(k_F^0,\bk)\Vert\nn,
\eea
for some positive constants $K_1$, $K_2$. So we gain a convergence factor $\g^{-[(s_{j_e}^{(a)}+s_{j_e}^{(b)})-(s_{j_r}^{(a)}+s_{j_r}^{(b)})]}$.
%%%%%%%%%%%%%%%%%%%%%%%%%%%%%%%%%%%%%%%%%%%%%%%%%%%%%%%%%%%%%%%%%%%%%%%%%%%%%%%%%%%%
%%%%%%%%%%%%%%%%%%%%%%%%%%%%%%%%%%%%%%%%%%%%%%%%%%%%%%%%%%%%%%%%%%%%%%%%%%%%%%%%%%%%%
\subsection{The renormalization of the self-energy function}
In this part we consider the renormalization the self-energy function $\cE(k_0,\bk,\l)$, which reduces to the renormalization for $\cE_{11}=\tilde\Sigma(k_0,\bk,\l)_{11}$ and $\cE_{12}=\hat\Sigma(k_0,\bk,\l)_{12}$, by symmetry properties of the self-energy (cf. \eqref{inv1}). Recall that the renormalization conditions for $\tilde\Sigma(k_0,\bk,\l)_{11}$
are given by \eqref{rncd3} and the first equation of \eqref{rncd5}. While the former corresponds to the renormalization of the chemical potential, the latter corresponds to the renormalization of the non-local part of $\tilde\Sigma(k_0,\bk,\l)_{11}$. The renormalization conditions for $\hat\Sigma(k_0,\bk,\l)_{12}$ is given by the second equation of \eqref{rncd5}.
The renormalization analysis is to be performed recursively in the multi-scale representation, from lower scale index toward higher scale index, in which, at each scale $r$, we move the counter-terms from the interaction to the covariance so that the tadpoles as well as the self-energy at that scale can be compensated by the corresponding counter-terms. In this procedure the renormalized band function and the renormalized Fermi surface remain fixed. Consider first the renormalization of $\tilde\Sigma(k_0,\bk,\l)_{11}$
%%%
%%%%%%%%%%%%%%%%%%%%%%%%%%%%%%%%%%%%%%%%%%%%%%%%%%%%%%%%%%%%%%%%%%%%
\subsubsection{Renormalization of the bare chemical potential}
In this part we consider the renormalization of the bare chemical potential $\mu_{bare}=\mu+\delta\mu$, realized by compensation of the tadpoles $T(\lambda)$ with the counter-term $\delta\mu(\l)$, in the multi-scale representation. Define
\bea
&&\delta\mu(\lambda):=\delta\mu^{\le r_{max}}(\lambda)=\sum_{r=0}^{r_{max}} \delta\mu^r(\lambda),\\
&&T(\lambda):=T^{\le r_{max}}(\lambda)= \sum_{r=0}^{r_{max}} T^r(\lambda),
\eea
in which $T^r(\lambda)\in\RRR$ is the sliced tadpole whose {\it internal momentum} belongs to sectors of scale $r$. At each scale $r$, the chemical potential counter-term $\delta\mu^r$ is compensated with $T^r$ and the renormalized chemical potential $\mu$ remain fixed. By locality, the compensations are $exact$. Before proceeding, it is useful to calculate explicitly the amplitude of a tadpole term. 
\begin{lemma}\label{tadmain1}
Let $T^r$ be the amplitude of a tadpole of scale $r$, let $\lambda\in\RR_T\subseteq\RR^\cQ_T$ be the coupling constant. There exist two positive constants $c_1$ and $c_2$, with $c_1<c_2$, which are dependent on the model but are independent of $j$ and $\l$, such that:
\be\label{tad01}
c_1|\lambda|j\g^{-j}\le\vert T^r\vert\le c_2|\lambda|j\g^{-j}.
\ee
\end{lemma}

\begin{proof}
For any scale index $0\le j\le j_{max}$, we have:
\be
\vert T^r\vert\le|\lambda|\sum_{\s=(s^{(a)},s^{(b)})}\ \int dk_0 dk^{(a)}dk^{(b)}\Big|\tilde C_{j,\sigma}(k_0,k^{(a)},k^{(b)})\ \Big|\le c_1 |\lambda| \sum_{(s^{(a)},s^{(b)})}\g^{-s^{(a)}-s^{(b)}}.
\ee
for some positive constant $c_1$. Using the constraint $s^{(a)}+s^{(b)}\ge j-1$, we have
\be
\sum_{(s^{(a)},s^{(b)})}\g^{-s^{(a)}-s^{(b)}}=\Big(\sum_{s^{(a)}=0}^{j}\g^{-s^{(a)}}\Big)\ \Big(\sum_{s^{(b)}=j-1-s^{(a)}}^{j} \g^{-s^{(b)}}\Big).
\ee
Now using the fact that
\be
\g^{-j+1+s^{(a)}}\le\sum_{s^{(b)}=j-1-s^{(a)}}^{j} \g^{-s^{(b)}}\le \g^{-j+1+s^{(a)}}\frac{1-\g^{-(1+s^{(a)})}}{1-\g^{-1}},
\ee
we conclude that there exists another positive constant $c_2>c_1$, also independent of $j$, such that
\bea\label{bdtj}
c_1|\lambda|j\g^{-j}\le\vert T^j\vert\le c_2|\lambda|j\g^{-j}.
\eea
\end{proof}
%%%%%%%%%%%%%%%%%%%%%%%%%%%
\begin{lemma}\label{tad05}
Let $T=\sum_{j=0}^{j_{max}}T^j$ be the full amplitude of a tadpole. There always exist two positive constants $c_1'$ and $c_2'$, with $c_1'<c_2'$, such that:
\be
c_1'|\lambda|\le\vert T\vert\le c_2'|\lambda|.
\ee
\end{lemma}
\begin{proof}
Since $|T|=\sum_{j=0}^{j_{max}}|T^j|$, we can prove this lemma directly by summing over the indices $j$, using \eqref{bdtj}.
\end{proof}

\begin{figure}[htp]
\centering
\includegraphics[width=.5\textwidth]{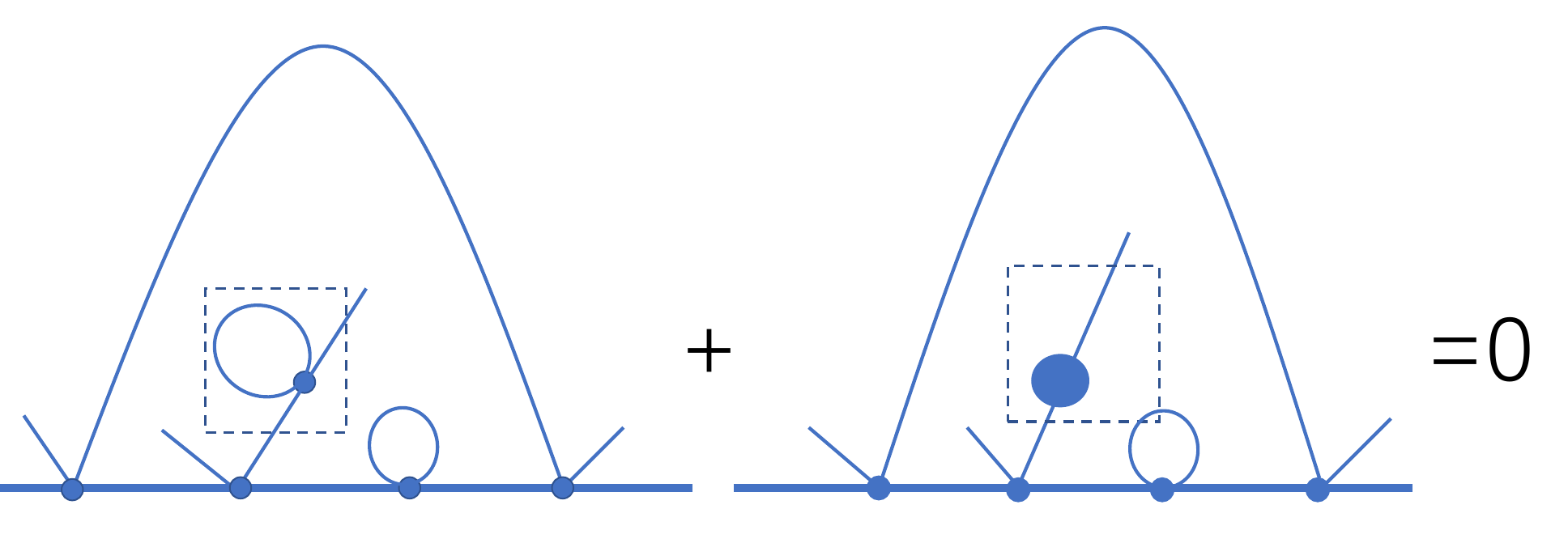}
\caption{\label{rtad}
Cancellation of a tadpole with the corresponding counter-term (the blue dot). }
\end{figure}

%%%%%%%%%%%%%%%%%%%%%%%%%%%%%%%%%%%%%%%%%%%%
In order that Equation \eqref{rncd3} can be valid, we have: $T^r+\delta\mu^r(\lambda)=0$,
$r=0,\cdots, r_{max}$, which implies that:
\be\delta\mu^{\le r}(\lambda)+\delta T^{\le r}=0,\ {\rm for}\ r=0,\cdots, r_{max}.\ee
%%%%%%%%%%%%%%%%%%%%%%%%%%%%%%%%%%%%%%%%%%%%%%%%%%%%%%%%%%%%%%%%%%%%%%%%%%%%%%%%%%%%%%%%
To remember that the cancellation is between a pair of GN trees. Let $F_{2,n}=F'_{2,n-1}\vert T^r\vert$ be the amplitude of a graph with $n$ vertices which contains a tadpole $T^r$. Let $F'_{2,n}=F'_{2,n-1}\delta\mu^r$ be the amplitude of another graph which contain counter-term, which is located at the same position in the GN tree as the tadpole. Then we have $F_{2,n}+F'_{2,n}=0$. See Figure \ref{rtad} for an illustration of the cancellation. 
%%%%%%%%
%Remark that, since the cancellation between a tadpole term with a counter-term is exact, we consider in the rest of this section only the graphs that are tadpole free.
In this way we can fix the renormalized chemical potential at all scales.
%%%%%%%%%%%%%%%%%%%%%%%%%%%%%%%%%%%%%%%%%%%%%%%%%%%%%%%%%%%%%%%%%%%%%%%%%%%%%%%%%%%%%%%%%%5
We have the following theorem concerning the coefficient $\delta\mu(\lambda)$ of the tadpole counter-terms.
\begin{theorem}\label{flowmu}
There exists a positive constant $K$ independent of the scale indices such that 
the tadpole counter-term can be bounded as follows:
\be\label{bdbare}
|\delta\mu(\lambda)|:=|\delta\mu^{\le r_{max}}(\lambda)|\le K|\lambda|,\ {\rm for}\ \lambda\in\RR_T.
\ee
\end{theorem}
The proof of this theorem will be provided in Section \ref{secflow}. 
\begin{remark}
This Theorem states that the counter-terms are bounded, therefore will not cause any divergence in the analysis of the two point functions. So we can replace $\delta\mu$ by some constants in the rest of this paper, except in Section \ref{secflow}.
\end{remark}
%%%%%%%%%%%%%%%%%%%%%%%%%%%%%%%%%%%%%%%%%%%%%%%%%%%%%%%%%%%%%%%%%%%%%%%%%%%%%%%%%%%%%%%%%%%%%%%%%%%%
%%%%%%%%%%%%%%%%%%%%%%%%%%%%%%%%%%%%%%%%%%%%%%%%%%%%%%%%%%%%%%%%%%%%%%%%%%%%%%%%%%%%%%%%%%%%%%%%%%%%
\subsubsection{Renormalization of $\hat\Sigma(k,\lambda)_{11}$} 
Now we consider the renormalization of $\hat\Sigma(k,\lambda)_{11}$, the non-local part of the diagonal term of the self-energy function. Since the cancellation between tadpoles and counter-terms are exact, we assume that the self-energy function $\hat\Sigma_{11}$ is tadpole free. Rewrite the counter-term and the self-energy in the multi-scale representation as
\bea
[\hat\nu(\lambda,\bk)]_{11}&:=&\big[\hat\nu^{\le r_{max}}(\lambda,\bk)\big]_{11}=\sum_{r=0}^{r_{max}}\sum_{\sigma=s^{(a)},s^{(b)}}\big[\hat\nu_{s^{(a)},s^{(b)}}^r(\lambda,\bk)\big]_{11},\\
\big[\hat\nu_{s^{(a)},s^{(b)}}^r(\lambda,\bk)\big]_{11}&=&\hat\nu(\lambda)(\bk)_{11}\chi_j(4k_0^2+e^2(\bk,1))\cdot v_{s^{(a)}}[t^{(a)}(\bk)]
\cdot v_{s^{(b)}}[t^{(b)}(\bk)],\nn
\eea
and
\be
\big[\hat\Sigma(k,\hat\nu,\lambda )\big]_{11}:=\big[\Sigma^{\le r_{max}}(k,\hat\nu^{\le r_{max}},\lambda )\big]_{11}=\sum_{r=0}^{r_{max}}\sum_{\sigma=(s^{(a)},s^{(b)})}\big[\hat\Sigma_{s^{(a)},s^{(b)}}^r(k,\hat\nu^{\le (r)},\lambda )\big]_{11},
\ee
\be\big[\hat\Sigma_{s^{(a)},s^{(b)}}^r(k,\hat\nu^{\le (r)},\lambda )\big]_{11}=\big[\hat\Sigma(k,\hat\nu^{\le (r)},\lambda )\big]_{11}\chi_j(4p_0^2+e^2(\bk,1))\cdot v_{s^{(a)}}[t^{(a)}(\bk)]
\cdot v_{s^{(b)}}[t^{(b)}(\bk)].\nn
\ee
The localization for the self-energy is defined as: 
\be\big[\tau \hat\Sigma_{s^{(a)},s^{(b)}}^{r}\big(k,\hat\nu^{\le r},\lambda \big)\big]_{11}=
\big[\hat\Sigma_{s^{(a)},s^{(b)}}^{r}\big((2\pi T,P_F(\bk)_{s^{(a)},s^{(b)}}),\hat\nu^{\le r},\lambda \big)\big]_{11},
\ee
in which $P_F(\bk)_{s^{(a)},s^{(b)}}\in\cF\cap \Delta^r_{{s^{(a)},s^{(b)}}}$ is a projection of the vector $\bk\in\Delta^r_{{s^{(a)},s^{(b)}}}$ on the Fermi surface. 
To perform the renormalization, at each scale $r$ we move the counter-term $\hat\nu^r$ from the interaction potential to the covariance, so that the localized self-energy term $\hat\Sigma^r$ can be compensated. The renormalization condition becomes:
\be\label{rs1}
\big[\hat\Sigma_{s^{(a)},s^{(b)}}^{r}\big((2\pi T,P_F(\bk)_{s^{(a)},s^{(b)}}),\hat\nu^{\le r},\lambda \big)\big]_{11}+
\big[\hat\nu^{r}_{s^{(a)},s^{(b)}}(P_F(\bk)_{s^{(a)},s^{(b)}},\lambda)\big]_{11}=0.
\ee
%\be\label{rs1}
%\hat\Sigma_{s^{(a)},s^{(b)}}^{r}\big[(2\pi T,P_F(\bk))_{s^{(a)},s^{(b)}},\hat\nu^{\le (r-1)},\lambda \big]+
%\hat\nu^{r}_{s^{(a)},s^{(b)}}(P_F(\bk)_{s^{(a)},s^{(b)}},\lambda)=0.
%\ee
%%%%%%%%%%%%%%%%%%%%%%
When the external momentum is not restricted on the Fermi surface, the compensation between the two terms needs not to be exact, due to the non-locality of the proper self-energy $\hat\Sigma^r(k)$ and the counter-term $\hat \nu^r(\bk)$. Then the renormalization is defined as:
\be\label{rs2}
\big[\hat\Sigma_{s^{(a)},s^{(b)}}^r((k_0,\bk),\hat\nu^{\le (r)},\lambda )\big]_{11}+\big[\hat\nu^{r}_{s^{(a)},s^{(b)}}(\bk,\l)\big]_{11}=\big[\hat R\hat\Sigma_{s^{(a)},s^{(b)}}^{r}((k_0,\bk),\hat\nu^{\le (r+1)},\lambda )\big]_{11},
\ee
%%%%%%%%%%%%%%%%%%%%%%%%%%%%%%%%%%%%
in which the remainder term $\hat R\hat\Sigma_{s^{(a)},s^{(b)}}^{r}((k_0,\bk),\hat\nu^{\le (r+1)},\lambda )$ is bounded by (cf. \eqref{rmd11}-\eqref{rmd13}) $\g^{-\delta^r}\Vert \hat\Sigma_{s^{(a)},s^{(b)}}^{r+1}(k,\hat\nu^{\le (r+1)},\lambda )\Vert_{L^\infty}$,
with
\be\label{rmd14}
\g^{-\delta^r}=\max \{\g^{-(r_e-r_r)},  \g^{-[(s_{j_e}^{(a)}+s_{j_e}^{(b)})-(s_{j_r}^{(a)}+s_{j_r}^{(b)})]}\}<1.\ee 
%%%%%%%%%%%%%%%%%%%%%%%%%%%%%%%%%%%%%%%%%%%%%%%%%%%%%%%%%%%%%%%%%%
%%%%%%%%%%%%%%%%%%%%%%%%%%%%%%%%%%%%%%%%%%%%%%%%%%%%%%%%%%%%%%%%%%%%%%%%
From the renormalization conditions \eqref{rs1} and \eqref{rs2}, we have:
\be\label{rmd15}
\Vert[\hat\nu(\bk,\lambda)]_{11}\Vert:=\sup_{\bk}|[\hat\nu(\bk,\lambda)]_{11}|\le \sup_{\bk}\sum_{r=0}^{r_{max}}|[(1+\hat R)\hat\Sigma^r(\bk,\lambda)]_{11}|\le 2\sup_{\bk}\sum_{r=0}^{r_{max}}|[\hat\Sigma^r(\bk,\lambda)]_{11}|.\nn
\ee
%%%%%%%%%%%%%%%%%%%%%%%%%%%%%%%%
\subsubsection{Renormalization of $\hat\Sigma(k,\lambda)_{12}$} 
By \eqref{rncd4} or \eqref{rncd5}, the renormalization condition for $\hat\Sigma(k,\lambda)_{12}$
in the multi-scale representation can be formulated as:
\be\label{rs2a}
\big[\hat\Sigma_{s^{(a)},s^{(b)}}^r((k_0,\bk),\hat\nu^{\le (r)},\lambda )\big]_{12}+\big[\hat\nu^{r}_{s^{(a)},s^{(b)}}(\bk,\l)\big]_{12}=\big[\hat R\hat\Sigma_{s^{(a)},s^{(b)}}^{r}((k_0,\bk),\hat\nu^{\le (r+1)},\lambda )\big]_{12}.\nn
\ee
%%%%%%%%%%%%%%%%%%%%%%%%%%%%%%%%%%%%
in which $\big[\hat R\hat\Sigma_{s^{(a)},s^{(b)}}^{r}((k_0,\bk),\hat\nu^{\le (r+1)},\lambda )\big]_{12}$ is
the remainder term. By Proposition \ref{mat0}, it has the same upper bound as \eqref{rmd14}, up to some numerical constant. Repeating the analysis in the last subsection we have
\be\label{rmd15a}
\Vert[\hat\nu(\bk,\lambda)]_{12}\Vert:=\sup_{\bk}|[\hat\nu(\bk,\lambda)]_{12}|\le 2\sup_{\bk}\sum_{r=0}^{r_{max}}|[\hat\Sigma^r(\bk,\lambda)]_{12}|.
\ee
Combining these results, we have:
\begin{theorem}\label{flownu}
There exists a positive constant $K$ independent of the scale indices, such that
\be\label{bdbare1}
\Vert\big[\nu^{\le r_{max}}(\bk,\lambda)\big]_{\a\a'}\Vert\le K|\lambda|,\ \a,\a'=1,2,\  \forall\ \lambda\in\RR_T.
\ee
\end{theorem}
Remark that, in order to prove this theorem, it is enough to prove such upper bound for $\big[\hat\Sigma^{\le r_{max}}\big]_{11}$, since by Proposition \ref{mat0}, the upper bound for $\big[\hat\Sigma^{\le r_{max}}\big]_{12}$ is the same, up to some numerical constant. This theorem will be proved in Section \ref{multiarch}. 
%%%%%%%%%%%%%%%%%%%%%%%%%%%%%%%%%%%%%%%%%%%%%%%%%%%%
\begin{remark}
Remark that, in order to obtain the upper bounds for the $2$-point Schwinger function and self-energy function, we have to consider not only the biped trees $\cG^{r_{max}}_\cB$, but also the full Gallavotti-Nicol\`o tree structure. However, since the contributions from the quadrupeds and convergent ones are convergent, which only set constraints to the analytic domain of the coupling constants, we can safely drop these contributions but take into account the fact that the analytic domain for the biped Schwinger functions can't be larger than the one for the quadruped Schwinger functions. 
\end{remark}
%%%%%%%%%%%%%%%%%%%%%%%%%%%
%%%%%%%%%%%%%%%%%%%%%%%%%%%%%%%%%%%%%%%%%%%%%%%%%%%%%%%%%%%%%%%%%%%%%%%%%%%%
\section{Construction of the Self-energy function}\label{multiarch}
In this section we shall establish the optimal upper bounds for the self-energy function and its derivatives w.r.t. the external momenta. The main tool for the construction of the self-energy function is the multi-arch expansion for the determinant, from which we can obtain perturbation terms labeled by the 1PI graphs between any {\it two external vertices} of a graph, without generating any divergent combinatorial factor. In order to obtain the optimal bounds for the self-energy function, we have to perform a second multi-arch expansion on top of the first one. The resulting graphs are now {\it two-particle irreducible} between the two external vertices, which means that there are two line-disjoint paths between the two external paths, namely, one can't disconnect this graph by deleting two paths between the external vertices. Then by Menger's theorem (cf. eg. \cite{graph}), there exists {\it a third} line-disjoint path joining the two external vertices. Among the three paths, one can choose two line-and-vertex disjoint paths, from which we can obtain the optimal power counting. The union of the two paths is called a {\it ring} and the propagators in the ring are called the {\it ring propagators}. The optimal upper bound for the self-energy is obtained by integrating out the ring propagators as well as summation over the corresponding scale and sector indices. This part follows closely \cite{AMR1}. Some technical details are omitted if they can be found in \cite{AMR1} for a similar setting.
%%%%%%%%%%%%%%%%%%%%%%%%%%%%%%%%%%%%%%%%%%%%%%%%%
\subsection{The multi-arch expansions}
\begin{figure}[htp]
\centering
\includegraphics[width=.8\textwidth]{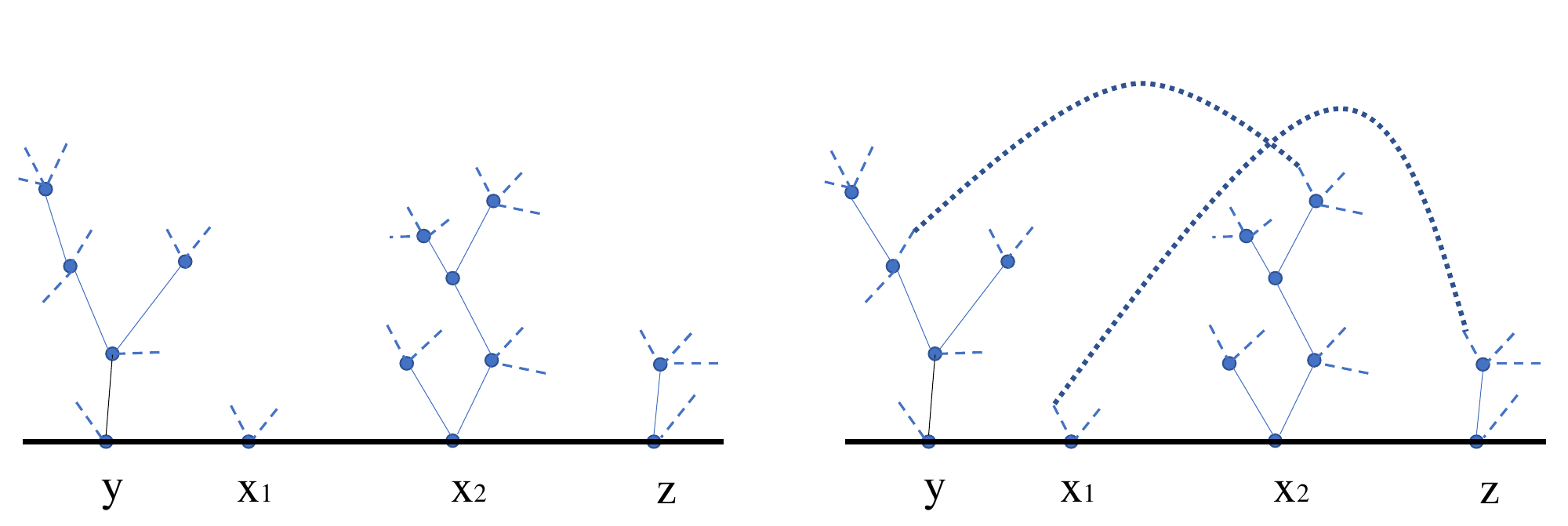}
\caption{\label{mtarch}
An illustration of the multi-arch expansion. The graph on the l.h.s. is a tree graph with $4$ branches, in which the dash lines are the half-lines, corresponding to the unexpanded matrix elements in the determinant. The graph on the r.h.s. is the one-particle irreducible graph constructed from the tree graph, by extracting two arches (the dotted lines) from the determinant.}
\end{figure}

Let $\cT$ be a spanning tree connecting the $n+2$ vertices $\{y,z,x_1\cdots,x_n\}$. The integrand of the connected two-point function labeled by $\cT$ is
\be\label{main0}
F(\{C_\ell\}_\cT, \{C(f_i,g_j)\})=\Big[\prod_{\ell\in\cT}C_{\s(\ell)}(f(\ell),g(\ell))\Big]\ \det(\{C(f_i,g_j)\})_{left,\cT},
\ee
in which we keep the matrix form of the propagators $C$ (cf. \eqref{prop2qx}). Let $P(y,z,\cT)$ be the unique path in $\cT$ such that $y$ and $z$ are the two ends of the path. Suppose that there are $p+1$ vertices in the path $P(y,z,\cT)$, $p\le n+1$. Then we can label each vertex in the path with an integer, starting from the label $0$ for the vertex $y$ in an increasing order towards $p+1$, which is the label for the vertex $z$. Let $\mathfrak{B}_i$ be a branch in $\cT$ at the vertex $i$, $1\le i\le p+1$, which is defined as the
subtree in $\cT$ whose root is the vertex $i$. See Figure \ref{mtarch} for an illustration of a tree graph with $4$ branches rooted at the four vertices $y, x_1,x_2$ and $z$.
We fix two half lines, also called the external fields, each is attached to one end vertex. Since each tree line in $P(y,z,\cT)$ contracts $2$ fields, there are $2(n+2)$ fields to be contracted from the determinant $\det_{left}$. We also call these fields the {\it remaining fields}, and denote the set of the remaining fields by $\mathfrak{F}_{left}$. A packet $\mathfrak{F}_i$ is defined as the set of the remaining fields restricted to a branch, $\mathfrak{B}_i$. By definition we have: $\mathfrak{F}_i\cap\mathfrak{F}_j=\emptyset$ for $i\neq j$, and $\mathfrak{F}_{left,\cT}=\mathfrak{F}_1\sqcup\cdots\sqcup\mathfrak{F}_p$, in which $\sqcup$ means disjoint union. Among all pairs of fields and anti-fields, we select the pair which has a contraction between an element of $\mF_1$ and an element of $\sqcup_{k=2}^{p}\mF_k$, through an explicit Taylor expansion with interpolating parameter $s_1$, as follows. Let $\{C(f_i,g_j)\}$ be the loop propagator in the remaining determinant for any loop line $\{(f_i,g_j)\}$, define \bea
C(f_i,g_j)(s_1)&:=&s_1C(f_i,g_j)\quad {\rm if}\ f_{1}\in \mF_1, g_{1}\notin \mF_1,\\
&:=&C(f_i,g_j)\quad\quad {\rm otherwise}.
\eea
We have
\bea
&&\det(\{C(f_i,g_j)\})_{left,\cT}=\det(\{C(f_i,g_j)(s_1)\})_{left,\cT}\ \big|_{s_1=1}\\
&&=\det(\{C(f_i,g_j)(s_1)\})_{left,\cT}\ \big|_{s_1=0}
+\int_0^1 ds_1\frac{d}{ds_1}\det(\{C(f_i,g_j)(s_1)\})_{left,\cT}\nn,
\eea
in which the first term means that there is no loop line connecting $\mF_1$ to its complement. The second term means that there is a contraction between a half-line $f_1\in\mF_1$ and $g_1\in\mF_{k_1}$, with $1\le k_1\le p$. Graphically this means that we add to $\cT$ an explicit line $\ell_1=(f_1, g_1)$, which joins the packet $\mF_1$ to $\mF_{k_1}$. The newly added line $\ell_1$ is called a loop line or an arch. $\mF_1$ is called the starting packet of the contraction, and the index $1$ is called the starting index of $\ell_1$. Similarly $\mF_{k_1}$ is called the arriving packet of the contraction and $k_1$ is called the arriving index of $\ell_1$. These definitions can be generalized to an arbitrary contraction between pairs of packets and the associated arches. The new graph $\cT\cup\{\ell_1\}$ becomes 1-PI between the vertices $y$ and $x_{k_1}$. If $k_1=p$, then the whole 1PI graph is generated and we are done. Otherwise we test whether there is a contraction between an element of $\sqcup_{k=1}^{k_1}\mF_k$ and its complement, by introducing a second interpolation parameter $s_2$ to the propagator. Define the interpolated propagator as:
\bea\label{mul1}
C(f_i,g_j)(s_1, s_2)&:=&s_2C(f_i,g_j)(s_1)\quad {\rm if}\quad f_{i}\in\sqcup_1^{k_1} \mF_{k},\  g_{j}\in\sqcup_{k_1+1}^{p}\mF_k\ ,\\
&:=&C(f_i,g_j)(s_1)\quad {\rm otherwise}.
\eea
We have:
\bea
&&\det(\{C(f_i,g_j)(s_1)\})_{1,left,\cT}=\det(\{C(f_i,g_j)(s_1,s_2)\})_{1,left,\cT}\})|_{s_2=1}\nn\\
&&=\det(\{C(f_i,g_j)(s_1,s_2)\})_{1,left,\cT}\})|_{s_2=0}\nn\\
&&\quad\quad\quad+\int_0^1 ds_2\frac{d}{ds_2}\det(\{C(f_i,g_j)(s_1,s_2)\})_{1,left,\cT}\}).
\eea
Again, the first term means that the block $\sqcup_{i=1}^{k_1}\mF_{k}$ is not linked to its complement by any arch. The second term can be written as
\bea\label{link1}
&&\int_0^1 ds_2\frac{d}{ds_2}\det(\{C(f_i,g_j)(s_1,s_2)\})_{1,left,\cT}\})\\
&&\quad=
\int_0^1 ds_2\ \frac{\partial}{\partial s_2}C(f_{i_1},g_{j_1})(s_1,s_2)\cdot \frac{\partial}{\partial C(f_{i_1},g_{j_1})}\det(\{C(f_i,g_j)(s_1,s_2)\})_{1,left,\cT}\}),\nn
\eea 
in which $f_{i_1}\in \sqcup_1^{k_1} \mF_{k}$, $g_{j_1}\in\sqcup_{k_1+1}^{p}\mF_k$, and we have
\bea\label{rprop0}
\frac{\partial}{\partial s_2}C(f_{i_1},g_{j_1})(s_1,s_2)&=&C(f_{i_1},g_{j_1})\quad\quad{\rm if}\quad f_{i_2}\in\sqcup_{i=2}^{k_1}\mF_{k}\nn\\
&=&s_1C(f_{i_1},g_{j_1})\quad{\rm if}\quad f_{i_2}\in\mF_{1}.
\eea
The graphical meaning is that there exists a contraction between the loop field $f_{i_1}$ and $g_{i_1}$, hence we add to $\cT$ another line $\ell_2=(f_{i_1},g_{i_1})$ joining the two packets. Now the new graph $\cT\cup\{\ell_1,\ell_2\}$ becomes 1-PI between the vertices $y$ and $x_{k_2}$. Similarly, for an arch ends at the packet $\mF_{k_u}$, the corresponding interpolated propagator is defined as
\be\label{rprop}
C(f_{u}, g_u)(s_1,\cdots, s_u)=s_uC(f_u,g_u)(s_1,\cdots, s_{u-1}).
\ee
We continue this interpolation process until the graph becomes 1-PI in the $y-z$ channel. 
Suppose that we have generated $m$ arches to form a 1PI graph, the set of arches 
\bea
&&\Big\{ \ell_1=(f_1,g_1),\cdots,\ell_m=(f_m,g_m)\ \Big|\ f_1\in\mF_1, g_1\in\mF_{k_1};\ f_2\in\sqcup_{u=1}^{k_1}\mF_u, g_2\in\sqcup_{u=k_{1}+1}^{k_2}{\mF_u};\nn\\
&&\quad\quad\quad \cdots ;\ 
f_m\in\sqcup_{u=1}^{k_{m-1}}\mF_u,\ g_m\in\mF_{k_m}=\mF_{p};\  k_1\le\cdots\le k_m;\ m\le p\ \Big\}
\eea
is called an {\it m-arches system}.

Finally, we have the following expression for the determinant:
\bea\label{main1}
&&\det\big(\{C(f_i,g_j)\}\big)_{left,\cT}=\sum_{\substack{{\rm m-arch-systems}\\(f_1,g_1),\cdots, (f_m,g_m)\\ {\rm with}\ m\le p}}\ \int_0^1 ds_1\cdots \int_0^1 ds_m  \nn\\
&&\Bigg[\frac{\partial}{\partial{s_1}}C(f_1,g_1)(s_1)\cdot\frac{\partial}{\partial{s_2}}C(f_2,g_2)(s_1,s_2)\cdots\frac{\partial}{\partial{s_m}}C(f_m,g_m)(s_1,s_2,\cdots,s_m)\cdot\nn\\
&&\quad\quad\cdot\frac{\partial^m det_{left,\cT}}{\prod_{u=1}^m \partial C(f_u, g_u)}\Big(\{s_u\}\Big)\Bigg]\ ,
\eea
where the sum runs over all the $m$-arch systems with $p$ vertices. It is useful to have a more explicit expression for the second line of the above formula. We have:
%%%%%%%%%%%%%%%%%%%%%%%%%%%%%%%%%%%
\begin{proposition}\label{prodpg}
Let $\ell_u$ be a loop line in an $m$-arch system introduced above. Let $q_u$ be the number of loop lines that fly over $\ell_u$, namely those loop lines whose starting indices are smaller than or equal to that of $\ell_u$ while whose arriving indices are greater than that of $\ell_u$. Let
$\prod_{u=1}^m C(f_u,g_u)(s_1,\cdots, s_{u-1})$ be the compact form corresponding to the second line of formula 
\eqref{main1}, then we have
\bea\label{indu0}
&&\prod_{u=1}^m C(f_u,g_u)(s_1,\cdots, s_{u-1}):=\prod_{u=1}^m\partial_{s_u}C(f_u,g_u)(s_1,\cdots,s_u)\nn\\
&&\quad=\big[\ \prod_{u=1}^m C(f_u,g_u)\ \big]\cdot \big[\ \prod_{u=1}^m
s_u^{q_u}\ \big]\ .
\eea
\end{proposition}

\begin{proof}
This proposition can be proved by induction, using the definition of the interpolated propagators (c.f. \eqref{rprop}, \eqref{rprop0}). Indeed, if the indices of the successive loop lines in a $m$-arch system are strictly increasing, then there is no interpolation parameter in the product
$\prod_{u=1}^m\partial_{s_u}C(f_u,g_u)(s_1,\cdots,s_u)$; A factor $s_u^{q_u}$ will be generated when there are exactly $q_u$ loop lines which completely fly over $\ell_u$. So we proved this proposition.
\end{proof}
Remark that, since each interpolations is performed between a subset of packets and its complement, the final interpolated covariance is a convex combination of block-diagonal covariances with positive coefficients. Hence the remaining matrix in \eqref{main1} is still positive and its determinant can be bounded by Gram's inequality.

Now we prove that no factorials will be generated in the multi-arch expansions, which is not trivial: while the arriving index for a successive arch is strictly increasing, once the arriving field is fixed, a factorial might be generated when choosing the departure fields. It has been proved in \cite{DR1,DR2,AMR1} that this factorial will not cause divergence, as it can be compensated by the integration over the interpolation parameters. We don't repeat the proof here but only collect some basic notions and results of \cite{DR1,DR2,AMR1}, for the reader's convenience. First of all, let's introduce some more notations concerning the multi-arch graphs.
\begin{definition}
Let $\cL_n=\{\ell_1,\cdots,\ell_n\}$ be a set of $n$ loop lines, $n\le m$, in an $m$-arch system such that the arriving indices of these lines are in the increasing order. The set $\cL_n$ is said to form a nesting system if the starting index of the last line $\ell_n$ is the lowest one among all the starting indices of the loop lines in $\cL_n$. In this case the loop lines $\{\ell_1,\cdots,\ell_{n-1}\}$ are said to be \emph{ineffective}, in that the graph remains 
1PI if we delete these loop lines. A loop line that is not ineffective is called an effective loop line.
%Define the number of loop lines that completely fly over a loop line $\ell_k$ be $q_k$. 
\end{definition}
%\begin{remark}
%Remark that a nesting system may contain a subset of loop lines that still form a nesting system.
%The nesting systems are indeed the source of the combinatorial factors.
%\end{remark}

By Proposition \ref{prodpg}, only the ineffective loop lines contribute to the interpolating factors in \eqref{indu0}. Let the number of loop lines that completely fly over a loop line $\ell_k$ be $q_k$, then the sum over all $m$-arch systems over $p$ vertices, which may result in combinatorial factors, should be weighted by the integral $\Big(\prod_{u=1}^m \int_0^1 s_u\Big)\ 
\prod_{u=1}^m s_u^{q_{u}}$. We have:
\begin{lemma}[cf. \cite{AMR1}, Lemma VI.1]
Let $n\ge1$ be the number of vertices of the 1-PI graph formed in the m-arch expansions. There exist some numerical constants $c_2$ and $K$, which are independent of the scale indices, such that:
\be
\sum_{m=1}^p\sum_{\substack{{\rm m-arch-systems}\\(f_1,g_1),\cdots, (f_m,g_m)\\ {\rm with}\ m\le p}}\ \Big(\prod_{u=1}^m \int_0^1 s_u\Big)\ 
\prod_{u=1}^m s_u^{q_{u}}\le c_2 K^n.
\ee
\end{lemma}
Remark that one can always optimize an $m$-arch system by choosing a {\it minimal} $m$-arch subsystem system which contains minimal number of loop lines whereas all the ineffective loop lines are deleted, and we assume that the $m$-arch systems are always minimal. Let $\Sigma(y,z,\l)$ be the $2\times2$ matrix in which $\Sigma(y,z,\l)_{\a\a'}$ is the Fourier transform of $\cE(k,\l)_{\a\a'}$,
$\a,\a'=1,2$, then in the multi-scale representation we have:
%%%%%%%
%%%%%%%%%%%%%%%%%%%
%%%%%%%%%%%%%%%
\begin{lemma}
The amplitude of the self-energy is given by:
\bea\label{sfe}
&&\Sigma^{\le r_{max}} (y,z,\l)= \sum_{n=2}^\infty\sum_{n_1+n_2+n_3=n} \frac{\lambda^{n_1}\delta\mu^{n_2}\hat\nu^{n_3}}{n_1!n_2!n_3!} \int_{\Lambda^n} d^3x_1 ... d^3x_n\sum_{\{ \underline\t \}}\sum_{\cG^{r_{max}}_\cB} \sum_{\text{external fields} \atop \mathcal{EB}}   \nn\\
&&\sum_{\text{spanning trees} \mathcal{T} }\sum_{\{ \sigma \}}
\sum_{{m-{\rm arch\ systems} \atop 
\big( (f_1,g_1,...,(f_m,g_m))\big) } \atop
{\rm with} \  m \leq p }
\left( \prod_{\ell \in \mathcal{T}} \int_0^1 dw_\ell \right) \left( \prod_{r = 1}^m \int_0^1 ds_r  \right)
\left( \prod_{\ell \in \mathcal{T}} C_{\sigma(\ell)} (f_\ell,g_\ell)\right)\nn
\\
&&\quad\left( \prod_{u=1}^m C(f_u,g_u) (s_1,...,s_{u-1})\right)
 \frac{\partial^m \det_{\text{left}, \mathcal{T}}}{\prod_{r=1}^m \partial C(f_u,g_u)} 
\big( \{ w_\ell\}, \{ s_u\}\big) \ .
\eea
\end{lemma}
%\vskip.3cm
\begin{remark}
Following Remark \ref{ctbd}, in the rest of this section we can simply replaced all the counter-terms by constants, except in the parts concerning the renormalization.
\end{remark}
Now we introduce a second multi-arch expansion, which completes the 1PI graphs into the 2PI graphs and one-vertex irreducible graphs (graphs that remain connected after deleting one vertex). We have:
\begin{lemma}
The amplitude for the corresponding 2PI biped graphs reads
\bea
&&\Sigma (y,z,\l)^{\le r_{max}}=\sum_{n=2}^\infty \frac{\lambda^{n}}{n!} \int_{(\Lambda_\b)^n} d^3x_1 ... d^3x_n 
\sum_{\{ \underline\t  \}} \sum_{\text{biped structures} \atop \mathcal{B}}\sum_{\text{external fields} \atop \mathcal{EB}}\nn\\
&&\quad\quad\sum_{\cG^{r_{max}}_\cB} \sum_{\text{spanning trees}\ \mathcal{T}} \sum_{\{\sigma \}}\sum_{ m-{\rm arch\ systems} \atop \bigl( (f_1,g_1), ... (f_m,g_m) \bigr)}
\sum_{m'-{\rm arch\ systems} \atop \bigl( (f'_1,g'_1), ... (f'_{m'},g'_{m'}) \bigr)}\nn\\
&&\quad\quad\left( \prod_{\ell \in \mathcal{T}} \int_0^1 dw_\ell \right) \left( \prod_{\ell \in \mathcal{T}} C_{\sigma(\ell)} (f_\ell,g_\ell)\right)\left( \prod_{u=1}^m \int_0^1 ds_u \right)\left( \prod_{u'=1}^{m' }\int_0^1 ds'_{u'} \right)\nn \\
&&\quad\quad\left( \prod_{u=1}^m C(f_u,g_u) (s_1,...,s_{u-1})\right) \left( \prod_{u' = 1}^{m'} C({f'}_{u'},{g'}_{u'}) (s'_1,...,s'_{u'-1})\right)\nn\\
&&\quad\quad\frac{\partial^{m+m'} \det_{\text{left}, \mathcal{T}}}{\prod_{u=1}^m \partial C(f_u,g_u)\prod_{u'=1}^{m'} \partial C(f'_{u'},g'_{u'}) }\big( \{ w_\ell\}, \{ s_u\} ,  \{ s'_{u'} \}  \big),
\eea
where we have summed over all the first multi-arch systems with $m$ loop lines and
the second multi-arch systems with $m'$ loop lines. The underlying graphs are two-line irreducible as well as one vertex irreducible.
\end{lemma}
\begin{proof}
The construction of the $2$-PI perturbation series from the 1PI one is analogous to the construction of the 1PI terms from the connected terms, except that the total ordering in the tree graph is lost; one has to consider the {\it partial ordering} of the various branches. This construction has discussed in great detail in \cite{AMR1}, so we don't repeat it here. Remark that the second multi-arch expansion respects again the positivity of the interpolated propagator at any stage, so the remaining determinant still satisfies the Gram's inequality.
\end{proof}
%%%%%%%%%%%%%%%%%%%%%%%%%%%%%%%%%%
%%%%%%%%%%%%%%%%%%%%%%%%%%%%%%%%%%%%%%%%%%%%%%%%%%%%%%%%%%%%%%%%%%%%%%%%%%%%%%%%%%%%%
\subsection{The ring sectors and the power-counting}
\begin{figure}[htp]
\centering
\includegraphics[width=.4\textwidth]{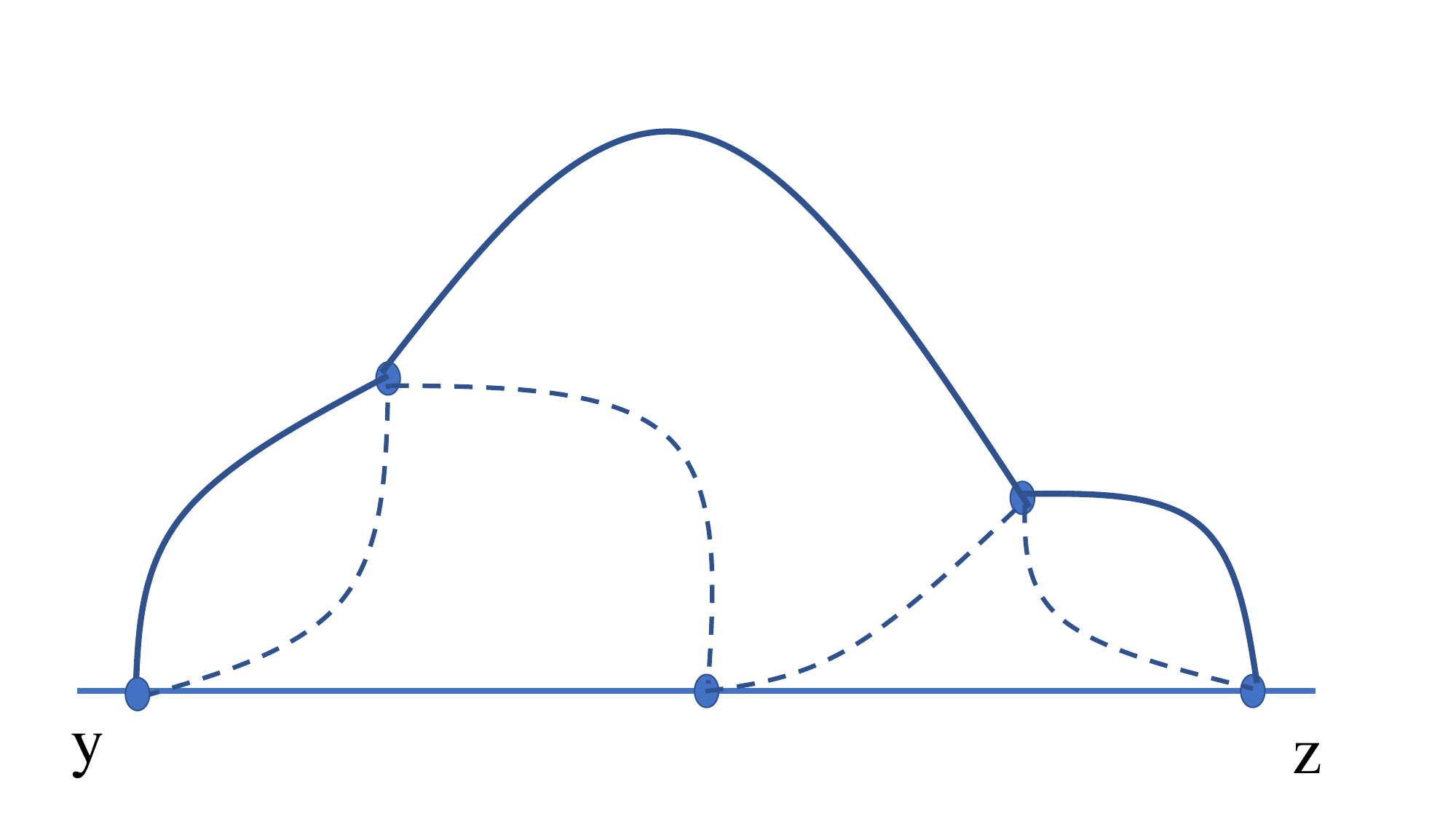}
\caption{\label{rin}
Ring structure in a $2$PI graph which connects two end vertices $y$ and $z$. The thick lines are the ring propagators and dash lines are the third path in the $2$PI graph.}
\end{figure}

Let $G=\cT\cup\cL$ be a 2PI graph generated by the two-level multi-arch expansions. Menger's theorem ensures that, any such graph $G$ has three line-disjoint independent paths and two internally vertex-disjoint paths joining the two external vertices \cite{AMR1} of $G$. Since different propagators in these paths may have different scale properties, in order to obtain the optimal bounds for the self-energy we need to choose the optimal integration paths from which we can obtain the best convergence factors. The optimal paths are called a ring structure. See Figure \ref{rin} for an illustration.
\begin{definition}
A ring $R$ is a set of two paths $P_{R,1}$, $P_{R,2}$ in $\cL\cup\cT$ connecting the two vertices $y$ and $z$ and satisfies the following conditions: Firstly, the two paths in $R$ don't have any intersection on the paths or on the vertices, except on the two external vertices $y$ and $z$. Secondly, let $b$ be any node in the biped tree $\cG^{r_{max}}_\cB$, then at least two external fields of $b$ are not contained in the ring. 
\end{definition}
%%%%%%%%%%%%%%%%%%%%%%%
%%%%%%%%%%%%%%%%%%%%%%%
The scale indices and the sector indices for the ring propagators are defined as follows. Let $P_{R,1}$ and $P_{R,1}$ be two paths in the ring and $n$ be the labeling of the ring propagators in the two paths. Let $r_T$ be the first scale at which $y$ and $z$ fall into a common connected component in the GN tree and $r_R$ be the first scale at which the ring connects $y$ and $z$, we have $r_R=\min_{a=1,2}\max_{n\in P_{R,a}}r(n)$ and $r_T\le r_R\le r_{max}$. In the same way, we can define the sectors indices for the ring propagators: 
\be
s_{+,R}:=\min_{a=1,2}s_{+,R,a}:=\min\max_{n\in P_{R,a}}s_+(n),\  \ s_{-,R}:=\min_{a=1,2}s_{+,R,a}:=\min\max_{n\in P_{R,a}}s_-(n),\nn
\ee
which are greater than the sector indices of the tree propagators. The corresponding sectors are called the {\it ring} sectors. The scale index for the ring propagators is defined by $j_{R,T}=\min\{j_R, j_\cT\}$, in which
$j_\cT=\max_{n\in P(y,z,\cT)}j_n$, $P(y,z,\cT)$ is the unique path in $\cT$ connecting $y$ and $z$, and $j_R=\min_{a=1,2}\max_{n\in P_{R,a}}j(n)$. Finally, the $r$-index for the ring sector is defined as
\be
r_{R,\cT}=\frac{j_{R,T}+s_{+,R}+s_{-,R}}{2},
\ee
and the rescaled distance for this ring sector is defined as $d_{R,T}(y,z)=d_{j_{R,T},s_{+,R},s_{-,R}}(y,z)$. With all these preparations, we can prove the following upper bound for the self-energy function:
%%%%%%%%%%%%%%%%%%%%%%%%%%%%%%%%
\begin{theorem}\label{maina}
For any $0\le r\le r_{max}$ and $\lambda\in\RR_T$, there exists a positive constant $K$, which is independent of $r$ and $\l$, such that:
\be\label{x2pta}
\Vert\big[\Sigma(y,z,\l)^{\le r}\big]_{11}\Vert_{L^\infty}\le \lambda^2 r^2\sup_{\sigma\in\sigma_{\cG^r} K\g^{-3r(\sigma)}}e^{-c[d_{j,\s}(y,z)]^\a},
\ee
\be\label{x2pt}
|z^{(a)}-y^{(a)}||z^{(b)}-y^{(b)}|\Vert\big[\Sigma(y,z,\l)^{\le r}\big]_{11}\Vert_{L^\infty}\le \lambda^2 r^2\sup_{\sigma\in\sigma_{\cG^r}} K\g^{-2r(\sigma)}e^{-c[d_{j,\s}(y,z)]^\a},
\ee
\be\label{x2ptc}
|y_0-z_0|\Vert\big[\Sigma_2(y,z,\l)^{\le r}\big]_{11}\Vert_{L^\infty}\le \lambda^2 r^2\sup_{\sigma\in\sigma_{\cG^r}} K\g^{-2r(\sigma)}e^{-c[d_{j,\s}(y,z)]^\a},
\ee
in which $\cG^r$ is the GN tree whose root is at scale $r$ and $\sigma_{\cG^r}$ is the set of sector indices that is compatible with $\cG^r$, $[d_{j,\s}(y,z)]^\a$ (cf. \eqref{dist0}) characterizes the decay of the propagator in position spaces. 
%The upper bounds presented in \eqref{x2pta}, \eqref{x2pt} and \eqref{x2ptc} are optimal.
\end{theorem}
Before proving this theorem, we consider the following lemma concerning sector counting for the biped graphs.
\begin{lemma}[Sector counting lemma for bipeds]\label{secbi}
Let $b_r$ be a 2PI biped with root scale index $r\in[0, r_{max}]$ and contains $n+2$ vertices. There exists a positive constant $C_1$, independent of the scale index $r$, such that the summation over all the sector indices is bounded by $C_1^{n+2}r^{2n+1}$. 
\end{lemma}
\begin{proof}
Let $b_r$ be a 2PI biped and $\cT_b=b_{r}\cap\cT$ be the set of tree lines in $b_r$. The two external fields are fixed. They have identical sector indices, by conservation of momentum. We choose a root field $f_r$ at each vertex, and among all the root fields, we choose the one with maximal scale index as the one for the whole biped. By conservation of momentum there can be at most $2n+1$ independent sectors to be summed. Since summing over each pair of sector indices is bounded by $r$, the total summation is the bound $C_1^n r^{2n+1}$, for some positive constant $C_1$. This concludes the lemma.
\end{proof}
%Remark that the bound proved in the above lemma is not optimal, as the number of sectors to be summed can be further reduced by taking into account the useful arches in the 2PI biped graph. But this bound is enough for our purpose. 

\begin{proof}[Proof of Theorem \ref{maina}]
This theorem for a similar setting has been proved in \cite{AMR1}, Section $VIII$. We only sketch the main idea for the proof and ask the interested readers to consult \cite{AMR1} for more details.
First of all, by integrating out the weakening factors for the tree expansions and multi-arch expansions, we have:
\bea
\Vert\big[\Sigma^{\le r}(y,z)\big]_{11}\Vert_{L^\infty}\le\sum_{n=2}^\infty\frac{(K\l)^{n}}{n!}\sum_{\{\underline\tau\}}\sum_{\cG^{r}_{\cB}}
\sum_{ \mathcal{EB}}  \sum_{ \mathcal{T} } \sum_{R}\sum_{\{ \sigma \}}\prod_p\chi_p(\sigma)\cI_{1,n}(y,z)\cI_{2,n}(y,z,x_{p,\pm}),\nn
\eea
in which $K$ is some positive constant that is independent of $r$ and $\l$.
\be
\cI_{2,n}(y,z,x_{p,\pm})=\int\prod_{v\in R,v\neq y,z} dx_{v,0}\prod_{v\notin R}d^3x_v
\prod_{f\notin R}\g^{-r_f/2-l_f/4}\prod_{p\in \cL}e^{-\frac{c}{2}[d_{j,\s(p)}(y,z)]^\a}\nn
\ee
is the factor in which we keep the position $y$, $z$ and the spatial positions of the ring vertices $x_{p,\pm}$ fixed but integrate out all the remaining positions (cf. Formula $VIII.86$, \cite{AMR1}). A fraction (one half) of the decaying factor from every loop line in $\cL$ and from the remaining determinant has been put here to compensate possible divergence from the integrations. And the factor
\be
\cI_{1,n}=\int\prod_{i=1}^p dx_{i,+}dx_{i,-}\prod_{k\in R}\g^{-(r+l/2)(k)}\prod_{p\in\cL}e^{-\frac{c}{2}[d_{j,\s(p)}(y,z)]^\a},
\ee
in which $x_i$ are the internal vertices other than $y$ and $z$ in the ring, contains the remaining terms and integrations. Then using the same techniques introduced in \cite{AMR1}, page 146, for the proof of Lemma $VIII.1$ and $VIII.2$, one can prove  that:
\be
\vert\cI_{2,n}\vert\le K_1^n \g^{-j_{\cT}}
\ee
and
\be
\vert\cI_{1,n}\vert\le K_2^p \g^{-s_{+,R,1}-s_{+,R,2}-s_{-,R,1}-s_{-,R,2}}e^{-\frac{c}{4}[d_{j,\s(p)}(y,z)]^\a}.
\ee
Combining these two factors and summing over all the tree structure $\cT$, the ring structure $R$ and the GN trees, we have
\be\label{finalsuma}
\Vert\big[\Sigma^{\le r}(y,z,\l)\big]_{11}\Vert_{L^\infty}\le \sum_n  \sum_{r'=0}^r\sum_{\{\sigma\}} C^n\lambda^{n} \g^{-j_{\cT}}\g^{-s_{+,R,1}-s_{+,R,2}-s_{-,R,1}-s_{-,R,2}}e^{-c'[d_{j,\s(p)}(y,z)]^\a},
\ee
for some positive constant $C$ and $c'$. For the sector indices in the exponential, we have: 
\be\label{finalsumb}
 \g^{-j_{\cT}}\g^{-s_{+,R,1}-s_{+,R,2}-s_{-,R,1}-s_{-,R,2}}\le\g^{-3/2(s_{+,R}+s_{-,R}+j_{R,\cT})}\le \g^{-3r_{R,T}}.
\ee
%%%%%%%%%%%%%%%%%%%%%%%%%%%%%%%%%%%%%%%%%%%%%%%%%%%%%%%%
Now we consider summation over the sector indices. For $n=N+2$ and using Lemma \ref{secbi}, we obtain
\bea\label{finalsum}
\sum_{N=0}^{\infty}  \sum_{r'=0}^r\sum_{\{\sigma\}} K_1^{N+2}\lambda^{N+2} \le
\sum_{N=0}^{\infty}  \sum_{r'=0}^rK_1^{N+2}C_1^{N}\lambda^{N+2} r'^{2N+1}\le 
\sum_{N=0}^{\infty} C_2^N(\lambda r^2)^N \l^2r^2,
\eea
for some positive constant $C_2$ depending on $K_1$ and $C_1$ but independent of the scale index and $\l$. Since $|\lambda r^2|\le |\lambda\log^2 T|\le C$ for $\lambda\in\RR_T$, summation over $N$
in \eqref{finalsum} is convergent provide that $C\cdot C_2<1$. This inequality can always be satisfied for $C$ small enough.
%%%%%%%%%%%%%%%%%%%%%
So we obtain:
\bea\label{taylorm1}
\Vert\big[\Sigma^{\le r}(y,z,\l)\big]_{11}\Vert_{L^\infty}&\le& \sum_{N=0}^{\infty} C_2^N(\lambda r^2)^N \l^2r^2\sup_{\sigma\in\sigma_{\cG^r}}\g^{-3r(\sigma)}e^{-c'[d_{j,\s}(y,z)]^\a}\\
&\le&
 K_3\lambda^2 r^2\sup_{\sigma\in\sigma_{\cG^r}}\g^{-3r(\sigma)}e^{-c'[d_{j,\s}(y,z)]^\a},\nn
\eea
for certain positive constants $K_3$. By choosing the ring structure, the convergence factors obtained from integrations over the spatial coordinates in $\cI_{1,n}$ and $\cI_{2,n}$ are optimal, hence the upper bounds we obtained in Theorem \ref{maina} are optimal.
Following exactly the same analysis for the derivatives of the self-energy, (see also \cite{AMR1}, pages 437-442) we can prove Formula \eqref{x2pt} and \eqref{x2ptc}. Hence we conclude Theorem \ref{maina}.
\end{proof}
Now we reformulate Theorem \ref{maina} in the momentum space, we have:
%%%%%%%%%%%%%%%%%%%%%%%%%%%%%%%%%%%%%%%%%%%%%%%%%%%%%%%%%%%%%%%%%%%%%%%%%%%%%%%%%%%%%5
\begin{theorem}[Bounds for the self-energy in the momentum space.]\label{mainb}
Consider $\big[\hat\Si^{r}(q,\l)\big]_{11}$ in which the external momentum $q=(k_0,q_+,q_-)$ is constrained to a sector with scale index $r$, and $\lambda\in\RR_T$. There exists a positive constant $K$, which is independent of $r$ and $\l$, such that:
\bea &&\sup_q|\big[\hat\Si^{r}(q,\l)\big]_{11}|\le K\l^2 r\g^{-r},\label{spa}
\\&& \sup_q\vert \Big[\frac{\partial}{\partial q_\mu } \hat\Si^{r} (q,\l)\Big]_{11} \vert\le K\lambda^2 r,\label{spb} 
\\&&\label{spc} \sup_q|\Big[ \frac{\partial^2}{\partial q_\mu \partial q_\nu}  \hat\Si^{r} (q,\l)\Big]_{11} | \le K\lambda^2 r  \g^{r}.
\eea
%These upper bounds are optimal.
\end{theorem}
%%%%%%%%%%%%%%%%%%%%%%%%%%%%%%%%
%%%%%%%%%%%%%%%%%%%%%%%%%%%%%%%%%%%%%%%%%%%%%%%%%%%%%%%%%%%%%%%
This theorem states that the self energy is uniformly $C^1$ in the external momentum for 
$\l\in\RR_T=\{\l\in\RRR,|\lambda|<C/|\log T|^2\}$ for some positive constant $C<1$, which is smaller than the one required by Salmhofer's criterion, which is $\{\l\in\RRR,\ |\lambda|<C/|\log T|\}$. What's more, for $r=r_{max}$, with $\g^{r_{max}}\sim \frac1T$, there exists some positive constant $K_1$, which is independent of the temperature and $\l$, such that
\be\sup_q|\Big[\frac{\partial^2}{\partial q_\mu \partial q_\nu}  \hat\Si^{r} (q,\l)\Big]_{11} |\le\frac{K_1\l^2}{T}.\ee
Hence $\hat\Si^{r}(q,\l)$ is not uniformly $C^2$ in $q$. This suggests that Salmhofer's criterion is violated and the ground state is not a Fermi liquid. We shall establish in the next section the following lower bound for the second derivative of the self-energy:
\be\label{lbd}\sup_q|\Big[\frac{\partial^2}{\partial q_\mu \partial q_\nu}  \hat\Si^{r} (q,\l)\Big]_{11} |\ge\frac{\tilde K_1\l^2}{T},\ee 
in which $\tilde K_1<K_1$ is a positive constant that is independent of $T$. With that lower bound we can conclude that the ground state of this model is not a Fermi liquid.

%%%%%%%%%%%%%%%%%%%%
\begin{remark}\label{rmkmain}
Since the self-energy function $\big[\hat\Sigma(q,\lambda)\big]_{11}$ is analytic in $\l$ for $\lambda\in\RR_T=\{\lambda\vert\ |\lambda|\log^2T\le C\}$, we can always choose the constant $C$ such that $\big[\hat\Sigma(q,\lambda)\big]_{11}$ can be Taylor expanded into the following convergent perturbation series:
\be\label{taylor2}
\big[\hat\Sigma(q,\lambda)\big]_{11}=\sum_{n=2}^\infty \lambda^n \big[\hat\Sigma_n(q,\lambda)\big]_{11},\ {\rm with}\ \lambda\in\RR_T,\ |\lambda\vert\cdot\Vert \hat\Sigma_n(q,\lambda)\Vert<0.1, \forall n\ge2, \nn
\ee
in which $\Vert \big[\hat\Sigma_n(q,\lambda)\big]_{11}\Vert:=\sup_q\vert \big[\hat\Sigma_n(q,\lambda)\big]_{11}\vert$.
Then from elementary mathematical analysis we know that
\be
\sum_{n=3}^\infty \lambda^n \Vert \big[\hat\Sigma_n(q,\lambda)\big]_{11}\Vert\le K_0\lambda^2\Vert\big[\hat\Sigma_2(q,\lambda)\big]_{11}\Vert,
\ee
for some positive constant $K_0>1$ that is independent of $\l$ and the scale indices. Here $\lambda^2\hat\Sigma_2(q,\lambda)$ is the amplitude of the lowest order perturbation term, namely, the amplitude of the sunset graph (see Figure \ref{sun} for an illustration). Therefore, in order to establish the lower bound for the self-energy and its derivatives, it is enough to study the corresponding quantities for the sunset graph.
\end{remark}
%%%%%%%%%%%%%%%%%%%%%%%%%%%%%%%5
\begin{proof}[Proof of Theorem \ref{mainb}]
Since $\big[\Sigma^{r}(y,z,\l)\big]_{11}$ is the Fourier transform of $\big[\hat\Sigma^{r}(q,\l)\big]_{11}$, we have
\be
\Vert\big[\Sigma^r(y,z)\big]_{11}\Vert_{L^\infty}=\Vert\int dq\ \big[\hat\Sigma(q,\l)\big]_{11} e^{ip(y-z)}\Vert_{L^\infty}=C \gamma^{-2r}\sup_q|\hat\Sigma^r(q)|,
\ee
for some positive constant $C$. By \eqref{finalsuma}, \eqref{finalsumb} and \eqref{finalsum},
we have
\be
\Vert\big[\Sigma^r(y,z,\l)\big]\Vert_{L^\infty}\le C\lambda^2 r \g^{-3r},
\ee
therefore
\be
\sup_q|\big[\hat\Sigma^r(q,\l)\big]_{11}|\le C\lambda^2 r \g^{-r}.
\ee
So we proved \eqref{spa}. Since any component of $q=(k_0,q_+,q_-)$ is bounded by $\g^{-r}$ when restricted to a sector of scale index $r$, the first order differentiation w.r.t. $q$ for the r.h.s. of \eqref{spa} gives the bound in \eqref{spb} and the second order differentiation gives the bound in \eqref{spc}. 
These bounds are optimal, since the bounds in \eqref{x2pta}-\eqref{x2ptc} are optimal.
\end{proof}
%%%%%%%%%%%%%%%%%%%%%%%%%%%%%%%%%%%%%%%%%%%%%%%%%%%%%%%%%%%%%%%%
\begin{remark}
Remark that what we have proved until now are the upper bounds for $\hat\Sigma_{11}(q,\l)$. 
Since all the matrix elements of the propagator can be bounded by some absolute constants (cf. Proposition \ref{mat0}), the upper bound for the other matrix elements of $[\Sigma(y,z,\l)]_{\a\a'}$ and $[\hat\Sigma(q,\l)]_{\a\a'}$, $\a,\a'=1,2$, can be obtained in the same way and they satisfy exactly the same upper bounds, up to some numerical constant.
\end{remark}
%%%%%%%%%%%%%%%%%%%%%%%%%%%%%%%%%%%%%%%%%%%%%%%%%%%%%%%%%%%%%%%%%%%
As a corollary of Theorem \ref{mainb}, we have the following result:
\begin{theorem}\label{mainc}
There exists a constant $C$ which is independent of the scale index and $\l$, such that the counter-term $[\nu^{ r}(\bk,\l)]_{\a\a'}$, $\a,\a'=1,2$, satisfies the following bound:
\be\label{cte1}
\sup_{\bk}|[\nu^{r}(\bk,\l)]_{\a\a'}|\le C\l^2 r\g^{-r}.
\ee
\end{theorem}
\begin{proof}
It is enough to prove the upper bound for $[\nu^{r}(\bk,\l)]_{11}$. 
By renormalization conditions, we have
$[\nu^{r}(\bk,\l)]_{11}:=-[\tau\Sigma^r(k_0,P_F\bk)]_{11}$, hence 
\be\sup_\bk|[\nu^{r}(\bk,\l)]_{11}|\le \sup_k|[\Sigma^r(k_0,\bk,\l)]_{11}|.\ee
By \eqref{spa}, this Theorem follows.
\end{proof}

We have the following theorem concerning the $2$-point Schwinger function:
\begin{theorem}\label{maine}
Let $\hat S_2(k,\lambda)$ be the two-point Schwinger function which is a $2\times2$ matrix. For any $\lambda$ such that $|\lambda\log^2T|\le K_1$, with $K_1$ some positive constant independent of $T$ and $\lambda$, we have
\be
\hat S_2(k,\lambda)=\hat C(k)[1+\hat R(k,\l)],
\ee
in which $\hat R(k,\l)$ is the remainder matrix such that $\sup_{k}\vert [\hat R(k,\l)]_{\a\a'}\vert\le K|\lambda|$, $\a,\a'=1,2$, for some positive constant $K$ that is independent of $k$ and $\l$.
\end{theorem}
\begin{proof}
By definition (cf.\eqref{sfe1}), the interacting 2-point Schwinger function can be written as the following geometric series:
\be
\hat S_2(k,\lambda)=\hat C(k)+\hat C(k)\cE(k,\l)\hat C(k)+\cdots+\hat C(k)[\cE(k,\l)\hat C(k)]^n+\cdots.
\ee
Consider the case in which the external momentum belongs to a sector of scale $r\le r_{max}\sim \log\frac1T$, then we have
\be
\hat S^r_2(k,\lambda)=\hat C^r(k)\big(1+\cE^r(k,\l)\hat C^r(k)+\cdots+[\cE^r(k,\l)\hat C^r(k)]^n+\cdots\big),
\ee
in which $\cE^r(k,\l)$ means that the external momentum of each element $\hat\cE^r(k,\l)_{\a\a'}$ (cf. \eqref{sf1}) is constrained to a sector of scale index $r$.
By Theorem \ref{mainb}, we have
\be
\Vert [\cE(k,\l)\hat C(k)]_{\a\a'}\Vert_{L^\infty}\le K\lambda^2 r \g^{-r}K_2\g^{-r}
\le K_3|\lambda|\g^{-r},
\ee
in which we have used the fact that $\Vert\hat C^r(k)\Vert_{L^\infty}\le K_2\g^{-r}$ for some constant $K_2>0$ (cf. Lemma \ref{bdsp}), and $r\g^{-r}\le1$ for $r\ge0$. Here $K_3=K\cdot K_2$. Define $\hat R(\lambda, k):=\hat R^{\le r_{max}}(k,\l)=\sum_{r=0}^{r_{max}} \hat R^r(k,\l)$, we have:
\bea
\Vert \big[\hat R^r(k,\l)\big]_{\a\a'}\Vert_{L^\infty}&:=& \Vert\big[\cE^r(k,\l)\hat C^r(k)\big]_{\a\a'}+\cdots+[\big[\cE^r(k,\l)\hat C^r(k)\big]_{\a\a'}]^n+\cdots\Vert_{L^\infty}\nn\\
&&\le 2K_3|\lambda|\g^{-r}.
\eea
Summing over $r$ and choosing $K=4K_3$, the conclusion follows.
\end{proof}
%%%%%%%%%%%%%%%%%%%%%%%%%%%%%%%%%%%%%%%%%%%%%%%%%%%%%
\subsection{Proof of Theorem \ref{conj2}}\label{contthm}
With all these preparations, we are ready to prove Theorem \ref{conj2}.
\begin{proof}\label{prfconj}
In order to prove this Theorem, it is enough to verify item (2) of the renormalization conditions (cf. \eqref{rncd2}) at any 
slice $r\le r_{max}$ and prove that the counter-term $\nu(\bk,\l)$ is $C^{1}$ in $\bk$. First of all, by the multi-slice renormalization condition \eqref{rs1}, the ratio in \eqref{rncd2} at any slice $r$ is bounded by
\bea
&&\Vert\big[\big(\hat\nu^{r}_{s^{(a)},s^{(a)}}+\Sigma^{r}_{s^{(a)},s^{(a)}}\big)\hat C_r(k)\big]_{11}\Vert_{L^\infty}\le K\cdot
\frac{\Vert\big[\Sigma^{r+1}_{s^{(a)},s^{(a)}}\big]_{11}\Vert_{L^\infty}}{\g^{-r}}\nn\\
&&\quad\le K_1\l^2\frac{(r+1)^2\g^{-r-1}}{\g^{-r}}=\g^{-1}\l^2 K'(r+1)^2
\le\l^2 K_2\log^2 T,
\eea
in which $K$, $K_1$ and $K_2$ are some positive constants and we have used the relation  $\gamma^{r_{max}}\sim\frac1T$. Since $\vert\l\log^2T\vert\le C$ for $\l\in\RR_T$ and for some positive constant $C$, we have
\be
\Vert\big[\big(\hat\nu^{r}_{s^{(a)},s^{(a)}}+\Sigma^{r}_{s^{(a)},s^{(a)}}\big)\hat C_r(k)\big]_{11}\Vert_{L^\infty}
\le K_3|\lambda|,
\ee
for some positive constant $K_3$ which is independent of $T$ and $\l$. Following the same analysis we can prove that each of the other elements $\big[\big(\hat\nu^{r}_{s^{(a)},s^{(a)}}+\Sigma^{r}_{s^{(a)},s^{(a)}}\big)\hat C_r(k)\big]_{\a\a'}$  satisfies the same upper bound, up to some numerical constant. By construction, $[\hat\nu^{r}_{s^{(a)},s^{(a)}}(\bk)]_{\a\a'}$ has the same regularity as $[\Sigma^{r}_{s^{(a)},s^{(a)}}(\bk)]_{\a\a'}$. By Theorem \ref{mainb} and Theorem \ref{thmain2}, the self-energy is uniformly $C^{1}$ w.r.t. $\bk$ but not uniformly $C^2$, so is the counter-term. This concludes Theorem \ref{conj2}.
%%%%%%%%%%%%%%%%%%%
%
%$[\Sigma^{r}_{s^{(a)},s^{(a)}}(\bk)]_{\a\a'}$
%
%\begin{remark}\label{rmknu}
%Remark that, since $[\hat\nu^r(\bk,\l)]_{\a\a'}$ has the same regularity as the self-energy function, which is not $C^2$ in $\bk$, (cf. Theorem \ref{thmain2}), $[\hat\nu(\bk,\l)]_{\a\a'}$ is not $C^{2}$ in $\bk$.
%\end{remark}
\end{proof}

%%%%%%%%%%%%%%%%%%%%%%%%%%%%%%%%%%%%%%%%%%%%%%%%%%%%%%%%%%%%%%%%%%%%%%%%%%%%%%%%%%%%%%%%%%%%%%%
%%%%%%%%%%%%%%%%%%%%%%%%%%%%%%%%%%%%%%%%%%%
%%%%%%%%%%%%%%%%%%%%%%%%%%%%%%%%%%%%%%%%%%%%%%%%%%%%%%%%%%%%%%%%%%%%%%%%%%%%%
\subsection{Proof of Theorem \ref{flowmu}.}\label{secflow}
\begin{figure}[htp]
\centering
\includegraphics[width=.4\textwidth]{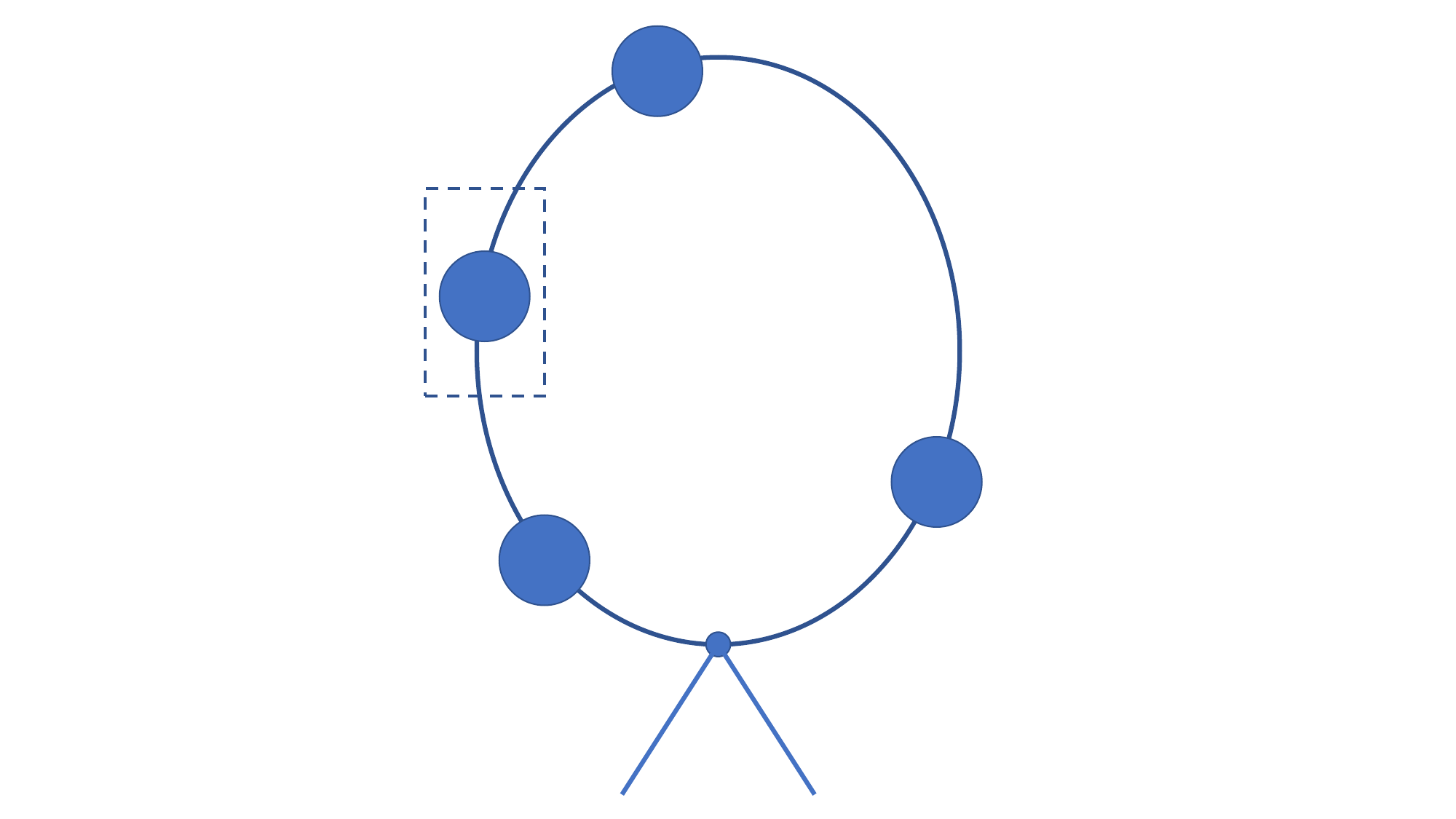}
\caption{\label{gtad1}
A generalized tadpole. Each big dot corresponds to the 2PI bipeds.}
\end{figure}

In this section we consider the upper bound for the counter-term $\delta\mu(\lambda)$. 
There are different kinds of terms that contribute to $\delta\mu(\lambda)$: the tadpole term, the integration of the self-energy function:
\be
\big[\Sigma^{\le r_{max}}(x,\l)\big]_{11}=\sum_{r=0}^{r_{max}}\int d^3y\ \big[\Sigma^r(x, y,\l)\big]_{11},
\ee
and the {\it the generalized tadpole term}, which is a tadpole term whose internal lines are decorated by 1PI bipeds. See Figure \ref{gtad1} for an illustration.
First of all, by Lemma \ref{tad05}, the amplitude of a tadpole is bounded by:
\be
\Vert T\Vert_{L^\infty}\le K_1|\lambda|,
\ee
where $K_1$ is a positive constant and $|\lambda|<C/|\log T|^2$. So the amplitude of a tadpole satisfies the bound in Theorem \ref{flowmu}. Secondly, we have:
\be\label{localmu1}
\Vert\big[\Sigma^{\le r_{max}}(x,\l)\big]_{11}\Vert_{L^\infty}\le \sum_{r=0}^{r_{max}}\int\ d^3y\ |\big[\Sigma^r(x, y,\l)\big]_{11}|,
\ee
in which the integrand satisfy the following bound: (cf. Theorem \ref{maina}, \eqref{x2pta} )
\be
|\big[\Sigma^r(x, y,\l)\big]_{11}|\le K_1|\lambda|^2 r^2 \sup_{\sigma\in\sigma_{\cG^r}}\g^{-3r(\sigma)}e^{-cd^\alpha_\sigma(x,y)},
\ee
for some constant $K_1>0$, so we can perform the integration in \eqref{localmu1} along the spanning tree in the 1PI graph, for which the spatial integration is bounded by 
\be|\int d^3y e^{-cd^\alpha_\sigma(x,y)}|\le K_2.\gamma^{2r_\cT},
\ee
for some constant $K_2>0$, in which $r_\cT$ is the maximal scale index among the tree propagators. Combine the above two terms, we find that, these exists another positive constant, $K_3$, which is independent of the scale indices, such that
\be\label{local3}
\sum_r\int\ dy_0 dy^{(a)}dy^{(b)}\ |\big[\Sigma^{r}(x,y,\l)\big]_{11}|\le K_3 |\lambda|^2.
\ee
 Since $|\lambda|^2\ll|\lambda|$ for $\lambda\in\RR_T$, the amplitudes of the localized term is also bounded by $K_3|\lambda|$.

Now we consider the amplitude for a generalized tadpole, which is formed by contracting a chain of bipeds with a bare vertex. Let $T^g_n$ be a generalized tadpole which contain $n$ irreducible and renormalized bipeds and $n+1$ propagators connecting these bipeds. Let the scale index of an external propagators be $r^e$ and the lowest scale index of the propagators in the biped be $r^i$. Then we can consider the generalized tadpole as a string of more elementary graphs, each corresponds
to the part contained in the dashed square in Figure \ref{gtad1}, whose amplitude is
\be 
T^\Sigma(r^e,r^i,\lambda,x)= \sum_{\a_1=1,2}\int d{x_1}\int dy_1 [C^{r^e}(x,x_1)]_{1\a_1}\big[(1-\tau)\Sigma^{r^i}(x_1,y_1,\l)\big]_{\a_11}.
\ee
Recall that the remainder term $\big[(1-\tau)\Sigma^{r^i}(x_1,y_1,\l)\big]_{\a\a'}$ is bounded by the sum of
\be\label{gtsm1}
\sum_{\a_2=1,2}\int dx_1 dy_1|x_{1,0}-y_{1,0}|\cdot|\big[\Sigma^{r^i}(x_1,y_1,\l)\big]_{\a\a_2}|\cdot|\frac{\partial}{\partial{x_{1,0}}}[ C^{r^e}(x,x_1)]_{\a_2\a}|,
\ee
and
\be\label{gtsm2}
\int dx_1 dy_1|x_1^{(a)}-y_1^{(a)}||x_1^{(b)}-y_1^{(b)}||\big[\Sigma^{r^i}(x_1,y_1,\l)\big]_{\a\a_2}|\cdot|
\frac{\partial}{\partial{x_1^{(a)}}}\frac{\partial}{\partial{x_1^{(b)}}}[ C^{r^e}(x,x_1)]_{\a_2\a}|.
\ee
Using the fact that $j^e+[s^{(a)}]^e+[s^{(b)}]^e=2r^e$ and $[s^{(a)}]^e+[s^{(b)}]^e\ge r^e$, we can prove that there exist positive constants $K_4$ and $K_5$ such that, $\forall\a,\a'=1,2$,
\bea
&&|\frac{\partial}{\partial{x_{1,0}}}[ C^{r^e}(x,x_1)]_{\a\a'}|\le K_4\g^{-2r^e}e^{-d^\a(x,x_1)},\nn\\
&&|\frac{\partial}{\partial{x_1^{(a)}}}\frac{\partial}{\partial{x_1^{(b)}}}[C^{r^e}(x,x_1)]_{\a\a'}|\le K_5\g^{-2r^e}e^{-d^\a(x,x_1)}.\nn
\eea

By Theorem \ref{maina}, Formula \eqref{x2pt} and \eqref{x2ptc}, we have
$$|x_{1,0}-y_{1,0}||\big[\Sigma^{r^i}(x_1,y_1,\l)\big]_{\a\a'}|\le K_4\g^{-2r^i}e^{-d^\alpha(x_1,y_1)},$$
$$|x_1^{(a)}-y_1^{(a)}||x_1^{(b)}-y_1^{(b)}||\big[\Sigma^{r^i}(x_1,y_1,\l)\big]_{\a\a'}|\le K_5\g^{-2r^i}e^{-d^\alpha(x_1,y_1)}.$$
We can easily find that, after performing the integration over $x_1, y_1$, \eqref{gtsm1} or \eqref{gtsm2} is bounded by 
\be
K_5\g^{2r^e}\g^{2r^i}\g^{-2r^e}\g^{-2r^i}\le \bar K,\quad \bar K=\max(K_4, K_5).\nn
\ee

Since there is a propagator in $T^g_n$ whose coordinates are not integrated,
we have
\be
\Vert T^g_n\Vert_{L^\infty}\le\prod_{i=1}^n\Vert\sum_{r^i}  T^\Sigma_i(r^e,r^i,\lambda,x)\Vert_{L^\infty}\cdot\Vert \big[C^{r^e}(x_n,x)\big]_{\a\a'}\Vert_{L^\infty}\le K^n |\lambda|^{2n+1} |\log T|^n\g^{-r^e}.
\ee
Summing over the indices $r^e$ and using the fact that $|\lambda\log^2 T|<C$, we can easily see that there exists a positive constant $K_6$, independent of the scale index and $\l$, such that
\be
\sum_{n=0}^\infty |T^g_n|\le 2 K_6 |\lambda|.
\ee
Summing up all the local terms and let $K'=K_1+K_3+2K_6$, the amplitudes of all the local terms are bounded by $K'|\lambda|$. Hence we proved Theorem \ref{flowmu}. 
%%%%%%%%%%%%%%%%%%%%%%%%%%%%%%%%%%%%%%%%%%%%%%%%%%%%%%%%%%%%%%%%%%%%%%%%%%%%

%%%%%%%%%%%%%%%%%%%%%%%%%%%%%%%%%%%%%%%%%%%%%%%%%%%%%%%%%%%%%%%%%%%%
\section{The Lower Bounds}
\begin{figure}[htp]
\centering
\includegraphics[width=.33\textwidth]{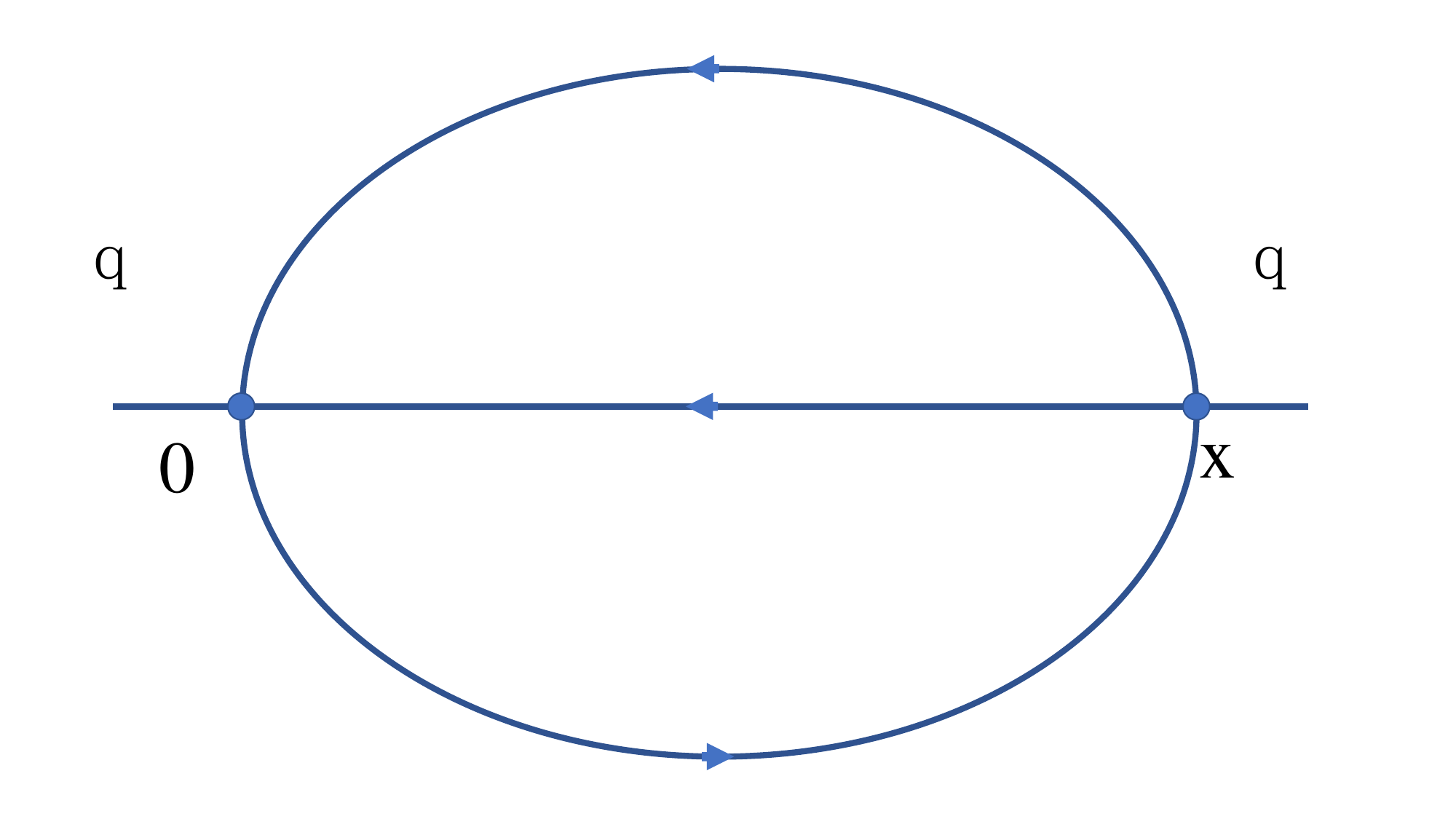}
\caption{\label{sun}
The sunset graph.}
\end{figure}

In this part we study the lower bound of the second derivative of the self-energy function. We  prove that a certain second derivative of the self-energy at a particular value of the external momentum is {\it not uniformly bounded} in the domain $\RR_T=\{\lambda\vert\ \vert\lambda\vert\cdot\vert\log^2T\vert\le C\}$, in which we have established analyticity. This domain being {\it smaller} than the Salmhofer's criterion, together with the non-uniformly boundedness of the second derivative of the self-energy, it suffices to show that the model does not satisfy this criterion and to conclude that the two-dimensional honeycomb Hubbard model is {\it not} a Fermi liquid. 
\begin{theorem}\label{thmain2}
Let $\hat\Sigma(k_e,\lambda)=\big(\hat\Sigma_{\a\a'}(k^e_0,k^e_+,k^e_-,\lambda)\big)$, $\a,\a'=1,2$, be the self-energy of the current model with external quasi-momentum $k_e=(k^e_0,k^e_+,k^e_-)$, in which $k^e_+=q^e_++1, k^e_-=q^e_--1$. There exists a positive constant $K$, which depends on the model but is independent of the temperature $T$ and $\l$, such that $\forall\l\in\RR_T$,
\be\label{sebd}
\bigg\vert\partial^2_{q^e_+}\hat\Sigma_{\a\a'}
(k_e,\lambda)\vert_{k_0=\pi T,\ q^e_+=0,\ q^e_-=0}\bigg\vert\ge K\frac{\lambda^2}{T}.\ 
\ee 
\end{theorem}
Define the amplitude of the sunset graph (see Figure \ref{sun}) as $\Sigma^{Sun}(k_e,\lambda)$.
By Remark \ref{rmkmain}, $\Sigma^{Sun}(k_e,\lambda)$ is the dominant contribution to $\hat\Sigma(k_e,\lambda)$.
\begin{remark}
In order to prove Theorem \ref{thmain2}, it is enough to prove the existence of a lower bound for {\it any} matrix element $\partial^2_{q^e_+}\Sigma^{Sun}_{\a\a'}(k_e,\lambda)$, $\a,\a'=1,2$.   In the following we shall concentrate on the lower bound for $\partial^2_{q^e_+}\Sigma^{Sun}_{11}(k_e,\lambda)$. The lower bounds for other matrix elements of $\partial^2_{q^e_+}\Sigma_{\a\a'}^{Sun}$, $\a,\a'=1,2$, can be proved in the same way.
\end{remark}
%\begin{notation}
%In the rest of this section we will denote the amplitude of sunset graph as $\Sigma_{\a\a'}^{Sun}(k_e,\l)$, namely, we drop the {\rm hat}, just for simplicity.
%
%\end{notation}

%%%%%%%%%%%%%%%%%%%%%%%%%%
%%%%%%%%%%%%%%%%%%%%%%%

The amplitude of the sunset graph with external momentum $k_e=(k^e_0,q^e_++1,q^e_--1)$ is
\be\label{ampsun}
\Sigma^{Sun}(k^e_0,q^e_+,q^e_-,\l)=\lambda^2\int_{\Lambda_\beta} d^3x  C(x)^2 C(-x)e^{-ix_0k^e_0-ix_+k^e_++ix_-k^e_-},
\ee
and the second derivative of $\Sigma^{Sun}(k^e_0,q^e_+,q^e_-)$ w.r.t.the external momentum,
at the point $\tilde k_e=(k^e_0=\pi T, q^e_+=0, q^e_-=0)$, is:
%%%%%%%%%%%%%%%%%%%%%%%%%%%%%%%%%%%%%%%%%%%%%%%%%%%%%%%
\be\label{partial1B}
\partial^2_+\Sigma^{Sun}(\tilde k_e,\l)=\lambda^2\int_{\Lambda_\beta} d^3x x^2_+ C(x)^2C(-x)e^{-i\pi Tx_0-ix_++ix_-},
\ee
in which $C(x)=C(x_0,x_+,x_-)$ is a $2\times2$ matrix given by \eqref{prop2qx}. Since we are mainly interested in the infrared behaviors, in which $k_{0}^2\ll k_{0}$, we can drop the term $k^2_{0}$ in \eqref{prop2q}. Define
\bea\label{propx1}
C^I(x)=\frac{2\pi^2}{3\sqrt3}\int_{\tilde\cD_\b} dk_0 dq_+dq_-e^{ik_0x_0+(q_++1) x_+ +(q_--1) x_-}\hat C^I(k_{i,0},\bq_i),
\eea 
\be\label{propq1}
\hat C^I(k_0,\bq)=\frac{1}{-2ik_0+e(\bq,1)}\begin{pmatrix}i k_{0}+1 &-\tilde\O^*(\bq) \\ -\tilde\O(\bq) &
ik_{0}+1\end{pmatrix}=:\tilde C(k_0,\bq)A(k_0,\bq),
\ee
in which $\tilde C(k_0,\bq)=(-2ik_0+e(\bq,1))^{-1}$ and $A(k_{0},\bq)=[A(k_{0},\bq)_{\a\a'}]$ is the remaining $2\times2$ matrix. We have:
%%%%%%%%%%%%%%%%%%%%%%%%%%%%%%%
\bea\label{lowerm}
&&\partial^2_+ \Sigma^{Sun}(\tilde k_e,\l)=\lambda^2\int_{\Lambda_\beta} d^3x x^2_+ e^{-i\pi Tx_0}\int_{\tilde\cD_\beta} dk_{1,0}dq_{1,+}dq_{1,-}\hat C^I(k_{1,0},\bq_1)\\
&&\times\int_{\tilde\cD_\beta} dk_{2,0}dq_{2,+}dq_{2,-}\hat C^I(k_{2,0},\bq_2)\int_{\tilde\cD_\beta} dk_{3,0}dq_{3,+}dq_{3,-}\hat C^I(k_{3,0},\bq_3).\nn
\eea
%%%%%%%%%%%%%
The main steps for proving Theorem \ref{thmain2} are as follows. First of all, we decompose $\partial^2_{+} \Sigma_{11}^{Sun}(\tilde k_e,\l)$ into a dominant contribution $\partial^2_{q^e_+} \Sigma_{11}^{Sun,dom}(\tilde k_e,\l)$ 
and an error term\\ $\partial^2_{q^e_+} \Sigma_{11}^{Sun,error}(\tilde k_e,\l)$, and then we provide a lower bound in $K\frac{\vert\lambda\vert^2}{T}$ for $\Sigma_{11}^{Sun,dom}(\tilde k_e,\lambda)$. It relies on a method based on residues for complex integrals. Finally, we prove that the upper bound for the error term is much smaller than the lower bound for the dominant term and complete the proof.
%%%%%%%%%%%%%%%%%%%%%%%%%%%%%%%%%%%%%%
\begin{proposition}\label{propdeco}
The amplitude $\partial^2_+\Sigma_{11}^{Sun}(\tilde k_e,\l)$ can be decomposed into the dominant contribution, $\partial^2_+\Sigma_{11}^{Sun,dom}(\tilde k_e,\l)$, which is defined
by replacing all propagators $C^I(x)$ in \eqref{lowerm} with $C^{II}(x)$, in which 
\be\label{propx2} 
C^{II}(x):=\int_{\tilde\cD_\b} dk_0dq_+dq_-e^{ik_0x_0+(q_++1) x_+ +(q_--1) x_-}\hat C^{II}(k_0,\bq),
\ee
\be\label{propq2}
\hat C^{II}(k_0,\bq)=\frac{1}{-2ik_0+2\pi q_+\sin\pi q_-}\begin{pmatrix}1 & 1 \\ 1 &
1\end{pmatrix},
\ee
and an error term, which is defined as
$$
\partial^2_+\Sigma_{11}^{Sun,error}(\tilde k_e,\l):=\partial^2_+\Sigma_{11}^{Sun}(\tilde k_e,\l)-
\partial^2_+\Sigma_{11}^{Sun,dom}(\tilde k_e,\l).$$
%%%%%%%%%%%%%%%
\end{proposition}
The rest of this part is devoted to an explicit construction of $\Sigma^{Sun,error}(\tilde k_e,\l)$. Let $1\ll N\ll j_{max}$ be a positive constant which is sufficiently large. Define the unrestricted summation  by:
\bea\label{cutbd}
\sum_{\{j_i\},\{s_{i,+}\},\{s_{i,-}\}}^N&:=&\sum_{\{j_i\},\{s_{i,+}\},\{s_{i,-}\}}{\bf1}(\inf\{s_{i,+}\}\le \min (j_i, j_{max}-N))\\
&=&\sum_{j_1,s_{1,-},\bar\sigma_{2},\bar\sigma_{3}}\sum_{s_{1,+}=0}^{\inf (j_1, j_{max}-N)}+\sum_{j_1,s_{1,-},j_2,s_{2,-},\bar\sigma_{3}}\sum_{s_{1,+}=j_{max}-N}^{j_1}
\sum_{s_{2,+}=0}^{\inf (j_2, j_{max}-N)}\nn\\
&+&\sum_{j_1,s_{1,-},j_2,s_{2,-},j_3,s_{3,-}}\sum_{s_{1,+}=j_{max}-N}^{j_1}
\sum_{s_{2,+}=j_{max}-N}^{j_2}\sum_{s_{3,+}=0}^{\inf (j_3, j_{max}-N)},\nn
\eea
which means that at least one sector index $s_{i,+}$ is {\it smaller} than $\min (j_i, j_{max}-N)$. Here ${\bf1}(E)$ is a characteristic function for the event $E$. 
%%%%%%%%%%%%%%
%For a sector with scale index $j$ and sector indices $\sigma=(s_+,s_-)$ with $0\le s_\pm\le j$, $s_++s_-\ge j-2$, it is useful to define the {\it extended sector index} $\bar\sigma=(j,s_+,s_-)$.
%%%%%%%%%%%%%%%%
Define
\bea
&&\partial^2_+\Sigma_{11}^{Sun,N,I}(\tilde k_e,\l)\\
&&\quad\quad\quad=\l^2\sum_{\{j_i\},\{s_{i,+}\},\{s_{i,-}\}}^N\int_{\Lambda_\beta} d^3x x^2_+
\big[C^{I}_{j_1,\sigma_{j_1}}(x) C^{I}_{j_2,\sigma_{j_2}}(x) C^{I}_{j_3,\sigma_{j_3}}(-x)\big]_{11}e^{-i\pi Tx_0-ix_++ix_-},\nn
\eea
then we can rewrite $\partial^2_+\Sigma_{11}^{sun}(\tilde k_e,\l)$ as
\be\label{sum0}
\partial^2_+\Sigma_{11}^{Sun}(\tilde k_e,\l)=
\partial^2_+\Sigma_{11}^{Sun,N,I}(\tilde k_e,\l)+\partial^2_+\tilde\Sigma_{11}^{Sun,N,I}(\tilde k_e,\l)
\ee
in which each sector index $s_{i,+}$ in $\partial^2_+\tilde\Sigma_{11}^{Sun,N,I}$ is greater than $j_{max}-N$. Let $\chi_N(q_+)=\chi(\g^{j_{max}-N}q_+)$ and define 
\bea\label{c1}
C^{I}_N(x)&=&\int_{\tilde\cD_\beta} dk_0dq_+dq_- e^{ik_{0}x_0+i(q_{+}+1) x_+ +i(q_{-}-1) x_-}\hat C^{I}_N(k_0,\bq),
\eea
in which $\hat C^{I}_N(k_0,\bq)=\chi_N(q_+)\hat C^{I}(k_0,\bq)$ (cf. \eqref{propq1}), then we have
\bea\label{sunmain1}
\partial^2_+ \tilde\Sigma^{Sun,N,I}_{11}(\tilde k_e,\l)=\l^2\int d^3x x^2_+
[C^{I}_N(x) C^{I}_N(x) C^{I}_N(-x)]_{11}e^{-i\pi Tx_0-ix_++ix_-}.
\eea 

Define also $\partial^2_+ \tilde\Sigma^{Sun,N,II}_{11}(\tilde k_e,\l)$ by replacing each propagator $\hat C^{I}_N(k_0,\bq)$ in \eqref{sunmain1} with
$\hat C^{II}_N(k_0,\bq)$, in which
\be
\hat C^{II}_N(k_0,\bq)=\frac{\chi_N(q_+)}{-2ik_0+2\pi q_+\sin\pi q_-}\begin{pmatrix}1 & 1 \\ 1 &
1\end{pmatrix}.\label{c2q}
\ee

The difference $\partial^2_+ \hat\Sigma^{Sun,N}_{11}(\tilde k_e,\l):=\partial^2_+ \tilde\Sigma^{Sun,N,I}_{11}(\tilde k_e,\l)-\partial^2_+ \tilde\Sigma^{Sun,N,II}_{11}(\tilde k_e,\l)$ is also an error term.
Notice that
\be\label{cutq2}
\chi_N(q_{+})=1+[\chi(q_{+})-1]+[\chi_N(q_{+})-\chi(q_{+})]:=1+0+\hat \chi_N(q_{+}),
\ee
in which the first term $1$ means that the cutoff is removed, for which we obtain the dominant contribution $\partial^2_+ \Sigma^{Sun,dom}$ (cf. Proposition \ref{propdeco}). The second term in \eqref{cutq2} is vanishing identically, since $q_{i+}\le1$ in $\tilde\cD$ (cf. \eqref{support}). So we obtain
\be
\partial^2_+ \tilde\Sigma^{Sun,N,II}_{11}(\tilde k_e,\l)=\partial^2_+ \Sigma_{11}^{Sun,dom}(\tilde k_e,\l)+\partial^2_+ \Sigma^{Sun,N,II}_{11}(\tilde k_e,\l),
\ee
in which
\bea\label{apt3}
&&\Sigma^{Sun,N,II}_{11}(\tilde k_e,\l)=4\l^2\int_{\Lambda_\beta} d^3x x^2_+ e^{-i\pi Tx_0} \\
&&\quad\times\prod_{i=1}^3\int_{\tilde\cD_\beta} dk_{0,i}dq_{i,+}dq_{i,-}\ \Bigg[\ 
\sum_{i=1}^3 \hat \chi_N(q_{i,+})\frac{e^{ik_{1,0}x_0+iq_{1,+} x_+ +iq_{1,-} x_-}}{-2ik_{1,0}-2\pi q_{1+}\sin\pi q_{1-}}\nn\\
&&\quad\quad\quad\quad\quad\times\frac{e^{ik_{2,0}x_0+iq_{2,+} x_+ +iq_{2,-} x_-}}{-2ik_{2,0}-2\pi q_{2+}\sin\pi q_{2-}}\frac{e^{-ik_{3,0}x_0-iq_{3,+} x_+ -iq_{3,-} x_-}}{-2ik_{3,0}-2\pi q_{3+}\sin\pi q_{3-}}\nn\\
&&\quad\quad\quad+\sum_{i,j=1,\cdots,3,\ i\neq j} \hat \chi_N(q_{i,+})\hat \chi_N(q_{j,+})\frac{e^{ik_{1,0}x_0+iq_{1,+} x_+ +iq_{1,-} x_-}}{-2ik_{1,0}-2\pi q_{1+}\sin\pi q_{1-}}\nn\\
&&\quad\quad\quad\quad\quad\times\frac{e^{ik_{2,0}x_0+iq_{2,+} x_+ +iq_{2,-} x_-}}{-2ik_{2,0}-2\pi q_{2+}\sin\pi q_{2-}}\frac{e^{-ik_{3,0}x_0-iq_{3,+} x_+ -iq_{3,-} x_-}}{-2ik_{3,0}-2\pi q_{3+}\sin\pi q_{3-}}\nn\\
&&\quad\quad\quad+\ \hat \chi_N(q_{1,+})\hat \chi_N(q_{2,+})\hat \chi_N(q_{3,+})\frac{e^{ik_{1,0}x_0+iq_{1,+} x_+ +iq_{1,-} x_-}}{-2ik_{1,0}-2\pi q_{1+}\sin\pi q_{1-}}\nn\\
&&\quad\quad\quad\quad\quad\times\frac{e^{ik_{2,0}x_0+iq_{2,+} x_+ +iq_{2,-} x_-}}{-2ik_{2,0}-2\pi q_{2+}\sin\pi q_{2-}}\frac{e^{-ik_{3,0}x_0-iq_{3,+} x_+ -iq_{3,-} x_-}}{-2ik_{3,0}-2\pi q_{3+}\sin\pi q_{3-}}\ \Bigg]\nn,
\eea
is also an error term.
Thus we have constructed the error terms:
\bea\label{apt4}
\partial^2_+ \Sigma_{11}^{Sun,error}=\partial^2_+ \Sigma_{11}^{Sun,N,I}+
\partial^2_+  \Sigma^{Sun,N,II}_{11}+\partial^2_+ \hat\Sigma^{Sun,N}_{11}.
\eea
%%%%%%%%%%%%%%%%%%%%%%%%%%%%%%%%%%%%%%%%%%%%%%%%%%%%%%%%%%%%%%%%%%%%%%%%%%%%%%%%%%%%%%%%%%%%%%%
%%%%%%%%%%%%%%%%%%%%%%%%%%%%%%%%%%%%%%%%%%%%%%%%%%%%%%%%%%%%%%%%%%%%%%%%%
%%%%%%%%%%%%%%%%%%%%%%%%%%%%%%%%%%%%%%%%%%%%%%%%%%%%%%%%%%%%%%%%%%%%%%%%%%%%%%%%%%%
%%%%%%%%%%%%%%%%%%%%%%%%%%%%%%%%%%%%%%%%%%%%%%%%%%%%%%%%%%%%%%%%%%%%%%%%%%%%%%%%%%%
\subsection{The dominant lower bounds}\label{bddom}
\begin{figure}[!htb]
\centering
\includegraphics[scale=.16]{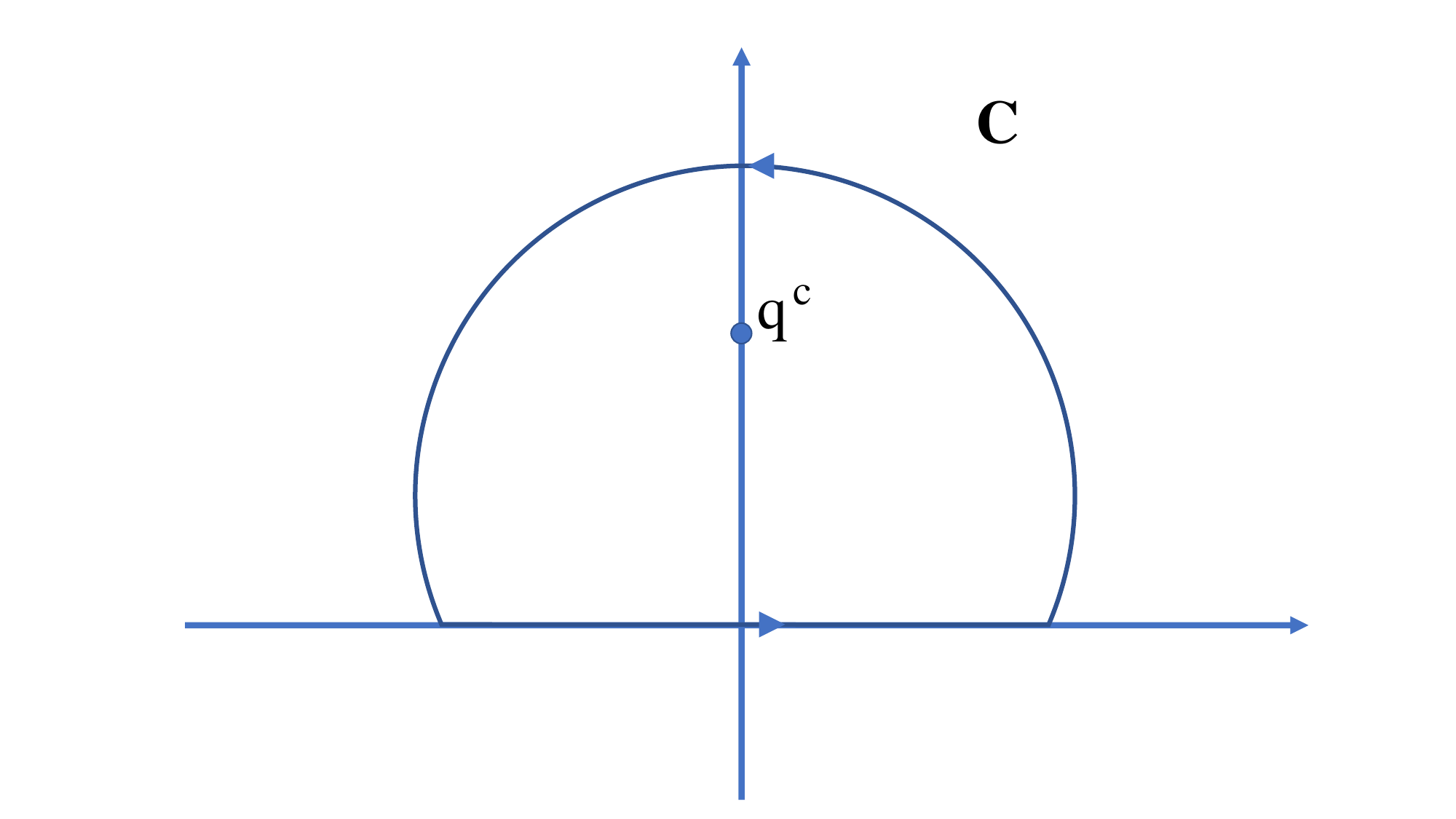}
\caption{The integral contour for $x_+>0$. The singularity $q^c_{i,+}=-\frac{ik_0}{\pi \sin\pi q_{i,-}}$ is on the upper half plane.}\label{contour}
\end{figure}

In this part we consider the lower bound for the dominant contribution $\partial^2_+ \Sigma_{11}^{Sun,dom}(\tilde k_e,\l)$, defined in Proposition \ref{propdeco}. First of all, consider the integration over $q_{i,+}$, $i=1,2,3$. Remark that the integration domain for each $q_{i,+}$ is $\cA=[-1,1]$, which is not very convenient for the analysis. We can enlarge the integration domain due to the presence of the ultraviolet cutoff function. Consider the following the integral:
\bea\label{cont1}
&&\int_{\cA} dq_{i,+}\frac{e^{iq_{i,+} x_+ }}{-2ik_{i,0}-2\pi q_{i,+}\sin\pi q_{i,-}}\\\
&&=
\int_{\RRR} dq_{i,+}\frac{e^{iq_{i,+} x_+ }}{-2ik_{i,0}-2\pi q_{i,+}\sin\pi q_{i,-}}-
\int_{\RRR\setminus\cA} dq_{i,+}\frac{e^{iq_{i,+} x_+ }}{-2ik_{i,0}-2\pi q_{i,+}\sin\pi q_{i,-}},\nn
\eea
in which the last term in the second line is the contribution from the momenta $\vert q_{i,+}\vert\ge1$, corresponding to the ultraviolet contributions. This term is vanishing due to the ultraviolet cutoff (cf. Definition \ref{uvq} or Definition \ref{wholeprop}), hence we can extend the integration domain in \eqref{cont1} from $\cA$ to $\RRR$. The integrand in \eqref{cont1} is a meromorphic function with a pole at $q^c_{i,+}=-\frac{ik_0}{\pi \sin\pi q_{i,-}}$. 
%%%%%%%%%%%%%%%
Remark that $\partial^2_+ \Sigma_{11}^{Sun,dom}$ is vanishing for $x_+=0$. If $x_+>0$, the integration contour is chosen on the the upper half plane (See Figure \ref{contour}), and if $x_+<0$, the integration contour is chosen on the lower half plane.
%%%%%%%%%%%%%%%%%%%%%%%%%%%
%%%%%%%%%%%%%%%%%%%
So we obtain:
\bea\label{cont4}
&&\int_{-\infty}^\infty dq_{i,+}\frac{e^{\pm iq_{i,+} x_+ }}{-2ik_{i,0}-2\pi q_{i,+}\sin\pi q_{i,-}}=
-\frac{i}{\sin\pi q_{i,-}}e^{\pm\frac{k_{i,0}x_+}{\pi \sin\pi q_{i,-}}}\\
&&\quad\quad\quad\quad\times\Big[\chi(x_+> 0)\chi(\pm\frac{k_{i,0}}{\pi \sin\pi q_{i,-}}<0)-\chi(x_+<0)\chi(\pm\frac{k_{i,0}}{\pi \sin\pi q_{i,-}}>0)\Big],\nn
\eea
%%%%%%%%%%%%%%%%%%%%%%%%%%%%%%%%%
for $i=1,2,3$, and we have:
\bea\label{cont5}
&&\partial^2_+ \Sigma_{11}^{Sun,dom}(\tilde k_e,\l)=\\
&&\quad 4i\l^2\int d^3x\int\frac{ dk_{1,0}dq_{1,-}dk_{2,0}dq_{2,-}
dk_{3,0}dq_{3,-}}{\sin\pi q_{1,-}\sin\pi q_{2,-}\sin\pi q_{3,-}}x_+^2e^{ix_0[k_{1,0}+k_{2,0}-k_{3,0}-\pi T]}\nn\\
&&\quad\times e^{(\frac{k_{1,0}}{\pi \sin\pi q_{1,-}}+\frac{k_{2,0}}{\pi \sin\pi q_{2,-}}  -\frac{k_{3,0}}{\pi \sin\pi q_{3,-}})x_+}e^{ix_-[q_{1,-}+q_{2,-}-q_{3,-}] }\nn\\
&&\quad\times\Big[\chi(x_+> 0)\chi(\frac{k_{1,0}}{\pi \sin\pi q_{1,-}}<0)
\chi(\frac{k_{2,0}}{\pi \sin\pi q_{2,-}}<0)\chi(\frac{k_{3,0}}{\pi \sin\pi q_{3,-}}>0)\nn\\
&&\quad-\chi(x_+<0)\chi(\frac{k_{1,0}}{\pi \sin\pi q_{1,-}}>0)
\chi(\frac{k_{2,0}}{\pi \sin\pi q_{2,-}}>0)\chi(\frac{k_{3,0}}{\pi \sin\pi q_{3,-}}<0)\Big].\nn
\eea 

\begin{figure}[!htb]
\centering
\includegraphics[scale=.22]{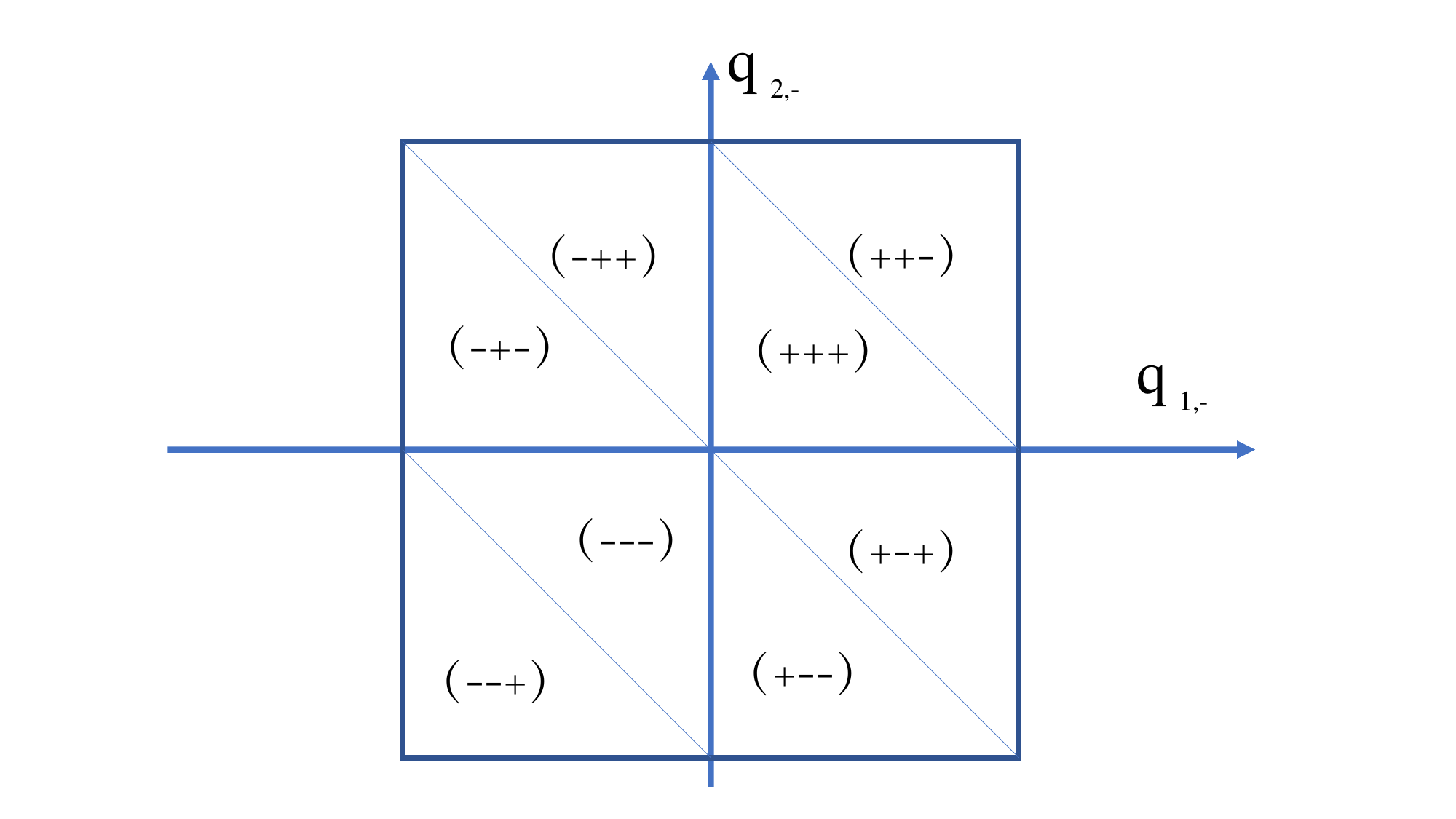}
\caption{The domain of integration for $(q_{1,-}, q_{2,-})$.}\label{intdom}
\end{figure}

Now we consider the integration over $x_0$ and $k_{3,0}$. While the former integration results in a
delta function $\frac{1}{T}\delta(k_{1,0}+k_{2,0}-k_{3,0}-\pi T),$
%%%%%%%%%%
%For $T$ fixed, we can write $x_0=a\beta=a\frac1T$, $a\in[-1,1)$, and $\int_{-\frac1T}^{\frac1T}dx_0=\frac1T\int_{-1}^1 da$.
%So we obtain:
%\be\label{cont6}
%\int dx_0e^{ix_0[k_{1,0}+k_{2,0}-k_{3,0}-\pi T]}=\frac{1}{T}\delta(k_{1,0}+k_{2,0}-k_{3,0}-\pi T).
%\ee
%Since $k_{3,0}=(2n+1)\pi T$, $n\in\ZZZ$,
%%%%%%%%%
the integration $\int d k_{3,0}$ is indeed the discrete sum $2\pi T\sum_{k_{3,0}\in \pi T+2\pi T\ZZZ}$. 
We obtain:
\bea\label{cont7}
&&\partial^2_+ \Sigma_{11}^{Sun,dom}(\tilde k_e,\l)=4i\l^2\int dx_+dx_-\int\frac{ dk_{1,0}dq_{1,-}dk_{2,0}dq_{2,-}
dq_{3,-}}{\sin\pi q_{1,-}\sin\pi q_{2,-}\sin\pi q_{3,-}}x_+^2\\
&&\quad\times e^{(\frac{k_{1,0}}{\pi \sin\pi q_{1,-}}+\frac{k_{2,0}}{\pi \sin\pi q_{2,-}}  +\frac{\pi T-k_{1,0}-k_{2,0}}{\pi \sin\pi q_{3,-}})x_+}e^{ix_-[q_{1,-}+q_{2,-}-q_{3,-}] }\nn\\
&&\quad\times\Big[\chi(x_+> 0)\chi(\frac{k_{1,0}}{\pi \sin\pi q_{1,-}}<0)
\chi(\frac{k_{2,0}}{\pi \sin\pi q_{2,-}}<0)\chi(\frac{\pi T-k_{1,0}-k_{2,0}}{\pi \sin\pi q_{3,-}}<0)\nn\\
&&\quad-\chi(x_+<0)\chi(\frac{k_{1,0}}{\pi \sin\pi q_{1,-}}>0)
\chi(\frac{k_{2,0}}{\pi \sin\pi q_{2,-}}>0)\chi(\frac{\pi T-k_{1,0}-k_{2,0}}{\pi \sin\pi q_{3,-}}>0)\Big].\nn
\eea 
%%%%%%%%%%%%%%%%%%%
The integration over $x_-$ is indeed the discrete summation $\sum_{x_-\in\pi\ZZZ}$ and the result is the delta function $\delta(q_{1,-}+q_{2,-}-q_{3,-}=0[2])$,
%\bea
%\sum_{x_-\in \pi\ZZZ}e^{ix_-[q_{1,-}+q_{2,-}-q_{3,-}] }=\delta(q_{1,-}+q_{2,-}-q_{3,-}=0[2]),
%\eea
in which $0[2]$ means $0$ modulo $2$. Integrating over $q_{3,-}$, performing the following change of variable in the fourth line in \eqref{cont7}:
%%%%%%%
\be\label{change1}
\begin{cases}
x_+\rightarrow -x_+\\
k_{1,0}\rightarrow -k_{1,0},\\
k_{2,0}\rightarrow -k_{2,0},
\end{cases}
\ee
and taking into account the characteristic function $\chi(x_+>0)$, we obtain:
\bea\label{aptm1}
&&\partial^2_+ \Sigma_{11}^{Sun,dom}(\tilde k_e,\l)\\
&&\quad=-4i\l^2\sum_{x_+\in \pi \ZZZ_+}\int dk_{1,0}dk_{2,0}
\int_{\cD_\bq}\frac{ dq_{1,-}dq_{2,-}
}{\sin\pi q_{1,-}\sin\pi q_{2,-}\sin\pi (q_{1,-}+q_{2,-})}x_+^2\nn\\
&&\quad\times  e^{\frac{k_{1,0}x_+}{\pi \sin\pi q_{1,-}}+\frac{k_{2,0}x_+}{\pi \sin\pi q_{2,-}} }e^{-\frac{k_{1,0}+k_{2,0}}{\pi \sin\pi (q_{1,-}+q_{2,-})}x_+}
\chi(\frac{k_{1,0}}{\pi \sin\pi q_{1,-}}<0)
\nn\\
&&\quad\times \chi(\frac{k_{2,0}}{\pi \sin\pi q_{2,-}}<0)\Big[\ e^{\frac{T x_+}{\sin\pi (q_{1,-}+q_{2,-}) }}\chi(\frac{k_{1,0}+k_{2,0}-\pi T}{\pi \sin\pi (q_{1,-}+q_{2,-})}>0)\nn\\
&&\quad\quad\quad\quad\quad\quad\quad\quad\quad\quad\quad -e^{-\frac{T x_+}{\sin\pi (q_{1,-}+q_{2,-}) }}\chi(\frac{k_{1,0}+k_{2,0}+\pi T}{\pi \sin\pi (q_{1,-}+q_{2,-})}>0)\ \Big]\nn,
\eea 
in which $\ZZZ_+$ is the set of positive integers. The integrations over $k_{1,0}, k_{2,0}, q_{1,-}, q_{2,-}$ are more involved, since these variables are not independent of each other, due to the characteristic functions in \eqref{aptm1}. In order to better analyze the integrations over $ q_{1,-}$ and $q_{2,-}$, we can divide the integration domain $\cD_\bq:=[-1,1]\times[-1,1]$  into different regions, according to the signs of $\sin\pi q_{1,-}$, $\sin\pi q_{2,-}$ and $\sin\pi (q_{1,-}+q_{2,-})$:
%\be\label{decom1}
%\cD_\bq=\cT^{+++}\cup \cT^{---}\cup\cT^{++-}\cup\cT^{--+}\cup\cT^{-++}\cup\cT^{+--}\cup\cT^{+-+}\cup\cT^{-+-},
%\ee
\bea\label{decom}
\cD_\bq=\cup_{i=1}^8\cT^{(i)},
\eea
in which $\cT^{(1)}=\cT^{+++}$, $\cT^{(2)}=\cT^{---}$, $\cT^{(3)}=\cT^{++-}$,
$\cT^{(4)}=\cT^{--+}$, $\cT^{(5)}=\cT^{+--}$, $\cT^{(6)}=\cT^{-++}$, $\cT^{(7)}=\cT^{+-+}$ and $\cT^{(8)}=\cT^{-+-}$. Here $\cT^{+++}$ is the region in which $\sin\pi q_{1,-}>0$, $\sin\pi q_{2,-}>0$ and $\sin\pi (q_{1,-}+q_{2,-})>0$. The same for the other regions. (See Figure \ref{intdom} for an illustration).
%%%%%%%%%%%%%%%%%%%%%
%$\cT^{(1)}=\cT^{+++}$, $\cT^{(2)}=\cT^{---},\cT^{(3)}=\cT^{++-},\\
%&&\cT^{(4)}=\cT^{--+}, \cT^{(5)}=\cT^{+--},\cT^{(6)}=\cT^{-++}, \cT^{(7)}=\cT^{+-+}, \cT^{(8)}=\cT^{-+-},\nn
%\eea
%in which $\cT^{+++}$ is the region in which $\sin\pi q_{1,-}>0$, $\sin\pi q_{2,-}>0$ and $\sin\pi (q_{1,-}+q_{2,-})>0$. The same for the other labeled triangles. 
%%It is useful to label these regions by natural numbers:
%%$\cT^{(1)}=\cT^{+++}$, $\cT^{(2)}=\cT^{---}$, $\cT^{(3)}=\cT^{++-}$,
%%$\cT^{(4)}=\cT^{--+}$, $\cT^{(5)}=\cT^{+--}$, $\cT^{(6)}=\cT^{-++}$, $\cT^{(7)}=\cT^{+-+}$ and $\cT^{(8)}=\cT^{-+-}$.
%%%%%%%%%%%%%%%%%%%%%%%%%%%%%%%%%%%%%%%%%%%%%%%%
Let
$\cA=\partial^2_+ \Sigma^{Sun,dom}(\tilde k_e,\l)$, then we have:
\bea
\cA&=&\sum_{a=1}^8\cA^{(a)},\label{aptm2}\\
\cA^{(a)}&=&-4i\l^2\sum_{x_+\in \pi \ZZZ_+}\int dk_{1,0}dk_{2,0}\int_{\cT^{(a)}}dq_{1,-}dq_{2,-}F(x_+, k_{1,0}, k_{2,0}, q_{1,-},  q_{2,-}),\nn
\eea
in which $F(x_+, k_{1,0}, k_{2,0}, q_{1,-},  q_{2,-})$ stands for the integrand in \eqref{aptm1}.
%%%%%%%%%%%%%%%%%%%%%%%%%%%%
%is the restriction of the integration \eqref{aptm1} to the triangle $\cT^{(a)}$, $a=1,\cdots,8$ with integrand $F(x_+, k_{1,0}, k_{2,0}, q_{1,-},  q_{2,-})$.
%%%%%%%%%%%%%55
%%%%%%%%%%%%%%%%%%%%%%%%%%%%%%%%%%%%%%%%%%%%%%%%%%%%%%%%%%%%%%%%%%%%%%%%%%%%%%
\subsubsection{The vanishing amplitudes}
First of all, we have:
\begin{lemma}
\be\cA^{(1)}=0,\quad\cA^{(2)}=0.\ee
\end{lemma}
\begin{proof}
We consider first $\cA^{(1)}$. The integration domain is $\cT^{(+++)}$, in which we have 
$\sin\pi q_{1,-}>0$, $\sin\pi q_{2,-}>0$ and $\sin\pi (q_{1,-}+q_{2,-})>0$. The characteristic functions in \eqref{aptm1} imply that
\be\label{cst1}
k_{1,0}<0,\quad k_{2,0}<0,\quad k_{1,0}+k_{2,0}>\pi T.
\ee
or
\be\label{cst1a}
k_{1,0}<0,\quad k_{2,0}<0,\quad k_{1,0}+k_{2,0}>-\pi T.
\ee
Since $k_{i,0}\in \pi T+2\pi T\ZZZ$, $i=1,2,3$, the inequalities in \eqref{cst1} or \eqref{cst1a} are not compatible with each other. Hence $\cA^{(1)}=0$. Repeat this analysis we prove that $\cA^{(2)}=0$.
\end{proof}
%All other terms in \eqref{decom2} are non vanishing, which will be studied in the next subsection.
%%%%%%%%%%%%%%%%%%%%%%%%%%%%%%%%%%%%%%%%%%%%%%%%%%%%%%%%%%%%%%%%%%%%%%
%%%%%%%%%%%%%%%%%%%%%%%%%%%%%%%%%%%%%%%%%%%%%%%%%%%%%%%%%%%%%%%%%%%5%
\subsubsection{The non-vanishing terms}
%%%%%%%%%%%%%%%%
In order to prove a lower bound for $\cA$, it is essential to determine if there can be any non-trivial cancellation among $\cA^{(a)}$, $a=3,\cdots,8$. We have:
\begin{proposition}\label{la0}
The amplitudes $\cA^{(3)},\cdots, \cA^{(8)}$ have the same sign, and we have  $\vert\cA\vert=2(\vert\cA^{(3)}\vert+\vert\cA^{(5)}\vert+\vert\cA^{(7)}\vert)$.
\end{proposition}
\begin{proof}
In the first step we shall prove that 
\be\label{la3}
\cA^{(3)}=\cA^{(4)},\quad {\rm and}\quad  i\cA^{(3)}\ge0.
\ee
The characteristic functions in \eqref{aptm1} set the following constrains:
\be\label{ca1b}
k_{1,0}<0,\quad k_{2,0}<0,\quad k_{1,0}+k_{2,0}<-\pi T.
\ee
Performing the following change of variables:
\be\label{change1a}
\begin{cases}
q_1\rightarrow -q_1,\\
q_2\rightarrow -q_2,\\
k_{1,0}\rightarrow-k_{1,0}\\
k_{2,0}\rightarrow -k_{2,0},
\end{cases}
\ee
the integration domain for $\cA^{(3)}$ becomes $\cT^{(--+)}$, which is the one for $\cA^{(4)}$, and the expression of $\cA^{(3)}$ becomes $\cA^{(4)}$. Hence we conclude that $\cA^{(3)}=\cA^{(4)}$.
Now we consider the sign of $\cA^{(3)}$. Taking into account the characteristic functions in \eqref{aptm1}, the integration over $k_{1,0}$ or $k_{2,0}$ in \eqref{aptm2} is constrained to the set $-\pi T-2\pi T\NNN$, where $\NNN$ is the set of non-negative integers. So we obtain:
\bea\label{apta1}
&&i\cA^{(3)}=4\l^2(\pi T)^2\sum_{x_+\in \pi \ZZZ_+}\sum_{k_{1,0}\in-\pi T-2\pi T\NNN  }\sum_{k_{2,0}\in-\pi T-2\pi T\NNN  }x_+^2\\
&&\quad\times
\int_{\cT^{(++-)}}\frac{ dq_{1,-}dq_{2,-}
}{\sin\pi q_{1,-}\sin\pi q_{2,-}\sin\pi (q_{1,-}+q_{2,-})} e^{k_{1,0}x_+\big(\frac{1}{\pi \sin\pi q_{1,-}}-\frac{1}{\pi \sin\pi (q_{1,-}+q_{2,-})}\big)}\nn\\
&&\quad\times 
e^{k_{2,0}x_+\big(\frac{1}{\pi \sin\pi q_{2,-}}-\frac{1}{\pi \sin\pi (q_{1,-}+q_{2,-})}\big)}
\Big[e^{\frac{ T}{ \sin\pi (q_{1,-}+q_{2,-})}x_+}
-e^{-\frac{T}{\sin\pi (q_{1,-}+q_{2,-})}x_+} \Big].\nn
\eea
Since $\sin\pi (q_{1,-}+q_{2,-})<0$ in $\cT^{(++-)}$, the integrand of
%%%%%%%%%%%%
\eqref{apta1} is positive. Hence we can change the order of summation and the integration over $q_{1,-}$ and $q_{2,-}$, by Fubini theorem. Performing the summations over $k_{1,0}$ and $k_{2,0}$, we obtain:
\bea\label{apta3}
&&i\cA^{(3)}=4\l^2(\pi T)^2\sum_{x_+\in \pi \ZZZ_+}
x_+^2\int_{\cT^{(++-)}}\frac{ dq_{1,-}dq_{2,-}
}{\sin\pi q_{1,-}\sin\pi q_{2,-}\sin\pi (q_{1,-}+q_{2,-})}\\
&&\frac{e^{-\big[\frac{T}{\sin\pi q_{1,-}}+\frac{T}{\sin\pi q_{2,-}}-\frac{T}{\sin\pi (q_{1,-}+q_{2,-})}\big]x_+}}
{1-e^{-2x_+\big[\frac{T}{\sin\pi q_{1,-}}-\frac{T}{\sin\pi (q_{1,-}+q_{2,-})}\big]}}\cdot
\frac{\big[e^{\frac{2Tx_+}{\sin\pi (q_{1,-}+q_{2,-})}}-1\big]}{1-e^{-2 x_+\big[\frac{T}{\sin\pi q_{2,-}}-\frac{T}{\sin\pi (q_{1,-}+q_{2,-})}\big]}}\nn.
\eea
We conclude that $i\cA^{(3)}\ge0$. 
%%%%%%%%%%%%%%%%%%%%%%%%%%%%%%
Then we prove that $\cA^{(5)}=\cA^{(6)}$, and $i\cA^{(5)}\ge0$.
First of all, performing the changing of variables on $\cA^{(5)}$ as in $(8.32)$, we obtain $\cA^{(6)}$. Hence we conclude that $\cA^{(5)}=\cA^{(6)}$. Now we consider the sign of $i\cA^{(5)}$.
In the integration domain of $\cA^{(5)}$, which is $\cT^{(+--)}$, we have:
$\sin\pi q_{1,-}>0$, $\sin\pi q_{2,-}<0$, $\sin\pi (q_{1,-}+q_{2,-})<0$, and the characteristic functions in $(8.24)$ set the following constrains:
\be\label{ca5a}
k_{1,0}<0,\quad k_{2,0}>0,\quad k_{1,0}+k_{2,0}\le0.\nn
\ee
%%%%%%%%%%
Let $s=k_{1,0}+k_{2,0}$ and consider $s$ and $k_{2,0}$ as independent variables, we obtain
\bea\label{apta5}
&&i\cA^{(5)}=4(\pi T)^2\sum_{x_+\in \pi \ZZZ_+}\sum_{s\in-2\pi T\NNN}\sum_{k_{2,0}\in\pi T+2\pi T\NNN  }x_+^2\\
&&\quad\times
\int_{\cT^{(+--)}}\frac{ dq_{1,-}dq_{2,-}
}{\sin\pi q_{1,-}\sin\pi q_{2,-}\sin\pi (q_{1,-}+q_{2,-})} e^{sx_+\big(\frac{1}{\pi \sin\pi q_{1,-}}-\frac{1}{\pi \sin\pi (q_{1,-}+q_{2,-})}\big)}\nn\\
&&\quad\times 
e^{-k_{2,0}x_+\big(\frac{1}{\pi \sin\pi q_{1,-}}-\frac{1}{\pi \sin\pi q_{2,-}}\big)}
\Big[e^{\frac{ T}{ \sin\pi (q_{1,-}+q_{2,-})}x_+}\chi(s<\pi T)\nn\\
&& \quad\quad\quad
-e^{-\frac{T}{\sin\pi (q_{1,-}+q_{2,-})}x_+}\chi(s<-\pi T) \Big].\nn
\eea
Performing the summation over $s$ and $k_{2,0}$, we obtain:
\bea\label{apt5}
&&i\cA^{(5)}=4(\pi T)^2\sum_{x_+\in \pi \ZZZ_+}
x_+^2\int_{\cT^{(+--)}}\frac{ dq_{1,-}dq_{2,-}
}{\sin\pi q_{1,-}\sin\pi q_{2,-}\sin\pi (q_{1,-}+q_{2,-})}\\
&&\frac{e^{-x_+\big[\frac{T}{\sin\pi q_{1,-}}-\frac{T}{\sin\pi q_{2,-}}\big]}}
{1-e^{-2x_+\big[\frac{T}{\sin\pi q_{1,-}}-\frac{T}{\sin\pi q_{2,-}}\big]}}\cdot
\frac{e^{\frac{2Tx_+}{\sin\pi (q_{1,-}+q_{2,-})}}\big[1-e^{-\frac{2Tx_+}{\sin\pi q_{1,-}}}\big]}{1-e^{-2 x_+\big[\frac{T}{\sin\pi q_{1,-}}-\frac{T}{\sin\pi (q_{1,-}+q_{2,-})}\big]}}\nn
\eea
Since the integrand is positive, we conclude that $i\cA^{(5)}\ge0$. Following the same analysis we can prove that
\be\label{la5}
\cA^{(7)}=\cA^{(8)}, 
\quad {\rm and}\quad i\cA^{(7)}\ge0.
\ee
Combining \eqref{la3} with \eqref{la5}, the conclusion of Proposition \ref{la0}
follows.
\end{proof}
%%%%%%%%%%%%%%%%%%%%%%%%%%%%%%%%%%%%%
%%%%%%%%%%%%%%%%%%%%%%

Now we consider the lower bound of $\cA$, we have:
%%%%%%%%%%%%%%%%%%%%%%%%%%%%%%%%%%%%%%%%%%%%%%%%%
\begin{lemma}\label{lmmain}
There exists a positive constant $K$, which is independent
of the temperature $T$ and the coupling constant $\l$, such that:
\be\label{bdmain}
\vert\cA\vert\ge\frac{K\l^2}{T}.
\ee
\end{lemma} 
By Proposition \ref{la0}, in order to estimate the lower bound for $\cA$, it is enough to obtain a lower bound for $\cA^{(3)}$, $\cA^{(5)}$ or $\cA^{(7)}$. We have:
\begin{lemma}\label{lmmain1}
There exists a positive constant $K_1$ which is independent of $T$ and $\l$, and $K_1\le K$, in which $K$ is the constant in \eqref{bdmain}, such that  
\be\label{bdm2}
\vert\cA^{(3)}\vert\ge\frac{K_1\l^2}{T}.
\ee
\end{lemma}
\begin{proof}
%%%%%%%%%%%%%%%%%%%%%%%%
%%%%%%%%%%%%%
By \eqref{apta3}, we have:
\bea\label{bdm1}
&&\vert\cA^{(3)}\vert\ge4\l^2(\pi T)^2\sum_{x_+\in \pi \ZZZ_+}
x_+^2\int_{\cT^{(++-)}}\frac{ dq_{1,-}dq_{2,-}
}{\sin\pi q_{1,-}\sin\pi q_{2,-}[-\sin\pi (q_{1,-}+q_{2,-})]}\nn\\
&&\quad\quad\times e^{-\big(\frac{T}{\sin\pi q_{1,-}}+\frac{T}{\sin\pi q_{2,-}}-\frac{T}{\sin\pi (q_{1,-}+q_{2,-})}\big)x_+}\cdot
\big[1-e^{\frac{2Tx_+}{\sin\pi (q_{1,-}+q_{2,-})}}\big].
\eea
Remark that, since the integrand in the r.h.s. of the above integral is positive, it is enough to prove a lower bound for the integral which is constrained to a smaller integration domain $\cT^{(++-)}_\e\subset \cT^{(++-)}$, in which we can better control the bounds for the trigonometric functions.
\begin{lemma}\label{edm}
For any $\e\in(0,\frac{1}{100})$, there exists a subset $\cT^{(++-)}_\e\subset\cT^{(++-)}$ with positive measure, such that $\forall (q_{1,-},q_{2,-})\in \cT^{(++-)}_\e$, we have:
\be\label{condm1}
\sin\pi q_{1,-}\ge\e,\  \sin\pi q_{2,-}\ge\e\ {\rm and}\ \sin\pi (q_{1,-}+q_{2,-})\le-\e.
\ee
\end{lemma}
%%%%%%%%%%%%%%%%%%%%%%%%%%
\begin{proof}
For any fixed $\e\in (0,\frac{1}{100})$, the solutions to the inequalities 
$ \sin\pi q_{1,-}\ge\e$ and $\sin\pi q_{2,-}\ge\e$ 
are
\be
q_{1,-}\in E_1:=\big(\ \frac{\arcsin\e}{\pi}, 1-\frac{\arcsin\e}{\pi}\ \big), 
\ q_{2,-}\in E_2:=\big(\ \frac{\arcsin\e}{\pi}, 1-\frac{\arcsin\e}{\pi}\ \big).
\ee
The solution to the inequality $\sin\pi (q_{1,-}+q_{2,-})\le-\e$
is
\be
q_{1,-}+q_{2,-}\in \big(\ 1+\frac{\arcsin\e}{\pi}, 2-\frac{\arcsin\e}{\pi}\ \big).
\ee
Hence it is enough to choose $\cT^{(++-)}_\e$ as 
\be\label{dmt2}
\cT^{(++-)}_\e=\Big\{\ (q_{1,-},q_{2,-})\in E_1\times E_2\ \vert\ 
q_{1,-}+q_{2,-}\in \big(\ 1+\frac{\arcsin\e}{\pi}, 2-2\frac{\arcsin\e}{\pi}\ \big)\ \Big\}.
\ee
Obviously the set $\cT^{(++-)}_\e$ defined as above has a positive volume measure, and we proved this lemma.
\end{proof}
Define
\bea
&&\cA^{(3)}_\e:=4\l^2(\pi T)^2\sum_{x_+\in \pi \ZZZ_+}
x_+^2\int_{\cT^{(++-)}_\e}\frac{ dq_{1,-}dq_{2,-}
}{\sin\pi q_{1,-}\sin\pi q_{2,-}[-\sin\pi (q_{1,-}+q_{2,-})]}\nn\\
&&\quad\quad\times e^{-\big(\frac{T}{\sin\pi q_{1,-}}+\frac{T}{\sin\pi q_{2,-}}-\frac{T}{\sin\pi (q_{1,-}+q_{2,-})}\big)x_+}\cdot
\big[1-e^{\frac{2Tx_+}{\sin\pi (q_{1,-}+q_{2,-})}}\big].
\eea
Now we consider the lower bound for $\cA^{(3)}_\e$. Obviously, the integrand of $\cA^{(3)}_\e$ is positive, hence we can change the order of integration and summation.
Since the measures in \eqref{bdm1} is positive, and since $\vert\sin x\vert\le1$, we have:
%%%%%%%%%%%%%%%%%%%%%%%%%
\bea\label{bdm22}
\vert\cA^{(3)}_\e\vert&\ge&4\l^2(\pi T)^2
\int_{\cT^{(++-)}_\e} dq_{1,-}dq_{2,-}
\sum_{x_+\in \pi \ZZZ_+}x_+^2 e^{-\frac{3T}{\e}x_+}
\big[1-e^{-2Tx_+}\big]\nn\\
&&=4\l^2(\pi T)^2 K_3 \sum_{x_+\in \pi \ZZZ_+}x_+^2 e^{-\frac{3T}{\e}x_+}
\big[1-e^{-2Tx_+}\big],
\eea 
in which $K_3=\vert \cT^{(++-)}_\e\vert$ is the volume measure of the set $\cT^{(++-)}_\e$. 
Performing the summation over $x_+$, and using the summation formula:
\be
\sum_{n=0}^\infty n^2e^{-an}=\frac{e^{-a}+e^{-2a}}{(1-e^{-a})^3},\quad {\rm for}\ a>0,
\ee
we obtain:
\bea\label{bdm7}
\vert\cA^{(3)}_\e\vert&\ge& 4K_3\l^2\pi^4 T^2 \Big[\frac{e^{-\frac{3\pi T}{\e}}+e^{-\frac{6\pi T}{\e}}}
{\big(1-e^{-\frac{3\pi T}{\e}}\big)^3}-\frac{e^{-\big(\frac{3\pi T}{\e}+2\pi T\big)}+e^{-\big(\frac{6\pi T}{\e}+4\pi T\big)}}
{\big(1-e^{-\big(\frac{3\pi T}{\e}+2\pi T\big)}\big)^3}\
\Big]\nn\\
&&\ge 4K_3\l^2\Big[\Big(\frac{2}{(\frac{3}{\e})^3}-\frac{2}{[(\frac{3}{\e})^3+2]}\Big)\frac{\pi}{T}-
\frac{3\pi^2}{(\frac{3}{\e})^2}\ \Big],
\eea
in which the second term in the last line is dominated by the first term, for $T$ sufficiently small. So we can always choose a positive constant $K_1$, say $K_1=4K_3\Big[\Big(\frac{\pi}{(\frac{3}{\e})^3}-\frac{\pi}{[(\frac{3}{\e})^3+2]}\Big)\Big]$, such that 
$
\vert\cA^{(3)}_\e\vert\ge\frac{K_1\l^2}{T}.$ 
So we have $$\vert\cA^{(3)}\vert\ge\vert\cA^{(3)}_\e\vert\ge\frac{K_1\l^2}{T}.$$
\end{proof}
\begin{proof}[Proof of lemma \ref{lmmain}]
Since $\cA^{(3)},\cdots, \cA^{(8)}$ have the same sign, we have
\be
\vert\cA^{(3)}+\cdots+\cA^{(8)}\vert\ge\vert\cA^{(3)}\vert\ge\frac{K_1\l^2}{T}.
\ee
Then this lemma follows by choosing $K=K_1$.
\end{proof}
%%%%%%%%%%%%%%%%%%%%%%%%%%%%%%%%%%%%%%%%%%%%%%%%%%%%%%%%%%%%%%%%%%%%
%%%%%%%%%%%%%%%%%%%%%%%%%%%%%%%%%%%%%%%%%%%%%%%%%%%%%%%%%%%%%%%%%%%%%%%%%%%%%%%%%%%%%%%%%%%%%%%
\subsection{Analysis of the Error Terms}\label{error}
\subsubsection{The Upper Bound for $\partial^2_+\Sigma_{11}^{Sun}(\pi T,0,0)$}\label{upall}
In this part we study the upper bound for the error terms. Before proceeding, it is instructive to consider first the upper bound for $\partial^2_+\Sigma_{11}^{Sun}(\pi T,0,0)$, as a useful preparation for the analysis of the error terms, and for completeness. The upper bound for the sunset graph in different settings have been proved in \cite{AMR2, FS2}.
%%%%%%%%%%%%%%%%LATER THIS SECTION%%%%%%%%%%%%%%%%%%%%%%55
\begin{lemma}\label{lerr1}
There exists some positive constant $K'_1>K$ which is independent of $T$ and $\l$, such that
\be\label{err1}
\vert\partial_+^2\Sigma_{11}^{Sun}(\tilde k_e,\l)\vert\le\frac{K'_1\l^2}{T},
\ee
in which $\tilde k_e=(k^e_0, q^e_+, q^e_-)=(\pi T,0,0)$.
\end{lemma}
\begin{proof}
%%%%%%%%
Using \eqref{propx1}, we obtain
\bea\label{ele11}
&&\partial^2_+ \Sigma_{11}^{Sun}(\tilde k_e,\l)=\lambda^2\int_{\Lambda_\beta} d^3x x^2_+ e^{-i\pi Tx_0}
\int_{\tilde\cD_\beta} dq_1\frac{e^{ik_{1,0}x_0+iq_{1,+} x_+ +iq_{1,-} x_-}}{-2ik_{1,0}+e(\bq_1,1)}\\
&&\times\int_{\tilde\cD_\beta} dq_2\frac{e^{ik_{2,0}x_0+iq_{2,+} x_+ +iq_{2,-} x_-}}{-2ik_{2,0}+e(\bq_2,1)}\int_{\tilde\cD_\beta} dq_3\frac{e^{-ik_{3,0}x_0-iq_{3,+} x_+ -iq_{3,-} x_-}}{-2ik_0+e(\bq_3,1)}\times E_{11}\nn,
\eea
in which
\bea\label{elee11}
&&E_{11}
=\big[A_{11}(k_{1,0})A_{11}(k_{2,0})A_{11}(k_{3,0})+ A_{12}(k_{1,0},\bq_1)A_{21}(k_{2,0},\bq_2)A_{11}(k_{3,0})\nn\\
&&+A_{11}(k_{1,0})A_{12}(k_{2,0},\bq_2)A_{21}(k_{3,0},\bq_3)+
A_{12}(k_{1,0},\bq_1)A_{22}(k_{2,0})A_{21}(k_{3,0},\bq_3)\big],\nn
\eea
and each matrix element $A_{\a,\a'}(k_{0,i},\bq_i)$ is bounded by some positive constant (cf. Proposition \ref{mat0}). Using the decay property of $C^I_{j,\sigma}(x)$ (cf. Lemma \ref{bdx1}), we
obtain:
\bea\label{upper1}
\vert\partial^2_{+} \Sigma_{11}^{Sun}(\tilde k_e,\l)\vert&\le& 4K_0^3\lambda^2\sum_{\substack{j_1,j_2,j_3\\ \sigma_1,\sigma_2\sigma_3}}\gamma^{-\sum_{i=1}^3 s_{+,i}-\sum_{i=1}^3 s_{-,i}}\\
&&\quad\times\int_{\Lambda_{\beta}} dx_0dx_+dx_-x_+^2\exp\Big[-c\sum_{i=1}^3 d_{j_i,\s_i}(x,y) \Big]\nn,
\eea
in which $K_0>0$ is the upper bound for all the matrix elements $\vert A_{\a\a'}(k_{i,0},\bq_i)\vert$, $\a,\a'=1,2$, $i=1,2$, $3$. Now we fix the notations for the scale and sector indices. Let $\{a_1,a_2,a_3\}$ be a set of three real numbers, define $\inf\{a_i\}:=\inf\{a_1, a_2,a_3\}$ and denote by $\inf_2\{a_i\}$ the next smallest element among $\{a_1,\cdots, a_3\}$. We have
\be
\inf\{a_i\}=\frac{1}{3}\sum_{i=1}^3a_i-\frac{1}{3}\big[\inf_2\{a_i\}-\inf\{a_i\}\big]
-\frac{1}{3}\big[\sup\{a_i\}-\inf\{a_i\}\big].
\ee
Performing the integration over $x_0$, $x_\pm$ in \eqref{upper1} we obtain:
%%%%%%%%%%%%%%%%%%%%%%%%%%%%%%%%%%%%%%%%%%%%%%%%%%%%%%%%%%%%%%%%%%%%%%%%%%%%%%%
\bea\label{partial4}
&&\vert\partial^2_{+} \Sigma_{11}^{Sun}(\tilde k_e,\l)\vert\le 4K_0^3\lambda^2\sum_{\substack{j_1,j_2,j_3\\ \sigma_1,\sigma_2\sigma_3}}\gamma^{-\sum_{i=1}^3 s_{+,i}-\sum_{i=1}^3 s_{-,i}}\g^{\inf\{j_i\}}
\g^{3\inf\{s_{+,i}\}}\g^{\inf\{s_{-,i}\}}\nn\\
&&\le 4K_0^3\lambda^2\sum_{\substack{j_1,j_2,j_3\\ \sigma_1,\sigma_2\sigma_3}}\gamma^{-\sum_{i=1}^3 s_{+,i}-\sum_{i=1}^3 s_{-,i}}\g^{\frac{1}{3}\sum_{i=1}^3j_i-\frac{1}{3}\delta\{j_i\}}\g^{\sum_{i=1}^3s_{+,i}-\delta\{s_{+,i}\}}
\g^{\inf\{s_{-,i}\}},\nn \\
\eea
in which $\delta\{j_i\}:=[\inf_2\{j_i\}-\inf\{j_i\}\big]
+\big[\sup\{j_i\}-\inf\{j_i\}\big]$. Now denote by $i_0$ the value of the index $i$ such that $s_{+,i_0}=\inf \{s_{+,i}\}$ and write $\inf \{s_{+,i}\}=j_{i_0}-(j_{i_0}-s_{+,{i_0}})$. Using the fact that 
$\inf\{s_{-,i}\}\le \frac{1}{3}\sum_{i=1}^3s_{-,i}$ and $s_{-,i}+s_{+,i}\ge j_i-1$, we obtain
\bea
\vert\partial^2_{+} \Sigma_{11}^{Sun}\vert&\le& 4K_0^3\lambda^2\g^2\sum_{s_{+,1},s_{+,2},s_{+,3}}
\gamma^{-(j_{i_0}-s_{+,{i_0}})-\frac{2}{3}\delta(s_{+,i})}\\
&&\quad\quad\quad\quad\times\sum_{s_{-,1},s_{-,2},s_{-,3}}\gamma^{-\frac{1}{3}\sum_{i=1}^3 s_{-,i}}\sum_{j_1,j_2,j_3}\g^{j_{i_0}-\frac{1}{3}\delta\{j_i\}}\nn.
\eea
First of all, each summation over the sector indices $s_{-,1},\cdots, s_{-,3}$ can be bounded by a positive constant $K=\frac{1}{1-\g^{-1/3}}$. For $i\neq {i_0}$, the decaying factor can be used to sum over $s_{+,i}$, which also costs a constant $K$. In the same way we can sum over the scale indices $j_i$ for $i\neq {i_0}$ and each summation costs a constant $K$. Summation over $s_{+,{i_0}}$ is bounded by $
\sum_{0\le s_{+,{i_0}}\le j_{i_0}}\g^{-(j_{i_0}-s_{+,{i_0}})}\le\frac{\g}{\g-1}.$
Finally, summing over $j_{i_0}$, we obtain
\be\label{stz}
\vert\partial^2_{+} \Sigma_{11}^{Sun}\vert\le\tilde K\lambda^2\sum_{j_{i_0}}\g^{j_{i_0}}=K'\lambda^2\frac{\g^{j_{max}+1}}{\g-1},
\ee
in which $\tilde K$ is the multiplication of all the previous positive constants generated in the estimation. Using the fact that $\g^{j_{max}}\sim\frac1T$, we prove this lemma.
\end{proof}
%%%%%%%%%%%%%%%%%%%%%%%%%%%%%%%%%%%%%%%%%%%%%%%%%%%%%%%%%%%%%%%%%%%%%%%%%%%%
\subsubsection{The Upper Bounds for the Error Terms}
%%%%%%%%%%%%%%%%%%%%%%%%%%%%%%%%%%%%%%%%%%%%%%%%%%%%%%%%%%%%%%%%%%%%%%%%%%%%%
Now we consider the upper bound for the error terms $\partial^2_+ \Sigma^{Sun,error}(\tilde k_e,\l)$ (cf. \eqref{apt4}).
First of all, we have: 
\begin{lemma}\label{lerr2}
Let $1\ll N\ll j_{max}$ be an integer that is sufficiently large, there exists a positive constant $K'_1$, independent of $T$, $\l$ and $N$, such that
\be\label{err2}
\vert\partial^2_+ \Sigma_{11}^{Sun,N,I}(\tilde k_e,\l)\vert\le(N+1)\frac{K'_1\l^2}{\g^N T},\ \vert\partial^2_+ \Sigma_{11}^{Sun,N,II}(\tilde k_e,\l)\vert\le(N+1)\frac{K'_1\l^2}{\g^N T}.
\ee
\end{lemma}
\begin{remark}
Remark that since $N\gg1$, the upper bound proved in this lemma is much smaller than the dominant lower bound.
\end{remark}
\begin{proof}
Observe that the cutoff function $\hat\chi_N(q_+)$ (cf.\eqref{cutq2}) in $\partial^2_+ \Sigma_{11}^{Sun,N,II}(\tilde k_e,\l)$ plays the same role as the restricted sum in $\partial^2_+ \Sigma_{11}^{Sun,N,I}(\tilde k_e,\l)$. By \eqref{propx1}, \eqref{propx2} and Lemma \ref{bdx1}, the sectorized propagator $C^I_{j,\sigma}(x)$ and $C^{II}_{j,\sigma}(x)$ have the same upper bounds. Hence the two terms in the l.h.s. of \eqref{err2} have the same upper bounds. Following the proof of Lemma \ref{lerr1} and by \eqref{partial4}-\eqref{stz}, we obtain:
%%%%%%%%%%%%%%%%%%%%%%%%%%%%%%%%%%%%%%%%%%%%%%%%%
%\bea
%&&\vert\partial^2_+ \Sigma_{11}^{Sun,N,I}(\tilde k_e,\l)\vert\le K_1' \sum_{j_{i_0}}\g^{j_{i_0}}\sum_{s_{+,{i_0}}=0}^{j_{max}-N}
%\gamma^{-(j_{i_0}-s_{+,{i_0}})},\label{err3}\\
%&&\vert\partial^2_+ \Sigma_{11}^{Sun,N,II}(\tilde k_e,\l)\vert\le K_1' \sum_{j_{i_0}}\g^{j_{i_0}}\sum_{s_{+,{i_0}}=0}^{j_{max}-N}
%\gamma^{-(j_{i_0}-s_{+,{i_0}})}, \label{err3a}
%\eea
%%%%%%%%%%%%%%%%%%%%%%%%%%%%%%%%%%%%
%%%%%%%%%%%%%%%%%
%%%%%%%%%%%%%%%%%%%%%%%%%%%%%%%%%%%
\bea
&&\vert\partial^2_+ \Sigma_{11}^{Sun,N,I}(\tilde k_e,\l)\vert\le K_2'\l^2 \sum_{j_{i_0}=0}^{j_{max}}\g^{j_{i_0}}\sum_{s_{+,{i_0}}=0}^{j_{i_0}}
\gamma^{-(j_{i_0}-s_{+,{i_0}})},\label{err3}\\
&=&K_2'\l^2\Bigg[\sum_{j_{i_0}=0}^{j_{max}-N}\g^{j_{i_0}}\sum_{s_{+,{i_0}}=0}^{j_{i_0}}
\gamma^{-(j_{i_0}-s_{+,{i_0}})}+\sum_{j_{i_0}=j_{max}-N+1}^{j_{max}}\g^{j_{i_0}}\sum_{s_{+,{i_0}}=0}^{j_{max}-N}
\gamma^{-(j_{i_0}-s_{+,{i_0}})}\Bigg],\nn
\eea
for some absolute positive constant $K_2'$. The first term in the second line of \eqref{err3} is bounded by 
\be\label{eana1}
K_2'\l^2\frac{1}{1-\g^{-1}}\sum_{j_{i_0}=0}^{j_{max}-N}\g^{j_{i_0}}\le K'_1\frac{\g^{j_{max}}}{\g^N}\sim \frac{K'_1}{\g^NT},
\ee
in which $K'_1=K_2'\frac{1}{1-\g^{-1}}$. Similarly, the second term in \eqref{err3} is bounded by
%\be
%K_2'\l^2\frac{1}{1-\g^{-1}}\sum_{j_{i_0}=j_{max}-N}^{j_{max}}\g^{j_{i_0}}\le K'_1\frac{\g^{j_{max}}}{\g^N}\sim \frac{K'_1}{\g^NT},
%\ee
%%%%%%%%%%%%
\bea\label{eana2}
K'_2\l^2\sum_{j_{i_0}=j_{max}-N+1}^{j_{max}}\sum_{s_{+,{i_0}}=0}^{j_{max}-N}
\gamma^{s_{+,{i_0}}} \le N K'_1\l^2\g^{-N}\g^{j_{max}}\sim N\frac{ K'_1\l^2}{\g^{N}T}.
\eea
Combining \eqref{eana1} with \eqref{eana2} we prove the upper bound for $\partial^2_+ \Sigma_{11}^{Sun,N,I}(\tilde k_e,\l)$. Repeating the same analysis we obtain the one for
$\partial^2_+ \Sigma_{11}^{Sun,N,II}(\tilde k_e,\l)$. Hence we conclude this lemma.
\end{proof}
%%%%%%%%%%%%%%%%%%%%%%%%%%%%%%%%%%%%%%%%%%%%%%%%%%%%%%%%%%%%%%%%%%%
%%%%%%%%%%%%%%%%%%%%%%%%%%%%%%%%%%%%%%%%%%%%%%%%%%%%%%%%%%%%%%%%%%%%%%%%%%%%%%%%%%%%%%%
Now we consider the upper bound for the last error term in \eqref{apt4}.
\begin{lemma}\label{lerr3}
There exists a constant $\tilde K_1>0$, independent of $T$, such that
\be\label{lerr3a}
\vert \partial^2_+ \hat\Sigma^{Sun,N}_{11}(\tilde k_e,\l) \vert\le \frac{\tilde K_1\l^2}{\g^{-N}}.
\ee
\end{lemma}
\begin{remark}
Remark that since $\frac{1}{\g^{-N}}\ll\frac{1}{\g^{-j_{max}}}=\frac1T $, for $N\ll j_{max}$, the upper bound proved in this lemma is much smaller than the dominant lower bound.
\end{remark}

%Remark that, since $\frac{1}{\g^{-N}}\ll\frac{1}{\g^{-j_{max}}}=\frac1T $, for $N\ll j_{max}$, the r.h.s. of\eqref{lerr3a} is much smaller then the lower bound of the dominant term.

\begin{proof}
%%%%%%%%%%%%%%%%%%%%%%%%%%%%%%%%%%%%%%%%%%%%%%%%%%%%%%%%%%%

Before proceeding, it is useful to introduce another propagator
%%%%
\bea\label{c3}
C^{III}_N(x)&=&\int_{\tilde\cD_\beta} dq e^{ik_{0}x_0+i(q_{+}+1) x_+ +i(q_{-}-1) x_-}\hat C^{II}_N(k_0,\bq)\\
\hat C^{III}_N(k_0,\bq)&=&\frac{\chi_N(q_+)}{-2ik_0+e(\bq,1)}\begin{pmatrix}1 & 1 \\ 1 &
1\end{pmatrix},\label{c3q}
\eea
%%%%%%%%%%%%%%%%%%%%%%%
and define
\bea\label{apte3}
\partial^2_+ \tilde\Sigma^{Sun,N,III}_{11}=\l^2\int d^3x x^2_+
[C^{III}_N(x) C^{III}_N(x) C^{III}_N(-x)]_{11}e^{-i\pi Tx_0-ix_++ix_-}.
\eea 
Then we have:
\bea\label{diff12}
\partial^2_+ \hat\Sigma^{Sun,N}_{11}&=&\big[\partial^2_+ \tilde\Sigma^{Sun,N,I}_{11}-\partial^2_+ \tilde\Sigma^{Sun,N,III}_{11}\big]+\big[\partial^2_+ \tilde\Sigma^{Sun,N,III}_{11}-\partial^2_+ \tilde\Sigma^{Sun,N,II}_{11}\big]\nn\\
&:=&\partial^2_+ \tilde\Sigma^{I,III}_{11}+\partial^2_+ \tilde\Sigma^{III,II}_{11}.
\eea
%%%%%%%%%%%%%%%%%%%
Consider first the term $\partial^2_+ \tilde\Sigma^{I,III}_{11}(\tilde k_e,\l)$. By interpolation,we have:
%\bea\label{diff13}
%\partial^2_+ \tilde\Sigma^{I,III}_{11}(\tilde k_e,\l)=\l^2\int d^3x x^2_+e^{-i\pi Tx_0-ix_++ix_-}\Delta^N_{11}(x),
%\eea
\bea\label{diff13}
&&\partial^2_+ \tilde\Sigma^{I,III}_{11}(\tilde k_e,\l)=\l^2\int d^3x x^2_+e^{-i\pi Tx_0-ix_++ix_-}[\Delta^{N,(1)}_{11}(x)+\Delta^{N,(2)}_{11}(x)+\Delta^{N,(3)}_{11}(x)]\nn\\
&&:=\partial^2_+ \tilde\Sigma^{I,III,(1)}_{11}+\partial^2_+ \tilde\Sigma^{I,III,(2)}_{11}(\tilde k_e,\l)+\partial^2_+ \tilde\Sigma^{I,III,(3)}_{11}(\tilde k_e,\l),
\eea
in which, 
\bea\label{diff13a}
\Delta^{N,(1)}_{11}(x)&=&\big[ \big(C^{I}_{N}(x)-C^{III}_{N}(x)\big) C^{III}_{N}(x) C^{III}_{N}(-x)\big]_{11},\nn\\
\Delta^{N,(2)}_{11}(x)&=&\big[ C^{I}_{N}(x)\big( C^{I}_{N}(x)- C^{III}_{N}(x)\big) C^{III}_{N}(-x)\big]_{11},\nn\\
\Delta^{N,(3)}_{11}(x)&=&\big[ C^{I}_{N}(x) C^{I}_{N}(x)\big( C^{I}_{N}(-x)-C^{III}_{N}(-x)\big)\big]_{11}.
\eea
%%%%%%%%%%%%%%%%%%%%%%%
%So we have $$\partial^2_+ \tilde\Sigma^{I,III}_{11}=\partial^2_+ \tilde\Sigma^{I,III,(1)}_{11}+\partial^2_+ \tilde\Sigma^{I,III,(2)}_{11}(\tilde k_e,\l)+\partial^2_+ \tilde\Sigma^{I,III,(3)}_{11}(\tilde k_e,\l),$$ in which $\partial^2_+ \tilde\Sigma^{I,III,(a)}_{11}$, $a=1,2$ or $3$, is obtained by replacing $\Delta^N_{11}(x)$ in \eqref{diff13} with the corresponding term $\Delta^{N,(a)}_{11}(x)$. 
%%%%%%%%%%%%%%%%%%%%%%%%%%%%%%%%%%%%%%%%%%%%%%%%%%%%%%%%%%%%%%%%%%%
Observe the expression for $\partial^2_+ \tilde\Sigma^{I,III,(a)}_{11}$, $a=1,2, 3$, is very similar to the expression for $\partial^2_+ \Sigma_{11}^{Sun}(\tilde k_e,\l)$ (cf. \eqref{ele11}), except that one of the propagator in \eqref{ele11} is now replaced with a difference $C^{I}_{N}(x)-C^{III}_{N}(x)$ or $C^{I}_{N}(x)-C^{III}_{N}(-x)$. So we can obtain the upper bound for these terms by following the same analysis as in Section \ref{upall}. By \eqref{decay1}, in any sector with scale index $j$ and sector indices $\sigma=(s_+,s_-)$, one has
\bea\label{diff11}
\Vert [C_{j,\sigma,N}^{I}(x)-C_{j,\sigma,N}^{III}(x)]_{11}\Vert_{L^\infty}\le K \g^{-j}\g^{-s_+-s_-}\ e^{-c[d_{j,\s}(x,0)]^\a}, \ s_+\ge j_{max}-N,
\eea
and one has exactly the same upper bound for $ [C_{j,\sigma,N}^{I}(-x)-C_{k,\sigma,N}^{III}(-x)]_{11}$.
So we gain a convergence factor $\g^{-j}$ from each difference w.r.t. $\Vert C_{j,\sigma,N}^{I}\Vert_{L^\infty}$ or $\Vert C_{j,\sigma,N}^{III}\Vert_{L^\infty}$.
Now consider the matrix element $[C_{j,\sigma,N}^{I}(x)-C_{j,\sigma,N}^{III}(x)]_{12}$.
By \eqref{prop2q}, we have
\bea\label{bd12}
\vert\tilde\O^*(\bq)+1\vert=\vert\tilde\O(\bq)+1\vert&=&2\vert[1-e^{-i\frac{\pi}{2}(q_+-q_-)}\cos \frac{\pi}{2}(q_++q_-)]\vert\\
&\le& c_1(\vert q_+\vert+\vert q_-\vert)\le 2c_1\g^{-\min(s_+, s_-)}\nn,
\eea
for some positive constant $c_1$. So we obtain 
\bea\label{diff10}
&&\Vert [C_{j,\sigma,N}^{I}(x)-C_{j,\sigma,N}^{III}(x)]_{12}\Vert_{L^\infty}=\Vert [C_{j,\sigma,N}^{I}(x)-C_{k,\sigma,N}^{III}(x)]_{21}\Vert_{L^\infty}\\
&&\le K \g^{-\min(s_+, s_-)}\g^{-s_+-s_-}\ e^{-c[d_{j,\s}(x,0)]^\a},\ {\rm for} \ s_+\ge j_{max}-N,\nn
\eea
which means that we gain a convergence factor $\g^{-\min(s_+, s_-)}$ comparing to $C^I_N(x)$. Taking into account the cutoff function $\chi_N(q_+)$, we have
\be
\g^{-\min(s_+, s_-)}\le \begin{cases} \g^{-s_+}, {\rm for}\ j_{max}-N\le s_+\le j, N\le \EEE(\frac{j_{max}}{2})-1\\  \g^{-s_-},  {\rm for}\  N-2\le s_-\le j, N\ge E(\frac{j_{max}}{2})-1\end{cases},
\ee
in which $\EEE(\frac{j_{max}}{2})$ means the integer part of $j_{max}/2$. Since by construction, $1\ll N\ll E(\frac{j_{max}}{2})$, we have: $\g^{-\min(s_+, s_-)}\le  \g^{-s_+}.$
Comparing the expressions of $\partial^2_+ \tilde\Sigma^{I,III,(a)}_{11}$, $a=1,2,3$, and 
$\partial^2_+ \Sigma_{11}^{Sun}$ (cf. \eqref{ele11}), we find that exactly one of the factors $A_{\a\a'}$, $\a,\a'=1,2$, in each of the four terms in \eqref{elee11} needs to be replaced by $\max(\g^{-s_+},\g^{-j})=\g^{-s_+}$. Hence each amplitude $\partial^2_+ \tilde\Sigma^{I,III,(a)}_{11}$ is bounded by the upper bound $\vert\partial_+^2\Sigma_{11}^{Sun}(\tilde k_e,\l)\vert$ times a convergence factor $\g^{-s_+}$. Using the fact that $\g^{-s_{+}}\le \g^{-(j_{max}-N)}\sim T{\g^N}$ and $|\partial^2_+ \Sigma_{11}^{Sun}(\tilde k_e,\l)|\le\frac{K'_1\l^2}{T}$ (cf. \eqref{err1}), we obtain:
\be
\vert\partial^2_+ \tilde\Sigma^{I,III}_{11}\vert\le 4\frac{K'_1\l^2}{T}\cdot {T}{\g^N}= \tilde K_2\l^2\g^N,\label{gain13}
\ee
with $\tilde K_2=4K_1$.

%%%%%%%%%%%%%%%%%%%%%%
The upper bound for $\partial^2_+ \tilde\Sigma^{III,II}_{11}(\tilde k_e,\l)$ can be proved in a similar way. By \eqref{c2q} and \eqref{c3q}, we have
%\bea\label{diff0b}
%&&\partial^2_+ \tilde\Sigma^{III,II}_{11}(\tilde k_e,\l)=\l^2\int d^3x x^2_+e^{-i\pi Tx_0-ix_++ix_-}\nn\\
%&&\quad\quad\quad\quad\times\Big[\ C^{III}_{N}(x) C^{III}_{N}(x) C^{III}_{N}(-x)-
%C^{II}_{N}(x) C^{II}_{N}(x) C^{II}_{N}(-x)
%\Big]_{11}\\
%&&\quad\quad=4\l^2\int d^3x x^2_+e^{-i\pi Tx_0-ix_++ix_-}\nn\\
%&&\quad\quad\times\Big[\ [C^{III}_{N}(x)]_{11} [C^{III}_{N}(x)]_{11}[ C^{III}_{N}(-x)]_{11}-
%[C^{II}_{N}(x)]_{11}[ C^{II}_{N}(x)]_{11}[ C^{II}_{N}(-x)]_{11}\nn
%\Big],
%\eea
%%%%%%%%%%%%%%%%%%%%%%%%
\bea\label{diff0b}
&&\partial^2_+ \tilde\Sigma^{III,II}_{11}(\tilde k_e,\l)=4\l^2\int d^3x x^2_+e^{-i\pi Tx_0-ix_++ix_-}\\
&&\quad\quad\times\Big[\ [C^{III}_{N}(x)]_{11} [C^{III}_{N}(x)]_{11}[ C^{III}_{N}(-x)]_{11}-
[C^{II}_{N}(x)]_{11}[ C^{II}_{N}(x)]_{11}[ C^{II}_{N}(-x)]_{11}\nn
\Big].
\eea
%%%%%%%%%%%%%%%%%%%%%%%%%
The terms in the square bracket can be written as
\bea
&&[C^{III}_{N}(x)-C^{II}_{N}(x)]_{11}[ C^{III}_{N}(x)]_{11}[ C^{III}_{N}(-x)]_{11}\nn\\
&&+
[C^{II}_{N}(x)]_{11}[ C^{III}_{N}(x)- C^{II}_{N}(x)]_{11}[ C^{III}_{N}(-x)]_{11}\nn\\
&&\quad+[ C^{II}_{N}(x)]_{11} [C^{II}_{N}(x)]_{11}[ C^{III}_{N}(-x)-C^{II}_{N}(-x)]_{11}.\nn
\eea
Now we consider the bound for the difference term
\bea
[C^{III}_{N}(x)-C^{II}_{N}(x)]_{11}&=&\int_{\tilde\cD_\beta} dk_0dq_+dq_-{e^{ik_{0}x_0+i(q_{+}+1) x_+ +i(q_{-}-1) x_-}}\\
&&\times
\chi_N(q_{+})[\hat C^{III}(k_{0},q_{\pm})-\hat C^{II}(k_{0},q_{\pm})]_{11}.\nn
\eea
By inserting and subtracting the term
\be
\frac{1}{-2ik_{0}-4\pi q_{+}\sin\frac\pi2 q_{-}\cos\frac{\pi}{2}(q_{+}+q_{-})}\nn,
\ee
and using \eqref{c2q} and \eqref{c3q}, we obtain
\bea
\big[ \hat C^{III}(k_{0},q_{\pm})-\hat C^{II}(k_{0},q_{\pm})\big]_{11}= \hat C_{a}(q)+\hat C_{b}(q),
\eea
%%%%%%%%%%%%%%%%%%%%%%%%%%%%%%%%%%%%%%%%%%%%%%%%%%%%%%%%%%%%%%%%
in which
\bea
&&\hat C_{a}(q)=\frac{\sin \frac\pi2q_{+}-\frac\pi2q_{+}}
{-2ik_{0}-8\pi \sin\frac{\pi}{2}q_{+}\sin\frac\pi2 q_{-}\cos\frac{\pi}{2}(q_{+}+q_{-})}\\
&&\qquad\qquad\qquad\times\frac{\sin\frac\pi2 q_{-}\cos\frac{\pi}{2}(q_{+}+q_{-})}
{-2ik_{0}-4\pi q_{+}\sin\frac\pi2 q_{-}\cos\frac{\pi}{2}(q_{+}+q_{-})},\nn
\eea
and 
\be
\hat C_{b}(q)=\frac{4\pi q_{+}\sin\frac\pi2 q_{-}[\cos\frac{\pi}{2}(q_{+}+q_{-})-\cos\frac{\pi}{2}q_{-}]}{[{-2ik_{0}-4\pi q_{+}\sin\frac\pi2 q_{-}\cos\frac{\pi}{2}(q_{+}+q_{-})}][
{-2ik_{0}-2\pi q_{+}\sin\pi q_{-}}]}.
\ee
%%%%%%%%%%%%%%%%%%%
Suppose that $(q_+,q_-)$ belongs to a sector with scale index $j$ and sector indices $\sigma=(s_+,s_-)$, with $0\le s_\pm\le j$, $s_++s_-\ge j-2$, we have:
\be
|\sin \frac\pi2q_{+}-\frac\pi2q_{+}|\le O_1q^2_{+}\sim O_1\g^{-2s_{+}},\quad 
\vert \sin\frac\pi2 q_{-}\vert\le O_2\g^{-s_{-}},
\ee
in which $O_1$, $O_2$ are some absolute positive constants. Then we obtain 
\be\label{gain1}
\Vert \hat C_{a}(q)\Vert\le O'\g^{-s_++j}.
\ee
Hence we gain an additional convergence factor $\g^{-s_+}$ w.r.t. $\Vert C^{II}_{j,\sigma,N}(q)\Vert$. Here $O'$ is another positive constant depending on $O_1$ and $O_2$.

Now we consider the bound for $\hat C_{b}(q)$. We have:
\bea
&&|[\cos\frac{\pi}{2}(q_{+}+q_{i,-})-\cos\frac{\pi}{2}q_{-}]|\le O'_1\g^{-2s_{+}}+ O'_2\g^{-s_{+}-s_-},
\eea
in which $O'_1$ and $O'_2$ are some absolute positive constants. We have
\be\label{gain2}
\vert \hat C_{b}(q)\vert\le O'[\g^{-2s_+}+\g^{-j}]\g^j,
\ee
hence we gain a convergence factor $[\g^{-2s_+}+\g^{-j}]$ w.r.t. $\Vert C^{II}_{j,\sigma,N}(q)\Vert$. Combining \eqref{gain1} and \eqref{gain2}, and using the fact that $s_+\le j$, we obtain
\bea\label{diff11g}
\Vert [C_{j,\sigma,N}^{III}(x)-C_{j,\sigma,N}^{II}(x)]_{11}\Vert_{L^\infty}\le O'' \g^{-2s_+-s_-}\ e^{-c[d_{j,\s}(x,0)]^\a}, \ s_+\ge j_{max}-N.
\eea
Here $O'$ in \eqref{gain2} and $O''$ are positive constants that are independent of the scale index. 
%%%%%%
Similarly we have:
\bea\label{diff2}
\Vert [C^{III}_{j,\sigma,N}(-x)-C^{II}_{j,\sigma,N}(-x)]_{11}\Vert_{L^\infty}\le O''\g^{-2s_{+}-s_{-}}e^{-cd^\alpha_{\sigma}(x)}.
\eea
Repeating the same analysis in the paragraph just before \eqref{gain13}, we obtain
\be
\vert\partial^2_+ \tilde\Sigma^{III,II}_{11}(\tilde k_e,\l)\vert\le \tilde K_3\l^2\g^N,
\ee
for some positive constant $\tilde K_3$ that is independent of the scale index. Let
$\tilde K_1=\tilde K_2+\tilde K_3$, we conclude this lemma.
\end{proof}
%%%%%%%%%%%%%%%%%%%%%%%%%%%%%%%%%%%
\begin{proof}[Proof of Proposition \ref{propdeco}]
By Lemma \ref{lerr2}, Lemma \ref{lerr3} and Lemma \ref{lmmain} we conclude that the upper bounds for the error terms is bounded by $2(N+1)\frac{ K'_1\l^2}{\g^NT}+\tilde K_1\l^2\g^N$, for some positive constant $K'_1$ and $\tilde K_1$ that are independent of $T$. Since both terms are much smaller than $\frac{K_1\l^2}{T}$, which is the lower bound of the dominant term, for $N$ large enough and $T$ small enough, the conclusion of this proposition follows. 
\end{proof}
%%%%%%%%%%%%%%%%%%%%%%%%%%%%%%%%%%%%%%%%

\section{Conclusions and Perspectives}
In this paper we construct the $2p$-point Schwinger functions in the $2$-dimensional honeycomb Hubbard with renormalized chemical potential $\mu=1$ and establish the upper and lower bounds for the self-energy as well as its second derivatives. We prove that this model is not a Fermi liquid in the mathematical precise sense of Salmhofer. In \cite{GM} the authors studied the honeycomb Hubbard model at half-filling, in which the $2$ point Schwinger functions are proved to be analytic down to zero temperature. It is therefore important to study the crossover between the cases of $\mu=0$ and $\mu=1$ and consider the honeycomb Hubbard model with $0<\mu<1$, in which the Fermi surfaces are neither pair of points nor exact triangles, but are disjoint union of convex curves with $\ZZZ_3$ symmetry. In this case the sector counting lemma of the current paper may not be valid and a different one \cite{wang20} is required. It is also important to notice that Lifshitz phase transitions may happen at the van Hove filling \cite{rosen}, which also deserve a rigorous study.

%%%%%%%%%%%%%%%%%%%%%%%%%%%%%%%%%%%%%%%%%%%%%%%%%%%%%%%%%%%%%%%%%%%%%%%%%%%%%%%%
\medskip
\noindent{\bf Acknowledgments}
Zhituo Wang is very grateful to Horst Kn\"orrer for useful discussions and encouragements, and to Alessandro Giuliani and Vieri Mastropietro for useful discussions. Part of this work has been finished during Zhituo Wang's visit to the Institute of Mathematics, University of Zurich. He is also very grateful to Benjamin Schlein for invitation and hospitality. We are grateful to the anonymous referee for his comments and suggestions, which lead to significant improvements of the first manuscript. Zhituo Wang is supported by NSFC No.12071099 and No.11701121. Vincent Rivasseau is supported by Paris-Saclay University and the IJCLab of the CNRS.


\thebibliography{0}


\bibitem{AR}
A.~Abdesselam and V.~Rivasseau,
  ``Trees, forests and jungles: A botanical garden for cluster expansions,''  in Constructive Physics, Lecture Notes in Physics 446, Springer Verlag, 1995,

\bibitem{AMR1} S. Afchain, J. Magnen and V. Rivasseau: {\it
Renormalization of the 2-Point Function of the Hubbard Model at Half-Filling},
Ann. Henri Poincar\'e {\bf 6}, 399-448 (2005). 

\bibitem{AMR2} S. Afchain, J. Magnen and V. Rivasseau:{\it
The Two Dimensional Hubbard Model at Half-Filling, part III: The Lower Bound on the Self-Energy},
Ann. Henri Poincar\'e {\bf 6}, 449-483 (2005).

\bibitem{and} P. W. Anderson: {\it "Luttinger-Liquid" behavior of the normal metallic state of the 2D Hubbard model}, Phys. Rev. Lett. {\bf 64}, 1839-1841 (1990).

%
%\bibitem{BCS} T. Bardeen, L. N. Cooper and J. R. Schrieffer: {\it Theory of
%Superconductivity}, Phys. Rev. {\bf 108}, 1175-1204 (1957).

\bibitem{BG} G. Benfatto and G. Gallavotti: {\it
Perturbation theory of the Fermi surface in a quantum liquid.
A general quasiparticle formalism and one dimensional systems},
Jour. Stat. Phys. {\bf  59}, 541-664 (1990).

\bibitem{BG1} G. Benfatto and G. Gallavotti: {\it
Renormalization Group},
Physics Notes, Vol. 1, Princeton University Press (1995).

\bibitem{BFM} G. Benfatto, P. Falco and V. Mastropietro, Universality of one-dimensional Fermi systems, II. The Luttinger liquid structure. Comm. Math. Phys. 330 {\bf 1}, 217–282, (2014)
%\bibitem{BM2} Benfatto, G.; Mastropietro, V. Ward identities and chiral anomaly in the Luttinger liquid. Comm. Math. Phys. 258 (2005), no. 3, 609–655

%\bibitem{BGM1} G. Benfatto, A. Giuliani and V. Mastropietro: {\it
%Low Temperature Analysis of Two-Dimensional Fermi Systems with Symmetric Fermi Surface},
%Ann. Henri Poincar\'e {\bf 4}, 137-193 (2003).

\bibitem{BGM2} G. Benfatto, A. Giuliani and V. Mastropietro: {\it
Fermi liquid behavior in the 2D Hubbard model at low temperatures},
Ann. Henri Poincar\'e {\bf 7}, 809-898 (2006).

\bibitem{BM} G. Benfatto and V. Mastropietro: {\it
Ward identities and chiral anomaly in the Luttinger liquid},
Comm. Math. Phys. {\bf 258}, 609-655 (2005).

\bibitem{graph} J.A. Bondy and U.S.R. Murty: {\it
Graph theory with applications}, North-Holland (1976).

%
%\bibitem{B} D. Brydges: {\it A short course on cluster expansions}, Ph\'enom\`enes critiques,
%syst\`emes al\'eatoires, th\'eories de jauge, Les Houches, 1984, North-Holland, Amsterdam, 1986.

%\bibitem{BrF} D. C. Brydges and P. Federbush: {\it A new form of the Meyer expansion
%in classical statistical mechanics}, Jour. Math. Phys. {\bf 19}, 2064-2067 (1978).

\bibitem{BK} D. Brydges and T. Kennedy:
{\it Mayer expansions and the Hamilton-Jacobi equation}, J. 
Statistical Phys. {\bf 48}, 19 (1987).

\bibitem{BR1} O. Bratteli, D. W. Robinson: {\it Operator Algebras and Quantum Statistical Mechanics 2}, second edition, Springer-Verlag, 2002.

\bibitem{lieb} D. Baeriswyl,  D. Campbell, J. Carmelo, F. Guinea, E. Louis, 
{\it The Hubbard Model}, Nato ASI series, V. 343, 
Springer Science+Business Media New York, 1995
%
%\bibitem{jito} S. Becker, R. Han and S. Jitomirskaya
%{\it Cantor spectrum of graphene in magnetic fields}, Invent. 
%Math. {\bf 218}, 979 (2019).

\bibitem{review1} A. H. Castro Neto, F. Guinea, N. M. R. Peres, K. S. Novoselov,
A. K. Geim: {\it The electronic properties of graphene}, Rev. Mod. Phys. {\bf 81}, 109-162 (2009)


\bibitem{review3} S. Das Sarma, S. Adam, E. H. Hwang, E. Rossi
: {\it Electronic transport in two-dimensional graphene}, Rev. Mod. Phys. {\bf 83}, 407-470 (2011)

%\bibitem{DR1} M. Disertori and V. Rivasseau: {\it Interacting Fermi liquid in
%two dimensions at finite temperature, Part I - Convergent attributions} and
%{\it Part II - Renormalization},
%Comm. Math. Phys. {\bf 215},  251-290 (2000) and
%291-341 (2000).

\bibitem{DR1} M. Disertori and V. Rivasseau: {\it Interacting Fermi liquid in
two dimensions at finite temperature, Part I - Convergent attributions} 
Comm. Math. Phys. {\bf 215},  251-290 (2000)

\bibitem{DR2} M. Disertori and V. Rivasseau: 
{\it Interacting Fermi liquid in
two dimensions at finite temperature, Part II - Renormalization},
Comm. Math. Phys. {\bf 215},  291-341 (2000).


%\bibitem{S2} W. De Roeck, M. Salmhofer: {\it
%Persistence of exponential decay and spectral gaps for interacting fermions. Comm. Math. Phys. {\bf 365},773–796 (2019). }

\bibitem{Feff1} C. Fefferman and M. Weinstein: {\it Honeycomb lattice potentials and Dirac points}, J. Amer. Math. Soc. {\bf 25}, 1169–1220 (2012). 
%
%\bibitem{Feff2} C. Fefferman, J.P. Lee-Thorp, M. Weinstein: {\it Topologically protected states in one dimensional continuous systems and Dirac points}, Proc. Natl. Acad. Sci. USA {\bf 111}, 8759–8763 (2014).
%
\bibitem{Feff3} C. Fefferman, J.P. Lee-Thorp, M. Weinstein: {\it Honeycomb Schr\"odinger operators in the strong binding regime}, Comm. Pure Appl. Math. {\bf 71}, 1178–1270 (2018). 


\bibitem{FMRT} J. Feldman, J. Magnen, V. Rivasseau and E. Trubowitz:
{\it An infinite volume expansion for many fermions Freen functions},
Helv. Phys. Acta {\bf 65}, 679-721 (1992).

\bibitem{FKT} J. Feldman, H. Kn\"orrer and E. Trubowitz: {\it
A Two Dimensional Fermi Liquid},
Comm. Math. Phys {\bf 247}, 1-319 (2004).

\bibitem{FS1} J. Feldman, M. Salmhofer: {\it Singular Fermi surfaces. I. General power counting and higher dimensional cases}, Rev. Math. Phys. {\bf 20}, 233–274, (2008)
%(2008)

\bibitem{FS2} J. Feldman, M. Salmhofer: {\it Singular Fermi surfaces. II. The two-dimensional case}, Rev. Math. Phys. {\bf 20}, 275–334, (2008)


\bibitem{FST1} J. Feldman, M. Salmhofer and E. Trubowitz: {\it
Perturbation Theory Around Nonnested Fermi Surfaces.
I. Keeping the Fermi Surface Fixed}, J. Stat. Phys., {\bf 84}, 1209-1336 (1996). 

\bibitem{FST2} J. Feldman, M. Salmhofer and E. Trubowitz: {\it 
An inversion theorem in Fermi surface theory}, 
Comm. Pure Appl. Math. 53, 1350-1384. (2000)


\bibitem{FT} J. Feldman, E. Trubowitz: {\it
Perturbation theory for many fermion systems},
Helv. Phys. Acta {\bf 63}, 156-260 (1990).

%\bibitem{G} G. Gallavotti: {\it
%Renormalization group and ultraviolet stability
%for scalar fields via renormalization group methods},
%Rev. Mod. Phys. {\bf 57}, 471-562 (1985).

\bibitem{GN} G. Gallavotti and F. Nicol\`o: {\it
Renormalization theory for four dimensional scalar fields. Part I},
{\it II}, Comm. Math. Phys. {\bf 100}, 545-590 (1985),
{\bf 101}, 471-562 (1985).

\bibitem{GK} K. Gawedzki and A. Kupiainen: {\it Gross-Neveu model through
convergent perturbation expansions}, Comm. Math. Phys. {\bf 102}, 1-30 (1985).


\bibitem{GM} A. Giuliani and V. Mastropietro: {\it The two-dimensional
Hubbard model on the honeycomb lattice}, Comm. Math. Phys. {\bf 293}, 301-346
(2010).


%\bibitem{Gruner} G. Gruner: {\it Density waves in solids}, Perseus publishing, Cambridge, Massachusetts, (1994).

\bibitem{hubb} J. Hubbard, {\it Electron correlations in narrow energy bands}, Proc. Roy. Soc. (London), {\bf A276}, 238-257 (1963).

%\bibitem{HCM} C.-Y. Hou, C. Chamon and C. Mudry: {\it Electron Fractionalization in
%Two-Dimensional Graphenelike Structures}, Phys. Rev. Lett. {\bf 98}, 186809 (2007).

\bibitem{iz} C. Itzykson and J.-B. Zuber: {\it Quantum Field Theory}, Dover publications, 2005

%\bibitem{Jito} S. Becker, R. Han, S. Jitomirskaya: {\it Cantor spectrum of graphene in magnetic fields}, Invent. Math. {\bf 218},
%979-1041 (2019).

\bibitem{review4} V. N. Kotov, B. Uchoa, V. M. Pereira, F. Guinea and A. H. Castro Neto
: {\it Electron-Electron Interactions in Graphene: Current Status and Perspectives}, Rev. Mod. Phys. {\bf 84}, 1067-1125 (2012)

\bibitem{lifshitz} I. M. Lifshitz: {\it Anomalies of Electron Characteristics of a Metal in the High Pressure}, Sov. Phys. JETP {\bf 11}, 1130 (1960).

\bibitem{link} S. Lint, et el.: {\it Introducing strong correlation effects into graphene by gadolinium interaction}, Phys. Rev. B, {\bf 100}, 121407(R),
(2019).

\bibitem{tutte}  T.~Krajewski, V.~Rivasseau, A.~Tanasa and Zhituo~Wang,
  ``Topological Graph Polynomials and Quantum Field Theory, Part I: Heat Kernel
  Theories,''
  J.\ Noncommut.\ Geom.\  {\bf 4},  29  (2010)

%\bibitem{landau1}
%Landau, L.D.: {\it The Theory of a Fermi Liquid}, Sov. Phys. JETP 3, 920 (1956), {\it Oscillations in a Fermi Liquid}, Sov. Phys. JETP 5, 101 (1957), {\it On the Theory of the Fermi Liquid} Sov. Phys. JETP 8, 70 (1959)


\bibitem{Le} A. Lesniewski: {\it Effective action
for the Yukawa$_2$ quantum field theory}, Comm. Math. Phys. {\bf 108},
437-467 (1987).

\bibitem{Lu} J. M. Luttinger: {\it An exactly soluble model of a many fermions
system}, J. Math. Phys. {\bf 4}, 1154-1162 (1963).

\bibitem{M2} V. Mastropietro: {\it Non-Perturbative
Renormalization}, World Scientific (2008).

%\bibitem{M3} V. Mastropietro: {\it Localization in interacting fermionic chains with quasi-random disorder}. Comm. Math. Phys.  {\bf 351}, 283–309 (2017)

\bibitem{N}  K. S. Novoselov, A. K. Geim, S. V. Morozov,
D. Jiang, Y. Zhang, S. V. Dubonos, I. V. Grigorieva and
 A. A. Firsov: {\it Electric Field Effect in Atomically Thin Carbon Films},
Science {\bf 306}, 666 (2004).



\bibitem{Riv} V. Rivasseau: {\it The Two Dimensional Hubbard Model at Half-Filling. I. Convergent Contributions}, J. Statistical Phys. {\bf 106}, 693-722 (2002).


\bibitem{rivbook} V. Rivasseau: {\it From Perturbative Renormalization to Constructive Renormalization}, Princeton university press


%
%\bibitem{RW2} V. Rivasseau and Z. ~Wang: {\it Hubbard model on the Honeycomb lattice at van Hove filling, Part II: the Lower Bound on the Self-Energy}

\bibitem{exp2} J. L. McChesney et el.: {\it
Extended van Hove Singularity and Superconducting Instability in Doped Graphene}, Phys. Rev. Lett. {\bf 104}, 136803 (2010)

\bibitem{exp1} P. Rosenberg et el.: {\it
Tuning the doping level of graphene in the vicinity of the Van Hove singularity
via ytterbium intercalation}, Phys. Rev. B {\bf 100}, 035445 (2019)

\bibitem{rosen} P. Rosenberg et el.: {\it
Overdoping Graphene beyond the van Hove Singularity}, Phys. Rev. Lett. {\bf 125}, 176403 (2020)

\bibitem{RW1} 
  V.~Rivasseau and Z.~Wang,
``How to Resum Feynman Graphs,''
 Annales Henri Poincare {\bf 15} 11,  2069 (2014)

%
%\bibitem{RW2} 
%  V.~Rivasseau and Z.~Wang,
%``Hubbard model on the Honeycomb lattice at van Hove filling: II. The two point functions and the renormalization", work in progress.


%\bibitem{sach} S. Sachdev: {\it Quantum Phase Transitions}, Cambridge university press (2014)

\bibitem{salm} M. Salmhofer: {\it Continuous Renormalization for Fermions and Fermi Liquid Theory}, Comm. Math. Phys. {\bf 194}, 249-295 (1998).


%\bibitem{Sa} M. Salmhofer: {\it Renormalization: An Introduction},
%Springer (1999).

\bibitem{To} S. Tomonaga: {\it Remarks on Bloch's methods of sound waves
applied to many fermion systems}, Progr. Theo. Phys. {\bf 5}, 544-569 (1950).

\bibitem{W} P. R. Wallace: {\it The Band Theory of Graphite}, Phys. Rev. {\bf 71},
622-634 (1947).

\bibitem{wang20} Zhituo Wang: {\it On the sector counting lemma}, Lett Math Phys 111, 128 (2021). 
%\bibitem{GMW} A. Giuliani, V. Mastropietro, Z. Wang, preprint.
\endthebibliography
\end{document}